\documentclass[letterpaper,12pt]{article}

\usepackage[margin=1in,letterpaper]{geometry} 
\usepackage[utf8]{inputenc} 
\usepackage{hyperref}       
\usepackage{url}            
\usepackage{booktabs}       
\usepackage{amsfonts}       
\usepackage{nicefrac}       
\usepackage{microtype}      

\usepackage[numbers]{natbib}
\usepackage{amsmath}
\usepackage{graphicx}
\usepackage{adjustbox}
\graphicspath{{figs/}}
\usepackage{xargs}
\usepackage[colorinlistoftodos]{todonotes}
\usepackage{amsthm}
\usepackage{amssymb}
\usepackage{mathtools}
\usepackage{bbm}
\usepackage{ dsfont }
\usepackage{subcaption}

\usepackage{xcolor}
\usepackage[normalem]{ulem}

\usepackage{tikz}
\usetikzlibrary{arrows,automata}
\usepackage{pgfplots}
\usepackage{array}
\usepackage{ifthen}
\usepackage{enumitem}

\usepackage{theoremref}
\usepackage{Newcommands}

\begin{document}

\title{The Central Spanning Tree Problem}
\author{\normalsize Enrique Fita Sanmartín$^1$,
\hspace{0.5cm} Christoph Schnörr$^{1 \ 2}$, \hspace{0.5cm} Fred A. Hamprecht$^1$\\
\normalsize $^1$IWR, $^2$IPA at Heidelberg University, 69120 Heidelberg, Germany\\
\footnotesize \texttt{\{enrique.fita.sanmartin@iwr,schnoerr@math fred.hamprecht@iwr\}}\texttt{.uni-heidelberg.de}}
\date{}
\maketitle
\begin{abstract}
	Spanning trees are an important primitive in many data analysis tasks, when a data set needs to be summarized in terms of its ``skeleton'', or when a tree-shaped graph over all observations is required for downstream processing. Popular definitions of spanning trees include the minimum spanning tree and the optimum distance spanning tree, a.k.a.~the minimum routing cost tree. When searching for the shortest spanning tree but admitting additional branching points, even shorter spanning trees can be realized: Steiner trees. Unfortunately, both minimum spanning and Steiner trees are not robust with respect to noise in the observations; that is, small perturbations of the original data set often lead to drastic changes in the associated spanning trees. In response, we make two contributions when the data lies in a Euclidean space: on the theoretical side, we introduce a new optimization problem, the ``(branched) central spanning tree'', which subsumes all previously mentioned definitions as special cases. On the practical side, we show empirically that the (branched) central spanning tree is more robust to noise in the data, and as such is better suited to summarize a data set in terms of its skeleton. We also propose a heuristic to address the NP-hard optimization problem, and illustrate its use on single cell RNA expression data from biology and 3D point clouds of plants. 
\end{abstract}

\section{Introduction}
Many data analysis tasks call for the summary of a data set in terms of a spanning tree, or use tree representations for downstream processing. Examples include the inference of trajectories in developmental biology~\citep{saelens_comparison_2019,chizat},  generative modeling in chemistry \citep{ahn_spanning_2021}, network design \citep{wong_worst-case_1980} or skeletonization in image analysis~\citep{BAI202311,wang2019deepflux}. The problem is akin to, but more complicated than, the estimation of principal curves because good recent methods such as \citep{lyu_manifold_denoising} cannot account for branched topologies. For a spanning tree representation to be meaningful, it is of paramount importance that the tree structure be robust to minor perturbations of the data, e.g. by measurement noise. In this work, we address the geometric stability of spanning trees over points lying in an Euclidean space.

\def\backgroundcolortext{cyan }
\def\backgroundcolor{cyan!10}
\def\semiringcolor{red}
\def\semiringcolortext{red }

\ifcsname theadd\endcsname
\renewcommand{\theadd}[1]{%
	\begin{tabular}{@{}c@{}}
		\vrule height 1.2\ht\strutbox width 0pt\ignorespaces
		##1
		\unskip\vrule depth 1.2\dp\strutbox width 0pt
	\end{tabular}%
}%
\else
\newcommand{\theadd}[1]{%
	\begin{tabular}{@{}c@{}}
		\vrule height 1.2\ht\strutbox width 0pt\ignorespaces
		##1
		\unskip\vrule depth 1.2\dp\strutbox width 0pt
	\end{tabular}%
}%
\fi

\def\colortableCST{lime}
\def\scalefig{0.25}
\begin{figure}[t]
	\begin{center}
		\scalebox{0.84}{\begin{tabular}{ c|c|c|c|c| } 
                \multicolumn{2}{c}{}&\multicolumn{3}{c}{\footnotesize Edge Centrality Exponent $\alpha$}\\\cline{2-5}
				&\diagbox{SP}{$\alpha$} &$0$& $(0,1)$&$1$ \\\cline{2-5}
				
				\multirow{2}{*}{\rotatebox[origin=c]{90}{\footnotesize Add Steiner points}}
                &YES& \makecell{\footnotesize\textbf{Steiner Tree} \\\tiny{\cite{gilbert_steiner_1968}} \\\small$\displaystyle\min_{X_B,\tree}\sum_{(i,j) \in E_{\tree}}||x_i-x_j||$\\ 
					\\  
					\begin{minipage}{\scalefig\textwidth}
						\includegraphics[width=\linewidth]{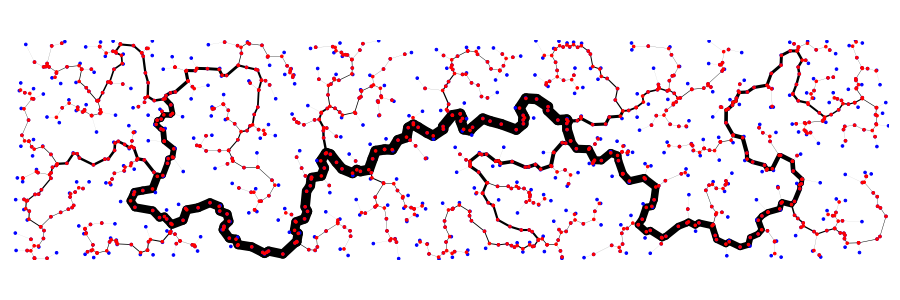}
				\end{minipage}}
				&\makecell{ \colorbox{\colortableCST}{\footnotesize\textbf{Branched Central Spanning Tree}} \\   
					\small$\displaystyle\min_{X_B,\tree}\sum_{(i,j) \in E_{\tree}}\big(m_{ij}(1-m_{ij})\big)^\alpha ||x_i-x_j||$
					\\
					\begin{minipage}{\scalefig\textwidth}
						\includegraphics[width=\linewidth]{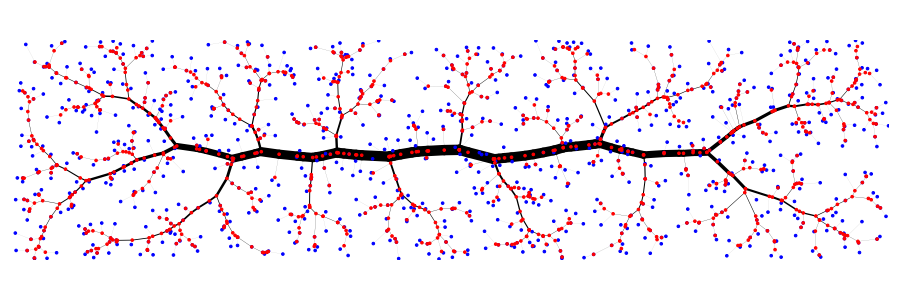}
				\end{minipage}}
				& \makecell{
                \colorbox{\colortableCST}{\footnotesize\textbf{Branched Minimum}}\\\footnotesize\colorbox{\colortableCST}{\textbf{Routing Cost Tree}}
                \\\small$\displaystyle\min_{X_B,\tree}\sum_{i,j \in V\times V}d_{\tree}(i,j)$
					\\
					\begin{minipage}{\scalefig\textwidth}
						\includegraphics[width=1\linewidth]{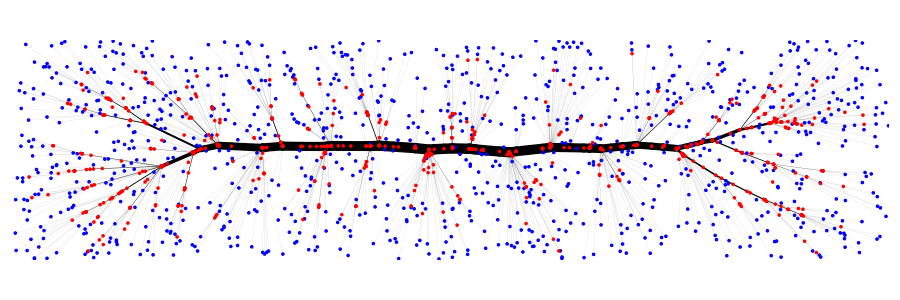}
				\end{minipage}} \\ \cline{2-5}

				&NO& \makecell{\footnotesize \textbf{Minimum Spanning }\textbf{Tree} \\\tiny{\cite{kruskal_shortest_1956}}
                \\\small$\displaystyle\min_{\tree}\sum_{(i,j) \in E_{\tree}}||x_i-x_j||$\\ 
					\begin{minipage}{\scalefig\textwidth}
						\includegraphics[width=\linewidth]{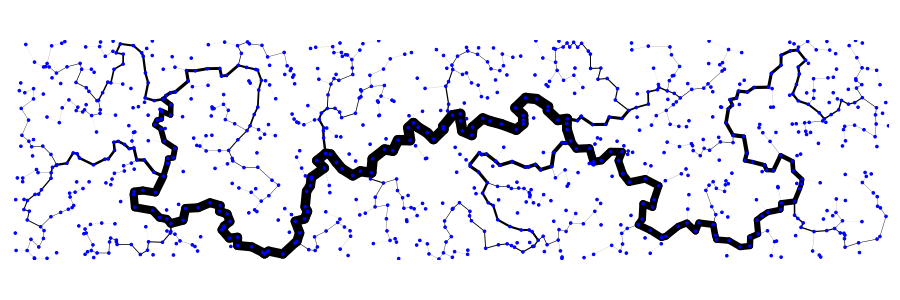}
				\end{minipage}}
				& \makecell{\colorbox{\colortableCST}{\footnotesize\textbf{Central Spanning Tree}} \\   
					\small$\displaystyle\min_{\tree}\sum_{i,j \in E_{\tree}}\big(m_{ij}(1-m_{ij})\big)^\alpha ||x_i-x_j||$\\
					\begin{minipage}{\scalefig\textwidth}
						\includegraphics[width=\linewidth]{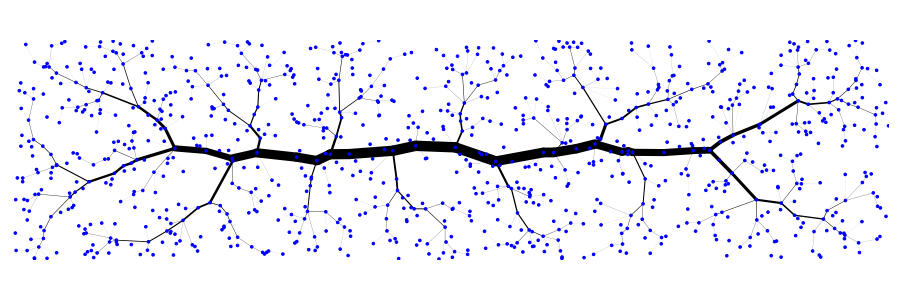}
				\end{minipage}}
				& \makecell{ 
                    \footnotesize \textbf{Minimum Routing}\textbf{ Cost Tree} \\\tiny{\cite{masone_minimum_2019}} \\\small$\displaystyle\min_{\tree}\sum_{i,j \in V\times V}d_{\tree}(i,j)$\\
					\begin{minipage}{\scalefig\textwidth}
						\includegraphics[width=\linewidth]{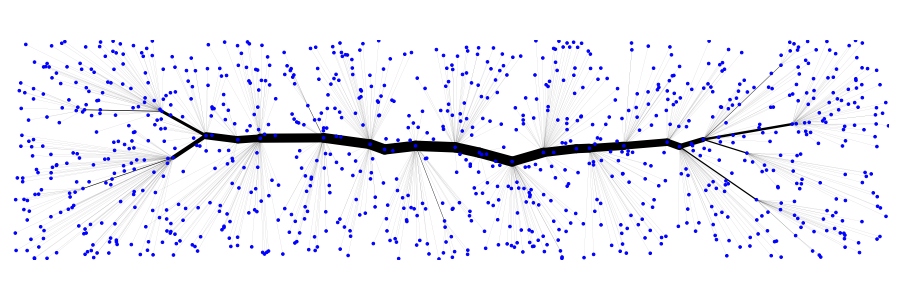}
				\end{minipage}} \\ \cline{2-5}
				
		\end{tabular}}
	\end{center}
	\caption[Euclidean Central Spanning Tree Family of Problems with and without Steiner Points]{\textbf{Euclidean Central Spanning Tree Family of Problems with and without Steiner Points}. The central spanning tree weighs the costs of the edges, given by node distances, with the centrality of the edges, $m_{ij}(1-m_{ij})$. The influence of the centrality is regulated by the parameter $\alpha\in [0,1]$. For lower $\alpha$ values the centrality becomes insignificant, and the tree tends to contain short edges. For higher $\alpha$ values, the tree encourages central edges of low cost, at the expense of peripheral edges of higher cost. We study central spanning trees with and without additional Steiner points (shown in red). The widths of all edges are proportional to their centrality. The central spanning tree problems includes well-known and novel spanning tree problems, the latter highlighted in green.}
	\label{tab:CST_problems}
\end{figure}%

The minimum spanning tree (\mST) is surely the most popular spanning tree, owing to its conceptual simplicity and ease of computation. For a graph $G=(V,E)$ with edge costs, the \mST is a tree that spans $G$ while minimizing the total sum of edge costs. It prioritizes shorter edges that connect closely located nodes enhancing data faithfulness. Unfortunately, its greedy nature makes the \mST susceptible to small data perturbations that may lead to drastic changes in its structure, see \figurename{} \ref{fig:CST_noiserobustness_toydata}. An alternative, the minimum routing cost tree (\MRCT), minimizes the sum of pairwise shortest path distances ~\citep{masone_minimum_2019}. Unlike the \mST, solving the \MRCT is NP-hard~\citep{wu_spanning_2004}. Despite this, the \MRCT exhibits a more stable geometric structure compared to the \mST, as it tends to be more "star-shaped" (see \figurename{} \ref{fig:CST_noiserobustness_toydata}). Nevertheless, this star-shaped tendency inclines towards connecting nodes that are spatially distant, introducing a risk of information loss and compromised data fidelity. This effect becomes particularly pronounced in high-dimensional spaces, potentially rendering the \MRCT approach unusable (see \figurename{s} \ref{sfig9:stability_paul}-\ref{sfig12:stability_paul})). Achieving a balance between data fidelity and geometric robustness is crucial for an effective spanning tree.

\textbf{Central spanning trees (CST)} In this paper, we propose a novel parameterized family of spanning trees that interpolate and generalize all the aforementioned ones. Unless otherwise stated, we will assume a complete Euclidean graph, where the nodes are embedded in the Euclidean space with coordinates $X_V=\{x_1,\dots,x_{|V|}\} \subset \mathbb{R}^n$ and the edge costs are given by the distances between the vertices, $c_{ij}=||x_i-x_j||$. We define the \CST as the spanning tree of $G$ that minimizes the objective
\begin{equation}\label{eq:def_CST}
	\operatorname{CST} \coloneqq \underset{\tree}{\arg \min}\sum_{e\in E_{\tree}} \big(m_e(1-m_e)\big)^{\alpha}c_e=\underset{\tree}{\arg \min}\sum_{(i,j)\in E_{\tree}}\big(m_{ij}(1-m_{ij})\big)^{\alpha}||x_i-x_j||,
\end{equation}%
where $m_e$ and $(1-m_e)$ are the normalized cardinalities of the components resulting from the removal of the edge $e$ from $\tree$. Thus, $m_e(1-m_e)$ is the product of the number of nodes on both sides of the edge divided by $|V|^2$. Because this value is proportional to the ``edge betweeness centrality''\footnote{The edge betweenness centrality measures an edge's frequency in shortest paths between nodes, with more traversed edges being deemed more central. In trees, it's the product of nodes on opposite sides of the edge.} of $e$ in $\tree$~\citep{brandes_variants_2008}, we call the problem the ``central spanning tree'' (\CST) problem.  The exponent $\alpha$ is the interpolating parameter that modulates the effect of the edge centrality. For $\alpha$ close to $0$, the centrality becomes insignificant, so the tree tends to contain lower cost edges overall. For $\alpha=0$ we retrieve the $\mST$. On the other hand, as $\alpha$ increases, the centrality becomes more relevant, leading the tree to favor topologies with low centrality edges, thus promoting a higher branching effect. For $\alpha=1$, the resulting expression is proportional to the \MRCT. Here, each edge cost is multiplied by the number of shortest paths it belongs to, leading the total sum to represent the sum of shortest path distances (see Appendix \ref{sec:app_CST_MRCT_equivalence}).

As will be seen in Section \ref{sec:stability}, the $\alpha$ parameter has an effect on the geometric stability of the spanning tree, with higher $\alpha$ resulting in greater robustness. The robustness of spanning trees has been explored in various contexts. For instance, researchers have investigated the robustness of the \mST cost under edge weight uncertainty \citep{kasperski2011approximability,sharma2022robust} and the robustness against node or edge failure in networks \citep{liu2019dynamic}. Additionally, studies have delved into the stability regions of \mST, under which any change in vertex location does not alter the \mST \cite{monma1991transitions,niendorf2016robustness}. The central tree problem \citep{bezrukov1996central}, related by name to ours, focuses on computing a tree that minimizes the maximal distance to a set of given trees. To our knowledge, we are the first to propose a spanning tree whose geometric structure is stable and resilient to data perturbations such as noise.

Finally, we remark the connection between the the \CST problem and the Minimum Concave Cost Network Flow (MCCNF) problem \cite{zangwill_minimum_1968,guisewite_minimum_1990}. The MCCNF problem aims to minimize the cost of distributing a certain commodity from source to sink nodes. Such a problem models the cost of an edge as a concave function of the transportation flow. The \CST can be reinterpreted as MCCNF where a commodity with mass equal to $|V|-1/|V|$, concentrated into a single source node, must be transported to the rest of nodes. In our case, the term $m_e$ in \eqref{eq:def_CST} can be interpreted as the flow of such problem. Since the function $(m_e(1-m_e)\big)^{\alpha}c_e$ is concave with respect to $m_e$ for $\alpha\in[0,1]$, we deduce that the \CST is an instance of the MCCNF. A more detailed discussion of the interpretation of the \CST as a MCCNF problem is offered in Appendix \ref{sec:app_CST_MCCNF}.

Considering the CST from the perspective of an MCCNF problem, it becomes clear that it falls into the NP-hard category. Indeed, the authors of \cite{guisewite1991algorithms} showed that single-source MCCNF problems with strictly concave functions are NP-hard. Consequently, we conclude that the CST problem is NP-hard for $\alpha \in ]0,1]$ due to the strictly concave nature of the edge cost function $\left(m_e(1-m_e)\right)^\alpha c_e$.\footnote{The same argument applies to the NP-hardness of the branched version of the \CST problem, which is explained next.}

\textbf{Branched central spanning trees (BCST)} Inspired by~\citet{lippmann_theory_2022}, we also study the variant of the \CST problem which allows for the introduction of additional nodes,  known in the literature as branching points or Steiner points (\BPs). Formally, we distinguish between two types of nodes. On the one hand, we have the terminal nodes with fixed coordinates given by $X_V$. On the other hand, we allow for an extra set of points, $B$, whose coordinates $X_B$ must be jointly optimized with the topology $\tree$. In this case, $\tree$ is a spanning tree defined over the  nodes $V\cup B$. Accordingly, the objective of the \CST problem becomes 
\begin{equation}
	\label{eq:BCST}
	\min_{\tree, X_B}\sum_{(i,j)\in E_{\tree}}\big(m_{ij}(1-m_{ij})\big)^{\alpha}||x_i-x_j||.
\end{equation}%
In this generalization, which we refer to as the branched \CST (\BCST), the well-known Steiner tree problem~\citep{hwang_steiner_1992,warme_exact_2000} arises when $\alpha=0$. \figurename{} \ref{tab:CST_problems} summarizes (B)\CST and its limiting cases, only some of which have been studied in the literature so far.

\textbf{Contributions.} 1) We present the novel (B)\CST problem and provide empirical evidence for its greater stability on toy and real-world data; 2) We propose an iterative heuristic to estimate the optimum of the \BCST problem.\footnote{Code available at \url{https://github.com/sciai-lab/CST}} By exploiting the connection between the branched and unbranched versions of the \CST problem, we are able to use the heuristic defined for the \BCST to also approximate the Euclidean \CST without Steiner points. We benchmark this heuristic and show its competitiveness. 3) On the theoretical side, we prove that for large $\alpha$ or large $|V|$ and $\alpha>1$, (B)\CST converges to a star-tree (hinting modeling limitations when $\alpha>1$), and for $\alpha\to-\infty$, it tends towards a path graph. Additionally, we show analytically that if the terminal points lie on a plane, then for $\alpha\in[0,0.5]\cup\{1\}$ the Steiner points of the optimal solution can have up to degree $3$, and we provide empirical evidence that this holds also for $\alpha\in \ ]0.5,1[$. 

\textbf{Outline} In Section \ref{sec:CSTlimits}, we explore the limiting cases of the (B)\CST optimum as $\alpha$ approaches $\pm\infty$, along with the scenario where the number of terminals tends to infinity for $\alpha>1$. Section \ref{sec:stability} demonstrates empirically the stability of the (B)\CST as $\alpha$ increases. In Section \ref{sec:relation-between-bcst-and-cst}, we establish a relationship between feasible \CST and \BCST topologies. Section \ref{sec:geometry-of-optimal-bcst-topologies} analyzes the geometry of optimal \BCST topologies, providing analytical expressions for the branching angles at the Steiner points. Moreover, it discusses the feasibility of 4-degree \BPs when the \BCST is restricted to the Euclidean plane. Section \ref{sec:Gen_optimization} presents a heuristic to approximate the (B)\CST optimal solution, while Section \ref{sec:benchmark} benchmarks this heuristic on small toy datasets. The conclusions of this work are given in section \ref{sec:CST_conclusion}.

\section{Limit Cases of the \CST/\BCST Problems Beyond the Range $\alpha\in[0,1]$}

The \CST problem, as defined in \eqref{eq:def_CST}, as well as its branched version are parameterized by $\alpha$. Throughout the manuscript our attention will be on the $\alpha$-range of $[0,1]$, nonetheless it is worth studying the problem beyond this range. We will show that when $\alpha\to\infty$ or the number of terminals approaches infinity when $\alpha>1$, the (B)\CST tends to a star graph centered on the medoid of the graph, i.e. the node that minimizes the distance to the rest of nodes. Consequently, the case with $\alpha>1$ becomes inadequate for modeling data structure, as the tree becomes increasingly trivial with a growing number of terminals. Conversely, as $\alpha\to-\infty$, the \CST tends towards the path graph that minimizes the \CST objective. These scenarios prove inadequate for modeling data structure, thus we restrict our focus exclusively to the $\alpha$-range of $[0,1]$.

\label{sec:CSTlimits}
\subsection{Limit Cases Where the Optimum (B)CST Transforms into a Star-Tree}\label{sec:limitCST_alpha>1_n_infty}

In this section, we delve into scenarios at the limits where the optimal solutions for both the  \CST and \BCST problems converge to a star-tree configuration. Specifically, we demonstrate that this outcome occurs as $\alpha$ approaches infinity and as the number of terminals, denoted by $N$, tends to infinity when $\alpha>1$. The later limit case is of special relevance, as it indicates that when the parameter $\alpha$ exceeds 1, both \CST and \BCST exhibit inadequacy in extracting meaningful structural information from the data. In this situation, an increasing number of data points lead to the formation of a star-tree, which, lacks the capacity to convey pertinent information about the underlying dataset.

Before studying the limit cases, we will analyze the optimum star-tree that minimizes the \CST and \BCST costs. Note that in a star-tree, all edges are adjacent to a leaf node and, therefore all have the same normalized centrality value, which is equal to $(N-1)/N^2$, where $N$ is the number of terminals. In this scenario, the centrality of the edges in the cost function from Equation \eqref{eq:def_CST} can be factored out, simplifying the problem to the identification of the star graph with the minimum cost. Indeed, if $u$ denotes the center node of a star graph, then its \CST objective is equal to
\[\sum_{v\neq u }\frac{N-1}{N^2}c_{uv}=\frac{N-1}{N^2}\sum_{v\neq u }c_{uv}.\]
Consequently, the optimal solution for the \CST problem manifests as a star graph centered at the node that minimizes the total distance to all other nodes, effectively the medoid. In the context of the \BCST problem, where Steiner Points can be introduced, the star graph is centered at the geometric median. This is because the Steiner Points strategically position themselves to minimize the distance to all nodes. The next result formalizes this statement.

\begin{lemma}
	The tree-star that minimizes the \CST cost is the star-shaped tree centered at the medoid of the terminals, that is, centered at the terminal which minimizes the sum of distances to all nodes. For the \BCST case, the tree is centered at the geometric median of all terminals.
\end{lemma}

As a consequence of this result, we infer that the limit cases wherein both \CST and \BCST converge to a star-tree will yield stable trees. This stability arises from the consistent output of star-trees, with their centers being the medoid and geometric median—both robust points resilient to noise.

In order to study limit cases where the star-tree emerges as the optimal solution, we establish first a sufficient condition for the \CST optimal solution to take the form of a star tree. This condition was first identified  by \citet{hu_optimum_1974} in the context of the Minimum Routing Cost Tree (MRCT), corresponding to the \CST with $\alpha=1$. \citeauthor{hu_optimum_1974} showed that if a ``stronger" variant of the regular triangle inequality holds, then the optimum solution of the MRCT is a star tree. The following theorem extends and generalizes this result for arbitrary values of $\alpha$.
\begin{theorem}\thlabel{lem:strong_triangle_inequality_(B)CST}
	Let $N$ be the number of terminals and  $c_{ij}$ be the edge-costs of any pair of points (Steiner or terminals) $i,j$. If there exists 
	$$t\leq \min_{\ell\in[2,N/2]} \frac{\left(\frac{\ell(N-\ell)}{N-1}\right)^{\alpha}-1}{\ell-1}$$
	such that 
	\begin{equation}\label{eq:strong_triangle_inequality_(B)CST_lemma}
	c_{kv}+tc_{uv}\geq c_{ku}
	\end{equation}
	for all triplets of nodes $u,k,v$, then there exists an optimum $(B)\CST$ evaluated at $\alpha$ which is a star tree.
\end{theorem}

We defer the complete proof to \appendixname{} \ref{sec:proof-threflemstrongtriangleinequalitybcst}, though we sketch briefly the key idea. The proof demonstrates that we can always iteratively increase the degree of certain node, by connecting all neighbors of one of its neighbors to it. If the ``stronger" variant of the triangle inequality holds, then this process does not increase the cost. Eventually, a node will reach maximum degree, indicating the formation of a star-tree structure.
\begin{remark}
	Note that \thref{lem:strong_triangle_inequality_(B)CST} states only a sufficient condition, which means that the optimum can be a star tree even if the strong triangle inequality does not hold. Additionally, it is worth to highlight that \thref{lem:strong_triangle_inequality_(B)CST} also holds true for the \CST problem even when the nodes lack embedding in any specific space, allowing for edge costs with arbitrary values.
\end{remark}

Let us define 
\begin{equation}
	h_1(\ell,N,\alpha)\coloneqq\frac{\left(\left(\frac{\ell(N-\ell)}{N-1}\right)^{\alpha}-1\right)}{\ell-1}=\frac{\left(1+\frac{(\ell-1)(N-\ell-1)}{N-1}\right)^{\alpha}-1}{\ell-1}.
\end{equation}
which characterizes the upper limit of the threshold $t$ in \thref{lem:strong_triangle_inequality_(B)CST}. Equation \eqref{eq:strong_triangle_inequality_(B)CST_lemma} represents a weighted version of the triangle inequality. Specifically, when $t=1$, equation \eqref{eq:strong_triangle_inequality_(B)CST_lemma} recovers the standard triangle inequality. Moreover, if $ h_1(\ell,N,\alpha) \geq 1$, there exists a value $t$ such that $ h_1(\ell,N,\alpha) \geq t \geq 1$. This implies that the relation is weaker than the triangle inequality. Therefore, if the triangle inequality holds, equation \eqref{eq:strong_triangle_inequality_(B)CST_lemma} will also hold. As a first consequence, is evident that
$$\lim_{\alpha\to\infty} h_1(l,N,\alpha)=\infty, \quad \forall N\geq 3, \ \ell \in[2,N/2].$$
This indicates that as $\alpha$ tends to infinity, the optimal tree tends to become a star-tree. The intuition behind this is clear: as $\alpha$ increases, the \CST/\BCST  aims to minimize the centrality of the edges, since the edge costs become relatively insignificant in comparison. Among all edges in a tree, those adjacent to a leaf have the least centrality. Thus, any star graph is a tree that minimizes simultaneously the edge centrality of all its edges

One can also show that $h_1$ is greater than $1$ when $\alpha>1$ and $N$ approaches infinity:
\[\lim_{N\to\infty} h_1(l,N,\alpha)\geq1, \quad \forall \alpha>1, \ \ell \in[2,N/2],\]
though we leave the technical proof for the \appendixname{} \ref{sec:proof-h1ellnalpha1-as-n-approaches-infinity-and-alpha1}. Consequently, we obtain the following corollary
\begin{corollary}\thlabel{cor:star_tree_limitN_and_alpha}
	As the parameter $\alpha$ approaches infinity, or $N$ approaches infinity and $\alpha>1$ the \CST/\BCST optimal solution is a star-shaped tree.
\end{corollary}

\begin{figure}[t!]
	\centering
	\begin{tikzpicture}
	\begin{axis}[
	xlabel={$N$},
	ylabel={$\alpha^{\ast}(N)$},
	xmin=100,
	xmax=5000,
	ymin=1,
	ymax=1.2,
	legend pos=south east,
	]
	\pgfmathdeclarefunction{g1}{1}{%
		\pgfmathparse{ln(2) / ln(1 + (#1 - 3) / (#1 - 1))}%
	}
	
	\pgfmathdeclarefunction{g2}{1}{%
		\pgfmathparse{ln(#1 / 2) / ln(1 + ((#1 / 2 - 1) ^ 2) / (#1 - 1))}%
	}
	
	\pgfmathdeclarefunction{h}{3}{%
		\pgfmathparse{((#1*(#2-#1)/(#2-1))^#3 - 1)/(#1 - 1)}%
	}
	
	\addplot[blue, domain=100:10000, samples=100] { max(g1(x), g2(x))};

	\end{axis}
	\end{tikzpicture}
	\caption[Threshold Function $\alpha^{\ast}(N)$ Guaranteeing Optimal (B)CST Star-Tree]{\textbf{Threshold Function $\alpha^{\ast}(N)$ Guaranteeing Optimal (B)CST Star-Tree}. The threshold function $\alpha^{\ast}(N)$, tied to the number of terminals $N$, defines the minimum $\alpha$ value ensuring the optimal solution for \CST/\BCST is a star-tree. For all $\alpha > \alpha^{\ast}(N)$, the optimum solution is a star-tree. The plot depicts the transition at which the optimum is ensured to be a star-tree is around $\alpha \approx 1$. As $N$ increases, $\alpha^{\ast}(N)$ approaches 1, implying that in the limit as $N$ tends to infinity, \CST/\BCST with $\alpha > 1$ converges to a star-tree.}
	\label{fig:threshold_alpha}
\end{figure}
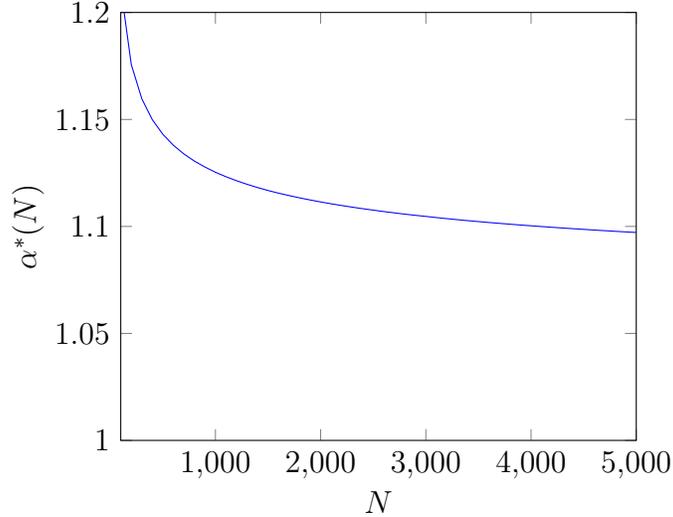

\begin{figure}[h!]
    \centering
    \begin{subfigure}{0.5\linewidth}
        \centering
		\includegraphics[width=\linewidth]{table_figs/BCST_1.00.png}
	\caption{\BCST, $\alpha=1.00$}
	\label{sfig1:rectangle_star}
	\end{subfigure}%
	\begin{subfigure}{0.5\linewidth}
	\centering
	\includegraphics[width=\linewidth]{table_figs/CST_1.00.png}
	\caption{\CST, $\alpha=1.00$}
	\label{sfig2:rectangle_star}
	\end{subfigure}
        \begin{subfigure}{0.5\linewidth}
    	\centering
    	\includegraphics[width=\linewidth]{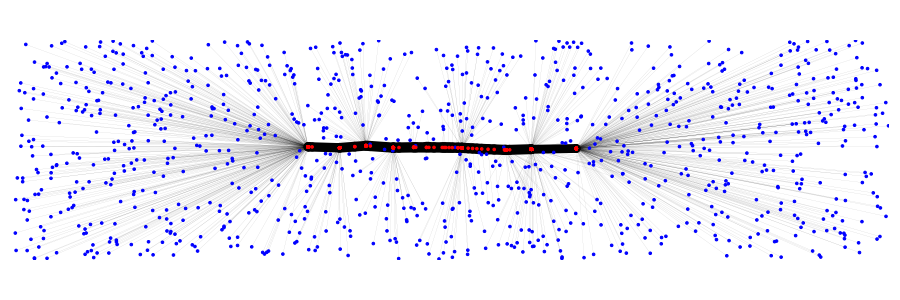}
    	\caption{\BCST, $\alpha=1.05$}
    	\label{sfig3:rectangle_star}
    \end{subfigure}%
    \begin{subfigure}{0.5\linewidth}
    	\centering
    	\includegraphics[width=\linewidth]{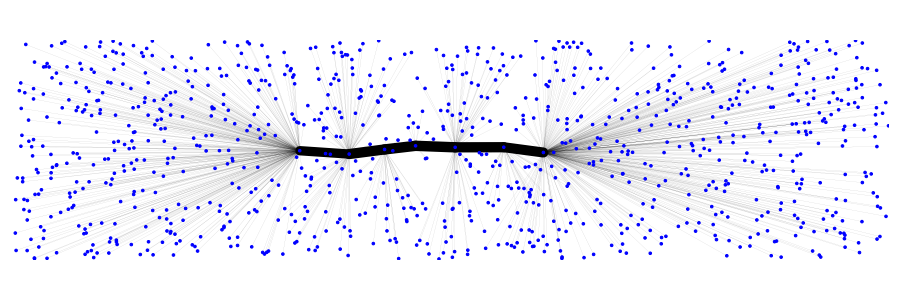}
    	\caption{\CST, $\alpha=1.05$}
    	\label{sfig4:rectangle_star}
    \end{subfigure}
    \begin{subfigure}{0.5\linewidth}
		\centering
		\includegraphics[width=\linewidth]{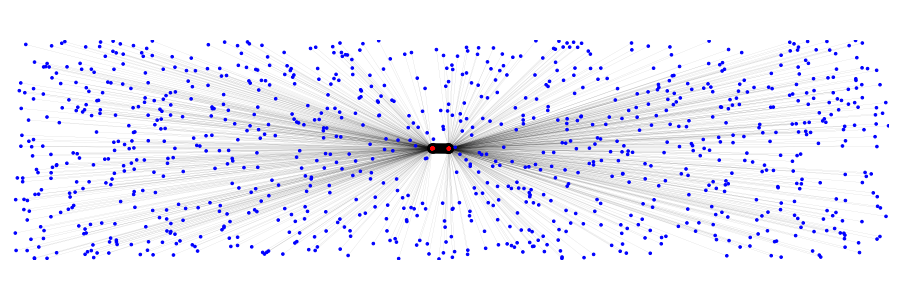}
		\caption{\BCST, $\alpha=1.10$}
		\label{sfig5:rectangle_star}
	\end{subfigure}%
	\begin{subfigure}{0.5\linewidth}
		\centering
		\includegraphics[width=\linewidth]{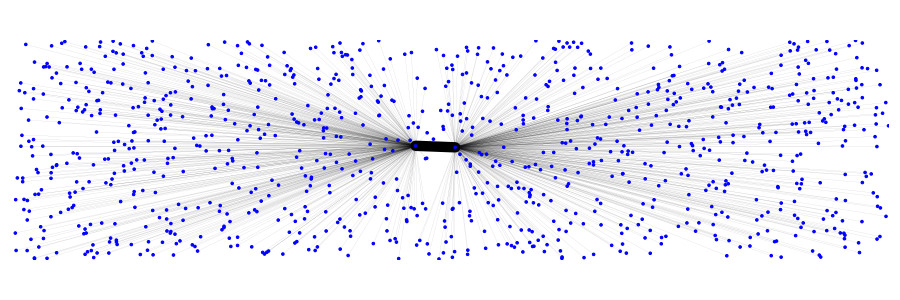}
		\caption{\CST, $\alpha=1.10$}
		\label{sfig6:rectangle_star}
	\end{subfigure}
	    \begin{subfigure}{0.5\linewidth}
		\centering
		\includegraphics[width=\linewidth]{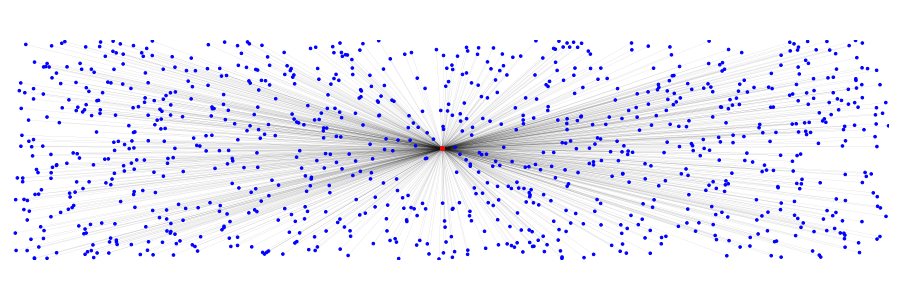}
		\caption{\BCST, $\alpha=1.15$}
		\label{sfig7:rectangle_star}
	\end{subfigure}%
	\begin{subfigure}{0.5\linewidth}
		\centering
		\includegraphics[width=\linewidth]{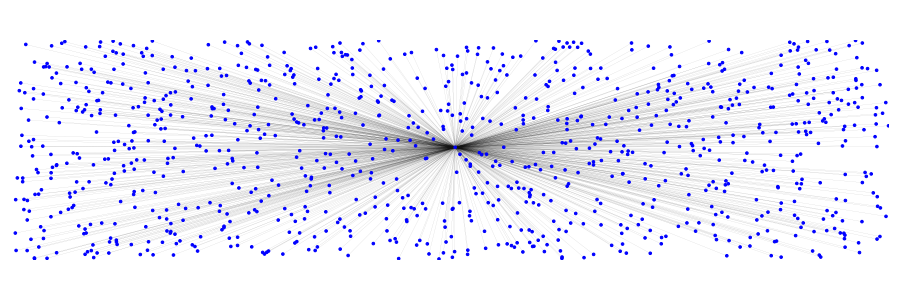}
		\caption{\CST, $\alpha=1.15$}
		\label{sfig8:rectangle_star}
	\end{subfigure}    
  \caption[(B)CST Star-Tree Optimality with Respect to $\alpha\gtrsim1$ in Samples Uniformly Drawn from a Rectangle]{\textbf{(B)CST Star-Tree Optimality with Respect to $\alpha\gtrsim1$ in Samples Uniformly Drawn from a Rectangle}. As $\alpha$ increases, both \CST and \BCST exhibit a transition towards a star graph. This effect may manifest relatively early. As anticipated in \figurename{} \ref{fig:threshold_alpha}, for a sample set with 1000 points, the value $\alpha=1.15$ transforms both \CST and \BCST into a star graph.}
	\label{fig:rectangle_star}	
\end{figure}

\thref{cor:star_tree_limitN_and_alpha} states  that the optimum tree is a star-tree when $\alpha>1$ and $N\to\infty$. What's intriguing is that this limit is reached at relatively low values of $\alpha\approx 1$ for moderate values of $N$. In \appendixname{} \ref{sec:computation-alphaastn}, we show that for $N$ nodes, the threshold $\alpha^{\ast}(N)$ at which $h_1(\ell,N,\alpha)\geq 1$ for all $\ell\in[2,N/2]$ and $\alpha\geq\alpha^*(N)$ is given by
\begin{equation}
\alpha^{\ast}(N)= \max\left(\frac{\log(2)}{\log\left(1+\frac{N-3}{N-1}\right)},\frac{\log(N/2)}{\log\left(1+\frac{\left(N/2-1\right)^2}{N-1}\right)}\right)
\end{equation}
The function $\alpha^{\ast}(N)$ serves as a threshold, ensuring that the optimal solution adopts a star-tree configuration. To illustrate this threshold, \figurename{} \ref{fig:threshold_alpha} depicts the function $\alpha^{\ast}(N)$. Indeed, we see that when $N=1000$, $\alpha = 1.15$ is enough to guarantee that the optimum is a star tree. A toy example is presented in \figurename{s} \ref{fig:rectangle_star}, with $N=1000$, showcasing an instance where the optimum is indeed a star tree.

\begin{remark}\thlabel{rem:more_star_in_high_dim}
	The star-shaped tendency becomes more prominent in higher dimensions. Empirical observations indicate that relatively small values of $\alpha$ ($\alpha \lesssim 1$) tend to produce an almost star-shaped tree, which compromises the preservation of data structure. This behavior can be attributed to the curse of dimensionality, where the majority of point pairs in a high-dimensional Euclidean space become nearly equidistant. Consequently, the strong triangle inequality derived in \thref{lem:strong_triangle_inequality_(B)CST} will hold for most triplets of points due to the approximate equality of $c_{ku}\approx c_{kv}$ in equation \eqref{eq:strong_triangle_inequality_(B)CST_lemma}. \figurename{s} \ref{sfig9:stability_paul}-\ref{sfig12:stability_paul}) exemplify this effect on a 50-dimnsional dataset.
\end{remark}

\subsection{Limit Cases Where the Optimum (B)CST Transforms into a Path-Tree}\label{sec:limit-cases-where-the-optimum-bcst-transforms-into-a-path-tree}
Negative values of $\alpha$ favour high central edges, as $(m_e(1-m_e)))^{\alpha}$ will be lower. Consequently, for sufficiently negative values of $\alpha$, the \CST/\BCST problem will prioritize minimizing the number of leaves since the centrality of its adjacent edges attain the minimum centrality. The tree that minimizes the number of leaves is a path. Therefore,  when $\alpha\to-\infty$ the optimum tree will be the Hamiltonian path that minimizes the \CST/\BCST objective function. A Hamiltonian path is a path that visits each node exactly once. 

Echoing \thref{lem:strong_triangle_inequality_(B)CST} we show that if a variant of the triangle inequality holds, then the optimum (B)\CST will be a tree.
\begin{theorem}\thlabel{th:tree_opt_alpha_-inf_reformulation}
	Let $N$ be the number of nodes and  $c_{ij}$ be the edge-costs of any pair of points (Steiner or terminals) $i,j$. If there exists
	$$t\leq \min_{\substack{1 \leq s\leq N-3\\1\leq \ell \leq \min(s,(N-s)/2-1)}} \frac{\left(\ell(N-\ell)\right)^{\alpha}-\left((\ell+s)(N-\ell-s)\right)^\alpha}{\left(s(N-s)\right)^\alpha}$$
	such that 
	\[c_{kv}+tc_{uv}\geq c_{ku}\]
	for all triangles in the graph, then there exists an optimum $(B)\CST$ evaluated at $\alpha$ which is a Hamiltonian path.
\end{theorem}
In contrast to the proof presented in \thref{lem:strong_triangle_inequality_(B)CST}, we demonstrate that we can systematically decrease the degree of nodes by iteratively connecting the neighbors of a specific node to one of its neighbors. This iterative reduction does not inflate the cost, provided the triangle inequality variant holds. Ultimately, all nodes will have at most degree 2, meaning that a path has been formed. We defer the complete proof to \appendixname{} \ref{sec:proof-threfthtreeoptalpha-infreformulation}. 

Let 
$$h_2(\ell,s,N,\alpha)=\frac{\left(\ell(N-\ell)\right)^{\alpha}-\left((\ell+s)(N-\ell-s)\right)^\alpha}{\left(s(N-s)\right)^\alpha}.$$
As a consequence of \thref{th:tree_opt_alpha_-inf_reformulation}, if $h_2(\ell,s)\geq1$ for $1 \leq s\leq N-3$ and $1\leq \ell \leq \min(s,(N-s)/2-1)$, then the satisfaction of the triangle inequality is a sufficient condition to ensure that the optimal (B)\CST topology is a path. We can easily check that as $\alpha$ approaches minus infinity the following limit holds
\begin{equation}\label{eq:limit_h2}
\begin{aligned}
\lim_{\alpha\to-\infty}h_2(\ell,s,N,\alpha)&=\lim_{\alpha\to-\infty}\left(\underbrace{\frac{\ell(N-\ell)}{s(N-s)}}_{\leq1}\right)^{\alpha}-\left(\underbrace{\frac{\left((\ell+s)(N-\ell-s)\right)}{\left(s(N-s)\right)}}_{>1}\right)^\alpha\\
&=\begin{cases}
1, \ &\text{if } \ell=s\\
\infty, \ &\text{if } \ell>s
\end{cases},
\end{aligned}
\end{equation}
where we have used the inequalities $1 \leq s\leq N-3$ and $1\leq \ell \leq \min(s,(N-s)/2-1)$. Consequently, as $\alpha$ approaches $-\infty$, the optimum tree will tend to a path tree. Note however, that if $l=s$, then $h_2(\ell,\ell,N,\alpha)<1$. In this case, according to \thref{th:tree_opt_alpha_-inf_reformulation}, the optimum tree can only be a path for $\alpha$ negative enough, if the triangle inequality holds strictly for all triplets of nodes. Nonetheless, in 
\thref{cor:tree_opt_alpha_-inf}, we show that when the points lie on a geodesic space, then the requirement of the strict triangle inequality is not necessary for the optimum to become a path as $\alpha$ approaches $-\infty$.

In the previous section, we demonstrated that for a  given $\alpha>1$ we can find $N$ large enough, such that for any configuration of terminals the optimal solution of the \CST/\BCST problem would be a star-tree. However, in this case, we cannot ensure that if $\alpha<0$, there exists $N$ large enough, where the optimum is a path. Indeed, given a fixed $\alpha\in\mathbb{R}$ and setting $\ell=N/4-1$ and $s=N/2$, we have that the limit of $h_2(N/4-1,N/2,N,\alpha)$  as $N$ increases is given by%
\[\lim_{N\to\infty}h_2\left(\frac{N}{4}-1,\frac{N}{2},N,\alpha\right)=\lim_{N\to\infty}\frac{\left((\frac{N}{4}-1)(\frac{3N}{4}+1)\right)^{\alpha}- \left((\frac{N}{4}+1)(\frac{3N}{4}-1)\right)^{\alpha}}{\left(\frac{N}{2}\right)^{2\alpha}}=0<1.\]
Consequently, \thref{th:tree_opt_alpha_-inf_reformulation} will not guarantee that the optimum is a path as $N$ increases, unless all edge-costs are equal.

\section{Stability of the CST Problem}\label{sec:stability}
In this section we argue in favour of the greater robustness of the \CST problem to small perturbations in data. We investigate first the robustness of the tree to noise in toy datasets. Next, we showcase two potential applications of the (B)\CST problem: single cell trajectory inference and skeletonization of 3D point clouds of plants.
\subsection{Toy Data}\label{sec:toydata}

\begin{figure}[h!]
\begin{minipage}{0.28\textwidth}
\center{\begin{subfigure}{1\linewidth}
            \centering
    		\includegraphics[width=\linewidth]{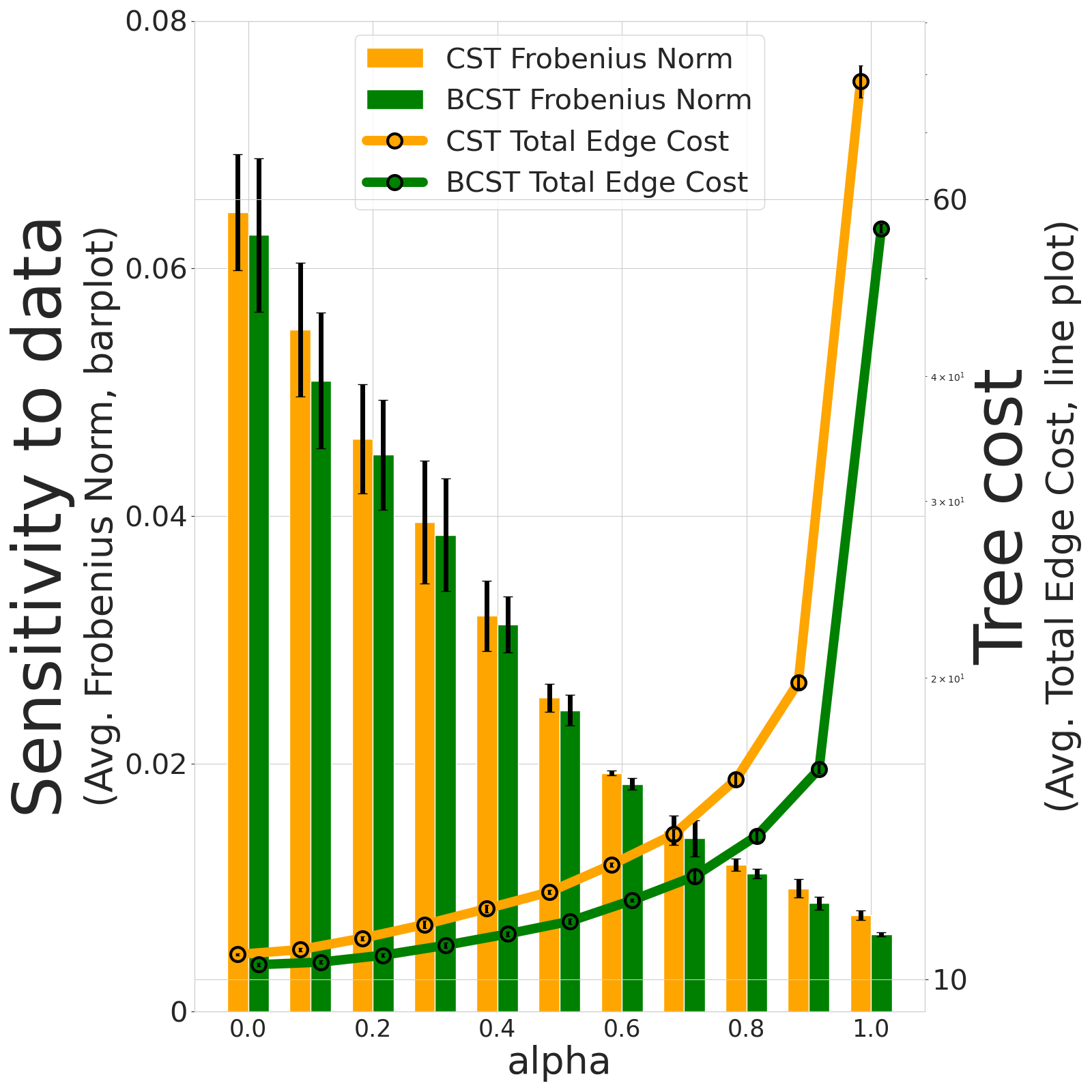}
    		\caption{Sensitivity vs. Tree cost}
    		\label{sfig0:CST_noiserobustness_toydata}
    \end{subfigure}}\end{minipage}
\hfill
\begin{minipage}{0.72\textwidth}
    \begin{minipage}[h]{0.5\linewidth}
    \center{\begin{subfigure}{1\linewidth}
        \includegraphics[width=\linewidth]{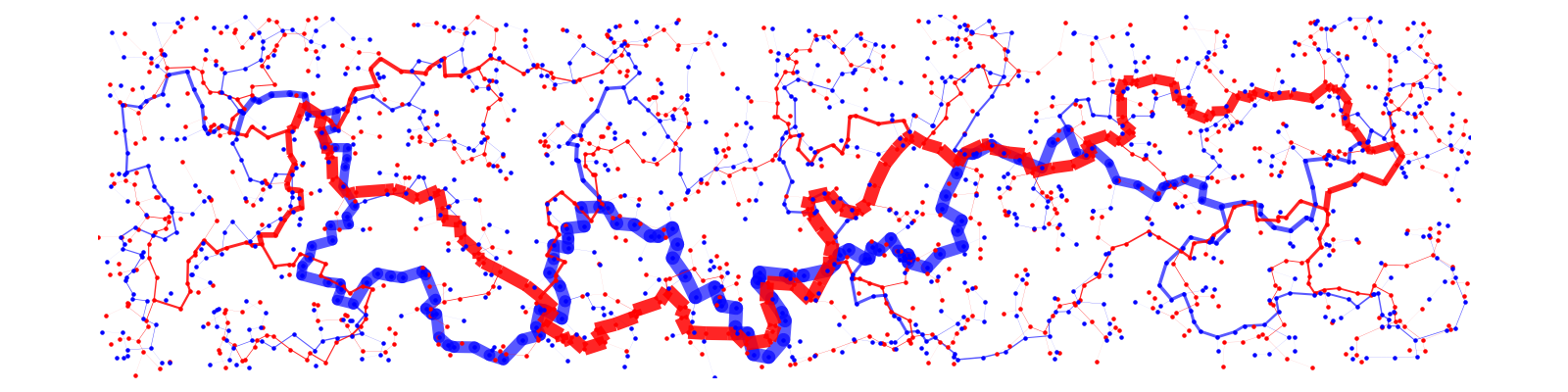}
         \caption{\mST (\CST, $\alpha=0)$}
		\label{sfig1:CST_noiserobustness_toydata}
	\end{subfigure}}
    \end{minipage}
    \hfill
    \begin{minipage}[h]{0.5\linewidth}
    \center{\begin{subfigure}{1\linewidth}
		\includegraphics[width=\linewidth]{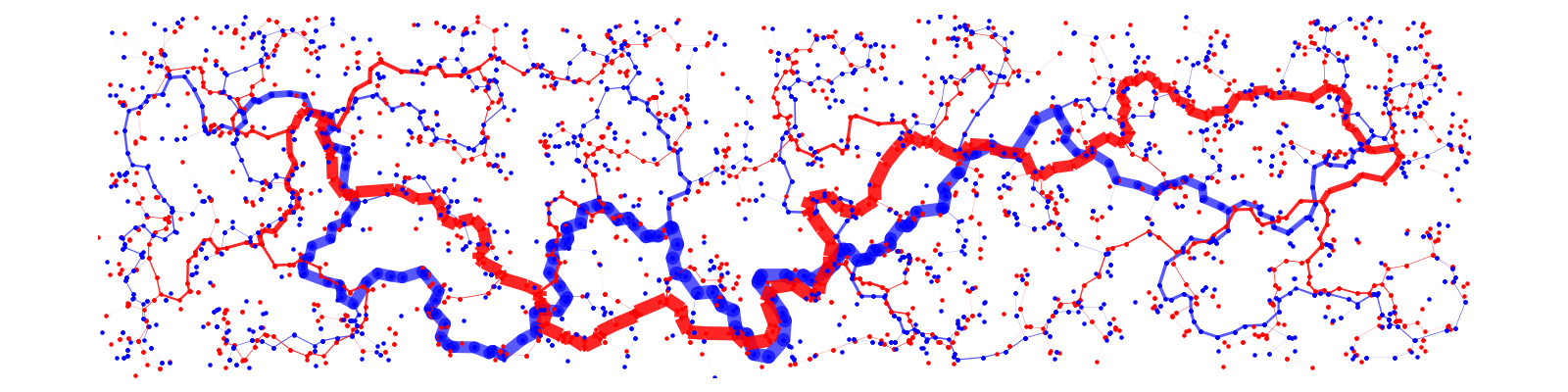}
		\caption{Steiner tree (\BCST, $\alpha=0$)}
		\label{sfig2:CST_noiserobustness_toydata}
	\end{subfigure}}
    \end{minipage}
    \vfill
    \begin{minipage}[h]{0.5\linewidth}
    \center{\begin{subfigure}{1\linewidth}
		\includegraphics[width=\linewidth]{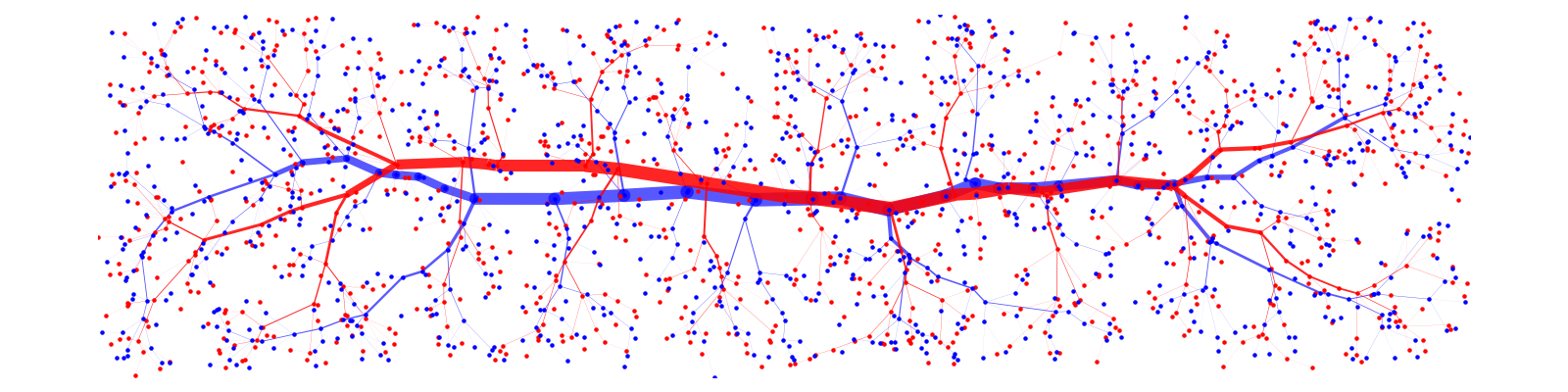}
		\caption{\CST, $\alpha=0.80$}
		\label{sfig3:CST_noiserobustness_toydata}
	\end{subfigure}}
    \end{minipage}
    \hfill
    \begin{minipage}[h]{0.5\linewidth}
    \begin{subfigure}{1\linewidth}
		\includegraphics[width=\linewidth]{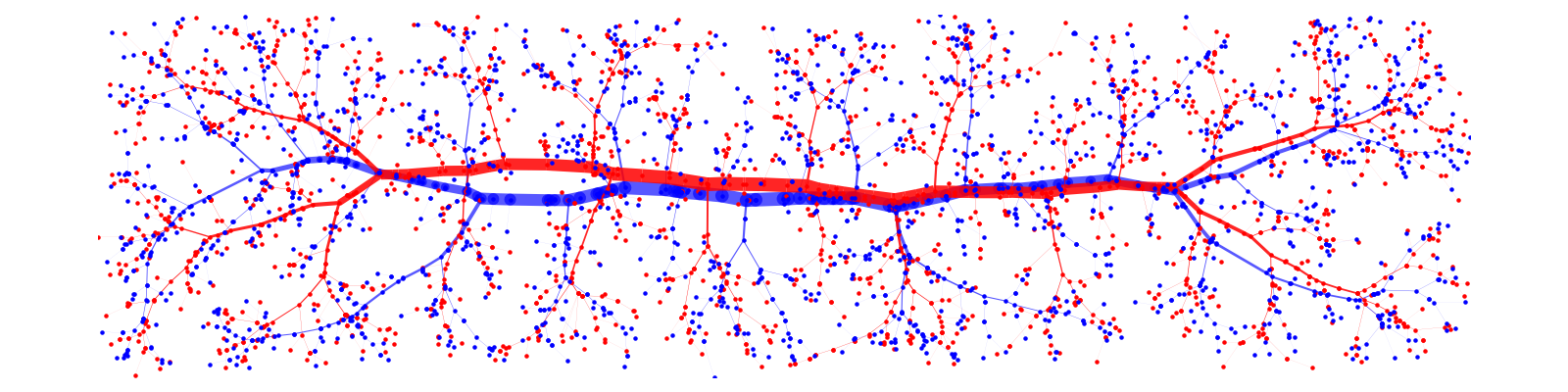}
		\caption{\BCST, $\alpha=0.80$}
		\label{sfig4:CST_noiserobustness_toydata}
	\end{subfigure}
    \end{minipage}
\end{minipage}
\caption[(B)CST Robustness Analysis]{\textbf{(B)CST Robustness Analysis}. (B)\CST for $\alpha > 0$ are more robust to noise and adhere to large scale structure in the data better than the \mST and Steiner tree. \textbf{Left}) When increasing $\alpha$, the sensitivity to random density fluctuations in the data decreases (good). At the same time, the total length of the tree increases (bad). This tradeoff can be adjusted with a single hyperparameter $\alpha$. More details in Section \ref{sec:toydata}. \textbf{Right}) \CST and \BCST of two samples, red and blue, drawn from the same distribution. The tree backbone reflects the global structure more accurately for $\alpha > 0$ than for $\alpha=0$. Edge widths are proportional to their centrality. All trees except for the \mST were computed using the heuristic proposed in Section \ref{sec:heuristic}. See Appendix \ref{sec:app_stability_examples} for more examples.}
	\label{fig:CST_noiserobustness_toydata}	
\end{figure}

We explore the robustness of the (B)\CST problem against data perturbations by comparing the (B)\CST topologies as small noise is introduced. In a toy example, three sets of 1000 points, uniformly sampled from a rectangle, are perturbed with Gaussian noise, yielding five perturbed sets per original set. We then compute the \CST and \BCST for $\alpha\in\{0, 0.1,\dots,0.9,1\}$, deeming a tree robust if minor data perturbations lead to minor geometrical changes in the tree. Formally, we consider a method $\delta$-robust, if, for any sets of points $P_1$ and $P_2$ and their respective trees $\tree_1$ and $\tree_2$,
$$d_{\tree}(\tree_1,\tree_2)\leq\delta d_P(P_1,P_2),$$
where $d_{\tree}$ and $d_P$ measure tree and set distances, respectively. The set distance $d_P$ quantifies the perturbations we aim to withstand, where lower $d_P$ values correspond to sets that are similar based on specific criteria. In our example, we define $d_P$ as the average distance between points and their perturbed counterparts. Since we apply the same noise to each point, the average distance between points approximates the Gaussian noise's standard deviation, making it nearly constant. To quantify structural tree changes, we set $d_{\tree}$ equal to the Frobenius norm of the shortest path distance matrices between the original and perturbed (B)\CST trees.

\figurename{} \ref{sfig0:CST_noiserobustness_toydata} shows the average Frobenius norm between the original and corresponding perturbed samples across various $\alpha$ values.  It is evident that as $\alpha$ increases, there is a noticeable decrease in the Frobenius norm. Since our $d_P$ is fairly constant, showcasing that the Frobenius norm decreases implies a reduction in $\delta$ as $\alpha$ rises, i.e. the trees become more robust. However, we also plot the average cost of the trees (sum of the individual edge costs), which increases with $\alpha$. Thus, the improvement in robustness comes at the expense of adding longer edges. This pattern is expected because, as $\alpha$ increases, the (B)\CST tends to a medoid-centered star graph (see Section \ref{sec:limitCST_alpha>1_n_infty}). This graph will have long edges but will also exhibit robustness to noise due to the medoid's inherent stability. According to our definition of $\delta$-robustness, the $\alpha\to\infty$ (B)\CST limiting case, which always outputs a star-graph, will be deemed robust despite its undesirability for describing the data structure.

We associate the data structure with the graph node interconnectivity, wherein shorter edges preserve it better. Thus, $\alpha$ serves as a parameter trading off stability vs.~data fidelity. Indeed, the \mST and Steiner tree ($\alpha=0$) on the right side of \figurename{} \ref{fig:CST_noiserobustness_toydata} are highly sensitive to minor data changes due to their greedy nature, prioritizing shorter edges. Conversely, the (B)\CST solutions at $\alpha=0.8$ are more stable, faithfully representing the data's overall layout, albeit with longer edges.

\subsection{Real-world data}\label{sec:real_word_data}
The ability to summarize the structure of data without excessive sensitivity to random jitter is important in many applications. In this section we briefly showcase some potential applications where
the (B)CST can be beneficial (see Appendix \ref{sec:app_applications} for more details):

\textbf{Trajectory inference of single cell data:} The high dimensional single cell RNA-sequencing data can be used to model the trajectories defined by the cell differentiation process. We show results on mouse bone marrow data \citep{paul_transcriptional_2015}, here denoted as Paul dataset. To study robustness we perturb the data by removing half of the samples and then compare how the backbone of different spanning trees is affected by this perturbation. For visualization, the data is embedded in 2D using PAGA~\citep{wolf_paga_2019}. If a spanning tree aligns well with the 2D embedding, this is an indication that the tree approximates the trajectory well. \figurename{} \ref{fig:stability_paul} shows the embedded trees. The \mST misses the highlighted bifurcation and it is more sensitive to the noise.
The \CST and \BCST are robust to the perturbation and align well with the PAGA embedding, though the \CST may not reconstruct well the finer details. The addition of \SPs enables the \BCST to follow the trajectory more closely. The case $\alpha=1$ results in a star graph, possibly due to the curse of dimensionality (see \thref{rem:more_star_in_high_dim}).




\def \bottomvspace{-1}
\def \topvspace{-0.}
\def \scalaabox{0.95}
\def \lowcornerx{2}
\def \lowcornery{2.3}
\def \topcornerx{3.8}
\def \topcornery{4.5}

\begin{figure}[]
	\captionsetup[subfigure]{justification=centering}
	\centering
	\scalebox{\scalaabox}{
		\begin{subfigure}{0.25\linewidth}
			\centering
			\vspace*{\topvspace cm}
			\begin{tikzpicture}
			\node[anchor=south west,inner sep=0] at (0,0) {\includegraphics[width=1\linewidth]{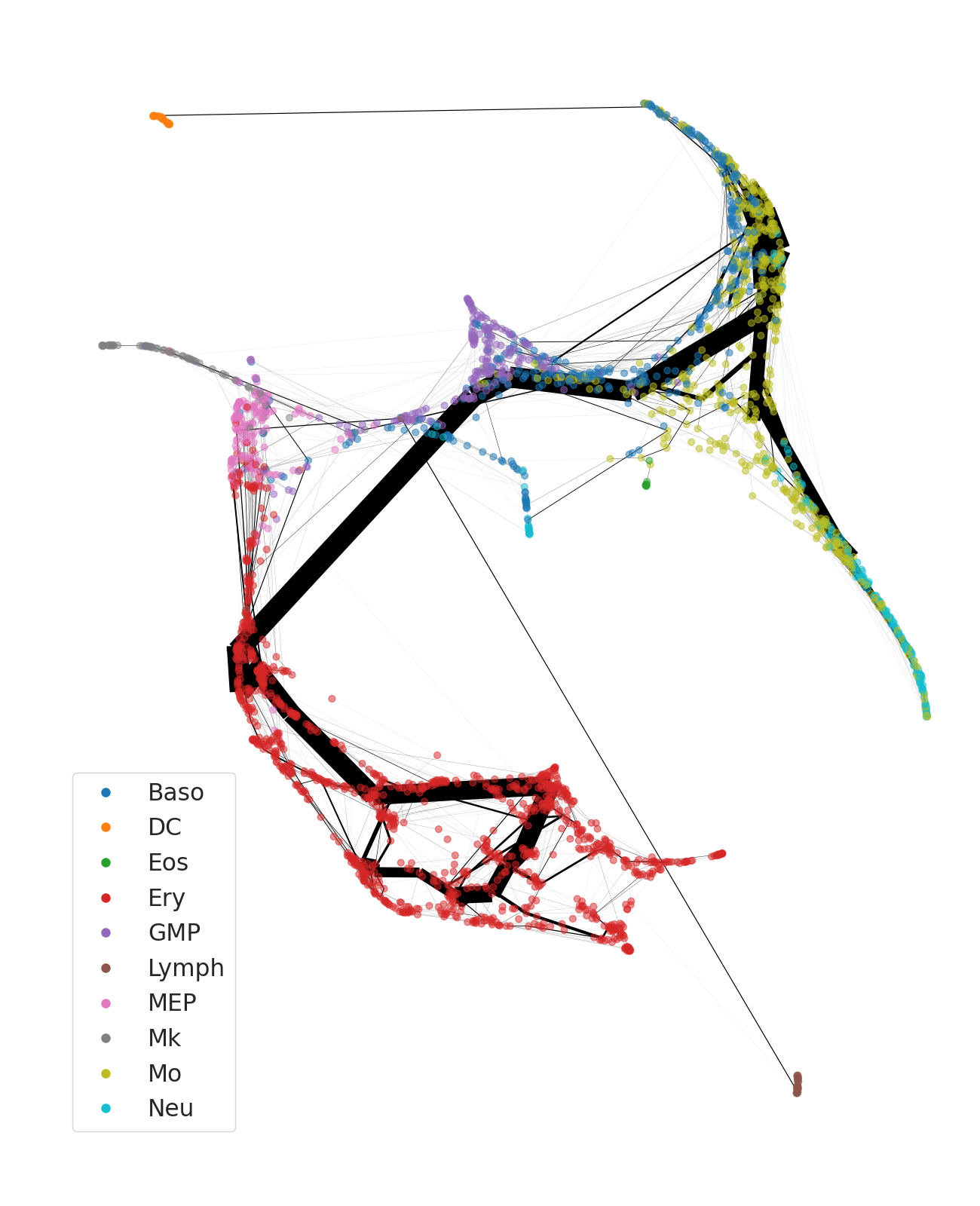}};
			\draw[red,ultra thick] (\lowcornerx,\lowcornery) rectangle (\topcornerx,\topcornery);
			\end{tikzpicture}
			\vspace{\bottomvspace cm}
			\caption{Original \\ \mST ($\alpha=0.0$)}
			\label{sfig1:stability_paul}
		\end{subfigure}%
		\begin{subfigure}{0.25\linewidth}
			\centering
			\vspace*{\topvspace cm}
			\begin{tikzpicture}
			\node[anchor=south west,inner sep=0] at (0,0) {\includegraphics[width=1\linewidth]{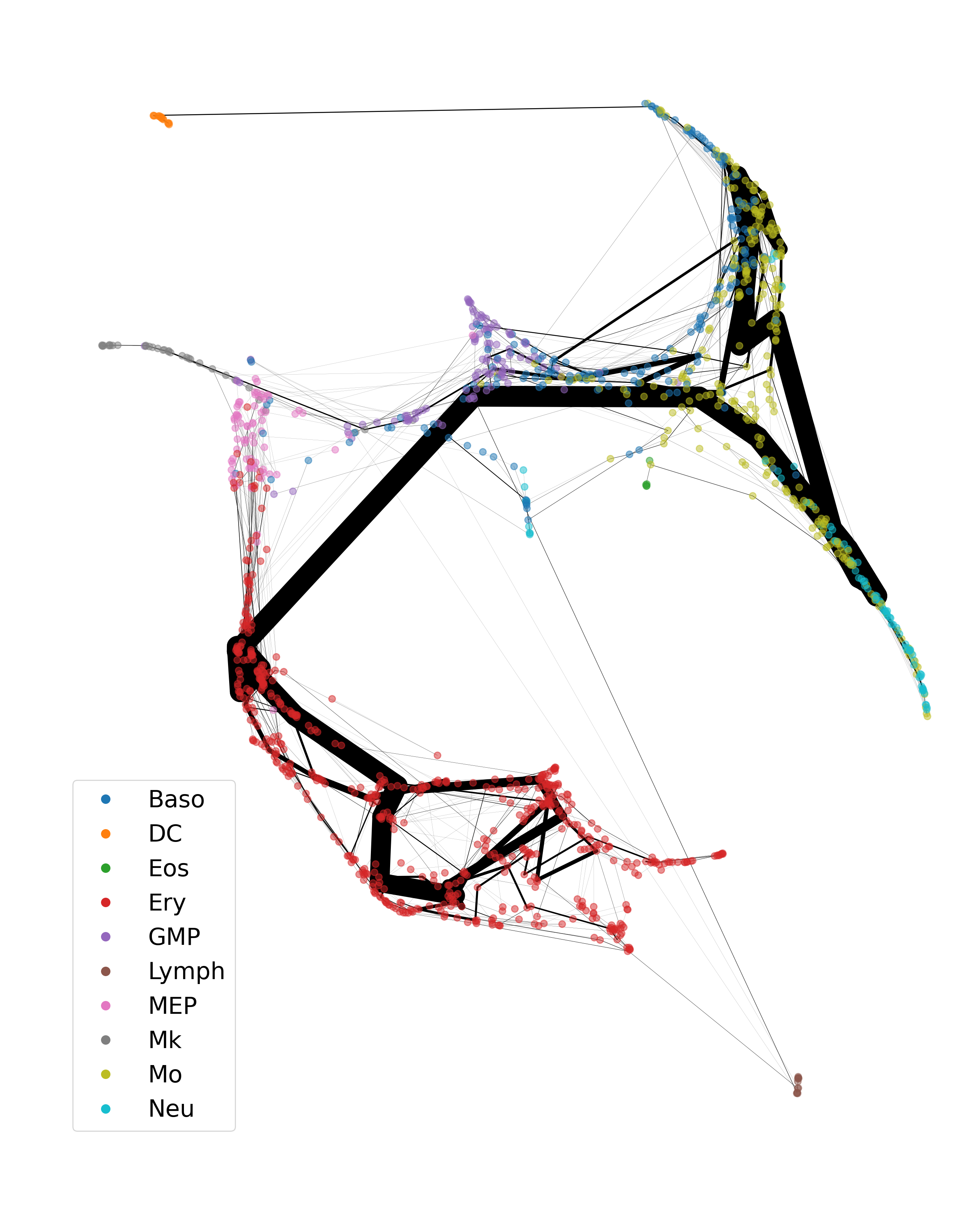}};
			\draw[red,ultra thick] (\lowcornerx,\lowcornery) rectangle (\topcornerx,\topcornery);
			\end{tikzpicture}
			\vspace{\bottomvspace cm}
			\caption{Subsample\\ \mST ($\alpha=0.0$)}
			\label{sfig2:stability_paul}
		\end{subfigure}%
		\begin{subfigure}{0.25\linewidth}
			\centering
			\vspace*{\topvspace cm}
			\begin{tikzpicture}
			\node[anchor=south west,inner sep=0] at (0,0) {\includegraphics[width=1\linewidth]{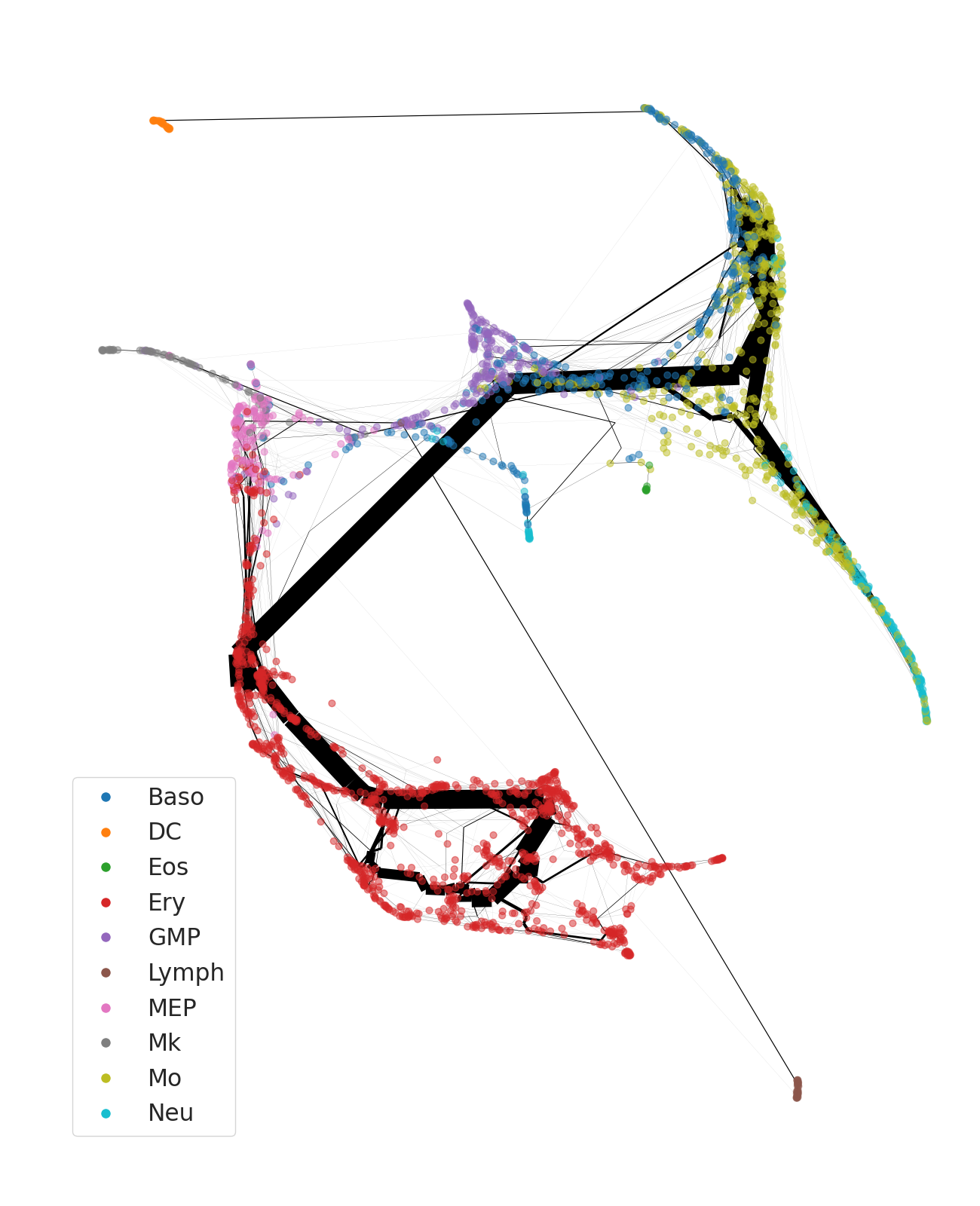}};
			\draw[red,ultra thick] (\lowcornerx,\lowcornery) rectangle (\topcornerx,\topcornery);
			\end{tikzpicture}
			\vspace{\bottomvspace cm}
			\caption{Original \\ Steiner ($\alpha=0.0$)}
			\label{sfig3:stability_paul}
		\end{subfigure}%
		\begin{subfigure}{0.25\linewidth}
			\centering
			\vspace*{\topvspace cm}
			\begin{tikzpicture}
			\node[anchor=south west,inner sep=0] at (0,0) {\includegraphics[width=1\linewidth]{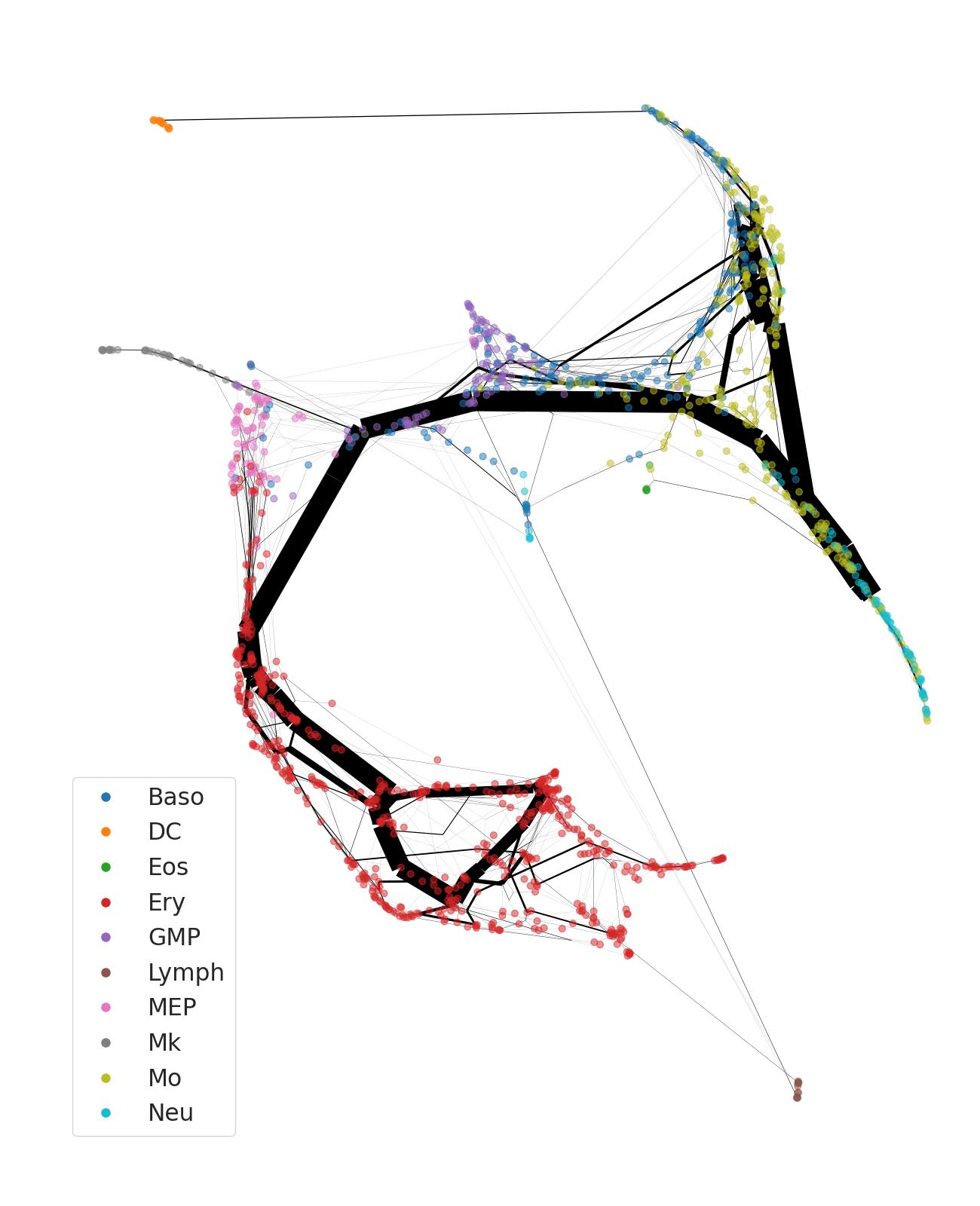}};
			\draw[red,ultra thick] (\lowcornerx,\lowcornery) rectangle (\topcornerx,\topcornery);
			\end{tikzpicture}
			\vspace{\bottomvspace cm}
			\caption{Subsample\\ Steiner ($\alpha=0.0$)}
			\label{sfig4:stability_paul}
		\end{subfigure}}
	\scalebox{\scalaabox}{
		\begin{subfigure}{0.25\linewidth}
			\centering
			\vspace*{\topvspace cm}
			\begin{tikzpicture}
			\node[anchor=south west,inner sep=0] at (0,0) {\includegraphics[width=1\linewidth]{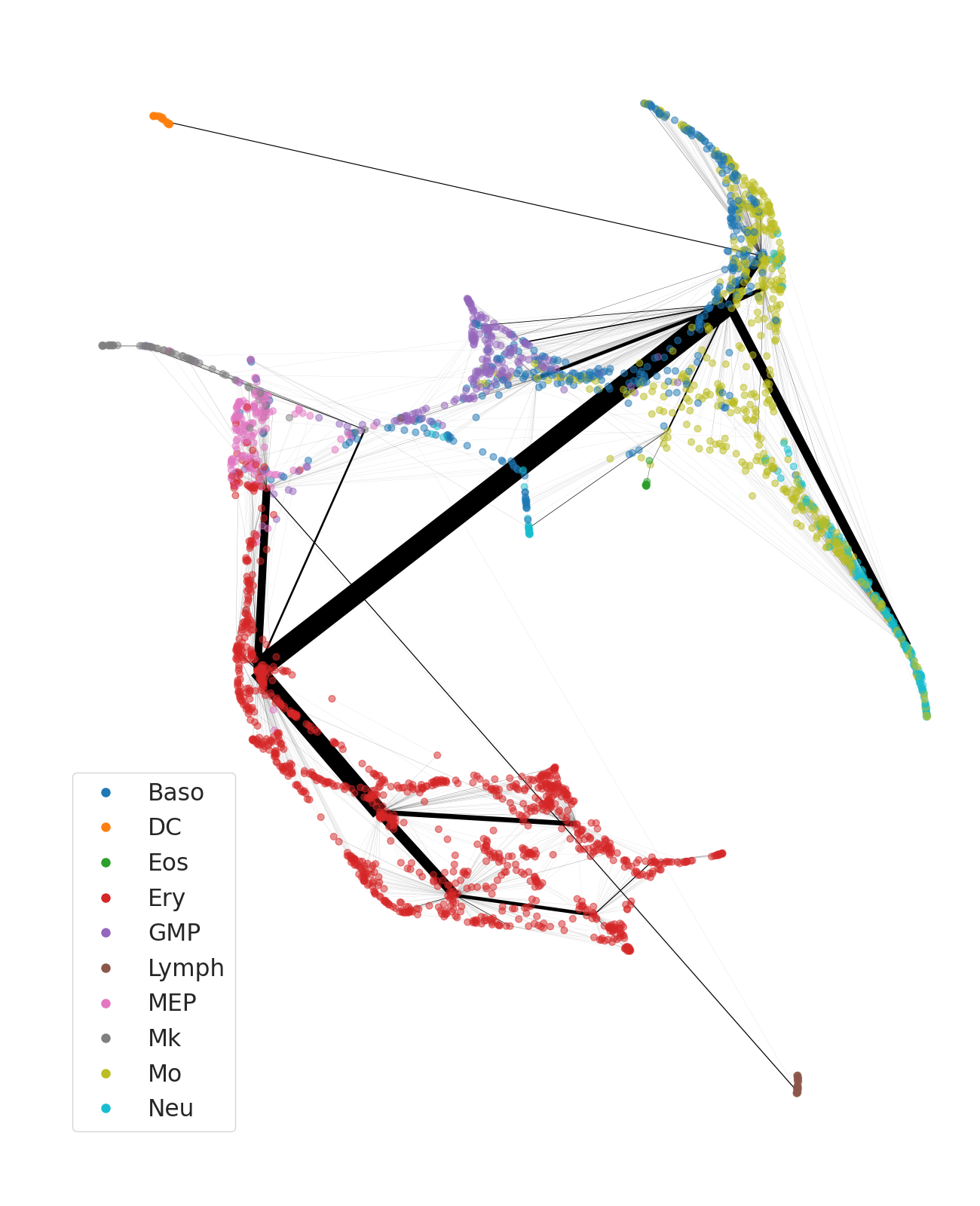}};
			\draw[red,ultra thick] (\lowcornerx,\lowcornery) rectangle (\topcornerx,\topcornery);
			\end{tikzpicture}
			\vspace{\bottomvspace cm}
			\caption{Original \\ \CST ($\alpha=0.5$)}
			\label{sfig5:stability_paul}
		\end{subfigure}%
		\begin{subfigure}{0.25\linewidth}
			\centering
			\vspace*{\topvspace cm}
			\begin{tikzpicture}
			\node[anchor=south west,inner sep=0] at (0,0) {\includegraphics[width=1\linewidth]{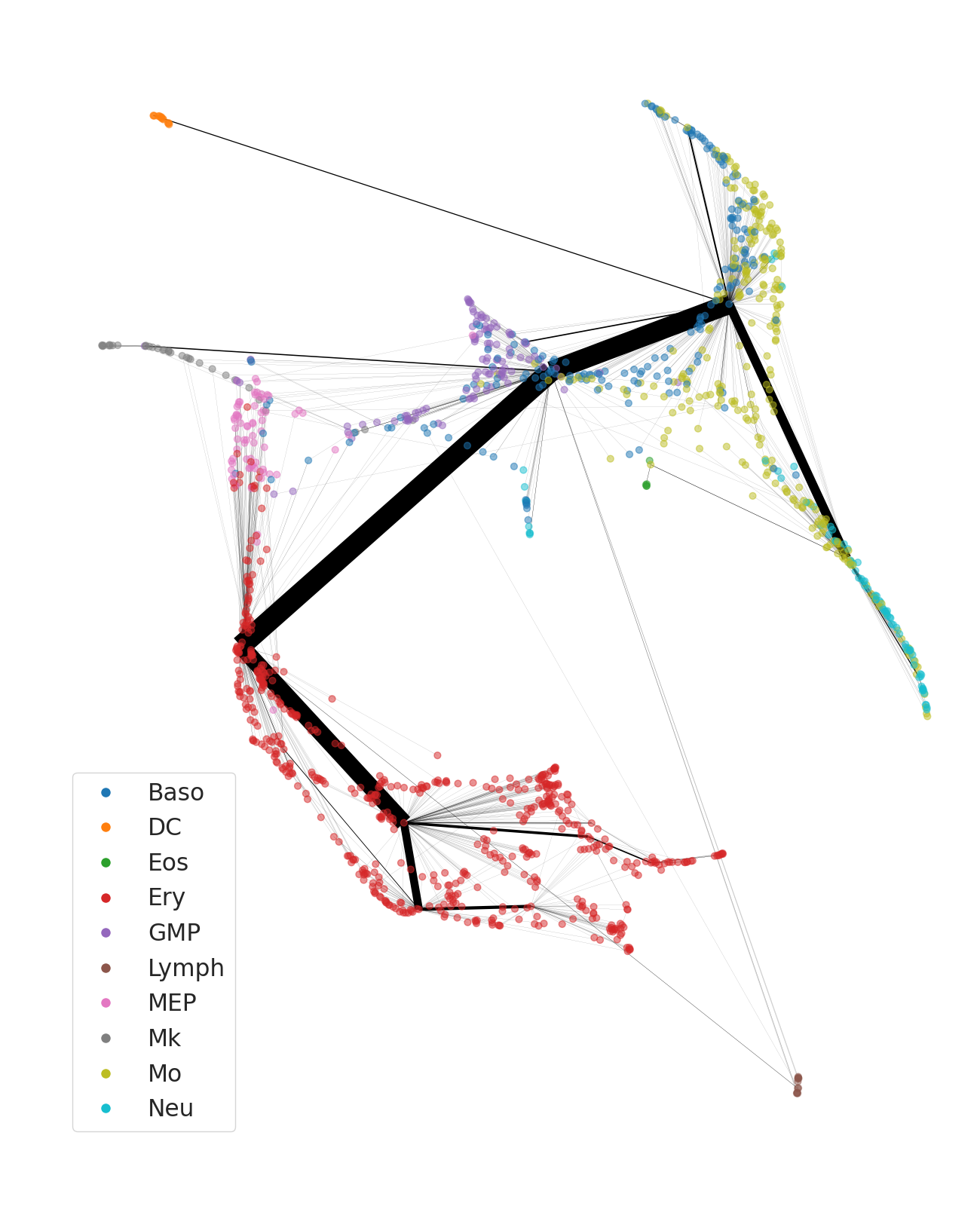}};
			\draw[red,ultra thick] (\lowcornerx,\lowcornery) rectangle (\topcornerx,\topcornery);
			\end{tikzpicture}
			\vspace{\bottomvspace cm}
			\caption{Subsample\\ \CST ($\alpha=0.5$)}
			\label{sfig6:stability_paul}
		\end{subfigure}%
		\begin{subfigure}{0.25\linewidth}
			\centering
			\vspace*{\topvspace cm}
			\begin{tikzpicture}
			\node[anchor=south west,inner sep=0] at (0,0) {\includegraphics[width=1\linewidth]{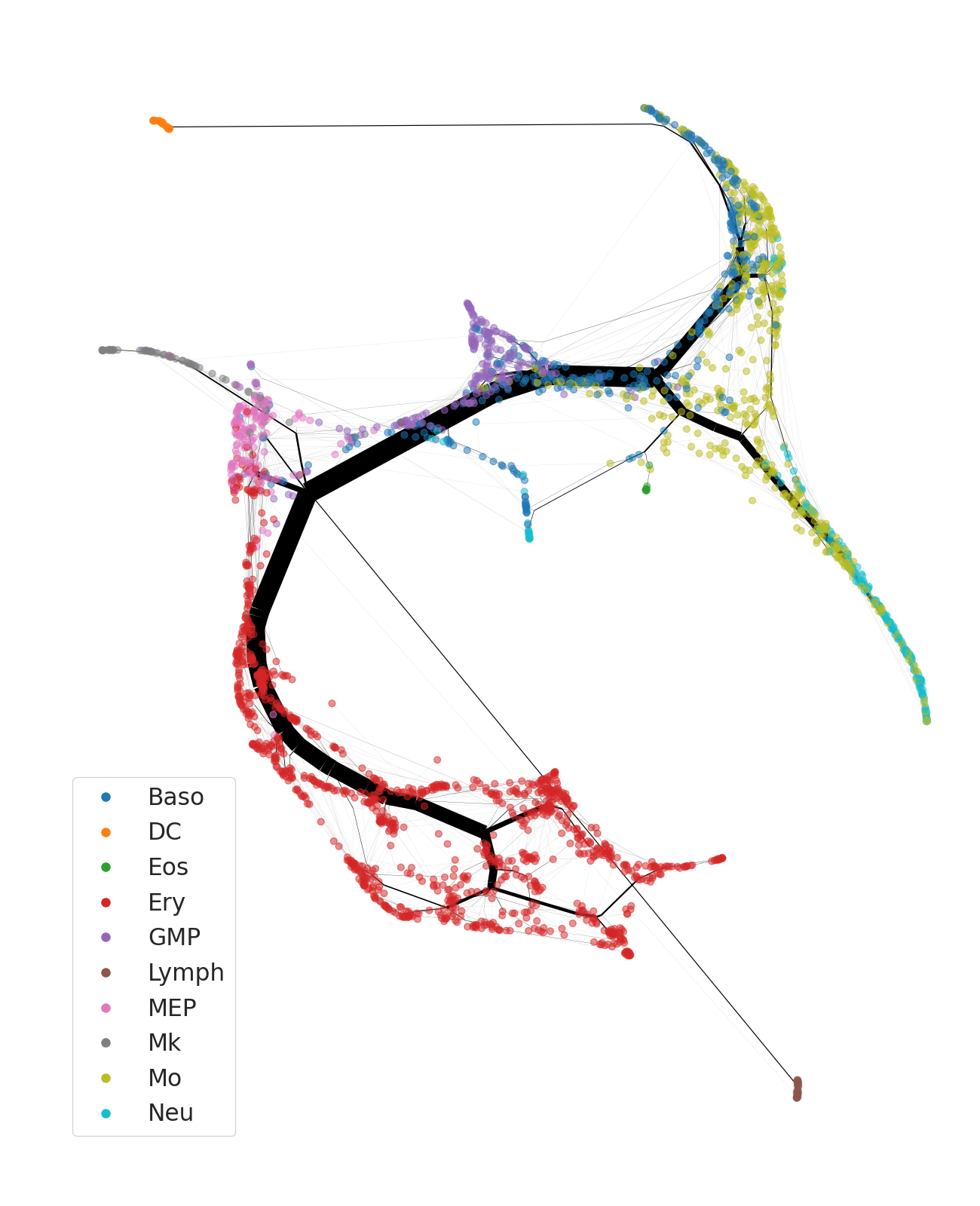}};
			\draw[red,ultra thick] (\lowcornerx,\lowcornery) rectangle (\topcornerx,\topcornery);
			\end{tikzpicture}
			\vspace{\bottomvspace cm}
			\caption{Original \\ \BCST ($\alpha=0.5$)}
			\label{sfig7:stability_paul}
		\end{subfigure}%
		\begin{subfigure}{0.25\linewidth}
			\centering
			\vspace*{\topvspace cm}
			\begin{tikzpicture}
			\node[anchor=south west,inner sep=0] at (0,0) {\includegraphics[width=1\linewidth]{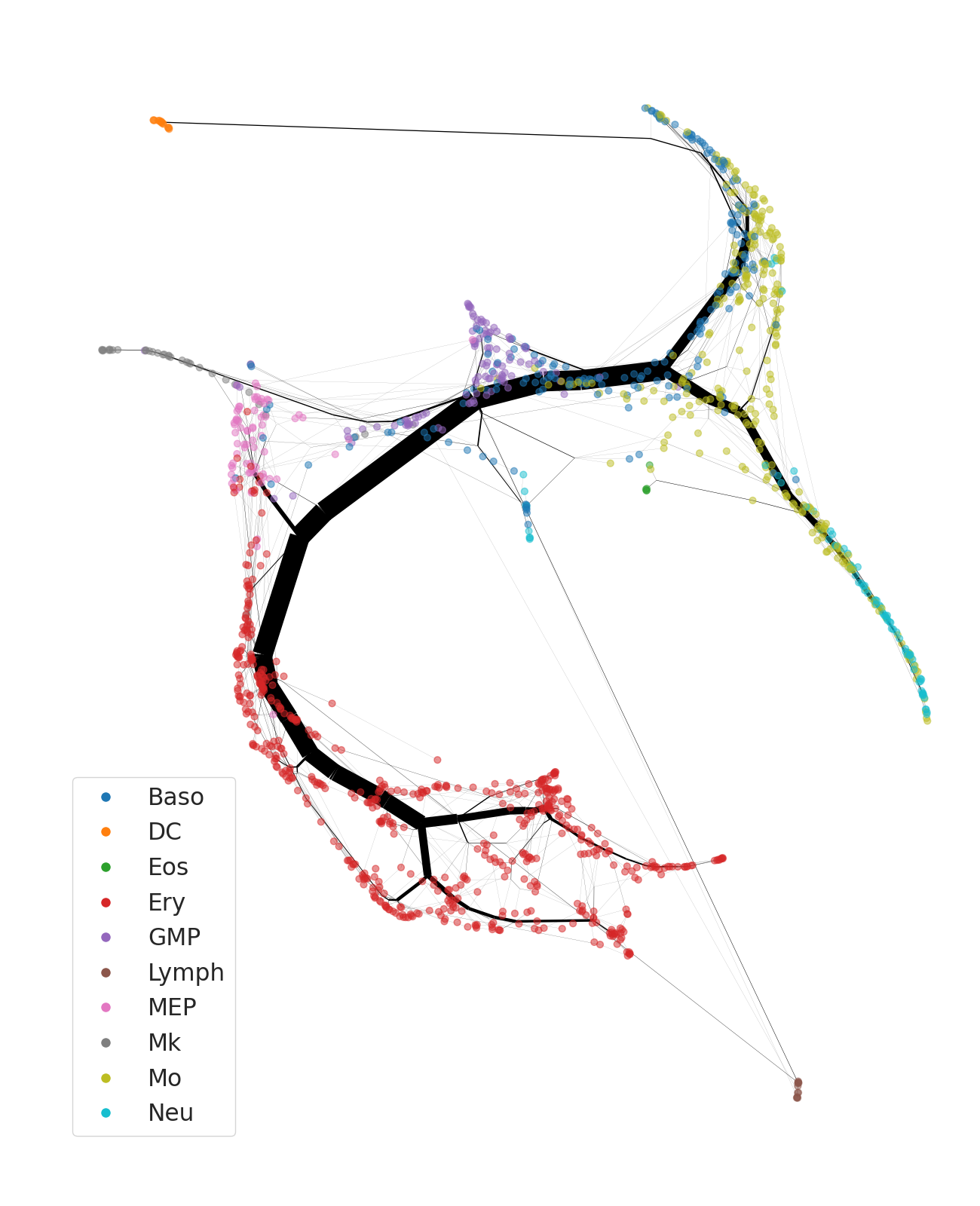}};
			\draw[red,ultra thick] (\lowcornerx,\lowcornery) rectangle (\topcornerx,\topcornery);
			\end{tikzpicture}
			\vspace{\bottomvspace cm}
			\caption{Subsample\\ \BCST ($\alpha=0.5$)}
			\label{sfig8:stability_paul}
		\end{subfigure}}
	\scalebox{\scalaabox}{		
		\begin{subfigure}{0.25\linewidth}
			\centering
			\vspace*{\topvspace cm}
			\includegraphics[width=1\linewidth]{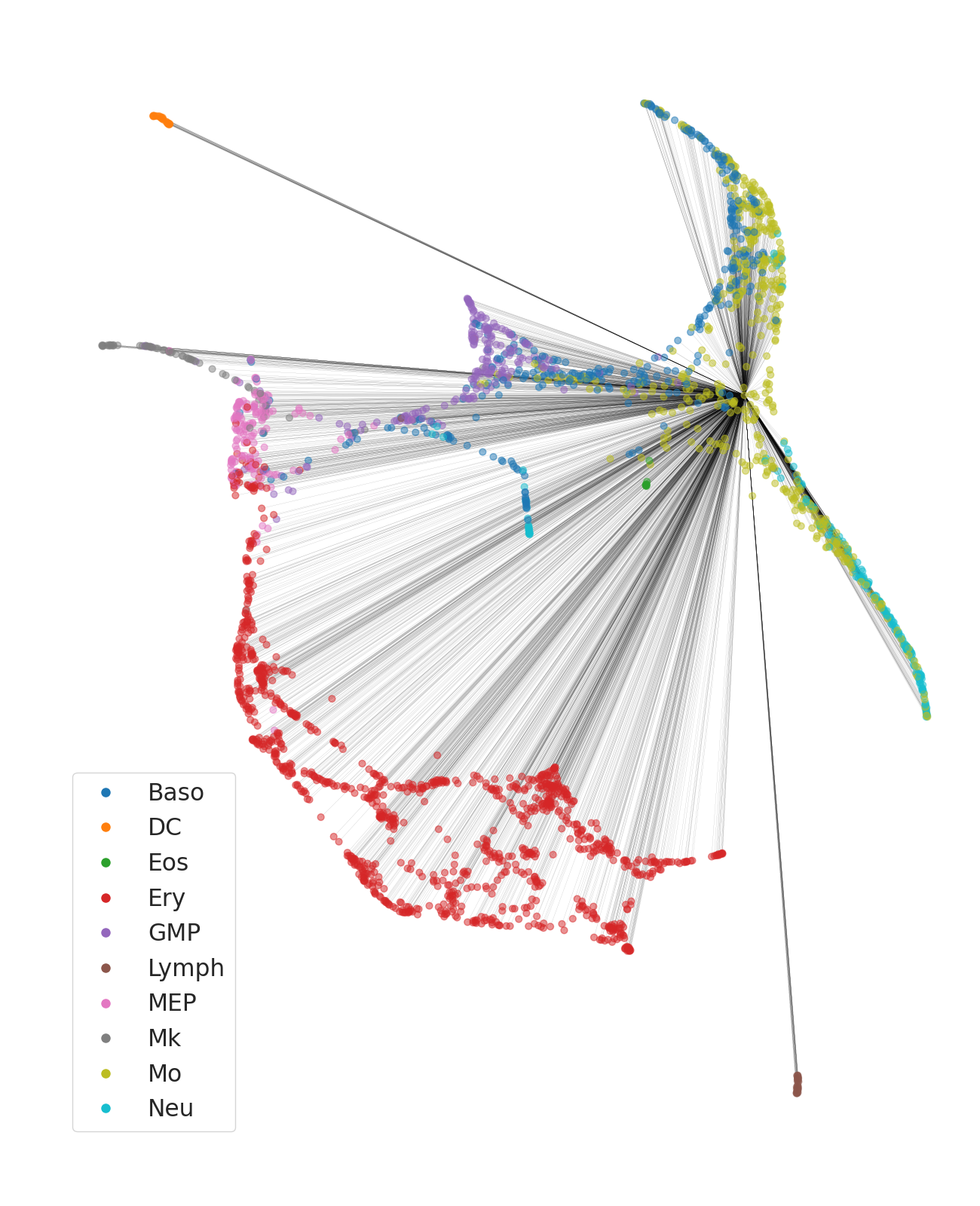}
			\vspace{\bottomvspace cm}
			\caption{Original\\ \MRCT ($\alpha=1.0$)}
			\label{sfig9:stability_paul}
		\end{subfigure}
		\begin{subfigure}{0.25\linewidth}
			\centering
			\vspace*{\topvspace cm}
			\includegraphics[width=1\linewidth]{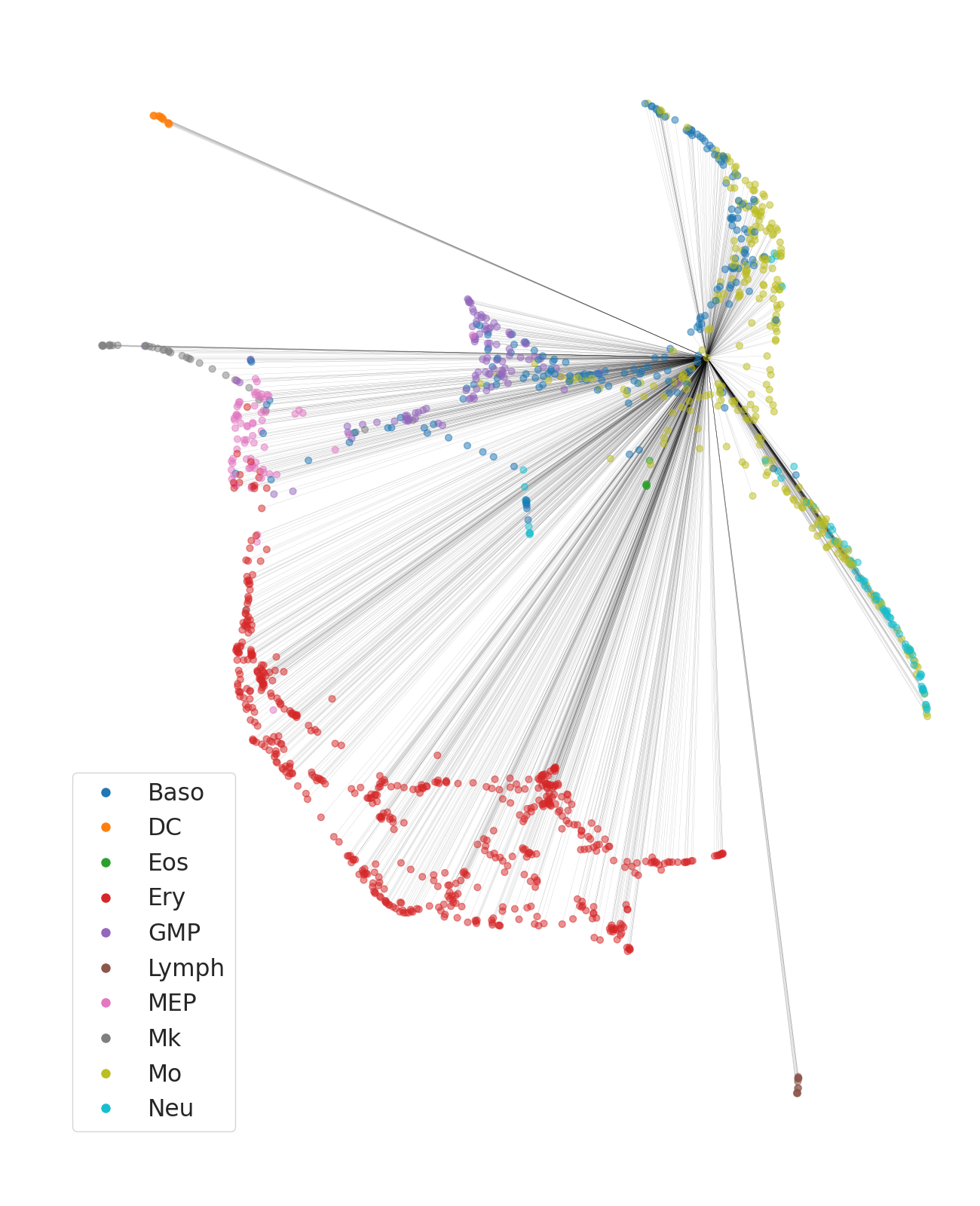}
			\vspace{\bottomvspace cm}
			\caption{Subsample \\ \MRCT ($\alpha=1.0$)}
			\label{sfig10:stability_paul}
		\end{subfigure}%
		\begin{subfigure}{0.25\linewidth}
			\centering
			\vspace*{\topvspace cm}
			\includegraphics[width=1\linewidth]{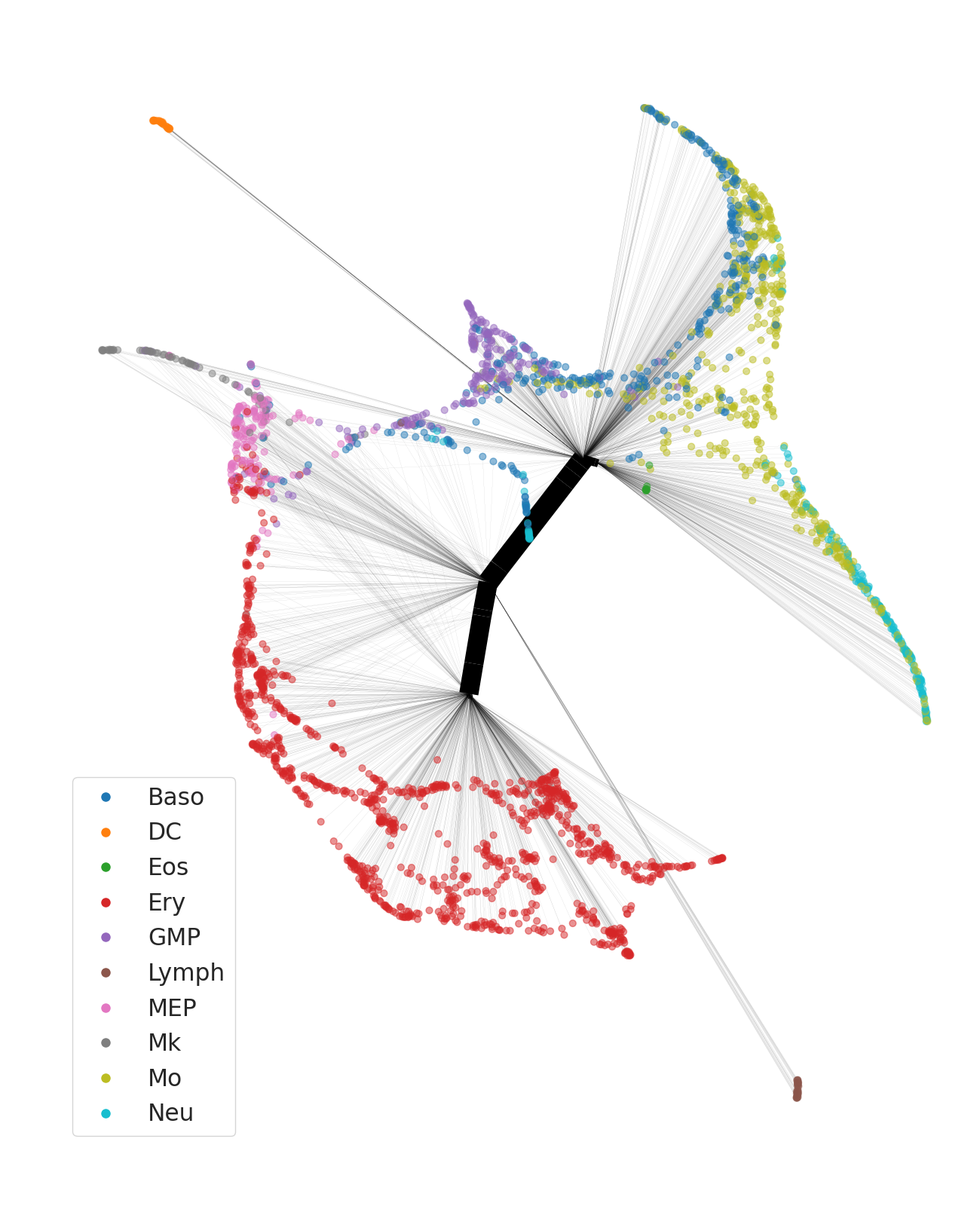}
			\vspace{\bottomvspace cm}
			\caption{Original\\ (B)\MRCT ($\alpha=1.0$)}
			\label{sfig11:stability_paul}
		\end{subfigure}
		\begin{subfigure}{0.25\linewidth}
			\centering
			\vspace*{\topvspace cm}
			\includegraphics[width=1\linewidth]{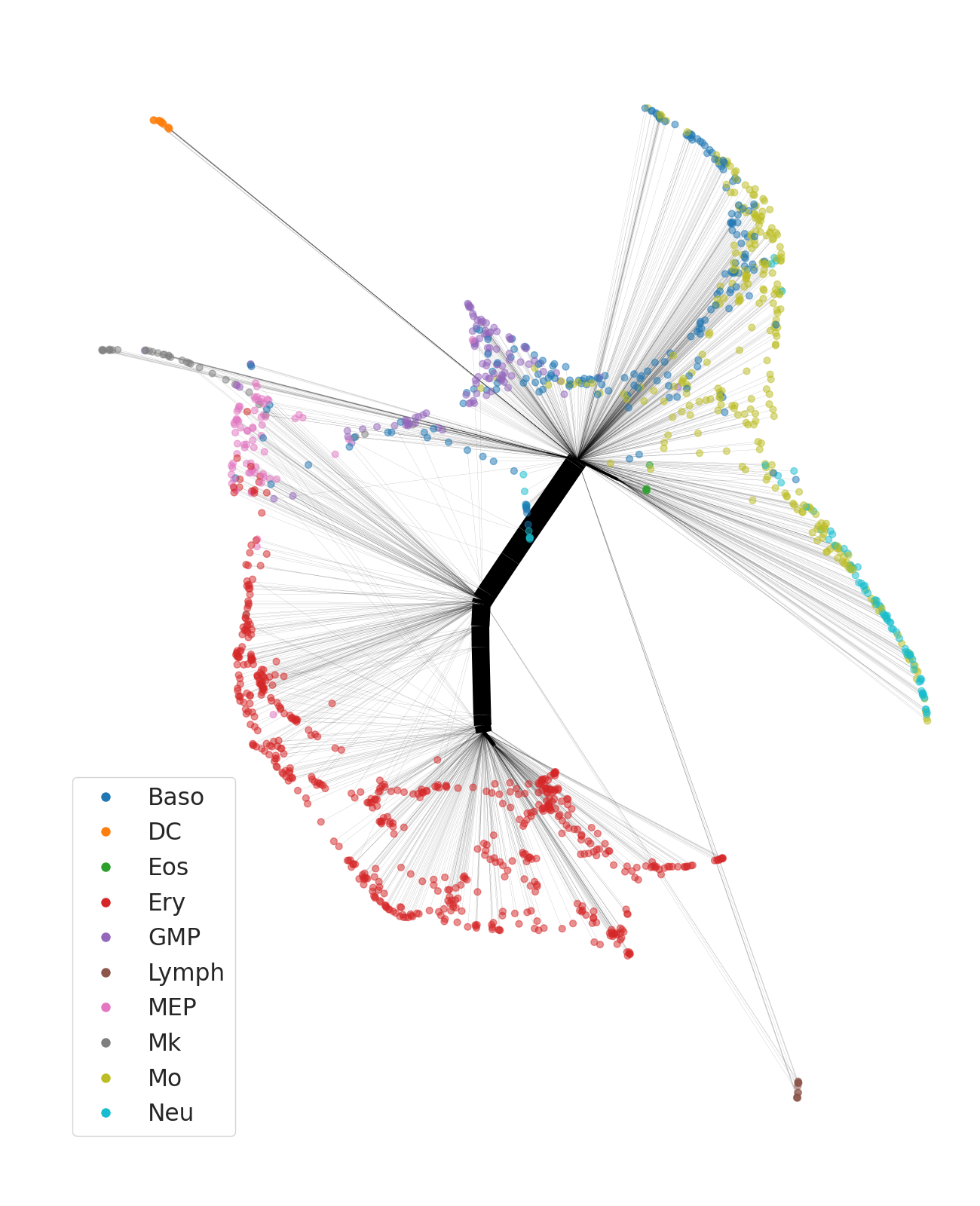}
			\vspace{\bottomvspace cm}
			\caption{Subsample \\ (B)\MRCT ($\alpha=1.0$)}
			\label{sfig12:stability_paul}
		\end{subfigure}}
	\caption[mST, (B)CST and B\MRCT of the Paul Dataset]{\textbf{mST, (B)CST and B\MRCT of the Paul Dataset}. We applied the algorithms to both the original data (top) and a perturbed version with half of the points randomly removed (bottom). PAGA was used for 2D visualization, while the trees were computed in a 50-dimensional PCA projection. Colors represent different cell populations. The width of the edges is proportional to their centralities. \ref{sfig1:stability_paul}-\ref{sfig4:stability_paul}) In the original data, the \mST and Steiner tree do not faithfully model the trajectory bifurcation highlighted by the rectangle. Moreover, the trajectory changes drastically at this point once a subset of the samples is removed. \ref{sfig5:stability_paul}-\ref{sfig8:stability_paul}) The \CST and \BCST at $\alpha=0.5$ are able to detect the bifurcation, and preserve the main backbone of the tree after the data has been perturbed. \ref{sfig9:stability_paul}-\ref{sfig12:stability_paul}) The \MRCT and its branched version are robust to perturbations due to its star shape structure, but this shape is also responsible for its incapability to model the data properly}
	\label{fig:stability_paul}
\end{figure}
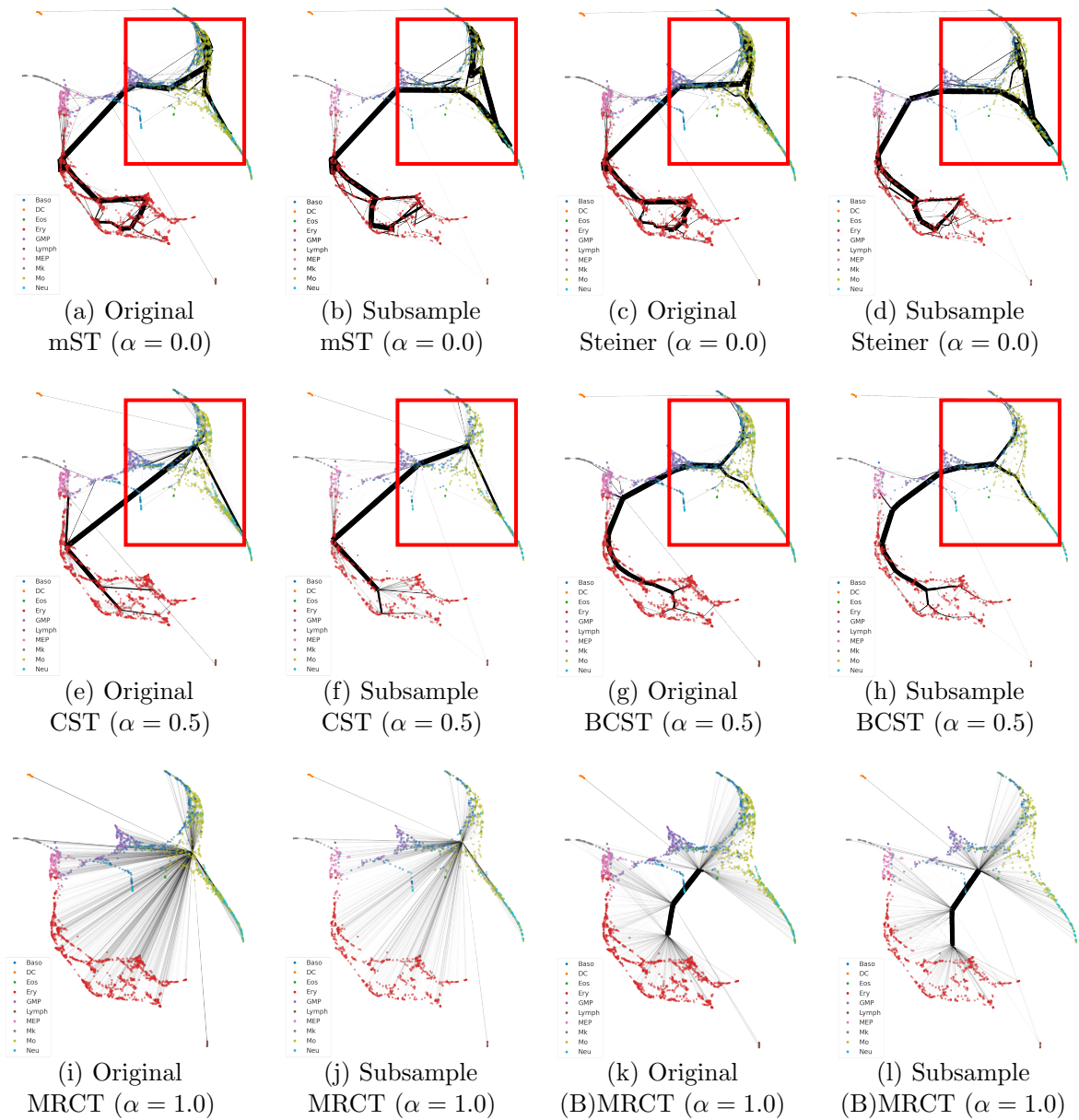%

\textbf{3D Plant skeletonization:} The skeletonization of plants is relevant for comprehending plant structure, with applications in subsequent tasks. The skeletons of a 3D point cloud of the surface of a plant \citep{schunck2021plosone}, obtained using the BCST for various $\alpha$ values, is shown in \figurename{} \ref{fig:plant_skeleton}. Intermediate $\alpha$ values are able to represent the plant's structure more accurately than extreme ones, producing qualitatively pleasing skeletons with fewer resources compared to dedicated handcrafted pipelines, which may depend on supplementary attributes like plant segmentation \citep{magistri2020segmentation}.

\def \bottomvspace{-.5}
\def \topvspace{-0.0}
\begin{figure}[h!]
	\centering
	\begin{subfigure}{0.25\linewidth}
		\centering
		\vspace*{\topvspace cm}
		\includegraphics[width=1\linewidth,trim=8cm 5cm 8cm 5cm ,clip]{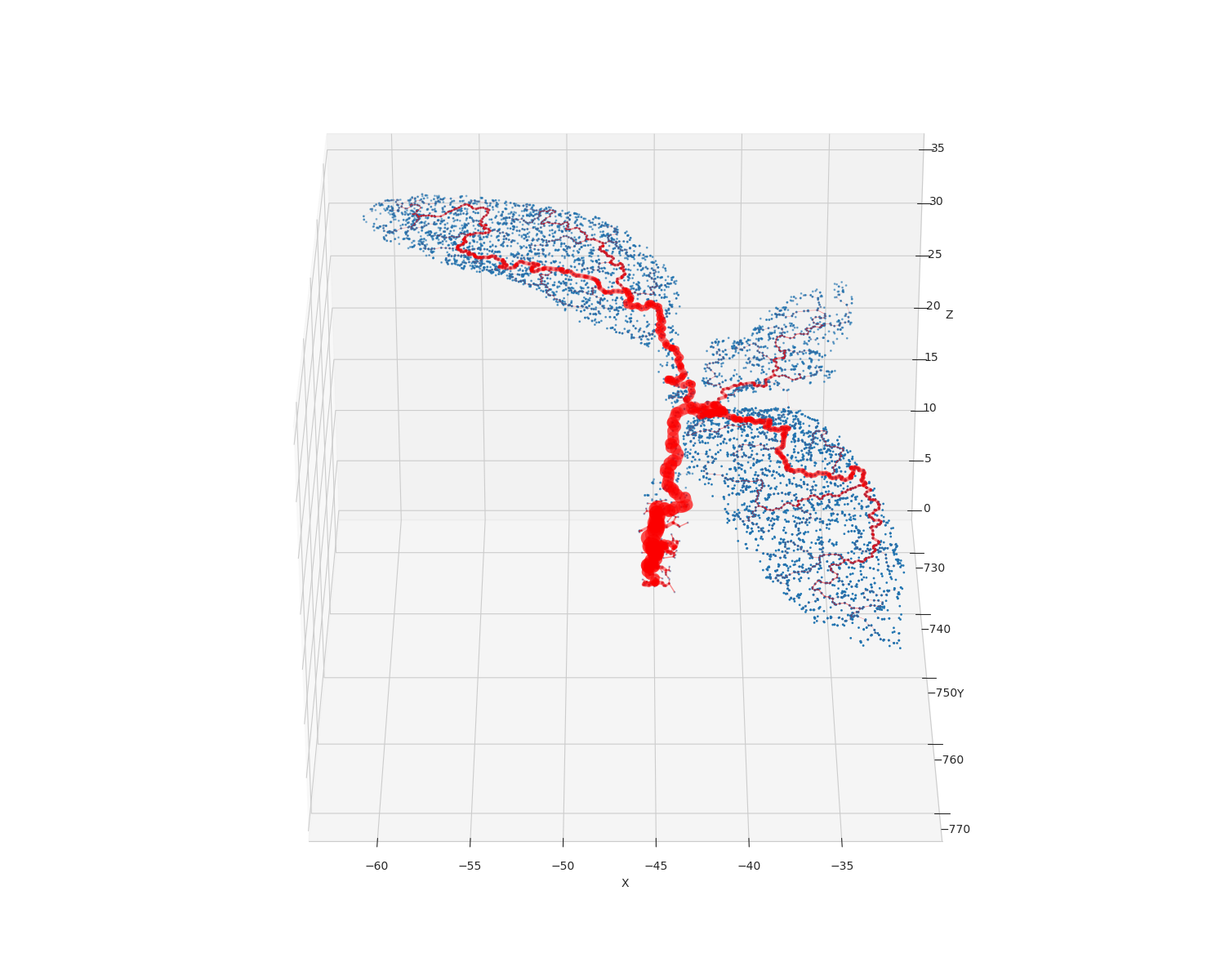}
		\vspace{\bottomvspace cm}
		\caption*{\BCST $\alpha=0.00$}
		\label{sfig1:plant_skeleton}
	\end{subfigure}%
	\begin{subfigure}{0.25\linewidth}
		\centering
		\vspace*{\topvspace cm}
		\includegraphics[width=1\linewidth,trim=8cm 5cm 8cm 5cm ,clip]{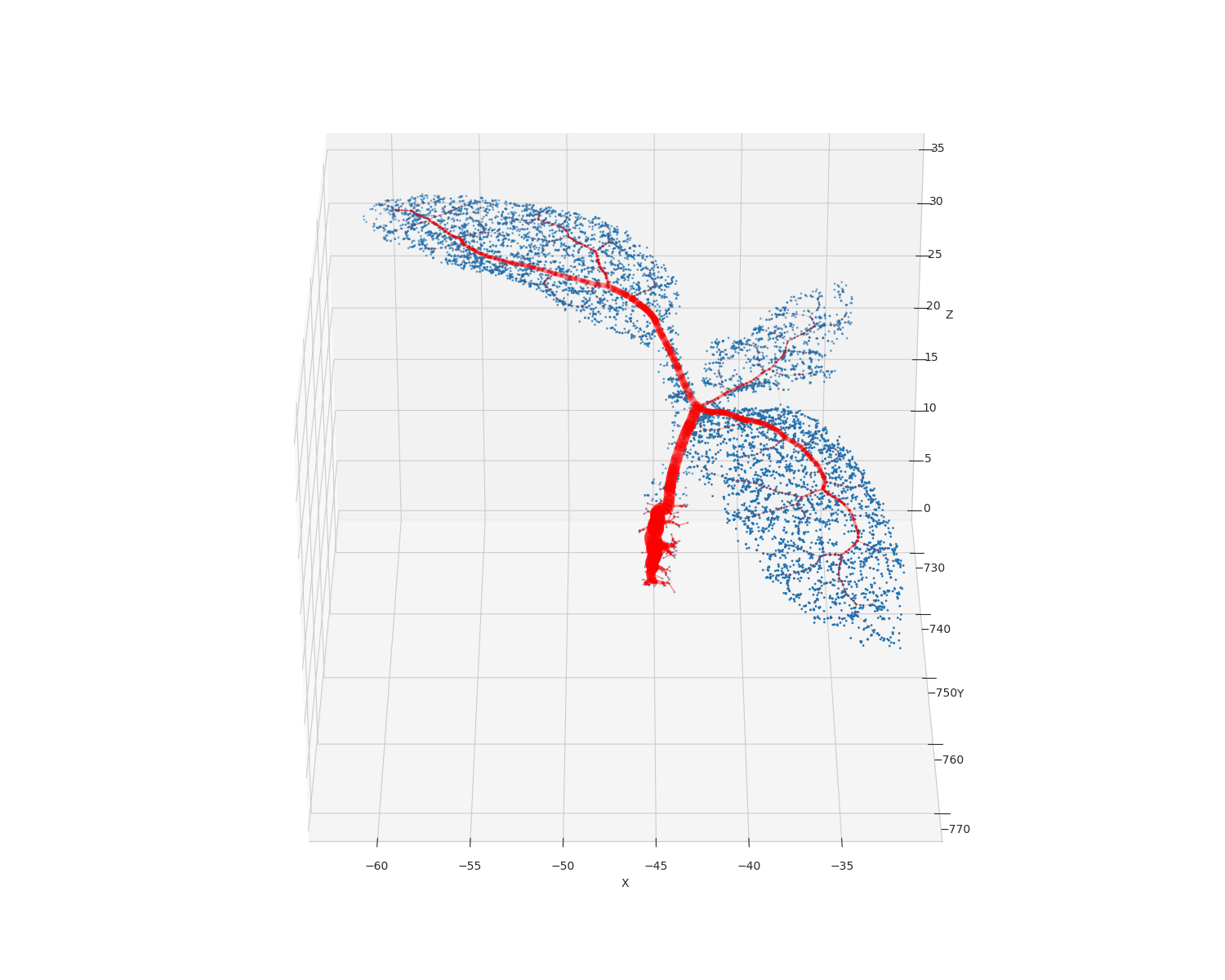}
		\vspace{\bottomvspace cm}
		\caption*{\BCST $\alpha=0.50$}
		\label{sfig2:plant_skeleton}
	\end{subfigure}%
	\begin{subfigure}{0.25\linewidth}
		\centering
		\vspace*{\topvspace cm}
		\includegraphics[width=1\linewidth,trim=8cm 5cm 8cm 5cm ,clip]{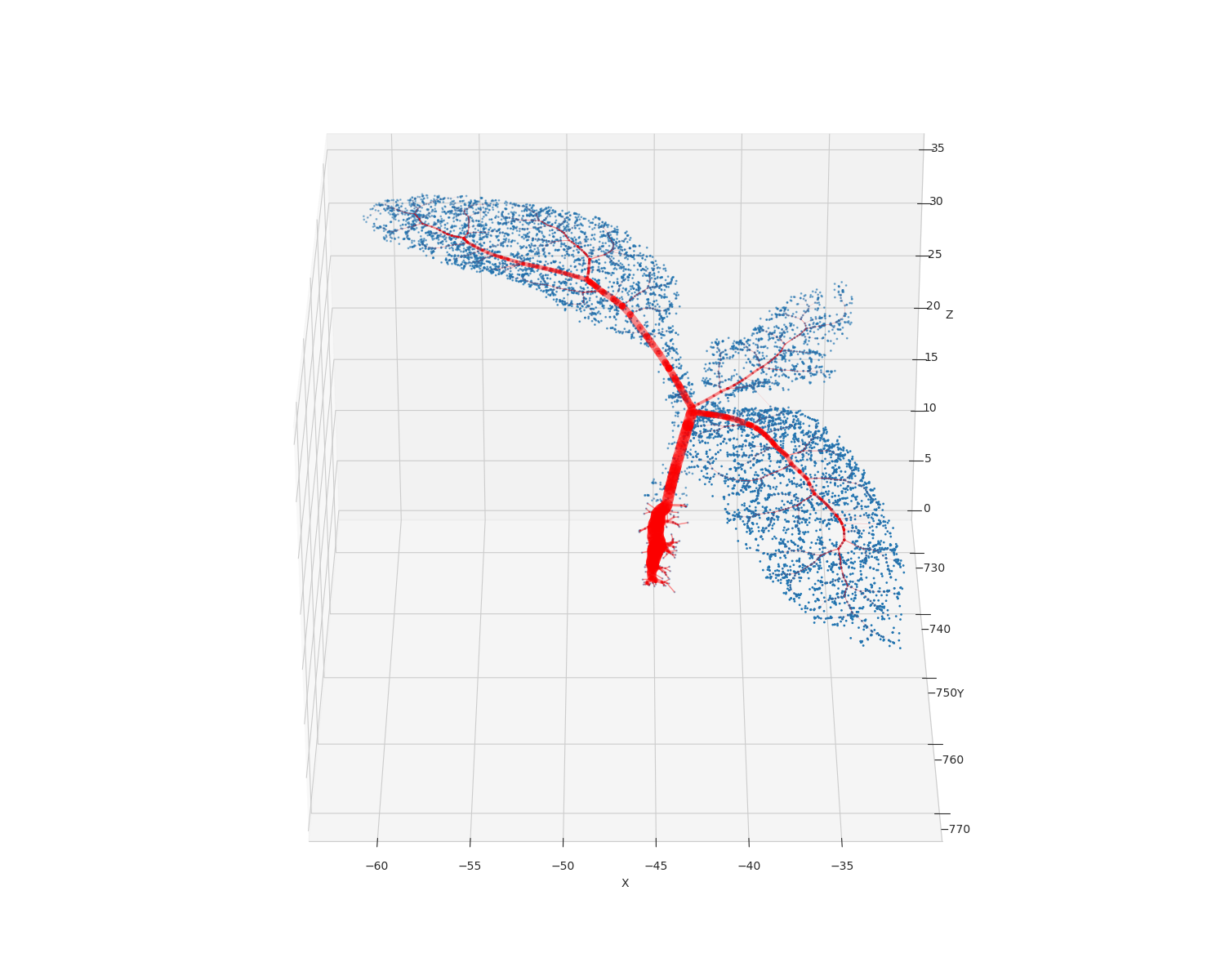}
		\vspace{\bottomvspace cm}
		\caption*{\BCST $\alpha=0.70$}
		\label{sfig3:plant_skeleton}
	\end{subfigure}%
	\begin{subfigure}{0.25\linewidth}
		\centering
		\vspace*{\topvspace cm}
		\includegraphics[width=1\linewidth,trim=8cm 5cm 8cm 5cm ,clip]{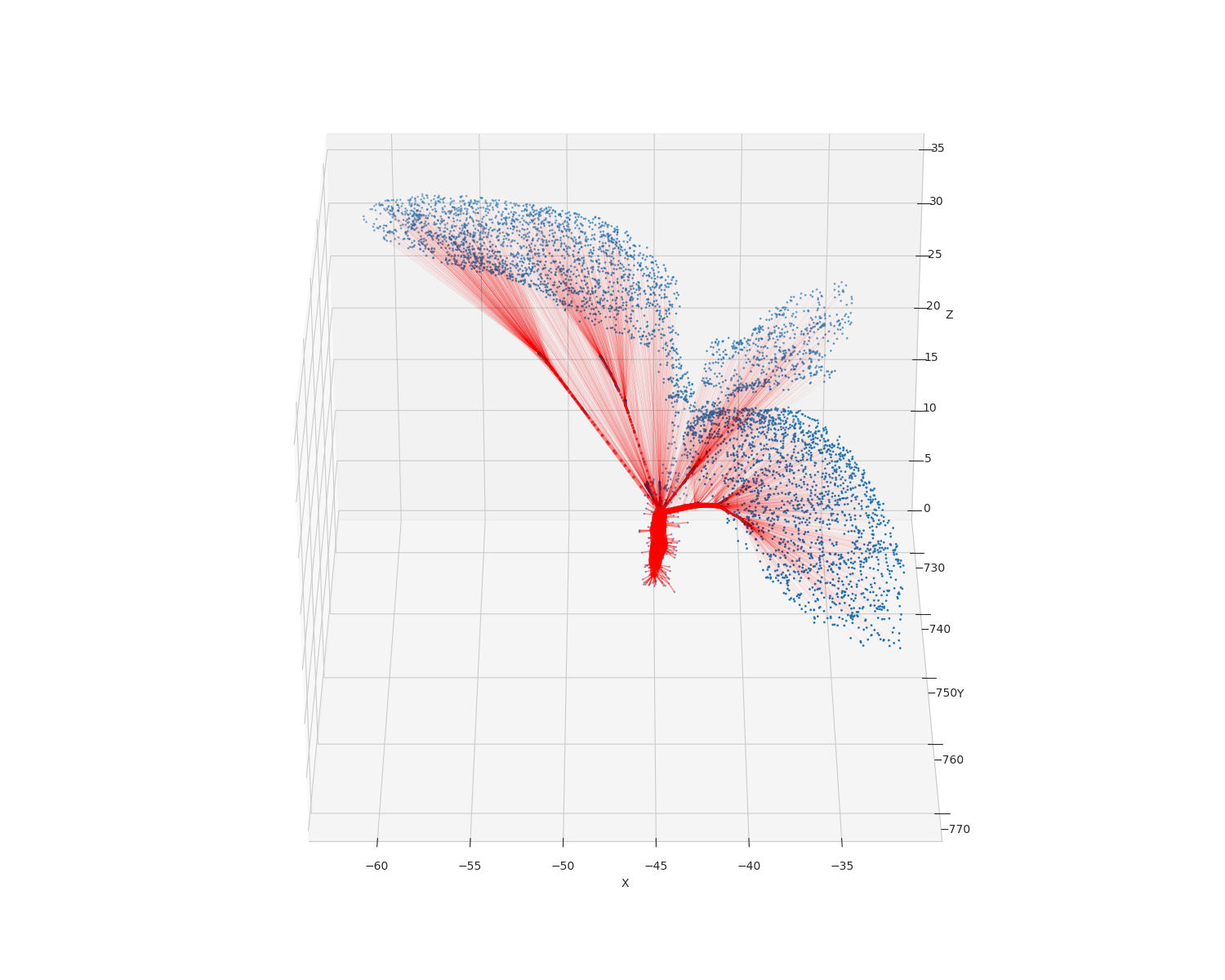}
		\vspace{\bottomvspace cm}
		\caption*{\BCST $\alpha=1.00$}
		\label{sfig4:plant_skeleton}
	\end{subfigure}%
	\caption[\BCST 3D plant skeletonization]{\textbf{\BCST 3D plant skeletonization}. \BCST for different $\alpha$ values of a 3D point cloud with 5000 samples capturing the surface of a plant. With $\alpha=0.00$, the tree branches exhibit greater irregularity, while at $\alpha=1.00$, finer details are obscured. Intermediate $\alpha$ values offer a more faithful representation of the plant's structure.}
	\label{fig:plant_skeleton}
\end{figure}

\section{Correspondence Between the BCST and CST Topologies}\label{sec:relation-between-bcst-and-cst}

Both the \CST and the \BCST problems have to be optimized over the set of feasible spanning tree topologies. This optimization is combinatorial in nature and turns out to be NP-hard in both cases. For the \CST case, Cayley's formula tells us that the number of feasible topologies is equal to $N^{N-2}$ \citep{cayley_theorem_1878}, which grows super-exponentially with the number of nodes $N$. For the \BCST case, w.l.o.g., we can represent any feasible topology as a full tree topology, i.e.~as a tree with $N-2$ Steiner points, each of degree $3$, and with leaves corresponding to the $N$ terminals. This representation is justified by the fact that any other feasible topology with \BPs of degree higher than~$3$ can be represented by a full tree topology by collapsing two or more \BPs, that is, when two or more \BPs have the same coordinates. \figurename{} \ref{fig:realizable_topos_fulltree} illustrates how a single full tree topology can realize different feasible topologies of the \BCST problem. For $N$ terminals, the number of possible full topologies is equal to $(2N-5)!!=(2N-5)\cdot(2N-7)\cdot\cdots3\cdot1$ \citep{schroder_vier_1870}, which also scales super-exponentially, but at a lower rate than the number of topologies of the \CST. Consequently, an exhaustive search through all trees is not feasible.

\tikzset{
	dot/.style 2 args={fill, circle, inner sep=0pt, label={#1:\scriptsize #2}},	
	fulldot/.style 2 args={circle,draw,minimum size=0.3cm,inner sep=0pt, label={#1:\scriptsize #2}},
	main node/.style={circle,draw,minimum size=0.3cm,inner sep=0pt]},
	mini node/.style={circle,draw,minimum size=0.3cm,inner sep=0pt]}
}

\def \scale{1 }
\def \scalemini{0.4*\scale}
\def \dispx{2.75}
\def \dispy{2}
\def \startarrow{0.58}
\def \endarrow{0.7}
\def \escalabox{0.6}

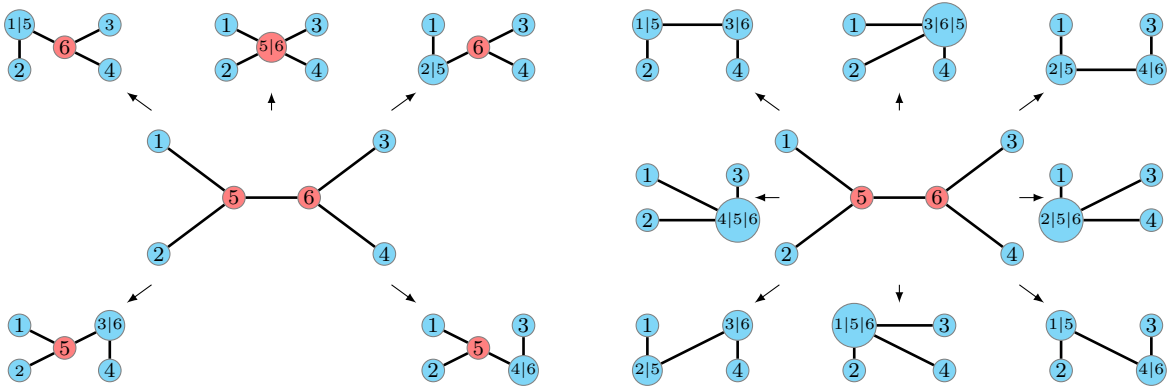
\begin{figure}
	\begin{subfigure}{0.5\textwidth}
		\centering
		\begin{tikzpicture}[]
			
			\node[main node,opacity=.5,black,fill=cyan,text opacity=1] (1) at (\scale*-1.5,\scale*0.75) {\scriptsize $1$};
			\node[main node,opacity=.5,black,fill=cyan,text opacity=1] (2) at (\scale*-1.5,\scale*-0.75) {\scriptsize $2$};
			\node[main node,opacity=.5,black,fill=cyan,text opacity=1] (3) at (\scale*1.5,\scale*0.75) {\scriptsize $3$};
			\node[main node,opacity=.5,black,fill=cyan,text opacity=1] (4) at (\scale*1.5,-\scale*0.75) {\scriptsize $4$};
			
			\node[main node,opacity=.5,black,fill=red,text opacity=1] (5) at (\scale*-0.5,\scale*0) {\scriptsize $5$};
			\node[main node,opacity=.5,black,fill=red,text opacity=1] (6) at (\scale*0.5,\scale*0) {\scriptsize $6$};			
			
			\path[-,draw,line width=1pt]

			(1) edge node[below] {} (5)
			
			(2) edge node[below] {} (5)
			
			(5) edge node[below] {} (6)
			
			(3) edge node[below] {} (6)
			
			(4) edge node[below] {} (6);
			

			\def \TOPLEFTx{-\dispx}
			\def \TOPLEFTy{\dispy}
			
			\draw[-latex](\scale*\startarrow*\TOPLEFTx,\scale*\startarrow*\TOPLEFTy) -- (\scale*\endarrow*\TOPLEFTx,\scale*\endarrow*\TOPLEFTy);
			\node[main node,opacity=.5,black,fill=cyan,text opacity=1] (1TOPLEFT) at (\scalemini*-1.5+\scale*\TOPLEFTx,\scalemini*0.75+\scale*\TOPLEFTy) {\tiny 1$|$5};
			\node[main node,opacity=.5,black,fill=cyan,text opacity=1] (2TOPLEFT) at (\scalemini*-1.5+\scale*\TOPLEFTx,\scalemini*-0.75+\scale*\TOPLEFTy) {\scriptsize 2};
			\node[main node,opacity=.5,black,fill=cyan,text opacity=1] (3TOPLEFT) at (\scalemini*1.5+\scale*\TOPLEFTx,\scalemini*0.75+\scale*\TOPLEFTy) {\tiny 3};
			\node[main node,opacity=.5,black,fill=cyan,text opacity=1] (4TOPLEFT) at (\scalemini*1.5+\scale*\TOPLEFTx,-\scalemini*0.75+\scale*\TOPLEFTy) {\scriptsize $4$};		
			\node[main node,opacity=.5,black,fill=red,text opacity=1] (6TOPLEFT) at (\scalemini*0.+\scale*\TOPLEFTx,\scalemini*0+\scale*\TOPLEFTy) {\scriptsize $6$};		
			
			\path[-,draw,line width=1pt]
			
			(1TOPLEFT) edge node[below] {} (2TOPLEFT)
			
			(1TOPLEFT) edge node[below] {} (6TOPLEFT)
			
			(3TOPLEFT) edge node[below] {} (6TOPLEFT)
			
			(4TOPLEFT) edge node[below] {} (6TOPLEFT);
			

			\def \TOPRIGHTx{\dispx}
			\def \TOPRIGHTy{\dispy}
			
			\draw [-latex](\scale*\startarrow*\TOPRIGHTx,\scale*\startarrow*\TOPRIGHTy) -- (\scale*\endarrow*\TOPRIGHTx,\scale*\endarrow*\TOPRIGHTy);
			\node[main node,opacity=.5,black,fill=cyan,text opacity=1] (1TOPRIGHT) at (\scalemini*-1.5+\scale*\TOPRIGHTx,\scalemini*0.75+\scale*\TOPRIGHTy) {\scriptsize 1};
			\node[main node,opacity=.5,black,fill=cyan,text opacity=1] (2TOPRIGHT) at (\scalemini*-1.5+\scale*\TOPRIGHTx,\scalemini*-0.75+\scale*\TOPRIGHTy) {\tiny 2$|$5};
			\node[main node,opacity=.5,black,fill=cyan,text opacity=1] (3TOPRIGHT) at (\scalemini*1.5+\scale*\TOPRIGHTx,\scalemini*0.75+\scale*\TOPRIGHTy) {\scriptsize 3};
			\node[main node,opacity=.5,black,fill=cyan,text opacity=1] (4TOPRIGHT) at (\scalemini*1.5+\scale*\TOPRIGHTx,-\scalemini*0.75+\scale*\TOPRIGHTy) {\scriptsize 4};

			\node[main node,opacity=.5,black,fill=red,text opacity=1] (6TOPRIGHT) at (\scalemini*0.+\scale*\TOPRIGHTx,-\scalemini*0+\scale*\TOPRIGHTy) {\scriptsize $6$};		
			
			\path[-,draw,line width=1pt]
			
			(1TOPRIGHT) edge node[below] {} (2TOPRIGHT)
			
			(2TOPRIGHT) edge node[below] {} (6TOPRIGHT)
			
			(3TOPRIGHT) edge node[below] {} (6TOPRIGHT)
			
			(4TOPRIGHT) edge node[below] {} (6TOPRIGHT);
			
			
			\def \TOPCENTERx{0}
			\def \TOPCENTERy{\dispy}
			\draw [-latex](\scale*\startarrow*\TOPCENTERx,\scale*\startarrow*\TOPCENTERy) -- (\scale*\endarrow*\TOPCENTERx,\scale*\endarrow*\TOPCENTERy);
			\node[main node,opacity=.5,black,fill=cyan,text opacity=1] (1TOPCENTER) at (\scalemini*-1.5+\scale*\TOPCENTERx,\scalemini*0.75+\scale*\TOPCENTERy) {\scriptsize 1};
			\node[main node,opacity=.5,black,fill=cyan,text opacity=1] (2TOPCENTER) at (\scalemini*-1.5+\scale*\TOPCENTERx,\scalemini*-0.75+\scale*\TOPCENTERy) {\scriptsize 2};
			\node[main node,opacity=.5,black,fill=cyan,text opacity=1] (3TOPCENTER) at (\scalemini*1.5+\scale*\TOPCENTERx,\scalemini*0.75+\scale*\TOPCENTERy) {\scriptsize 3};
			\node[main node,opacity=.5,black,fill=cyan,text opacity=1] (4TOPCENTER) at (\scalemini*1.5+\scale*\TOPCENTERx,-\scalemini*0.75+\scale*\TOPCENTERy) {\scriptsize 4};	
			\node[main node,opacity=.5,black,fill=red,text opacity=1] (6TOPCENTER) at (\scalemini*0.+\scale*\TOPCENTERx,\scalemini*.0+\scale*\TOPCENTERy) {\tiny $5|6$};		
			
			\path[-,draw,line width=1pt]
			
			(1TOPCENTER) edge node[below] {} (6TOPCENTER)
			
			(2TOPCENTER) edge node[below] {} (6TOPCENTER)
			
			(3TOPCENTER) edge node[below] {} (6TOPCENTER)
			
			(4TOPCENTER) edge node[below] {} (6TOPCENTER);

			\def \BOTTOMLEFTx{-\dispx}
			\def \BOTTOMLEFTy{-\dispy}
			\draw [-latex](\scale*\startarrow*\BOTTOMLEFTx,\scale*\startarrow*\BOTTOMLEFTy) -- (\scale*\endarrow*\BOTTOMLEFTx,\scale*\endarrow*\BOTTOMLEFTy);
			\node[main node,opacity=.5,black,fill=cyan,text opacity=1] (1BOTTOMLEFT) at (\scalemini*-1.5+\scale*\BOTTOMLEFTx,\scalemini*0.75+\scale*\BOTTOMLEFTy) {\scriptsize 1};
			\node[main node,opacity=.5,black,fill=cyan,text opacity=1] (2BOTTOMLEFT) at (\scalemini*-1.5+\scale*\BOTTOMLEFTx,\scalemini*-0.75+\scale*\BOTTOMLEFTy) {\tiny 2};
			\node[main node,opacity=.5,black,fill=cyan,text opacity=1] (3BOTTOMLEFT) at (\scalemini*1.5+\scale*\BOTTOMLEFTx,\scalemini*0.75+\scale*\BOTTOMLEFTy) {\tiny 3$|$6};
			\node[main node,opacity=.5,black,fill=cyan,text opacity=1] (4BOTTOMLEFT) at (\scalemini*1.5+\scale*\BOTTOMLEFTx,-\scalemini*0.75+\scale*\BOTTOMLEFTy) {\scriptsize 4};	
			\node[main node,opacity=.5,black,fill=red,text opacity=1] (5BOTTOMLEFT) at (\scalemini*0.+\scale*\BOTTOMLEFTx,\scalemini*.0+\scale*\BOTTOMLEFTy) {\scriptsize $5$};		
			
			\path[-,draw,line width=1pt]
			
			(1BOTTOMLEFT) edge node[below] {} (5BOTTOMLEFT)
			
			(2BOTTOMLEFT) edge node[below] {} (5BOTTOMLEFT)
			
			(3BOTTOMLEFT) edge node[below] {} (5BOTTOMLEFT)
			
			(3BOTTOMLEFT) edge node[below] {} (4BOTTOMLEFT);

			
			\def \BOTTOMRIGHTx{\dispx}
			\def \BOTTOMRIGHTy{-\dispy}
			\draw [-latex](\scale*\startarrow*\BOTTOMRIGHTx,\scale*\startarrow*\BOTTOMRIGHTy) -- (\scale*\endarrow*\BOTTOMRIGHTx,\scale*\endarrow*\BOTTOMRIGHTy);
			\node[main node,opacity=.5,black,fill=cyan,text opacity=1] (1BOTTOMRIGHT) at (\scalemini*-1.5+\scale*\BOTTOMRIGHTx,\scalemini*0.75+\scale*\BOTTOMRIGHTy) {\scriptsize 1};
			\node[main node,opacity=.5,black,fill=cyan,text opacity=1] (2BOTTOMRIGHT) at (\scalemini*-1.5+\scale*\BOTTOMRIGHTx,\scalemini*-0.75+\scale*\BOTTOMRIGHTy) {\scriptsize 2};
			\node[main node,opacity=.5,black,fill=cyan,text opacity=1] (3BOTTOMRIGHT) at (\scalemini*1.5+\scale*\BOTTOMRIGHTx,\scalemini*0.75+\scale*\BOTTOMRIGHTy) {\scriptsize 3};
			\node[main node,opacity=.5,black,fill=cyan,text opacity=1] (4BOTTOMRIGHT) at (\scalemini*1.5+\scale*\BOTTOMRIGHTx,-\scalemini*0.75+\scale*\BOTTOMRIGHTy) {\tiny 4$|$6};	
			
			\node[main node,opacity=.5,black,fill=red,text opacity=1] (5BOTTOMRIGHT) at (\scalemini*0.+\scale*\BOTTOMRIGHTx,\scalemini*.0+\scale*\BOTTOMRIGHTy) {\scriptsize $5$};	
			\path[-,draw,line width=1pt]
			
			(1BOTTOMRIGHT) edge node[below] {} (5BOTTOMRIGHT)
			
			(2BOTTOMRIGHT) edge node[below] {} (5BOTTOMRIGHT)
			
			(3BOTTOMRIGHT) edge node[below] {} (4BOTTOMRIGHT)
			
			(4BOTTOMRIGHT) edge node[below] {} (5BOTTOMRIGHT);
	
		\end{tikzpicture}
	\end{subfigure}
	\begin{subfigure}{0.5\textwidth}
		\centering
		\begin{tikzpicture}[]
			
			\node[main node,opacity=.5,black,fill=cyan,text opacity=1] (1) at (\scale*-1.5,\scale*0.75) {\scriptsize $1$};
			\node[main node,opacity=.5,black,fill=cyan,text opacity=1] (2) at (\scale*-1.5,\scale*-0.75) {\scriptsize $2$};
			\node[main node,opacity=.5,black,fill=cyan,text opacity=1] (3) at (\scale*1.5,\scale*0.75) {\scriptsize $3$};
			\node[main node,opacity=.5,black,fill=cyan,text opacity=1] (4) at (\scale*1.5,-\scale*0.75) {\scriptsize $4$};
			
			\node[main node,opacity=.5,black,fill=red,text opacity=1] (5) at (\scale*-0.5,\scale*0) {\scriptsize $5$};
			\node[main node,opacity=.5,black,fill=red,text opacity=1] (6) at (\scale*0.5,\scale*0) {\scriptsize $6$};			
			
			\path[-,draw,line width=1pt]

			(1) edge node[below] {} (5)
			
			(2) edge node[below] {} (5)
			
			(5) edge node[below] {} (6)
			
			(3) edge node[below] {} (6)
			
			(4) edge node[below] {} (6);
			

			\def \TOPLEFTx{-\dispx}
			\def \TOPLEFTy{\dispy}
			
			\draw[-latex](\scale*\startarrow*\TOPLEFTx,\scale*\startarrow*\TOPLEFTy) -- (\scale*\endarrow*\TOPLEFTx,\scale*\endarrow*\TOPLEFTy);
			\node[main node,opacity=.5,black,fill=cyan,text opacity=1] (1TOPLEFT) at (\scalemini*-1.5+\scale*\TOPLEFTx,\scalemini*0.75+\scale*\TOPLEFTy) {\tiny 1$|$5};
			\node[main node,opacity=.5,black,fill=cyan,text opacity=1] (2TOPLEFT) at (\scalemini*-1.5+\scale*\TOPLEFTx,\scalemini*-0.75+\scale*\TOPLEFTy) {\scriptsize 2};
			\node[main node,opacity=.5,black,fill=cyan,text opacity=1] (3TOPLEFT) at (\scalemini*1.5+\scale*\TOPLEFTx,\scalemini*0.75+\scale*\TOPLEFTy) {\tiny 3$|$6};
			\node[main node,opacity=.5,black,fill=cyan,text opacity=1] (4TOPLEFT) at (\scalemini*1.5+\scale*\TOPLEFTx,-\scalemini*0.75+\scale*\TOPLEFTy) {\scriptsize $4$};		
			
			\path[-,draw,line width=1pt]
			
			(1TOPLEFT) edge node[below] {} (2TOPLEFT)
			
			(3TOPLEFT) edge node[below] {} (1TOPLEFT)
			
			(4TOPLEFT) edge node[below] {} (3TOPLEFT);
			

			\def \TOPRIGHTx{\dispx}
			\def \TOPRIGHTy{\dispy}
			
			\draw [-latex](\scale*\startarrow*\TOPRIGHTx,\scale*\startarrow*\TOPRIGHTy) -- (\scale*\endarrow*\TOPRIGHTx,\scale*\endarrow*\TOPRIGHTy);
			\node[main node,opacity=.5,black,fill=cyan,text opacity=1] (1TOPRIGHT) at (\scalemini*-1.5+\scale*\TOPRIGHTx,\scalemini*0.75+\scale*\TOPRIGHTy) {\scriptsize 1};
			\node[main node,opacity=.5,black,fill=cyan,text opacity=1] (2TOPRIGHT) at (\scalemini*-1.5+\scale*\TOPRIGHTx,\scalemini*-0.75+\scale*\TOPRIGHTy) {\tiny 2$|$5};
			\node[main node,opacity=.5,black,fill=cyan,text opacity=1] (3TOPRIGHT) at (\scalemini*1.5+\scale*\TOPRIGHTx,\scalemini*0.75+\scale*\TOPRIGHTy) {\scriptsize 3};
			\node[main node,opacity=.5,black,fill=cyan,text opacity=1] (4TOPRIGHT) at (\scalemini*1.5+\scale*\TOPRIGHTx,-\scalemini*0.75+\scale*\TOPRIGHTy) {\tiny 4$|$6};		
			
			\path[-,draw,line width=1pt]
			
			(1TOPRIGHT) edge node[below] {} (2TOPRIGHT)
			
			(3TOPRIGHT) edge node[below] {} (4TOPRIGHT)
			
			(2TOPRIGHT) edge node[below] {} (4TOPRIGHT);
			
			
			\def \TOPCENTERx{0}
			\def \TOPCENTERy{\dispy}
			\draw [-latex](\scale*\startarrow*\TOPCENTERx,\scale*\startarrow*\TOPCENTERy) -- (\scale*\endarrow*\TOPCENTERx,\scale*\endarrow*\TOPCENTERy);
			\node[main node,opacity=.5,black,fill=cyan,text opacity=1] (1TOPCENTER) at (\scalemini*-1.5+\scale*\TOPCENTERx,\scalemini*0.75+\scale*\TOPCENTERy) {\scriptsize 1};
			\node[main node,opacity=.5,black,fill=cyan,text opacity=1] (2TOPCENTER) at (\scalemini*-1.5+\scale*\TOPCENTERx,\scalemini*-0.75+\scale*\TOPCENTERy) {\scriptsize 2};
			\node[main node,opacity=.5,black,fill=cyan,text opacity=1] (3TOPCENTER) at (\scalemini*1.5+\scale*\TOPCENTERx,\scalemini*0.75+\scale*\TOPCENTERy) {\tiny 3$|$6$|$5};
			\node[main node,opacity=.5,black,fill=cyan,text opacity=1] (4TOPCENTER) at (\scalemini*1.5+\scale*\TOPCENTERx,-\scalemini*0.75+\scale*\TOPCENTERy) {\scriptsize 4};		
			
			\path[-,draw,line width=1pt]
			
			(1TOPCENTER) edge node[below] {} (3TOPCENTER)
			
			(3TOPCENTER) edge node[below] {} (2TOPCENTER)
			
			(4TOPCENTER) edge node[below] {} (3TOPCENTER);

			
			\def \BOTTOMCENTERx{0}
			\def \BOTTOMCENTERy{-\dispy}
			\draw [-latex](\scale*\startarrow*\BOTTOMCENTERx,\scale*\startarrow*\BOTTOMCENTERy) -- (\scale*\endarrow*\BOTTOMCENTERx,\scale*\endarrow*\BOTTOMCENTERy);
			\node[main node,opacity=.5,black,fill=cyan,text opacity=1] (1BOTTOMCENTER) at (\scalemini*-1.5+\scale*\BOTTOMCENTERx,\scalemini*0.75+\scale*\BOTTOMCENTERy) {\tiny 1$|$5$|$6};
			\node[main node,opacity=.5,black,fill=cyan,text opacity=1] (2BOTTOMCENTER) at (\scalemini*-1.5+\scale*\BOTTOMCENTERx,\scalemini*-0.75+\scale*\BOTTOMCENTERy) {\scriptsize 2};
			\node[main node,opacity=.5,black,fill=cyan,text opacity=1] (3BOTTOMCENTER) at (\scalemini*1.5+\scale*\BOTTOMCENTERx,\scalemini*0.75+\scale*\BOTTOMCENTERy) {\scriptsize 3};
			\node[main node,opacity=.5,black,fill=cyan,text opacity=1] (4BOTTOMCENTER) at (\scalemini*1.5+\scale*\BOTTOMCENTERx,-\scalemini*0.75+\scale*\BOTTOMCENTERy) {\scriptsize 4};		
			
			\path[-,draw,line width=1pt]
			
			(1BOTTOMCENTER) edge node[below] {} (2BOTTOMCENTER)
			
			(1BOTTOMCENTER) edge node[below] {} (3BOTTOMCENTER)
			
			(1BOTTOMCENTER) edge node[below] {} (4BOTTOMCENTER);

			
			
			\def \BOTTOMLEFTx{-\dispx}
			\def \BOTTOMLEFTy{-\dispy}
			\draw [-latex](\scale*\startarrow*\BOTTOMLEFTx,\scale*\startarrow*\BOTTOMLEFTy) -- (\scale*\endarrow*\BOTTOMLEFTx,\scale*\endarrow*\BOTTOMLEFTy);
			\node[main node,opacity=.5,black,fill=cyan,text opacity=1] (1BOTTOMLEFT) at (\scalemini*-1.5+\scale*\BOTTOMLEFTx,\scalemini*0.75+\scale*\BOTTOMLEFTy) {\scriptsize 1};
			\node[main node,opacity=.5,black,fill=cyan,text opacity=1] (2BOTTOMLEFT) at (\scalemini*-1.5+\scale*\BOTTOMLEFTx,\scalemini*-0.75+\scale*\BOTTOMLEFTy) {\tiny 2$|$5};
			\node[main node,opacity=.5,black,fill=cyan,text opacity=1] (3BOTTOMLEFT) at (\scalemini*1.5+\scale*\BOTTOMLEFTx,\scalemini*0.75+\scale*\BOTTOMLEFTy) {\tiny 3$|$6};
			\node[main node,opacity=.5,black,fill=cyan,text opacity=1] (4BOTTOMLEFT) at (\scalemini*1.5+\scale*\BOTTOMLEFTx,-\scalemini*0.75+\scale*\BOTTOMLEFTy) {\scriptsize 4};		
			
			\path[-,draw,line width=1pt]
			
			(1BOTTOMLEFT) edge node[below] {} (2BOTTOMLEFT)
			
			(3BOTTOMLEFT) edge node[below] {} (4BOTTOMLEFT)
			
			(3BOTTOMLEFT) edge node[below] {} (2BOTTOMLEFT);

			
			\def \BOTTOMRIGHTx{\dispx}
			\def \BOTTOMRIGHTy{-\dispy}
			\draw [-latex](\scale*\startarrow*\BOTTOMRIGHTx,\scale*\startarrow*\BOTTOMRIGHTy) -- (\scale*\endarrow*\BOTTOMRIGHTx,\scale*\endarrow*\BOTTOMRIGHTy);
			\node[main node,opacity=.5,black,fill=cyan,text opacity=1] (1BOTTOMRIGHT) at (\scalemini*-1.5+\scale*\BOTTOMRIGHTx,\scalemini*0.75+\scale*\BOTTOMRIGHTy) {\tiny 1$|$5};
			\node[main node,opacity=.5,black,fill=cyan,text opacity=1] (2BOTTOMRIGHT) at (\scalemini*-1.5+\scale*\BOTTOMRIGHTx,\scalemini*-0.75+\scale*\BOTTOMRIGHTy) {\scriptsize 2};
			\node[main node,opacity=.5,black,fill=cyan,text opacity=1] (3BOTTOMRIGHT) at (\scalemini*1.5+\scale*\BOTTOMRIGHTx,\scalemini*0.75+\scale*\BOTTOMRIGHTy) {\scriptsize 3};
			\node[main node,opacity=.5,black,fill=cyan,text opacity=1] (4BOTTOMRIGHT) at (\scalemini*1.5+\scale*\BOTTOMRIGHTx,-\scalemini*0.75+\scale*\BOTTOMRIGHTy) {\tiny 4$|$6};		
			
			\path[-,draw,line width=1pt]
			
			(1BOTTOMRIGHT) edge node[below] {} (2BOTTOMRIGHT)
			
			(3BOTTOMRIGHT) edge node[below] {} (4BOTTOMRIGHT)
			
			(1BOTTOMRIGHT) edge node[below] {} (4BOTTOMRIGHT);

			
			\def \RIGHTx{\dispx}
			\def \RIGHTy{0}
			\draw [-latex](\scale*\startarrow*\RIGHTx,\scale*\startarrow*\RIGHTy) -- (\scale*\endarrow*\RIGHTx,\scale*\endarrow*\RIGHTy);
			\node[main node,opacity=.5,black,fill=cyan,text opacity=1] (1RIGHT) at (\scalemini*-1.5+\scale*\RIGHTx,\scalemini*0.75+\scale*\RIGHTy) {\scriptsize 1};
			\node[main node,opacity=.5,black,fill=cyan,text opacity=1] (2RIGHT) at (\scalemini*-1.5+\scale*\RIGHTx,\scalemini*-0.75+\scale*\RIGHTy) {\tiny 2$|$5$|$6};
			\node[main node,opacity=.5,black,fill=cyan,text opacity=1] (3RIGHT) at (\scalemini*1.5+\scale*\RIGHTx,\scalemini*0.75+\scale*\RIGHTy) {\scriptsize 3};
			\node[main node,opacity=.5,black,fill=cyan,text opacity=1] (4RIGHT) at (\scalemini*1.5+\scale*\RIGHTx,-\scalemini*0.75+\scale*\RIGHTy) {\scriptsize 4};		
			
			\path[-,draw,line width=1pt]
			
			(1RIGHT) edge node[below] {} (2RIGHT)
			
			(3RIGHT) edge node[below] {} (2RIGHT)
			
			(4RIGHT) edge node[below] {} (2RIGHT);

			
			\def \LEFTx{-\dispx}
			\def \LEFTy{0*\dispy}
			\draw [-latex](\scale*\startarrow*\LEFTx,\scale*\startarrow*\LEFTy) -- (\scale*\endarrow*\LEFTx,\scale*\endarrow*\LEFTy);
			\node[main node,opacity=.5,black,fill=cyan,text opacity=1] (1LEFT) at (\scalemini*-1.5+\scale*\LEFTx,\scalemini*0.75+\scale*\LEFTy) {\scriptsize 1};
			\node[main node,opacity=.5,black,fill=cyan,text opacity=1] (2LEFT) at (\scalemini*-1.5+\scale*\LEFTx,\scalemini*-0.75+\scale*\LEFTy) {\scriptsize 2};
			\node[main node,opacity=.5,black,fill=cyan,text opacity=1] (3LEFT) at (\scalemini*1.5+\scale*\LEFTx,\scalemini*0.75+\scale*\LEFTy) {\scriptsize 3};
			\node[main node,opacity=.5,black,fill=cyan,text opacity=1] (4LEFT) at (\scalemini*1.5+\scale*\LEFTx,-\scalemini*0.75+\scale*\LEFTy) {\tiny 4$|$5$|$6};		
			
			\path[-,draw,line width=1pt]
			
			(1LEFT) edge node[below] {} (4LEFT)
			
			(2LEFT) edge node[below] {} (4LEFT)
			
			(3LEFT) edge node[below] {} (4LEFT);

		\end{tikzpicture}
	\end{subfigure}
	\caption[Realizable \BCST Solutions Topologies from a Full Tree Topology with 4 Terminals]{\textbf{Realizable \BCST Solutions Topologies from a Full Tree Topology with 4 Terminals}. Terminal nodes and \BPs are represented in blue and red, respectively. Different topologies emerge from a single full tree topology depending on how the \BPs collapse with other nodes.}
	\label{fig:realizable_topos_fulltree}
\end{figure}

\input{Figures/CST2BCST_compendium.tex}

The heuristic presented in Section \ref{sec:Gen_optimization} exploits the correspondence between the feasible topologies of the \BCST and the \CST problems. Given a full tree topology $\tree_{\BCST}$, we say that a topology $\tree_{\CST}$ of the \CST problem can be derived from $\tree_{\BCST}$ if: 1) we can collapse the \BPs of $\tree_{\BCST}$ with the terminals such that  the resulting topology is $\tree_{\CST}$, and 2) for any \BP $s$ that is collapsed with terminal $t$, then all \BPs along the path connecting $s$ to $t$ must also collapse with $t$. In other words, a \BP cannot overtake any other \BP in the collapse process. \figurename{} \ref{sfig:stepsCST-BCST} (from top to bottom) shows the steps to transform a topology $\tree_{\BCST}$ into a topology $\tree_{\CST}$ by iteratively collapsing the \BPs. Analogously, we can derive a topology $\tree_{\BCST}$ from $\tree_{\CST}$ by spawning SPs from the terminals in $\tree_{\CST}$, i.e.,~introducing SPs connected to the terminals. Since in a full tree topology SPs have degree 3 and terminals have degree 1, we add one \BP per each pair of nodes adjacent to a common terminal node, so that the SP is connected to the triple of nodes.

The correspondence mapping between $\tree_{\CST}$ and $\tree_{\BCST}$, as shown in Figures \ref{sfig1:topo_correspondence} and \ref{sfig2:topo_correspondence}, is not unique. Multiple $\tree_{\CST}$ can be derived from a single $\tree_{\BCST}$ and vice versa. At most, any $\tree_{\BCST}$ can generate $O(3^{N-2})$ $\tree_{\CST}$ topologies because each \BP locally has 3 nodes to merge with. In concrete, the number of $\tree_{\CST}$ topologies  derivable from a singular full tree topology is given by the determinant of a submatrix of the Laplacian matrix, denoted as $L_{\BPs,\BPs}$. This submatrix is obtained by selecting the rows and columns indexed by the \BPs. Hence, the number of derivable topologies is precisely $\det L_{\BPs,\BPs}$. This assertion is proven in \thref{th:num_derivable_topos_from_BCST} (refer to \appendixname{} \ref{subsec:app_numtopos_CST} for details). The proof leverages the bijective relationship between the derivable \CST topologies and the terminal separating spanning forests of a full tree topology. By combining this bijectivity with the fact that the minors of the Laplacian matrix provide the count of forests separating the non-indexed rows and columns, the desired result is established.

Similarly, the number of $\tree_{\BCST}$ topologies derived from $\tree_{\CST}$ is given by
\begin{equation}
	\prod_{v\ :\ d_v\geq 2} (2d_v-3)!!,
\end{equation}
where $d_v$ is the degree of terminal $v$ in the $\tree_{\CST}$ topology. Higher-degree terminals can generate more topologies as they can spawn more pairs of nodes. For more details on the cardinalities of derivable topologies, see Appendix \ref{sec:app_num_topos}.

\begin{figure}[h]
    \centering
    \begin{subfigure}{0.5\textwidth}
        \centering
        \includegraphics[width=\linewidth]{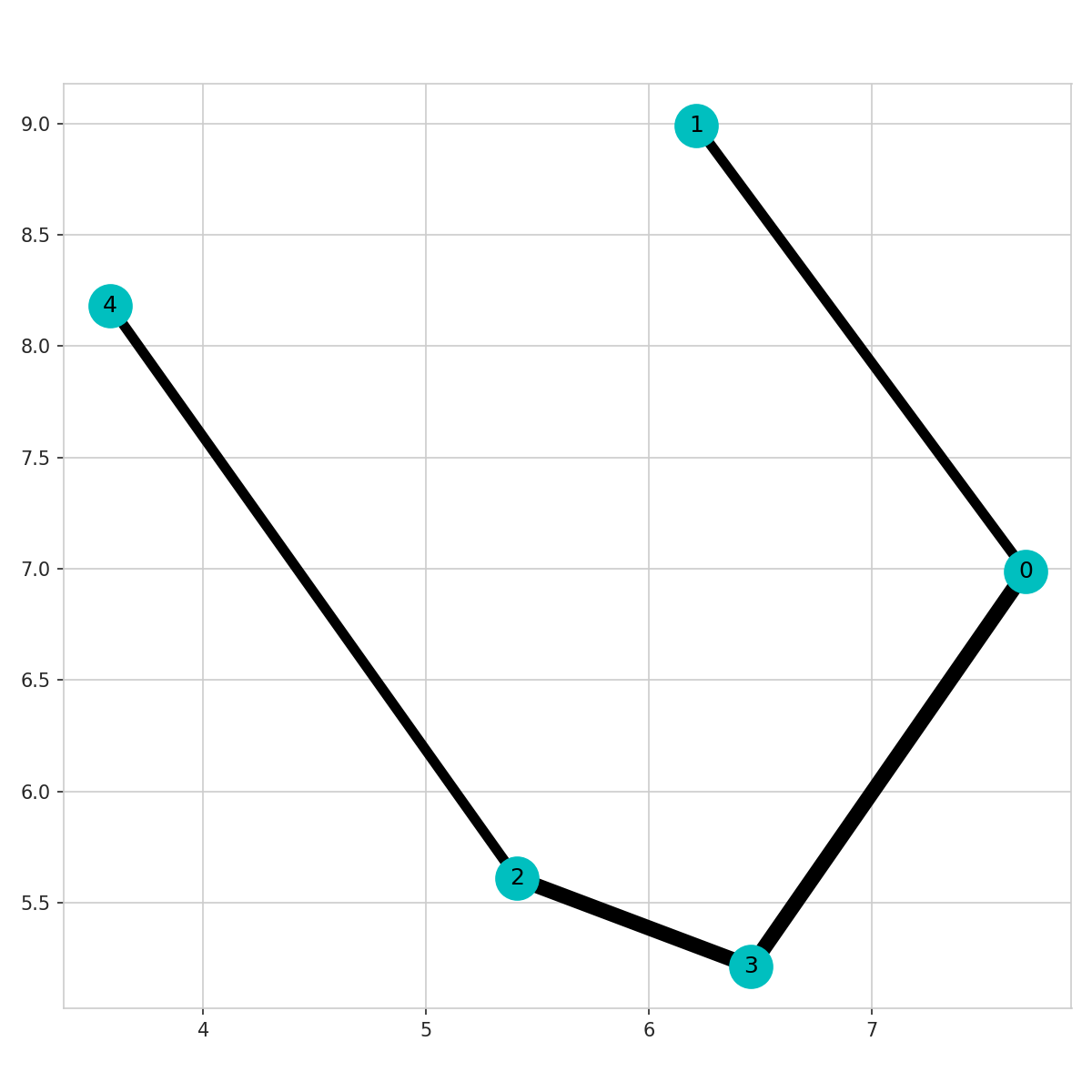}
        \caption{Optimal \CST solution}
        \label{sfig1:coutnerexapme_CST_BCST_topo_optimal}
    \end{subfigure}%
    \begin{subfigure}{0.5\textwidth}
        \centering
        \includegraphics[width=\linewidth]{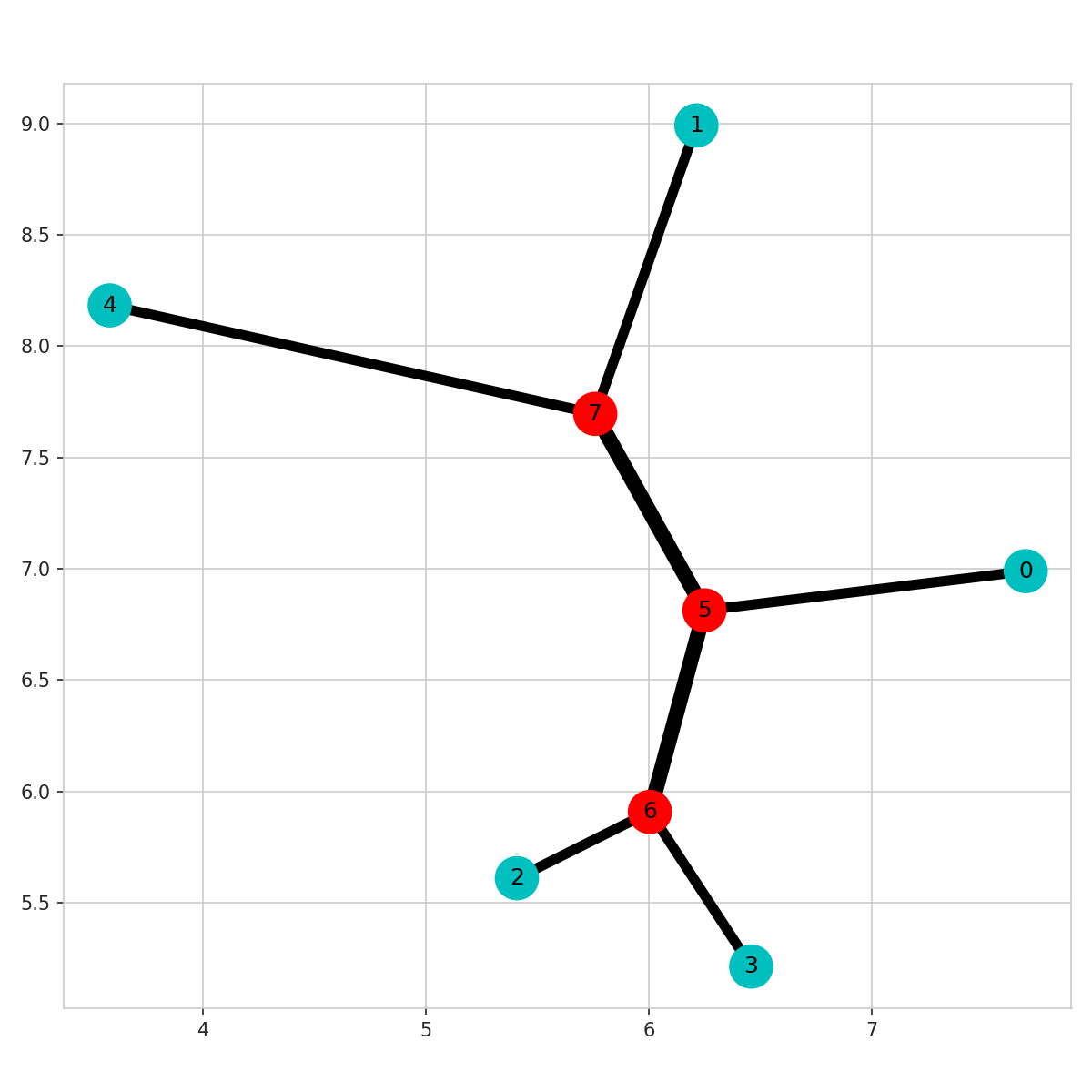}
        \caption{Optimal \BCST solution}
        \label{sfig2:coutnerexapme_CST_BCST_topo_optimal}
    \end{subfigure}
    \caption[Optimal \CST and \BCST Topologies May Not Be Derived from Each Other Given the Same Terminal Configuration]{\textbf{Optimal \CST and \BCST Topologies May Not Be Derived from Each Other Given the Same Terminal Configuration.} Left: Optimal \CST solution. Right: Optimal \BCST solution. The \CST topology cannot include nodes 4 and 1 as direct neighbors if derived from \BCST, as it would result in nodes adjacent to a common neighbor. Similarly, the optimal \BCST topology cannot be derived from the \CST topology, as nodes 1 and 4 would not be connected to a common \BP.}
        \label{fig:coutnerexapme_CST_BCST_topo_optimal}
\end{figure}

Despite the mapping ambiguity between the topologies, we can reduce the number of trees to explore in the \CST/\BCST given a \BCST/\CST topology. Although the optimum of one problem is not guaranteed to be derived from the optimum of the other (see \figurename{} \ref{fig:coutnerexapme_CST_BCST_topo_optimal}),
we show empirically that the heuristic proposed in Section \ref{sec:Gen_optimization} can  exploit the positions of the \BPs together with the correspondence between the sets of topologies of both problems to produce competitive results.

\section{Geometry of Optimal \BCST Topologies}\label{sec:geometry-of-optimal-bcst-topologies}
In this section, we will analyze the geometry of the optimal topologies of a \BCST problem. Concretely, we will determine an analytical formula of the angles that form the edges at a \BP. Using this relation, we will prove that when the terminal points lie in a plane, then \BPs with degree higher than $3$ are not realized in optimal solutions for $\alpha\in[0,0.5]\cup\{1\}$ unless they collapse with a terminal. For $\alpha\in \ ]0.5,1[$ we provide empirical evidence that the statement also holds in that case.

\subsection{Branching Angles at the Steiner Points}\label{sec:branching-angles-at-the-steiner-points}

In this section, we formulate the branching angles in terms of the centralities of the edges for a given topology of the \BCST problem. The derivation of the angles is based on previous works \citep{bernot_optimal_2008,lippmann_theory_2022}, which apply analogous arguments for the Branched Optimal Transport (BOT) problem. The main difference lies in the weighting factors that multiply the distances in the objective function \eqref{eq:BCST}, which in our case are the edge betweeness centralities, and in BOT are flows matching supply to demand. \appendixname{} \ref{subsec:app_relation_BOT_BCST} elaborates on the similarities and differences between the BOT and \BCST problems.

First and foremost, we emphasize the locality characteristic of the geometric optimization of \BPs of the \BCST problem. Because of the convexity of the \BCST objective \eqref{eq:BCST} given a fixed topology, it can be shown that the geometric optimization of the \BPs coordinates can be solved locally, meaning that the optimal position of a \BP is determined by its neighbors and weighting factors. \thref{lem:locality_BCST} formalizes this statement. For a proof, we refer to Lemma 2.1 of \cite{lippmann_theory_2022}, where the same statement was shown for the BOT problem. Since the proof is independent of the weighting factors multiplying the distances, the result applies to the \BCST problem as well.

\begin{lemma}\thlabel{lem:locality_BCST}
	Given a topology, its \BPs are in optimal position w.r.t.~the \BCST problem if and only if any individual \BP interconnects its neighbors at minimal cost. Moreover, the optimal topology of the \BCST is optimal if and only if for any subset of connected nodes the corresponding subtopology solves the respective subproblem.    
\end{lemma}
\begin{proof}
	See Lemma 2.1 of \cite{lippmann_theory_2022}.
\end{proof}

\tikzset{
	dot/.style 2 args={fill, circle, inner sep=0pt, label={#1:\scriptsize #2}},	
	fulldot/.style 2 args={circle,draw,minimum size=0.3cm,inner sep=0pt, label={#1:\scriptsize #2}},
	main node/.style={circle,draw,minimum size=0.4cm,inner sep=0pt]},
	mini node/.style={circle,draw,minimum size=0.3cm,inner sep=0pt]},
    invisible/.style={circle,minimum size=0.001cm,inner sep=0pt]},
}
\def \scale{2}

\begin{figure*}[h!]
	\centering

    \begin{tikzpicture}[]

        \node[invisible,text opacity=0] (i) at (\scale*-0.,\scale*1) {\small $1$};

        \node[main node,opacity=.5,black,fill=cyan,text opacity=1] (0) at (\scale*-
                      0.,\scale*-1) {\small $a_0$};
        \node[main node,opacity=.5,black,fill=cyan,text opacity=1] (1) at (\scale* -0.8,\scale*0.866) {\small $a_1$};
        \node[main node,opacity=.5,black,fill=cyan,text opacity=1] (2) at (\scale *1,\scale*0.75) {\small $a_2$};

        \node[main node,opacity=.5,black,fill=red,text opacity=1] (b) at (\scale*0,\scale*0) {\small b};
        
        \path[-,draw,line width=1pt]		
        
            (b) edge node[right] {$\centflow_0$} (0)
            
            (b) edge node[left] {$\centflow_1$} (1)
            
            (b) edge node[right] {$\centflow_2$} (2);

        \path[dashed,draw,line width=1pt]		
        
            (b) edge node[above] {} (i);

        \draw  pic["$\theta_1$", draw, -, angle eccentricity=1.2, angle radius=1.2cm]
        {angle=i--b--1};
        \draw  pic["$\theta_2$", draw, -, angle eccentricity=1.2, angle radius=1cm]
        {angle=2--b--i};


    \end{tikzpicture}
    \caption[Branching Angles at Steiner Point]{\textbf{Branching Angles at Steiner Point}. The symbols $\centflow_i$ represent the normalized centralities of the edges, that is $\centflow_i\coloneqq m_{ba_i}(1-m_{ba_i})$.}
    \label{fig:MAIN_angles_SP}
\end{figure*}
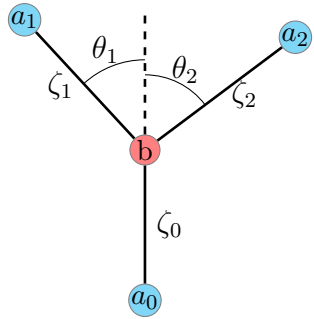
Recall that any feasible topology of the \BCST problem can be represented as a full tree topology where each \BP has degree 3. Thus, as a consequence of \thref{lem:locality_BCST}, it is enough to study the geometric optimization of 3 nodes connected by a single \BP. Consider the problem configuration depicted in \figurename{} \ref{fig:MAIN_angles_SP}, where node $b$ represents the branching point whose coordinates need to be optimized, nodes $\{a_i\}_i$ are the terminals with fixed positions and $\{\centflow_i\coloneqq m_{ba_i}(1-m_{ba_i})\}_i$ are the normalized centralities of the edges $\{(b,a_i)\}_i$. The objective to be minimized is
\begin{equation}\label{eq:MAIN_objective_3terminals}
C(b)=\centflow_0 ||b-a_0||+\centflow_1 ||b-a_1||+\centflow_2 ||b-a_2||
\end{equation}
\citeauthor{bernot_optimal_2008} showed that when the $b$ does not collapse with any terminal, then the angles $\theta_1$ and $\theta_2$ are given by
\begin{equation}\label{eq:MAIN_optimal_angles}
\begin{aligned}
\cos(\theta_1)&=&\frac{\centflow_0^{2\alpha}+\centflow_1^{2\alpha}-\centflow_2^{2\alpha}}{2\centflow_0^{\alpha}\cdot\centflow_1^{\alpha}}\\
\cos(\theta_2)&=&\frac{\centflow_0^{2\alpha}+\centflow_2^{2\alpha}-\centflow_1^{2\alpha}}{2\centflow_0^{\alpha}\cdot\centflow_2^{\alpha}}\\
\cos(\theta_1+\theta_2)&=&\frac{\centflow_0^{2\alpha}-\centflow_1^{2\alpha}-\centflow_2^{2\alpha}}{2\centflow_1^{\alpha}\cdot\centflow_2^{\alpha}}\\
\end{aligned}
\end{equation}

Alternatively, we can analyze when node $b$ collapses with one of the terminals. Assuming without loss of generality that $b$ collapses with $a_0$, let $\gamma\coloneqq\angle a_1a_0a_2$. According to \citeauthor{lippmann_theory_2022}, $b$ collapses with $a_0$ if
\begin{equation}
\label{eq:MAIN_branching angle}
\gamma\geq \arccos\left(\frac{\centflow_0^{2\alpha}-\centflow_1^{2\alpha}-\centflow_2^{2\alpha}}{2\centflow_1^{\alpha}\cdot\centflow_2^{\alpha}}\right)=\theta_1+\theta_2.
\end{equation}
Hence, $b$ collapses to $a_0$ if $\angle a_1a_0a_2$ exceeds the optimal angle specified by \eqref{eq:MAIN_optimal_angles}. This scenario results in the so-called $V$-branching.

For a comprehensive derivation of the angles, please refer to \appendixname{} \ref{sec:app_angles}, where we present the arguments from the works of \citet{bernot_optimal_2008} and \citet{lippmann_theory_2022}.

\subsection{Infeasibility of Degree-4 Steiner Points in the Plane}\label{sec:infeasibility-of-degree-4-steiner-points-in-the-plane}
In this section, we will prove the infeasibility of degree-4 \BPs in the optimal solution of the \BCST. Specifically, we will focus on the scenario where the terminal nodes lie in the plane and the value of $\alpha$ falls within the range $\alpha \in [0, 0.5]\cup\{1\}$. Moreover, we will provide compelling evidence to support the validity of the statement for the case where $\alpha \in \ ]0.5,1[$. We will divide the proof into two parts:  one for $\alpha\in[0,0.5]$, presented here, and the other for $\alpha=1$, which is detailed in \appendixname{} \ref{sec:infeasibility-of-degree-4-steiner-point-for-alpha1}.

\tikzset{
	dot/.style 2 args={fill, circle, inner sep=0pt, label={#1:\scriptsize #2}},	
	fulldot/.style 2 args={circle,draw,minimum size=0.3cm,inner sep=0pt, label={#1:\scriptsize #2}},
	main node/.style={circle,draw,minimum size=0.4cm,inner sep=0pt]},
	mini node/.style={circle,draw,minimum size=0.3cm,inner sep=0pt]},
    invisible/.style={circle,minimum size=0.001cm,inner sep=0pt]},
}
\def \scale{1.3}

\begin{figure*}[h!]
	\centering
    \begin{subfigure}{0.34\textwidth}
    	\centering

        \begin{tikzpicture}[]

            \node[main node,opacity=.5,black,fill=cyan,text opacity=1] (0) at (\scale*-
                          1.2,\scale*-0.8) {\small $a_3$};
            \node[main node,opacity=.5,black,fill=cyan,text opacity=1] (1) at (\scale* -1.2,\scale*0.8) {\small $a_1$};
            \node[main node,opacity=.5,black,fill=cyan,text opacity=1] (2) at (\scale *1.2,\scale*0.8) {\small $a_2$};
    
            \node[main node,opacity=.5,black,fill=cyan,text opacity=1] (3) at (\scale *1.2,-\scale*0.8) {\small $a_4$};
    
            \node[main node,opacity=.5,black,fill=red,text opacity=1] (b1) at (\scale*-0.4,\scale*0) {\small $b_1$};

            \node[main node,opacity=.5,black,fill=red,text opacity=1] (b2) at (\scale*0.4,\scale*0) {\small $b_2$};
            
            \path[-,draw,line width=1pt]		
            
                (b1) edge node[right] {} (0)
                
                (b1) edge node[left] {} (1)
                
                (b2) edge node[right] {} (2)
                
                (b2) edge node[right] {} (3)
                
                (b2) edge node[right] {} (b1);

            \draw  pic["\rotatebox{90}{\scriptsize$\psi_1=\theta_{1}+\theta_2$}", draw, -, angle eccentricity=1, angle radius=0.8cm,left]
            {angle=1--b1--0};
            \draw  pic["\rotatebox{90}{\scriptsize$\psi_3=\hat{\theta}_1+\hat{\theta}_2$}", draw, -, angle eccentricity=1, angle radius=0.8cm,right,]
            {angle=3--b2--2};

        \end{tikzpicture}
        \caption{Feasible Topology}
        \label{sfig1:degree4}
    \end{subfigure}\hfill
    \begin{subfigure}{0.33\textwidth}
        	\centering
        \begin{tikzpicture}[]

            \node[main node,opacity=.5,black,fill=cyan,text opacity=1] (0) at (\scale*-
                          1.2,\scale*-0.8) {\small $a_3$};
            \node[main node,opacity=.5,black,fill=cyan,text opacity=1] (1) at (\scale* -1.2,\scale*0.8) {\small $a_1$};
            \node[main node,opacity=.5,black,fill=cyan,text opacity=1] (2) at (\scale *1.2,\scale*0.8) {\small $a_2$};
    
            \node[main node,opacity=.5,black,fill=cyan,text opacity=1] (3) at (\scale *1.2,-\scale*0.8) {\small $a_4$};
    
            \node[main node,opacity=.5,black,fill=red,text opacity=1] (b1) at (\scale*0.,\scale*0.3) {\small $b_1$};

            \node[main node,opacity=.5,black,fill=red,text opacity=1] (b2) at (\scale*0.,\scale*-0.3) {\small $b_2$};
            
            \path[-,draw,line width=1pt]		
            
                (b1) edge node[right] {} (1)
                
                (b1) edge node[left] {} (2)
                
                (b2) edge node[right] {} (0)
                
                (b2) edge node[right] {} (3)
                
                (b2) edge node[right] {} (b1);

            \draw  pic["\scriptsize$\psi_2=\bar{\theta}_{1}+\bar{\theta}_2$", draw, -, angle eccentricity=0.9, angle radius=0.5cm,above]
            {angle=2--b1--1};
            \draw  pic["\scriptsize$\psi_4=\tilde{\theta}_1+\tilde{\theta}_2$", draw, -, angle eccentricity=0.8, angle radius=0.5cm,below]
            {angle=0--b2--3};

        \end{tikzpicture}
        \caption{Feasible Topology}
        \label{sfig2:degree4}
    \end{subfigure}%
    \begin{subfigure}{0.33\textwidth}
        \centering
        \begin{tikzpicture}[]

            \node[main node,opacity=.5,black,fill=cyan,text opacity=1] (0) at (\scale*-
                          1.2,\scale*-0.8) {\small $a_3$};
            \node[main node,opacity=.5,black,fill=cyan,text opacity=1] (1) at (\scale* -1.2,\scale*0.8) {\small $a_1$};
            \node[main node,opacity=.5,black,fill=cyan,text opacity=1] (2) at (\scale *1.2,\scale*0.8) {\small $a_2$};
    
            \node[main node,opacity=.5,black,fill=cyan,text opacity=1] (3) at (\scale *1.2,-\scale*0.8) {\small $a_4$};
    
            \node[main node,opacity=.5,black,fill=red,text opacity=1] (b) at (\scale*0,\scale*0) {\small b};
            
            \path[-,draw,line width=1pt]		
            
                (b) edge node[right] {} (0)
                
                (b) edge node[left] {} (1)
                
                (b) edge node[right] {} (2)
                
                (b) edge node[right] {} (3);

             \draw  pic["$\gamma_1$", draw, -, angle eccentricity=1.2, angle radius=1.2cm]
            {angle=1--b--0};
            \draw  pic["$\gamma_2$", draw, -, angle eccentricity=1.2, angle radius=1cm]
            {angle=2--b--1};
            \draw  pic["$\gamma_4$", draw, -, angle eccentricity=1.2, angle radius=1.2cm]
            {angle=3--b--2};
            \draw  pic["$\gamma_4$", draw, -, angle eccentricity=1.2, angle radius=1cm]
            {angle=0--b--3};

        \end{tikzpicture}
        \caption{Collapsed topology}
        \label{sfig3:degree4}

    \end{subfigure}
    \caption[Optimal Angles in Degree-4 $\BP$ Require $\psi\leq \gamma_i$]{\textbf{Optimal Angles in Degree-4 $\BP$ Require $\psi\leq \gamma_i$}. \figurename{s} \ref{sfig1:degree4}) and \ref{sfig2:degree4}) depict two topologies, where the optimal angles given by equation \eqref{eq:MAIN_branching angle} are represented by $\psi_i$. \figurename{} \ref{sfig3:degree4} illustrates the collapsed solution with the corresponding angles $\gamma_i$. An essential requirement for the optimality of \figurename{} \ref{sfig3:degree4} is that $\psi_i\leq \gamma_i$.}
    \label{fig:degree4}
\end{figure*}
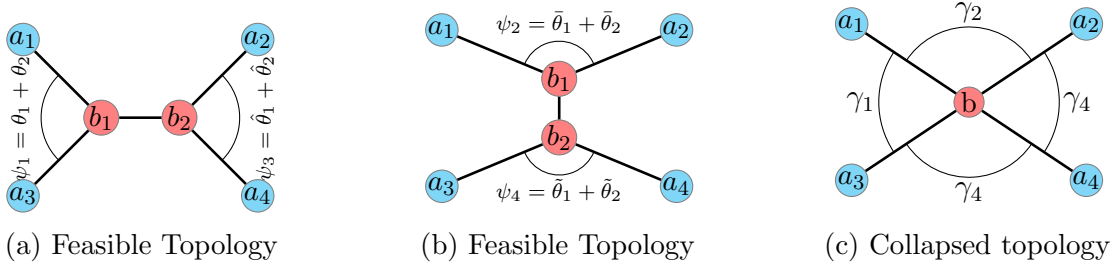

\begin{theorem}\thlabel{thm:infeasibility_4deg_alpha<0.5}
	Let $\alpha \in [0,0.5]$. Given a set of terminals which lie in the plane, then the \BPs of the optimal solution of the \BCST problem will not contain \BPs of degree-$4$ unless these collapse with a terminal.
\end{theorem}
\begin{proof}
The optimality of a solution in the \BCST problem relies on the locality characteristic, as stated in \thref{lem:locality_BCST}. Specifically, each subtopology within a connected subset must solve its respective problem for the overall solution to be optimal. Consequently, the realization of a degree-4 \BP, as depicted in \figurename{} \ref{fig:degree4}, requires the collapse of $b_2$ with $b_1$. Moreover, this collapse must occur in any topology. As we have discussed in Section \ref{sec:branching-angles-at-the-steiner-points}, both nodes $b_1$ and $b_2$ will collapse if a $V$-branching occurs, that is if the angle realized between the collapsed node and the two other nodes connected to it exceeds the optimal angle given by \eqref{eq:MAIN_branching angle}. Therefore, from \figurename{} \ref{fig:degree4} it follows that $\gamma_i \geq \psi$. We will demonstrate that the sum of $\displaystyle\sum_{i=1} ^4 \gamma_i$ is greater than $2\pi$, rendering a \BP of degree-4 infeasible. To do this we will prove that $\psi_i>\pi/2$ for all $i$.

W.l.o.g.~let us consider $i=1$ and denote $\psi_1$ as $\psi$. If $\cos(\psi)<0$, the angle $\psi\in[0,\pi]$ will be greater than $\pi/2$. Based on \eqref{eq:MAIN_branching angle}, we can derive the following:
\begin{equation}\label{eq:cospsi}
\begin{aligned}
\cos(\psi)&=&\frac{F\left(m_{a_3b_1}+m_{a_1b_1}\right)^{2\alpha}-F\left(m_{a_3b_1}\right)^{2\alpha}-F\left(m_{a_1b_1}\right)^{2\alpha}}{2F\left(m_{a_3b_1}\right)^{\alpha}F\left(m_{a_1b_1}\right)^{\alpha}}
\end{aligned}
\end{equation}
The function $F(x)^{2\alpha}=\left(x(1-x)\right)^{2\alpha}$ is strictly subadditive  in $\mathbb{R}^+$ for $\alpha \in [0,0.5]$,\footnote{In fact, all concave functions, $f$, with $f(0)\geq 0$ are subadditive on the positive domain. It is easy to see that $\left(x(1-x)\right)^{2\alpha}$ is concave for $\alpha \in [0,0.5]$.} that is $F(x+y)^{2\alpha}< F(x)^{2\alpha}+F(y)^{2\alpha}$ if $x,y>0$. Thus, the numerator of \eqref{eq:cospsi} is negative and therefore $\cos(\psi)<0$. Consequently, $\psi>\pi/2$. This argument applies to all $\psi_i$, hence their sum will be greater than $2\pi$.  Consequently, a \BP with degree-4 cannot be part of an optimal solution.
\end{proof}

In \cite{lippmann_theory_2022}, \citeauthor{lippmann_theory_2022} applied the same argument to establish the infeasibility of degree-4 \BPs in the context of the BOT problem for $\alpha\in[0,0.5]$. Our proof shows explicitly that this argument generalizes to other minimization problems of the same nature, where the edge-lengths are multiplied by weighting factors $F(m_{ij})$, with $F(\cdot)$ being a positive function dependent on $m_{ij}$ and exhibiting subadditivity when squared. For higher $\alpha>0.5$ the argument used in the previous theorem does not apply. Indeed, we can find values for the edge centralities for which the lower bounds of $\gamma_i$, given by \eqref{eq:MAIN_branching angle}, are all lower than $\pi/2$.\footnote{For instance for $\alpha=1$, $m_{a_3b_1}=m_{a_1b_1}=0.2$ and $m_{a_2b_2}=m_{a_4b_2}=0.3$ all $\psi_i$ angles are acute. Hence their sum is also lower than $2\pi$.} Now, we will take a more general approach that can rule out degree-4 branching for higher $\alpha$ values. In concrete, we will show the infeasibility for $\alpha=1$.

\tikzset{
	dot/.style 2 args={fill, circle, inner sep=0pt, label={#1:\scriptsize #2}},	
	fulldot/.style 2 args={circle,draw,minimum size=0.3cm,inner sep=0pt, label={#1:\scriptsize #2}},
	main node/.style={circle,draw,minimum size=0.4cm,inner sep=0pt]},
	mini node/.style={circle,draw,minimum size=0.3cm,inner sep=0pt]},
    invisible/.style={circle,minimum size=0.001cm,inner sep=0pt]},
}
\def \scale{1.8}

\begin{figure*}[h!]
	\centering
    \begin{subfigure}{0.5\textwidth}
    	\centering

        \begin{tikzpicture}[]
            \node[main node,opacity=.5,black,fill=cyan,text opacity=1] (0) at (\scale*-
                          1.2,\scale*-0.8) {\small $a_3$};
            \node[main node,opacity=.5,black,fill=cyan,text opacity=1] (1) at (\scale* -1.2,\scale*0.8) {\small $a_1$};
            \node[main node,opacity=.5,black,fill=cyan,text opacity=1] (2) at (\scale *1.2,\scale*0.8) {\small $a_2$};
    
            \node[main node,opacity=.5,black,fill=cyan,text opacity=1] (3) at (\scale *1.2,-\scale*0.8) {\small $a_4$};
    
            \node[main node,opacity=.5,black,fill=red,text opacity=1] (b) at (\scale*0,\scale*0) {\small $b$};

            \node[invisible] (i1) at (\scale*-1.2,\scale*0) {};
            \node[invisible] (i2) at (\scale*1.2,\scale*0) {};
            
            \path[-,draw,line width=1pt]		
            
                (b) edge node[right] {} (0)
                
                (b) edge node[left] {} (1)
                
                (b) edge node[right] {} (2)
                
                (b) edge node[right] {} (3);

            \path[dashed,draw,line width=1pt]		
                (b) edge node[right] {} (i1)
                
                (b) edge node[right] {} (i2);
            \draw  pic["$\gamma$", draw, -, angle eccentricity=1.2, angle radius=1cm]
            {angle=2--b--1};
            \draw  pic["$\theta_1$", draw, -, angle eccentricity=1, angle radius=1.2cm,left]
            {angle=1--b--i1};
            \draw  pic["$\theta_2$", draw, -, angle eccentricity=1, angle radius=1.2cm,right,]
            {angle=i2--b--2};
        \end{tikzpicture}
        \caption{}
        \label{sfig1:MAIN_degree4_alph1_approach}
    \end{subfigure}\hfill
    \begin{subfigure}{0.5\textwidth}
        \centering
        \begin{tikzpicture}[]

            \node[main node,opacity=.5,black,fill=cyan,text opacity=1] (0) at (\scale*-
                          1.2,\scale*-0.8) {\small $a_3$};
            \node[main node,opacity=.5,black,fill=cyan,text opacity=1] (1) at (\scale* -1.2,\scale*0.8) {\small $a_1$};
            \node[main node,opacity=.5,black,fill=cyan,text opacity=1] (2) at (\scale *1.2,\scale*0.8) {\small $a_2$};
    
            \node[main node,opacity=.5,black,fill=cyan,text opacity=1] (3) at (\scale *1.2,-\scale*0.8) {\small $a_4$};
    
            \node[main node,opacity=.5,black,fill=red,text opacity=1] (b1) at (\scale*-0.4,\scale*0) {\small $b_1$};

            \node[main node,opacity=.5,black,fill=red,text opacity=1] (b2) at (\scale*0.4,\scale*0) {\small $b_2$};

            \node[invisible] (i1) at (\scale*-1.2,\scale*0) {};
            \node[invisible] (i2) at (\scale*1.2,\scale*0) {};

            \node[invisible] (ia0) at (\scale*-1.2+-\scale*0.3,\scale*-0.8) {};
            \node[invisible] (ia1) at (\scale*-1.2+-\scale*0.3,\scale*0.8) {};
            \node[invisible] (ia2) at (\scale*1.2+\scale*0.3,\scale*0.8) {};
            \node[invisible] (ia3) at (\scale*1.2+\scale*0.3,-\scale*0.8) {};
            
            \path[-,draw,line width=1pt]		
            
                (b1) edge node[right] {} (0)
                
                (b1) edge node[left] {} (1)
                
                (b2) edge node[right] {} (2)
                
                (b2) edge node[right] {} (3)
                
                (b2) edge node[above ] {$\leftarrow \cdot \rightarrow$} (b1);
                
            \path[dashed,draw,line width=1pt]		
                (b1) edge node[right] {} (i1)
                
                (b2) edge node[right] {} (i2);

            \path[->,draw,line width=0.5pt]		
                (0) edge node[right] {} (ia0)
                
                (1) edge node[right] {} (ia1)
                
                (2) edge node[right] {} (ia2)
                
                (3) edge node[right] {} (ia3);

            \draw  pic["$\theta_1$", draw, -, angle eccentricity=1, angle radius=0.8cm,left]
            {angle=1--b1--i1};
            \draw  pic["$\theta_2$", draw, -, angle eccentricity=1, angle radius=0.8cm,right,]
            {angle=i2--b2--2};

        \end{tikzpicture}
        
        \caption{}
        \label{sfig2:MAIN_degree4_alph1_approach}
    \end{subfigure}
    \caption[Splitting Collapsed \BP while Preserving Optimal Angles]{\textbf{Splitting Collapsed \BP while Preserving Optimal Angles}. \ref{sfig1:MAIN_degree4_alph1_approach}) illustrates the collapsed solution of a 4-terminal configuration. (\ref{sfig2:MAIN_degree4_alph1_approach}) demonstrates that it is possible to move jointly the terminal points $\{a_1,a_3\}$ in a specific but opposite direction to the one of the terminals $\{a_2,a_4\}$, resulting in the splitting of the collapsed \BP $b$ into two distinct \BPs, $b_1$ and $b_2$. Remarkably, this split can be executed while preserving the angles $\theta_1$ and $\theta_2$. Importantly, these angles must correspond to the optimal angles given by \eqref{eq:MAIN_optimal_angles}.}
    \label{fig:MAIN_degree4_alph1_approach}
\end{figure*}

The optimal position of the \BPs is continuously dependent on the terminal positions and solely relies on the branching angles, as shown in Section \ref{sec:branching-angles-at-the-steiner-points}. Consequently, assuming that there exists a configuration such that the \BPs collapse, it is possible to find terminal positions that lead to an unstable collapse of the \BPs. Here, instability refers to a configuration where an infinitesimal translation of the terminals results in the splitting of the \BPs. This scenario is depicted in \figurename{} \ref{fig:MAIN_degree4_alph1_approach}. In such cases, the angles realized by the terminals and the \BPs will reach the upper bounds specified by \eqref{eq:MAIN_branching angle}. Therefore, the angles depicted in \figurename{} \ref{sfig1:MAIN_degree4_alph1_approach} fulfill the condition 
\begin{equation}\label{eq:MAIN_equality_angles_approach_alpha=1}
\gamma=\pi -\theta_1-\theta_2,
\end{equation}
where the angles satisfy
\begin{eqnarray}
\cos(\gamma)&=&\frac{F\left(m_{a_1b}+m_{a_2b}\right)^{2\alpha}-F\left(m_{a_1b}\right)^{2\alpha}-F\left(m_{a_2b}\right)^{2\alpha}}{2F\left(m_{a_1b}\right)^{\alpha}F\left(m_{a_2b}\right)^{\alpha}},\\
\cos(\theta_1)&=&\frac{F\left(m_{a_3b}+m_{a_1b}\right)^{2\alpha}+F\left(m_{a_1b}\right)^{2\alpha}-F\left(m_{a_3b}\right)^{2\alpha}}{2F\left(m_{a_3b}+m_{a_1b}\right)^{\alpha}F\left(m_{a_1b}\right)^{\alpha}},\\
\cos(\theta_2)&=&\frac{F\left(m_{a_2b}+m_{a_4b}\right)^{2\alpha}+F\left(m_{a_2b}\right)^{2\alpha}-F\left(m_{a_4b}\right)^{2\alpha}}{2F\left(m_{a_2b}+m_{a_4b}\right)^{\alpha}F\left(m_{a_2b}\right)^{\alpha}},
\end{eqnarray}
with $F(x)=x(1-x)$. By processing further equation \ref{eq:MAIN_equality_angles_approach_alpha=1}, we arrive at the following expression (see \appendixname{} \ref{sec:infeasibility-of-degree-4-steiner-point-for-alpha1} for further details)
\begin{equation}\label{eq:MAIN_equation2solve_alpha1}
\left(\cos(\gamma)+\cos(\theta_1)\cos(\theta_2)\right)^2-\left(1-\cos(\theta_1)^2\right)\left(1-\cos(\theta_2)^2\right)=0.
\end{equation}
Solving \eqref{eq:MAIN_equation2solve_alpha1} analitically for all $\alpha$ is difficult. Nonetheless, in \appendixname{} \ref{sec:infeasibility-of-degree-4-steiner-point-for-alpha1} we show analitically that for $\alpha=1$ \eqref{eq:MAIN_equality_angles_approach_alpha=1} can not hold given the constraints on the terms $m_{a_i,b}$, namely $\sum_{i=1}^4m_{a_i,b}=1$ and $0<m_{a_i,b}<1$ for all $i$. 

\begin{theorem}
	Let $\alpha=1$. Given a set of terminals which lie in the plane, then the \BPs of the optimal solution of the \BCST problem will not contain \BPs of degree-$4$ unless these collapse with a terminal.
\end{theorem}
\begin{proof}
	See \appendixname{} \ref{sec:infeasibility-of-degree-4-steiner-point-for-alpha1}
\end{proof}

\begin{figure}
	\centering
	\includegraphics[width=1\textwidth]{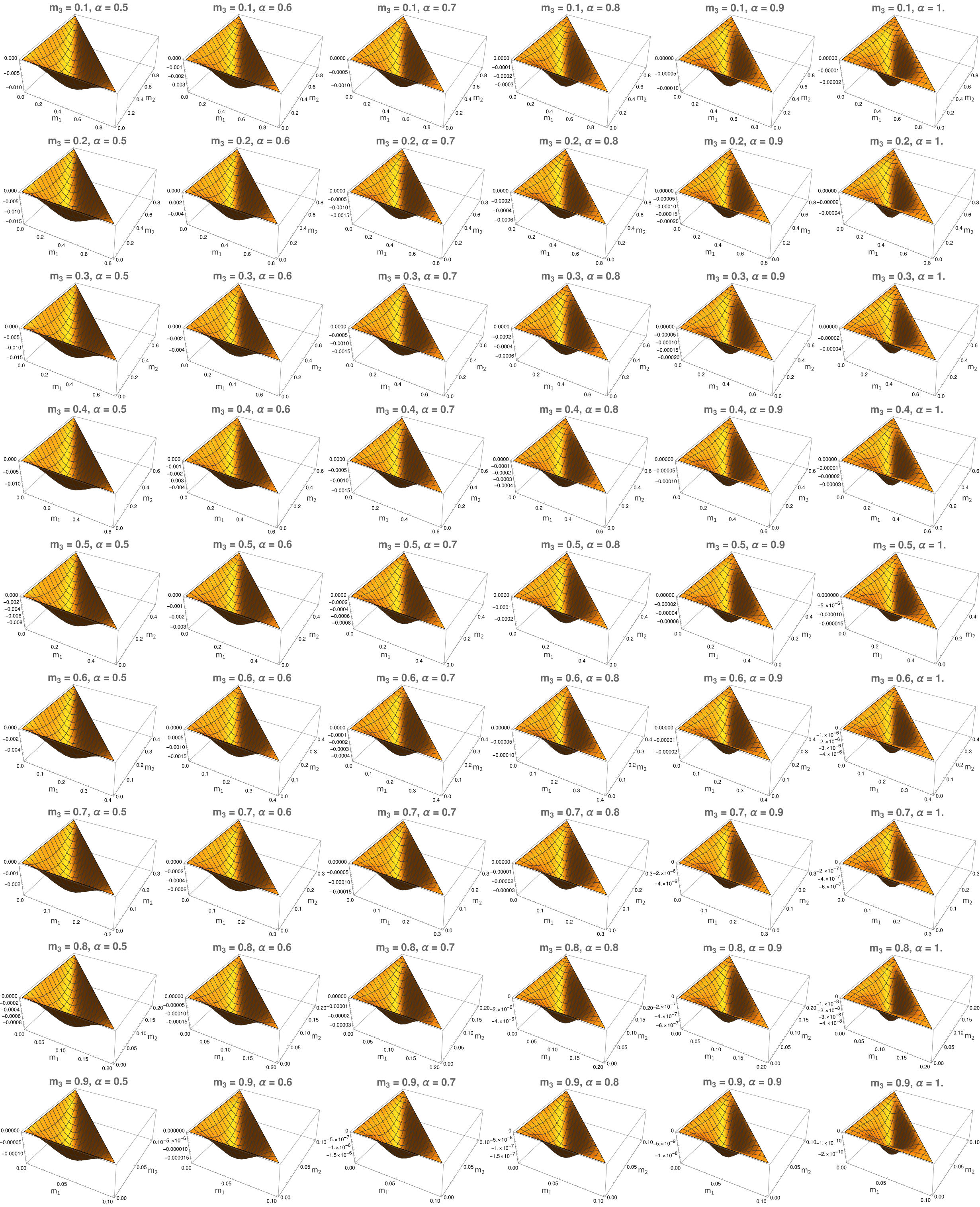}
	\def\AUXargs{]0.5,1[}
	\caption[\BPs of Degree-4 in the Plane are Likely not Feasible for $\alpha\in\AUXargs$]{\footnotesize \textbf{\BPs of Degree-4 in the Plane are Likely not Feasible for $\alpha\in\AUXargs$}. Surface plots are depicted, illustrating the left side of equation \eqref{eq:MAIN_equation2solve_alpha1}, as a function of $m_{1}$ and $m_{2}$, with different fixed values of $\alpha$ and $m_{3}$. From left to right: $m_3$ is fixed and $\alpha$ ranges over $\{0.5,0.6,0.7,0.8,0.9,1.0\}$. From top to bottom: $\alpha$ is fixed and $m_3$ ranges over $\{0.1,0.2,\dots,0.8,0.9\}$. For $m_3$ fixed, $m_1$ and $m_2$ range over the domain defined by $\{(x,y) \ : \ 0<x+y+m_3<1\}$. In all plots, the function values are negative and tend towards 0 as $m_1$, $m_2$, or $m_3$ approaches 0. We can observe that for fixed $m_1$, $m_2$ and $m_3$, the function seems to be increasing with respect to $\alpha$ (from right to left). Since we have previously demonstrated that the left side of equation \eqref{eq:MAIN_equation2solve_alpha1} does not equal zero in the desired domain for $\alpha=0.5$ and $\alpha=1$, the plots suggest that this is also the case for $\alpha \in ]0.5,1[$.}
	\label{fig:3dplot_inequality}
\end{figure}

Though we have not been able to prove analytically the infeasibility of degree-4 \BPs for $\alpha \in ]0.5,1[$, we strongly believe that the statement still holds. \figurename{} \ref{fig:3dplot_inequality} shows the surface plots of the numerator of the equality \ref{eq:MAIN_equation2solve_alpha1} (once expanded) w.r.t. $m_1$ and $m_2$ for different fixed values of $\alpha$ and $m_{3}$. Upon analysis, it appears that the numerator exhibits an increasing trend with respect to $\alpha$ within the interval $[0.5,1]$. This observation leads us to hypothesize that if the equality holds for $\alpha=0.5$ and $\alpha=1$, it is likely to hold for intermediate values as well. However, due to the complexity of the formula, it is challenging to verify this hypothesis analytically. 

In section \ref{sec:limitCST_alpha>1_n_infty}, we have demonstrated that as $\alpha>1$ and $N\to\infty$, the \BCST tends to converge to a star graph centered at the geometric median. Consequently, for $\alpha>1$, a degree-4 \BP becomes feasible. 

\floatpagestyle{empty} 

\begin{remark}
	In \cite{lippmann_theory_2022}, it was shown for the BOT problem that if degree-4 \BPs are not feasible then higher degree \BPs are not possible either. The same reasoning applies for the \BCST, since the proof does not depend on the weighting factors. Thus for $\alpha\in[0, 0.5]\cup\{1\}$, only degree-$3$ \BPs are feasible unless they do collapse with a terminal node. Due to the compelling evidence shown, we also believe this is the also the case for $\alpha\in ]0.5,1[$. \citet{lippmann_theory_2022} also showed that some of the results of the BOT problem obtained on the plane can be extended to other 2-dimensional manifolds. Again, this is also the case for the \BCST problem. Among these properties, we emphasize the optimal angles formulae exposed in section \ref{sec:branching-angles-at-the-steiner-points} and the infeasibility of degree-4 \BPs for appropriate $\alpha$ values. We refer to Appendix F of \cite{lippmann_theory_2022} for more details.
\end{remark}

\section{CST and BCST Optimization Algorithm}\label{sec:Gen_optimization}
This section details the proposed heuristic for optimizing the \BCST and \CST problems. We will first focus on the \BCST. The heuristic iterates over two steps: First, given a fixed topology, the algorithm finds the geometric positions of the Steiner points (\BPs) that exactly minimize the cost conditioned on the topology. Given the optimal coordinates of the \BPs, we then update the topology of the tree by computing an \mST over the terminals and SPs. This procedure is iterated until convergence or until some stopping criterion is met.

\subsection{Geometry Optimization}\label{sec:geometry-optimization}
The \BCST problem can be divided into two subproblems: combinatorial optimization of the tree topology and geometric optimization of the coordinates of the  \BPs, $X_B$. When conditioning on a topology $T$, the \BCST objective \eqref{eq:BCST} is a convex problem w.r.t.~$X_B$. Despite its convexity, the objective is not everywhere differentiable. We build on the iteratively reweighted least squares (IRLS) approach from \citet{smith_how_1992} and \citet{lippmann_theory_2022} to efficiently find the positions of the \BPs.

Starting from arbitrary \BPs coordinates, denoted as $X^{(0)}=\{x_i^{(0)}\}_{i=1}^{2N-2}$, the algorithm iteratively solves the following linear system of equations.
\begin{equation}\label{eq:MAINiteration_IRLS}
x_i^{(k+1)}=\frac{\displaystyle\sum_{j:(i,j)\in E}\centflow_{ij}^\alpha\frac{x_j^{(k+1)}}{||x_i^{(k)}-x_j^{(k)}||}}{\displaystyle\sum_{j:(i,j)\in E}\frac{\centflow_{ij}^\alpha}{||x_i^{(k)}-x_j^{(k)}||}}, \qquad \forall N+1\leq i\leq 2N-2.
\end{equation}
where $\centflow_{ij}=m_{ij}(1-m_{ij})$. We assume, without loss of generality, that the coordinates corresponding to the \BPs are indexed from $N+1$ to $2N-2$, where $N$ is the number of terminals. The coordinates for the terminals, which remain fixed throughout all iterations, are represented by the other indices. Thanks to the tree structure of the graph, the linear systems can be efficiently solved in linear time.

In Appendix \ref{sec:app_IRLS}, we show that the algorithm is agnostic to the weighting factors that multiply the distances, and can therefore be applied to compute any weighted geometric mean.

\subsection{Heuristic Optimizer for the (B)CST Problem}\label{sec:heuristic}
\begin{figure*}[t]
	\centering
	\includegraphics[width=\linewidth]{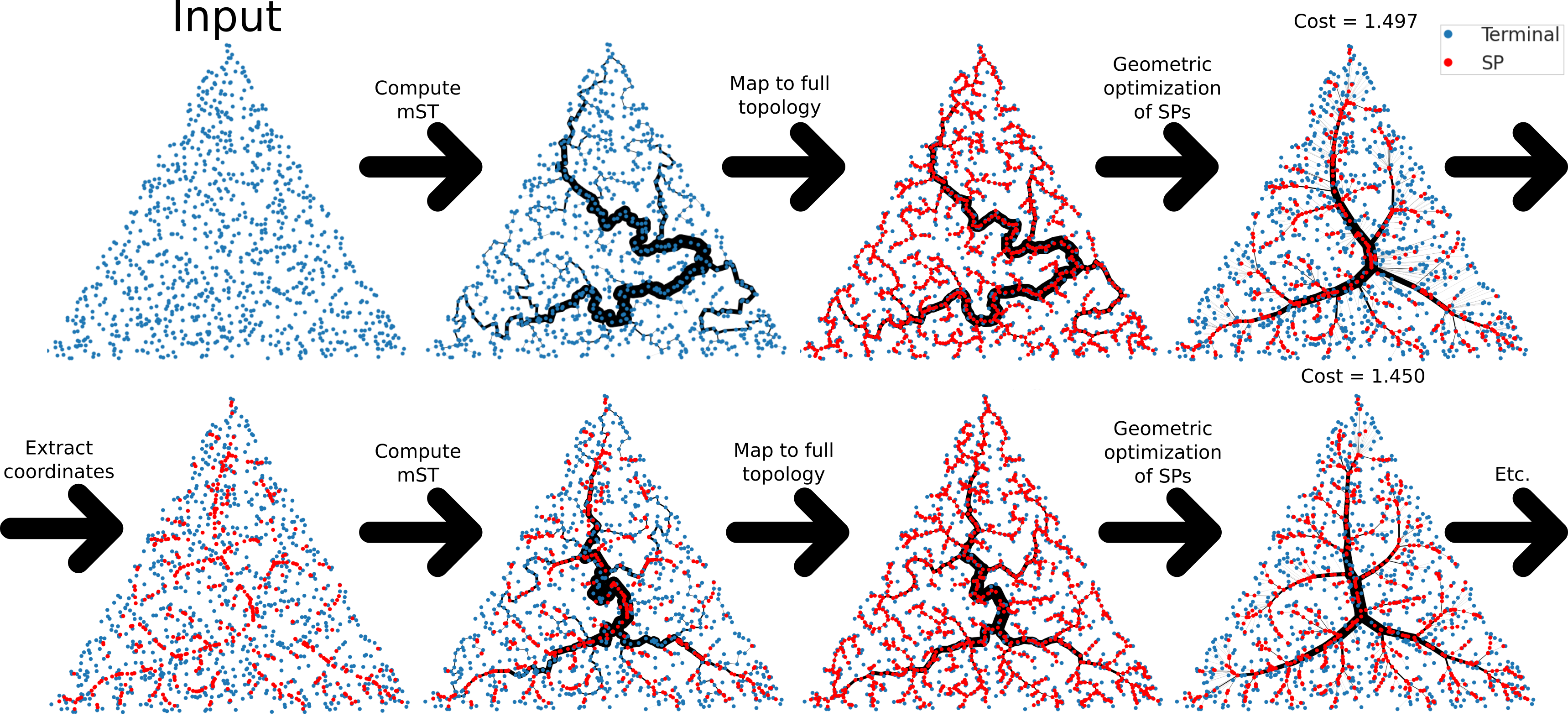}
	\caption[\mSTreg Heuristic]{\textbf{\mSTreg Heuristic}. The mSTreg heuristic iteratively transforms an mST to an approximate BCST. Given a set of points, the heuristic first computes the \mST over all points and transforms it into a full tree topology by adding Steiner points (\BPs). Next, the optimal positions of the \BPs are computed using iteratively reweighted least squares. Given the updated SPs coordinates, the heuristic recomputes the \mST over the union of the terminals nodes and the previous \BPs. This \mST produces a new topology, which is again transformed into a full tree topology by adding \BPs whose coordinates are optimized. This process is repeated until some stopping criterion is satisfied.}
	\label{fig:mSTreg_scheme}
\end{figure*}

We now present a heuristic which alternates between the \BP geometric coordinate optimization (convex) and a topology update (combinatorial). The heuristic's main characteristic is how it exploits the location of the Steiner points given an initial topology guess. The heuristic's underlying assumption is that the optimum position of the Steiner points may suggest a more desirable topology.

Unless otherwise stated, the heuristic we propose starts from the \mST over all terminal nodes. At this point, the \mST does not contain any \BPs and is therefore not a full tree topology. Thus, we need to transform the \mST into a full tree topology. As mentioned in Section \ref{sec:relation-between-bcst-and-cst}, and highlighted in  \figurename{} \ref{fig:topo_correspondence}, this process is not unambiguous. In particular, for each terminal node $v$ with degree $d_v\geq 2$, we have to add $d_v-1$ \BPs. Consequently, there are $(2d_v-3)!!$ ways to connect these \BPs to the neighbors of $v$. Among all possible subtopologies connecting the \BPs with $v$ and its neighbors, we choose the one given by the dendrogram defined by the hierarchical single linkage clustering algorithm applied to $v$ and its neighbors. In practice, this choice tends to work relatively well since nearby terminals are also closer in the subtopology.

Once we have a full tree topology, we can apply the geometry optimization step to obtain the optimal coordinates of the \BPs. We assume that the optimal positions of the \BPs indicate which connections between nodes might be more desirable, since they may be biased to move closer to other nodes than the ones to which they are connected. Therefore, we propose to recompute a mST over the terminals together with the \BPs. This new mST defines a new topology that needs to be transformed into a full tree topology for the geometry optimization. Once we have a valid full tree topology, we recompute the optimal positions of the \BPs. This process is repeated iteratively until convergence or until some stopping criterion is met. We refer to this algorithm as the \mST regularization (\mSTreg) heuristic. The algorithm's steps are illustrated in \figurename{} \ref{fig:mSTreg_scheme}, and its pseudocode is provided in Algorithm \ref{alg:CST_mSTreg}. The algorithm's complexity is $\mathcal{O}(dn\log(n)^2)$. A detailed complexity analysis is available in \appendixname{} \ref{sec:complexity-mstreg-heuristic}. We remark that the \mSTreg heuristic is independent of the weighting factors that multiply the distances, thus it can also be used to approximate other problems as well, for instance a generalized version of the optimum communication tree with \BPs.

\IncMargin{1.6em}
\begin{algorithm}[h!]
	\caption{\mSTreg Heuristic}\label{alg:CST_mSTreg}
	\DontPrintSemicolon
	\Indm  
	\KwInput{$X$, num\_iterations, sampling\_frequency, optimize\_\CST}
	\KwOutput{Tree}
	\Indp  
	
	$mST_{init}=$minimum\_spanning\_tree($X$)\ \tcp*{Define initial topology as mST}
	
	$\tree_{\BCST}=$transform2fulltopo($mST_{init}$) \ \tcp*{transform topology to full tree topology}

	$\BP=$compute\_SP($\tree_{init}$)\ \tcp*{compute optimal $\BP$ coordinates}

	bestcost\_\BCST=$\infty$\ 
	\If{\upshape optimize\_\CST}
	{
		bestcost\_\CST=$\infty$\;
	}
	
	\BlankLine \BlankLine 
	\While{\upshape $it<$num\_iterations}{
		\If{\upshape sampling\_frequency$>$2}
		{
			\tcc{sample extra points from edges}
			$Y=$sample\_from\_edge($\tree_{\BCST}$,$X\cup \BP$,sampling\_frequency)\;
			$\BP=\BP\cup Y$
		}
		$mST_{X\cup \BP}=$minimum\_spanning\_tree($X\cup \BP$)\;
		\BlankLine
		$\tree_{\BCST}=$transform2fulltopo($mST_{X\cup \BP}$)\ \tcp*{transform $mST_{X\cup \BP}$ to full tree topology}
		\BlankLine 
		$\BP=$compute\_SP($\tree_{reg}$)\ \tcp*{compute optimal $\BP$ coordinates}
		\BlankLine  
		\If{\upshape cost($\tree_{\BCST})<$bestcost\_\BCST}
		{
			bestcost\_\BCST=cost($\tree_{\BCST})$\;
			$\tree_{\BCST best}=\tree_{\BCST}$\;
		}	
		
		\BlankLine 
		\If{\upshape optimize\_\CST}
		{
			$\tree_{\CST}=$remove\_SP($\tree_{\BCST}$)\ \tcp*{Derive \CST topology from \BCST topology}
			\BlankLine 

			\If{\upshape cost($\tree_{\CST})<$bestcost\_\CST}
			{
				bestcost\_\CST=cost($\tree_{\CST})$\;
				$\tree_{\CST best}=\tree_{\CST}$\;
			}	
		}	
	}
\end{algorithm}
 \DecMargin{1.6em}

Optionally, before the mST step is computed over the terminals and previous \BPs, we can add intermediate points along the edges of the output generated by the geometry optimization step. These additional points will allow the \mST to more reliably follow the edges of the geometry-optimized tree from previous step. Moreover, in case the initial topology was poor, these extra points may help to detect and correct edge crossings, which are known to be suboptimal. An illustration of the effect of these extra points can be found in Appendix \ref{sec:app_effect_freq_sampling}. 

The heuristic designed for the \BCST problem can also be applied to the \CST problem by transforming \BCST topologies at each iteration into \CST topologies. While this transformation isn't unique, we found that iteratively collapsing one \BP at a time with the neighbor that leads the smallest increase in cost produces compelling results. Additionally, when collapsing \BPs together, centering the new node at the weighted geometric median of its new neighbors improves results slightly. Further details can be found in Appendix \ref{sec:app_SP_removal_strategies}.

\section{Benchmark}\label{sec:benchmark}
\subsection{Brute Force Benchmark}
To assess the quality of the \mSTreg heuristic, we compare the cost of the trees computed by the \mSTreg algorithm with the globally optimal solutions obtained by brute force of configurations with up to nine terminals. We generate 200 instances with $N \in \{5,6,7,8,9\}$ terminals sampled from a unit square. Both \CST and \BCST problems are solved for each $\alpha\in\{0,0.1,\dots,0.9,1\}$. In Figure \ref{fig:bruteforce}, the relative error, calculated as $100 (c_{h}-c_{o})/c_{o}$ where $c_{h}$ and $c_{o}$ are heuristic and optimal costs, is shown for different $N$. We also show how the heuristic solution ranks, once the costs of all topologies are sorted. The heuristic attains the optimum in the majority of cases, with slightly better performance in the \BCST problem than in the \CST one. Appendix \ref{sec:app_toydata_experiments} provides $\alpha$-based results.

\def \bottomvspace{-.5}
\def \topvspace{-0.0}
\begin{figure}
	\centering
	\begin{subfigure}{0.25\linewidth}
		\centering
		\vspace*{\topvspace cm}
		\includegraphics[width=1\linewidth]{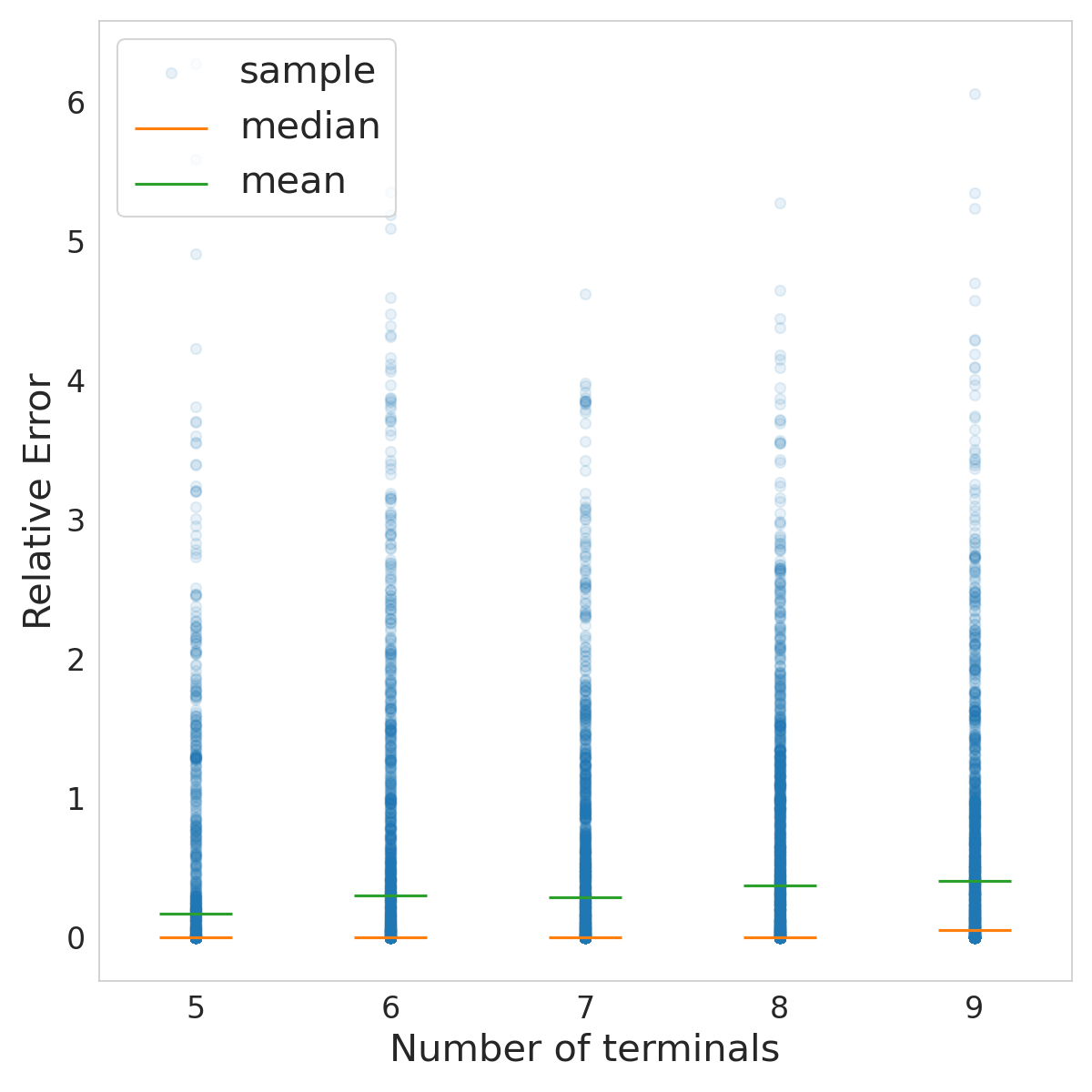}
		\vspace{\bottomvspace cm}
		\caption{\footnotesize\BCST relative error}
		\label{sfig1:bruteforce}
	\end{subfigure}%
	\begin{subfigure}{0.25\linewidth}
		\centering
		\vspace*{\topvspace cm}
		\includegraphics[width=1\linewidth]{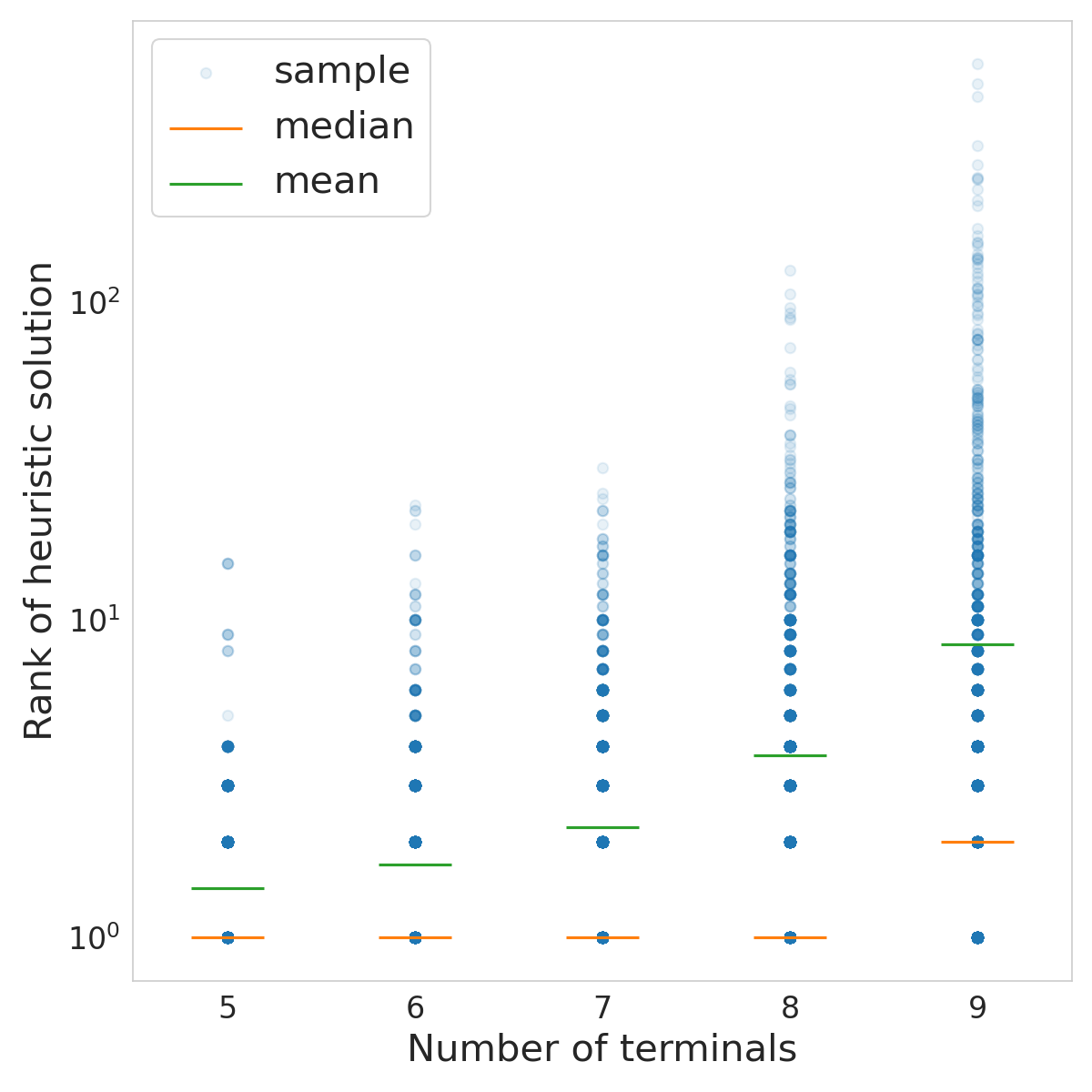}
		\vspace{\bottomvspace cm}
		\caption{\footnotesize\BCST rank of heuristic}
		\label{sfig2:bruteforce}
	\end{subfigure}%
	\begin{subfigure}{0.25\linewidth}
		\centering
		\vspace*{\topvspace cm}
		\includegraphics[width=1\linewidth]{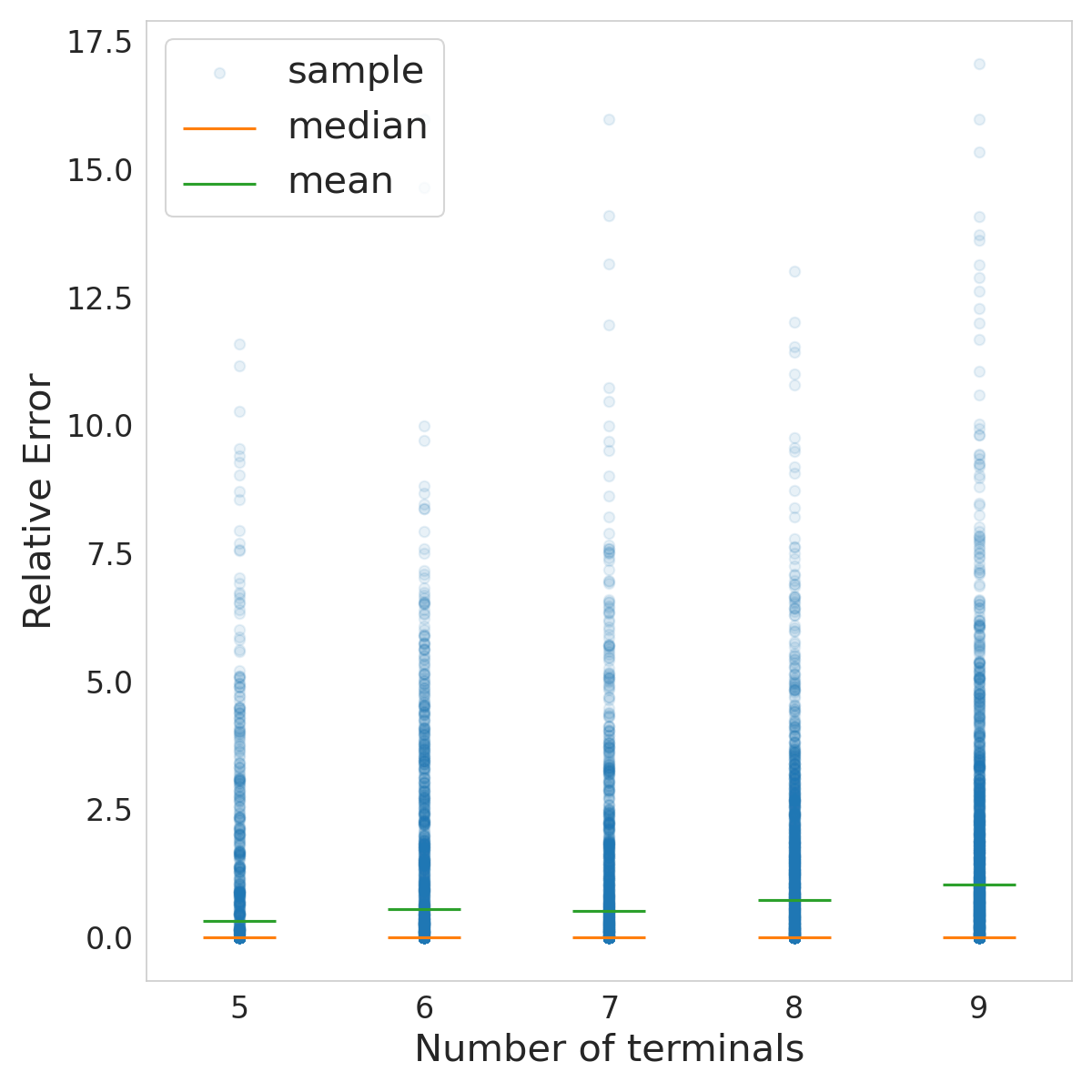}
		\vspace{\bottomvspace cm}
		\caption{\footnotesize\CST relative error}
		\label{sfig3:bruteforce}
	\end{subfigure}%
	\begin{subfigure}{0.25\linewidth}
		\centering
		\vspace*{\topvspace cm}
		\includegraphics[width=1\linewidth]{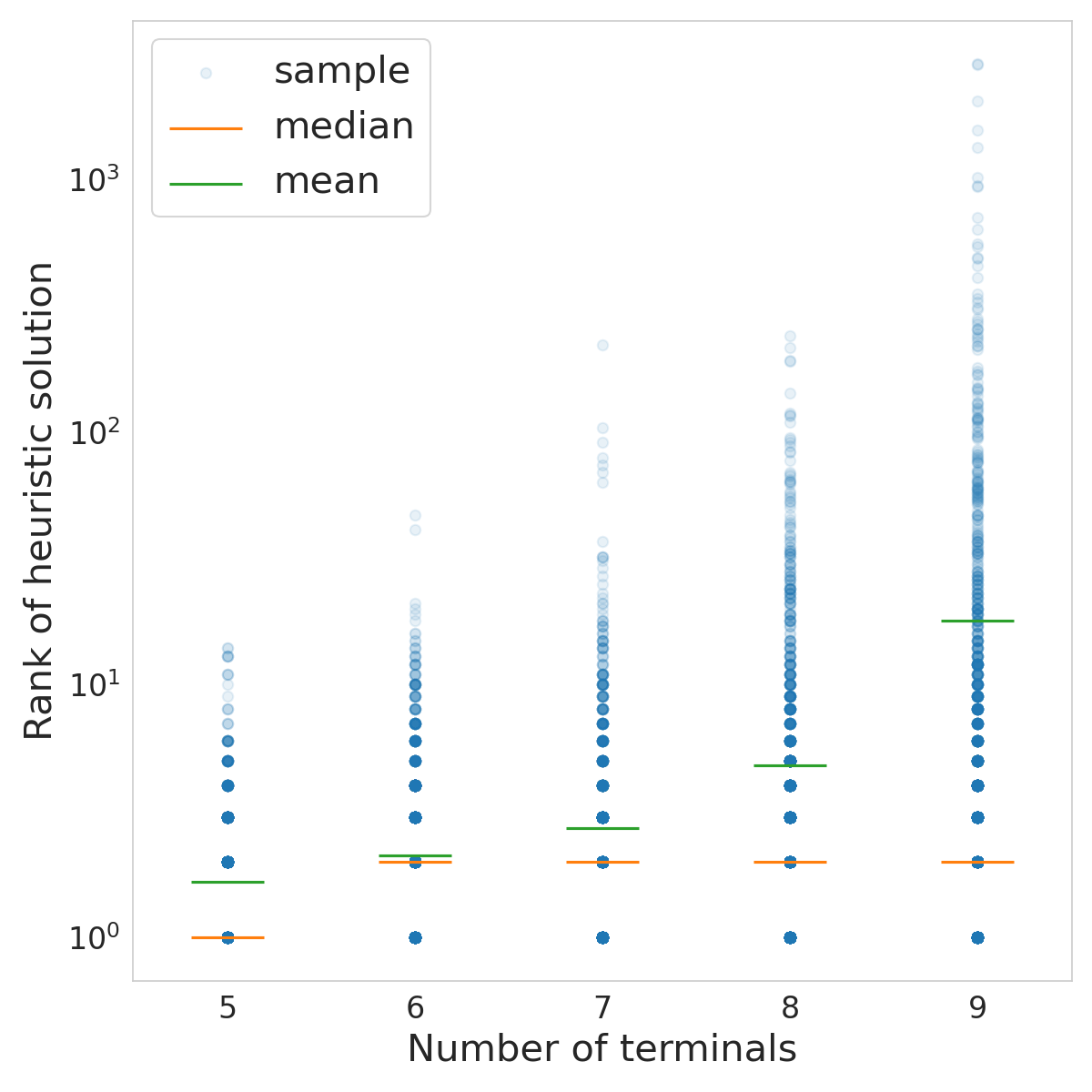}
		\vspace{\bottomvspace cm}
		\caption{\footnotesize\CST rank of heuristic}
		\label{sfig4:bruteforce}
	\end{subfigure}%
	\caption[Bruteforce \mSTreg Benchmark]{\textbf{Bruteforce \mSTreg Benchmark}. Relative cost errors between the \mSTreg heuristic and optimal solutions; and sorted position of the heuristic tree for different number of terminals, $N$. For each $N$, we uniformly sampled 200 different terminal configuration and solved them for different $\alpha$ values. Most runs ended up close to the global optimum, though the heuristic is slightly better for the \BCST problem.}
	\label{fig:bruteforce}
\end{figure}

\subsection{Steiner and MRCT Benchmark}
In addition, we evaluate \mSTreg on bigger datasets from the OR library\footnote{\url{http://people.brunel.ac.uk/~mastjjb/jeb/orlib/esteininfo.html}} \citep{beasley_heuristic_1992} for the Steiner tree ($\alpha$=0) and the \MRCT ($\alpha$=1) problems. This dataset includes exact solutions of Steiner problem instances of up to 100 nodes randomly distributed in a unit square. The used instances are labeled as e{$n$}.{$k$}, where $n$ denotes the number of terminals, and $k$ represents the instance number. \figurename{} \ref{sfig1:benchmark_gap} compares the cost of our heuristic with the optimal cost. We also provide for reference the costs of the \mST and the topology obtained by transforming the \mST into a full tree topology with its \BP coordinates optimized (referred to as mST\_fulltopo). Though our heuristic does not reach the optimal cost, it produces good topologies. The average relative error is lower than 1\%. 

For the \MRCT, we compare our heuristic with the Campos \citep{campos_fast_2008} and GRASP\_PR \citep{sattari_metaheuristic_2015} algorithms. Campos modifies Prim's algorithm with heuristic rules, while GRASP\_PR conducts local search by exchanging one edge at a time. We test the algorithms on the OR library datasets for problem instances with 50, 100 and 250 terminals. \figurename{} \ref{sfig2:benchmark_gap} shows the relative errors. In this case, we do not have access to the optimal cost, therefore we use GRASP\_PR costs cited from \citet{sattari_metaheuristic_2015} as reference. Campos costs are obtained from our own implementation. For reference, we also show the 2-approximation \citep{wong_worst-case_1980} given by the star graph centered at the data centroid. While \mSTreg proves competitive (surpassing Campos but falling short of GRASP\_PR by a modest average relative error of 1.16\%), it is worth noting that GRASP\_PR relies on a time-consuming local search. Leveraging the competitive solution provided by \mSTreg as an initial step can enhance the performance and convergence of local search based algorithms, such as GRASP\_PR.

\def \scalafig{1}
\begin{figure}[t]
	\centering
	\begin{subfigure}{0.5\linewidth}
		\includegraphics[width=\scalafig\linewidth]{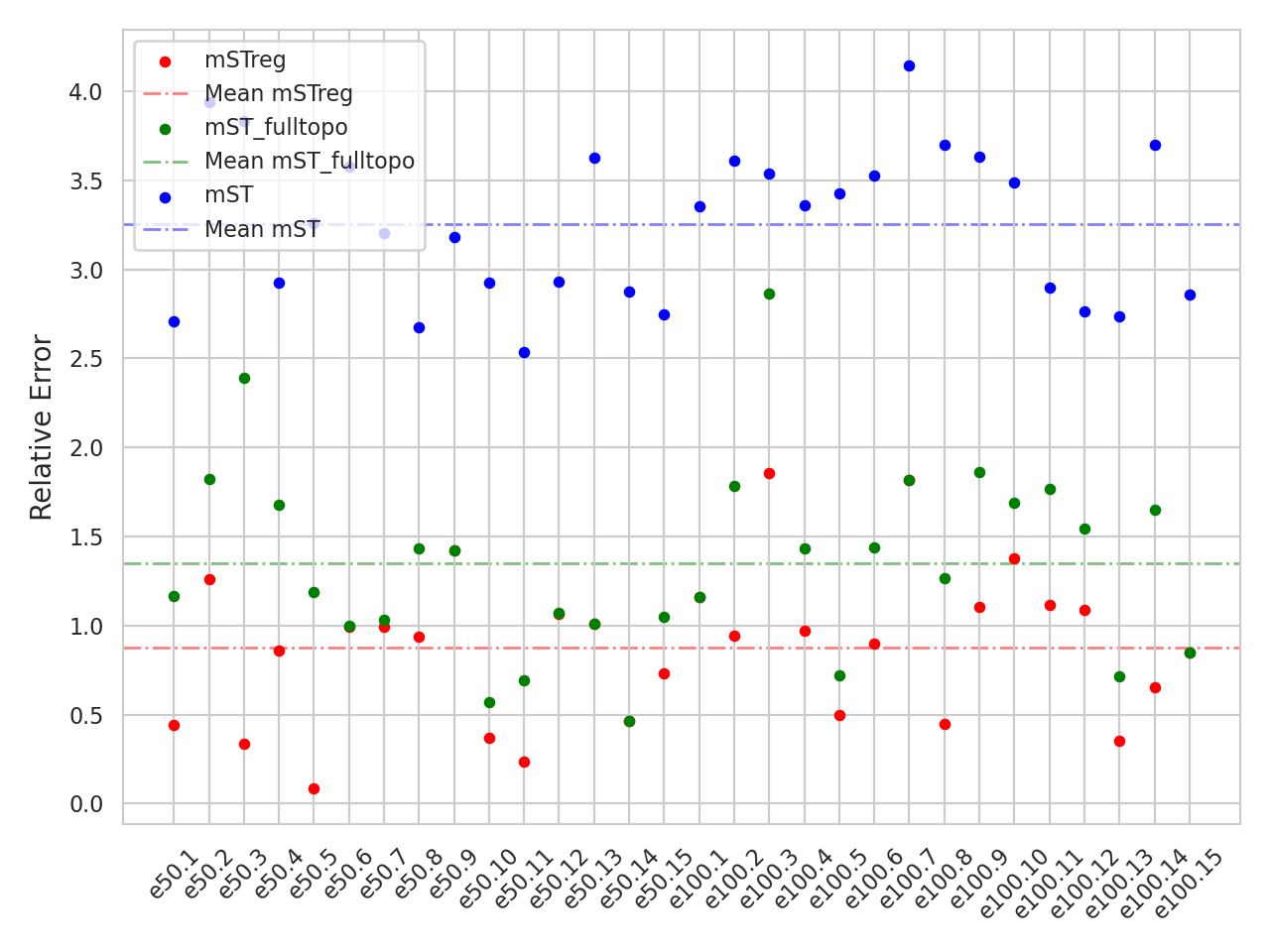}
		\caption{Steiner}
		\label{sfig1:benchmark_gap}
	\end{subfigure}%
	\begin{subfigure}{0.5\linewidth}
		\includegraphics[width=\scalafig\linewidth]{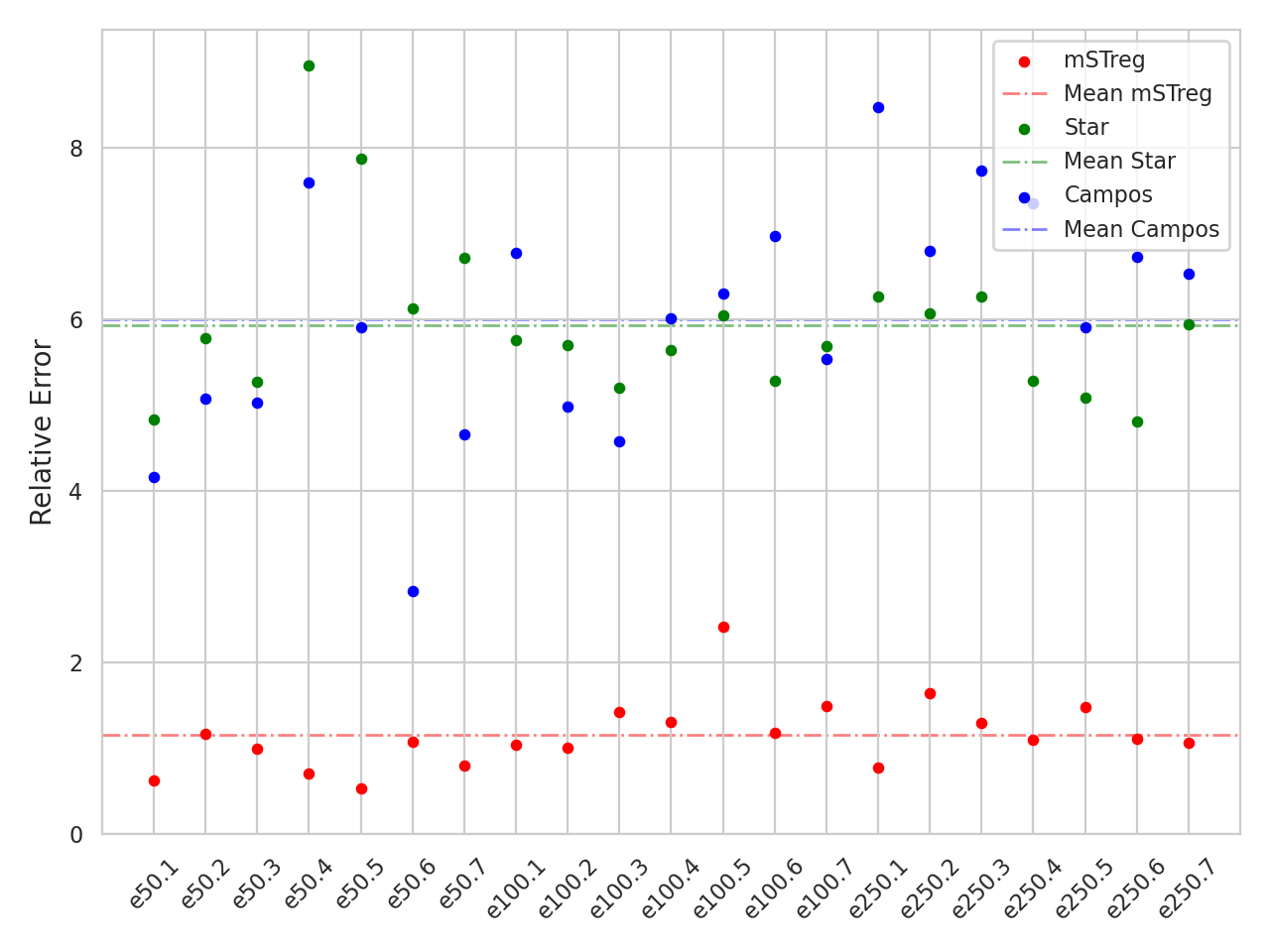}
		\caption{\MRCT}
		\label{sfig2:benchmark_gap}
	\end{subfigure}
	\caption[Steiner and MRCT OR Library Dataset Benchmark]{\textbf{Steiner and MRCT OR Library Dataset Benchmark}. Relative cost error with respect to a reference cost for the Steiner and \MRCT problems for different instances and different methods (lower is better). Left: For the Steiner problem, the reference cost is the optimal cost. \mSTreg finds good solutions and improves over mST\_fulltopo. Right: For MRCT the reference cost is given by the GRASP\_PR algorithm. The heuristic beats all other methods, but is still sightly worse than GRASP\_PR algorithm.}
	\label{fig:benchmark_gap}	
    
\end{figure}%

\subsection{Comparing GRASP\_PR and \mSTreg for the \CST Problem}
In the previous section, the GRASP\_PR algorithm by \cite{sattari_metaheuristic_2015} outperformed the \mSTreg heuristic when solving the \MRCT problem. However, it's important to note that the complexity of each iteration of the GRASP\_PR scales quadratically with the number of nodes for complete graphs. This is due to its local search algorithm, which involves swapping edges. In contrast, the complexity per iteration of the \mSTreg algorithm is $\mathcal{O}(dn\log(n)^2)$ (refer to \appendixname{} \ref{sec:complexity-mstreg-heuristic}), making it more efficient.

Additionally, GRASP\_PR requires an initial random guess to initiate the local search. The quality of this guess impacts the number of iterations needed for the local search to converge. We will show that initializing the local search with a solution generated by the \mSTreg allows for initializations that enhance the performance of GRASP\_PR. To ensure variability in the mSTreg output, we initialize it with random topologies instead of the \mST. Despite this modification, as the \mST initialization yielded favorable results, we opt to sample trees similar to it. To achieve this, we construct a tree by randomly sampling edges, prioritizing shorter ones. In concrete, we sample edge $(i,j)$ with probability proportional to $\exp(-\mu||x_i-x_j||)$ for some inverse temperature $\mu$. This sampling approach is akin to the one used in Karger's algorithm for approximating the minimum cut \citep{karger1993global,jenner2021extensions} and differs from the GRASP\_PR construction phase in that it doesn't require that one end of the edges belongs to the current tree.

While the previous section relied on the MRCT costs reported in \cite{sattari_metaheuristic_2015} for the GRASP\_PR  algorithm, we now validate our claims using our implementation of GRASP\_PR. Given that the GRASP\_PR is a versatile algorithm, we used it to compute the \CST with alternative $\alpha$ values besides 1.  Acknowledging that for lower $\alpha$ values, the optimum trees will be more similar to the \mST, we adapted the construction phase of GRASP\_PR. Specifically, for $\alpha<0.7$, it generates the random tree based on the Karger's sampling method mentioned above. For $\alpha\geq 0.7$, it uses the construction proposed by \cite{sattari_metaheuristic_2015}. Due to the relatively slow performance of our Python implementation of GRASP\_PR, we imposed a 5-minute time threshold. If exceeded, the algorithm returns the best solution at the end of the path relinking (PR) phase of the GRASP\_PR algorithm.

\begin{figure}
	\centering         
    \begin{subfigure}{0.33\linewidth}
    		\centering
    		\vspace*{\topvspace cm}
    		\includegraphics[width=1\linewidth]{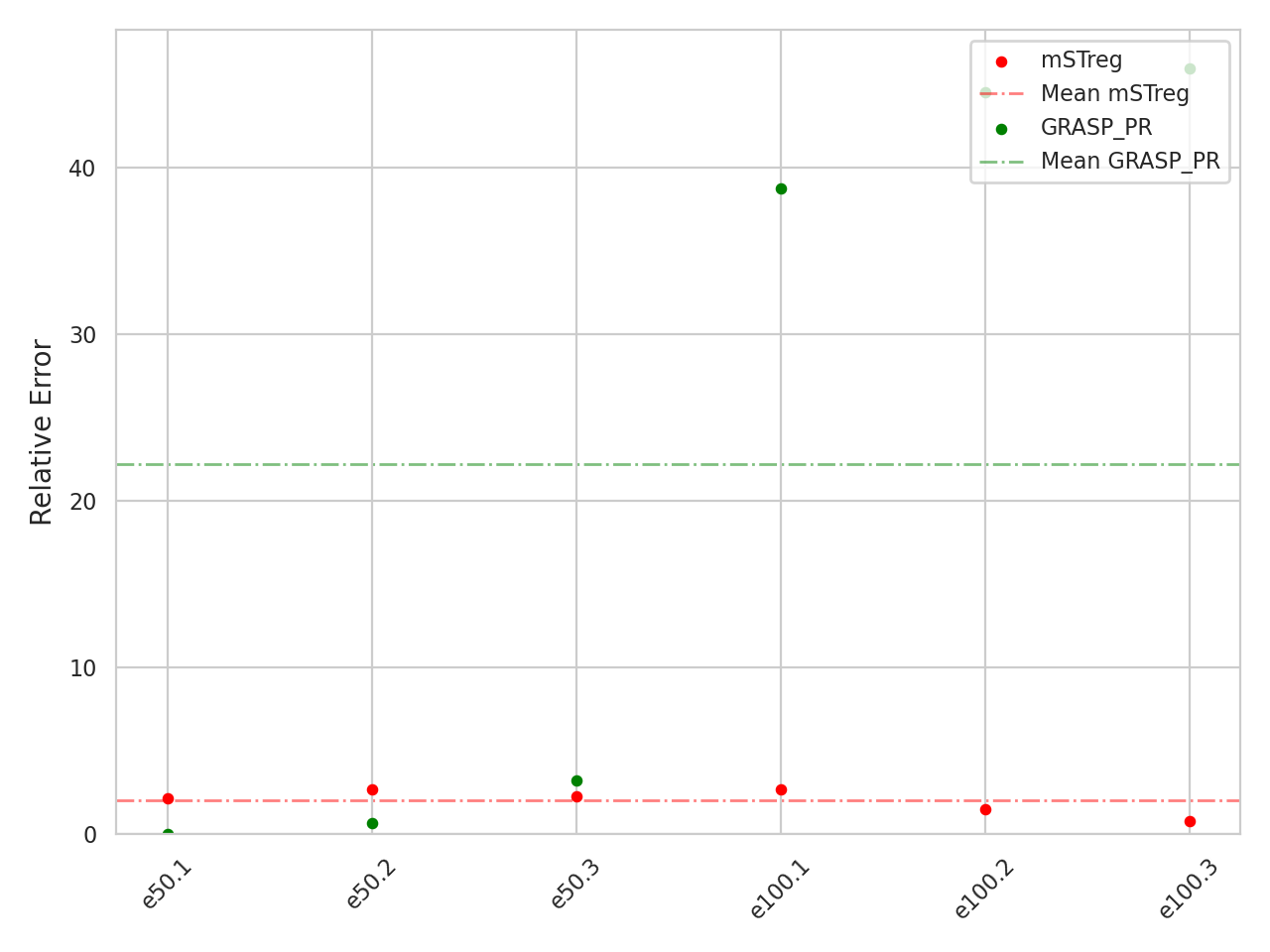}
    		\vspace{\bottomvspace cm}
    		\caption{$\alpha=0.2$}
    		\label{sfig1:GRASPPR_CST}
    	\end{subfigure}%
    	\begin{subfigure}{0.33\linewidth}
    		\centering
    		\vspace*{\topvspace cm}
    		\includegraphics[width=1\linewidth]{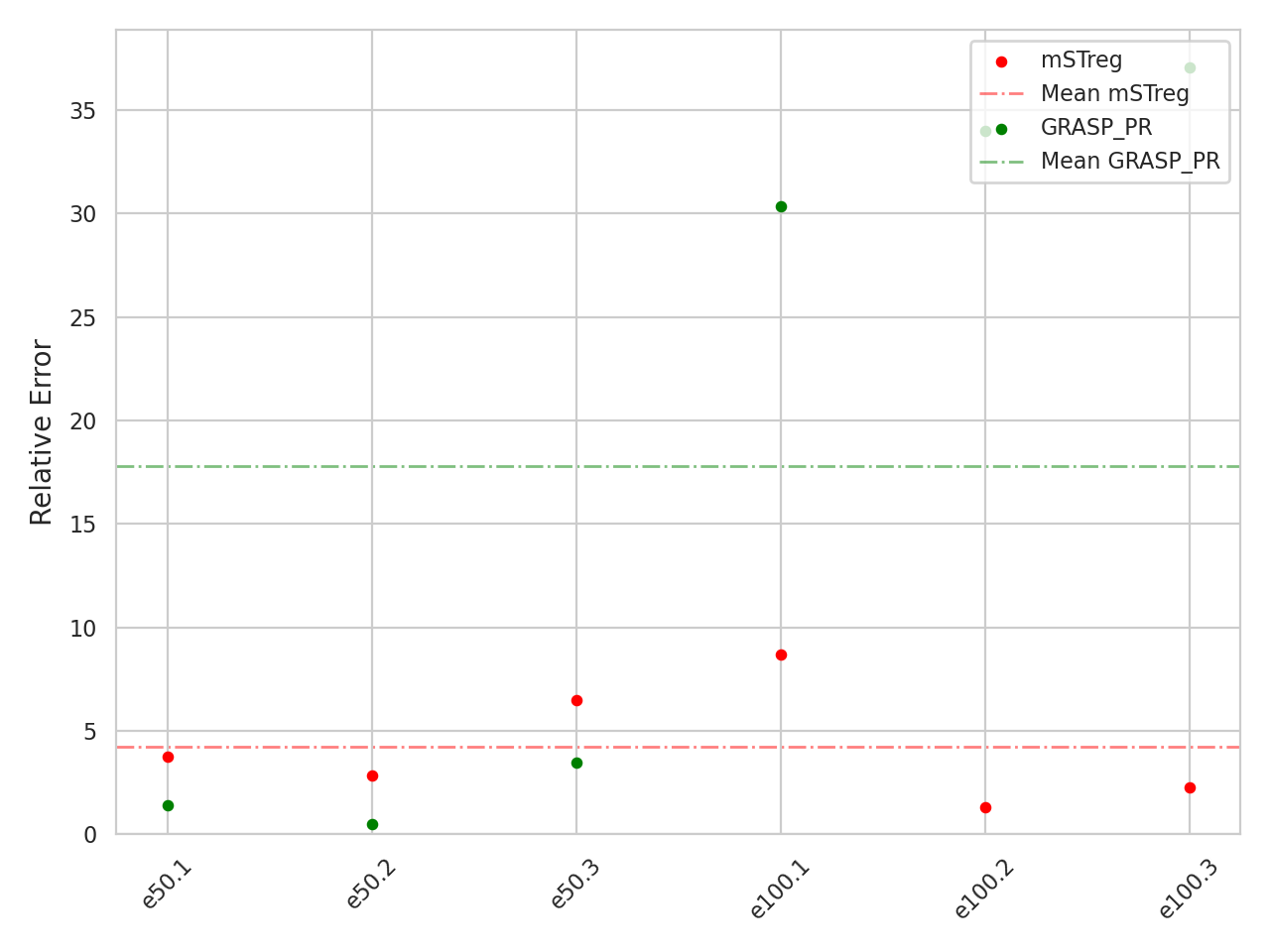}
    		\vspace{\bottomvspace cm}
    		\caption{$\alpha=0.4$}
    		\label{sfig2:GRASPPR_CST}
    	\end{subfigure}%
    	\begin{subfigure}{0.33\linewidth}
    		\centering
    		\vspace*{\topvspace cm}
    		\includegraphics[width=1\linewidth]{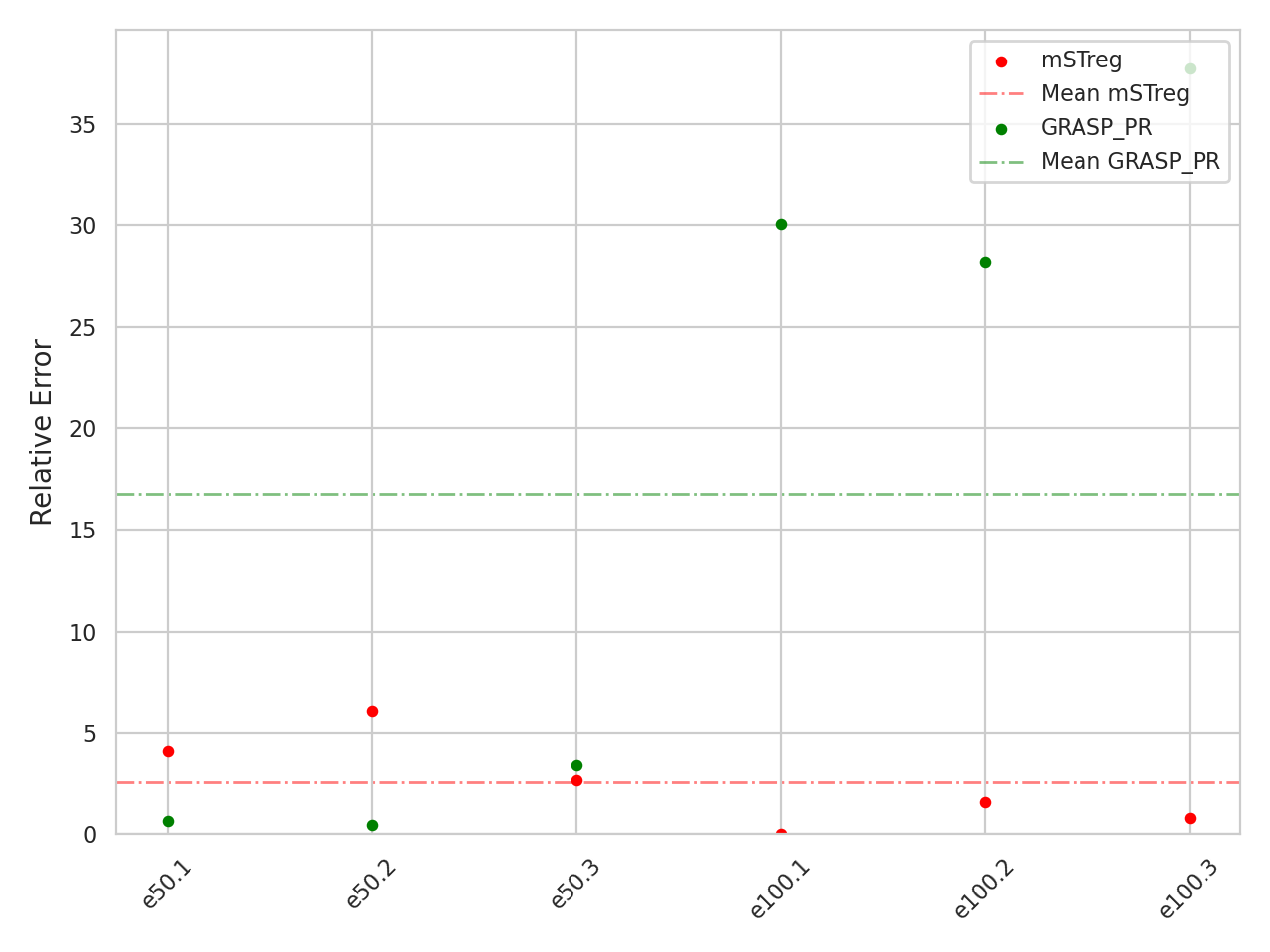}
    		\vspace{\bottomvspace cm}
    		\caption{$\alpha=0.6$}
    		\label{sfig3:GRASPPR_CST}
    	\end{subfigure}
        \begin{subfigure}{0.33\linewidth}
    		\centering
    		\vspace*{\topvspace cm}
    		\includegraphics[width=1\linewidth]{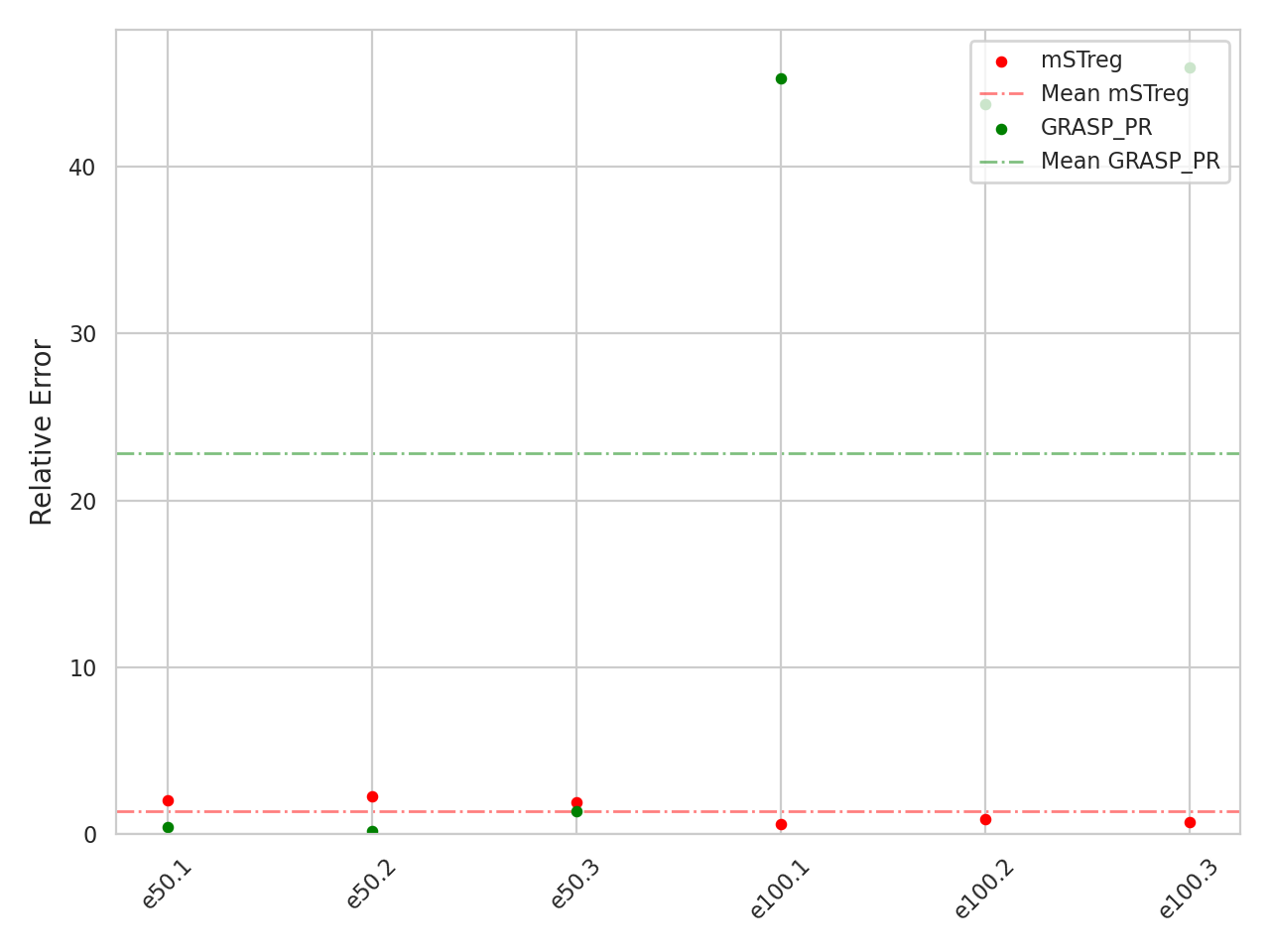}
    		\vspace{\bottomvspace cm}
    		\caption{$\alpha=0.8$}
    		\label{sfig4:GRASPPR_CST}
    	\end{subfigure}%
        \begin{subfigure}{0.33\linewidth}
    		\centering
    		\vspace*{\topvspace cm}
    		\includegraphics[width=1\linewidth]{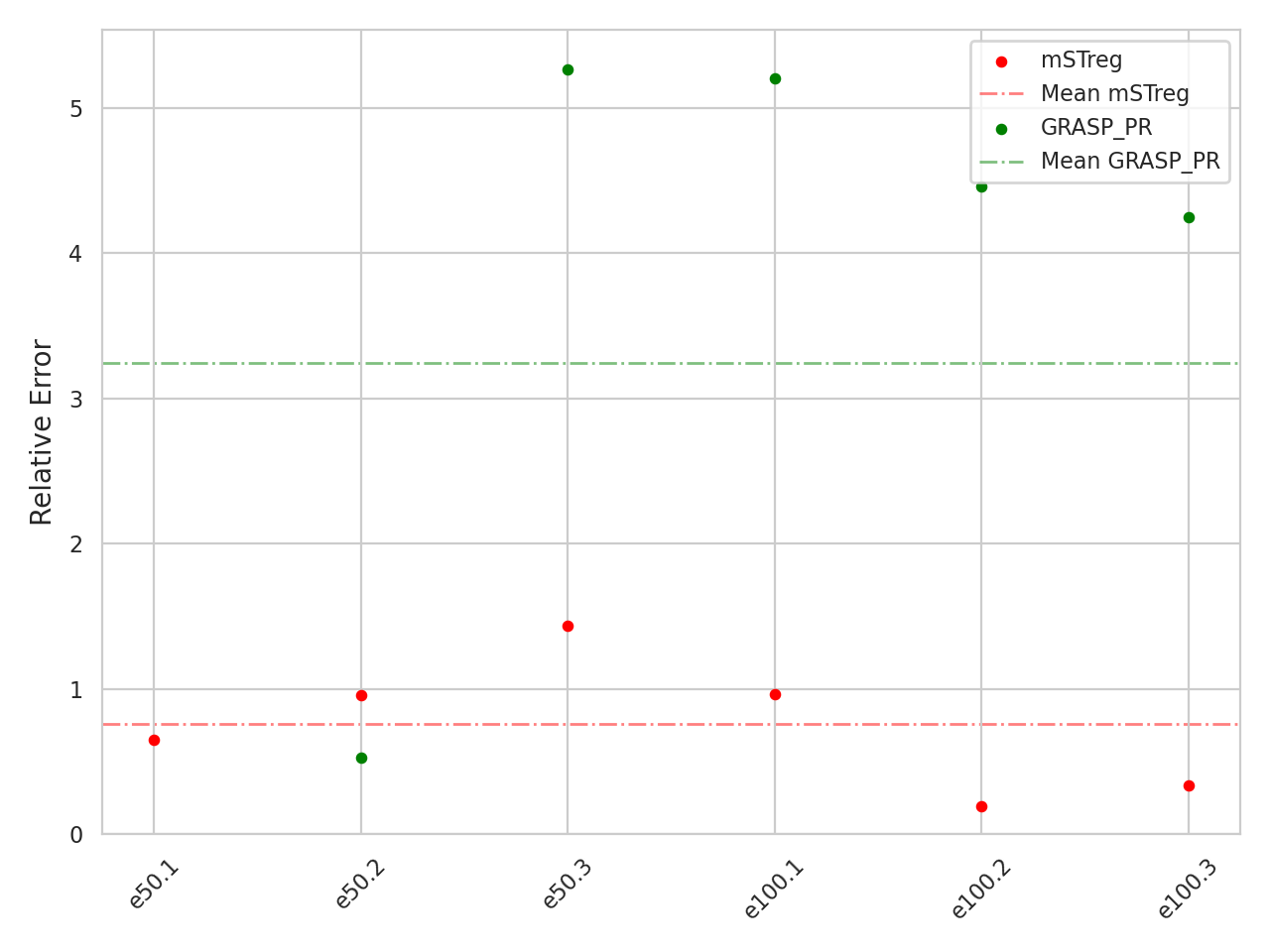}
    		\vspace{\bottomvspace cm}
    		\caption{$\alpha=1.0$}
    		\label{sfig5:GRASPPR_CST}
    	\end{subfigure}%
     	\caption[Comparing \mSTreg and GRASP\_PR]{\textbf{Comparing \mSTreg and GRASP\_PR}. Relative cost error concerning the \CST problem at different $\alpha$ values in the set $\{0.2, 0.4, 0.6, 0.8, 1\}$ for various instances of the OR library dataset (lower is better). The comparison involves \mSTreg, GRASP\_PR, and the combination of GRASP\_PR initialized with the \mSTreg solution (referred to as GRASP\_PR\_\mSTreg). The relative error is computed using the cost from GRASP\_PR\_\mSTreg as a reference. The combination of the mSTreg heuristic with GRASP\_PR consistently achieved the lowest cost, with all relative costs above $0$, demonstrating the enhancement of GRASP\_PR performance by mSTreg. To manage time constraints, a threshold of 5 minutes was imposed for methods utilizing GRASP\_PR.}
	 
	\label{fig:GRASPPR_CST}
\end{figure}

To assess performance, we conducted tests using GRASP\_PR, GRASP\_PR initialized with \mSTreg (GRASP\_PR\_\mSTreg), and the \mSTreg algorithms on OR library datasets for problems with 50 and 100 terminals. Analogously to the plots shown in the previous section, \figurename \ref{fig:GRASPPR_CST} displays the relative errors using GRASP\_PR\_\mSTreg cost as a reference for different $\alpha$ values in the set $\{0.2, 0.4, 0.6, 0.8, 1\}$. Notably, the combination of the mSTreg heuristic with GRASP\_PR consistently achieved the lowest cost, as all relative costs are above $0$, proving that the mSTreg can enhance GRASP\_PR performance.

Across 30 runs (combining 5 $\alpha$ values, 2 problem sizes, and 3 instances), GRASP\_PR outperformed mSTreg in only 12 instances. Moreover, achieving a lower cost took minutes for GRASP\_PR, while mSTreg run in the order of seconds. Consequently, mSTreg emerges as a favorable alternative, delivering a descent solution quickly.
\section{Conclusion}\label{sec:CST_conclusion}

We introduced the novel problem of the (branched) central spanning tree, which encompasses the minimum spanning, the Steiner and minimum routing cost trees as particular cases. The \CST weighs the edge-costs with the edge-centralities, whose influence are regulated by the parameter $\alpha$. We have focused on the Euclidean version of the problem, where the nodes are embedded in an Euclidean space. Moreover, we presented a variant of the problem allowing the addition of extra nodes, referred to as Steiner points (\BPs). In addition, we provided empirical evidence for the robustness of the (B)\CST tree structure to perturbations in the data, which increases with $\alpha$. In this regard, $\alpha$ serves as a parameter that trades-off between data-fidelity and stability. We also provided examples of potential applications such as 3D plant skeletonization or single cell trajectory inference.

On the theoretical side, we showed that as $\alpha\to\infty$ or the number of terminals approaches infinity when $\alpha>1$, (B)\CST converges to a star-tree, indicating inadequacy in extracting structural information when $\alpha>1$. Conversely, as $\alpha\to-\infty$, the (B)\CST tends towards a path graph. Additionally, thanks to the closed formulae of the branching angles, we proved the infeasibility of degree-4 \BPs when the terminals lie on a plane and $\alpha\in[0,0.5]\cup\{1\}$. We also provided evidence that suggests a similar case for $\alpha\in \ ]0.5,1[$.

Based on an efficient algorithm to compute the optimal locations of the \BPs, we have proposed the \mSTreg heuristic, which exploits the optimal position of the \BPs and the correspondence between the \CST and \BCST topologies to find approximate solutions for either. We benchmarked this algorithm and showed its competitiveness on small toy data sets. Since the proposed heuristic is agnostic to the weighting factors that multiply the distances, we leave as future work to test whether it is equally competitive for other problems, like the general optimum communication tree. Another open question is whether the algorithm can be adapted to perform well on non-complete or non-Euclidean graphs.

\section*{Acknowledgements}
This work is supported by the Deutsche Forschungsgemeinschaft (DFG, German Research Foundation) under Germany's Excellence Strategy EXC 2181/1 - 390900948 (the Heidelberg STRUCTURES Excellence Cluster).

\bibliographystyle{thesis_bibstyle}
\bibliography{CST_non_zotero,CST}

\newpage
\appendix
\section*{Overview of Appendix Contents}
The appendix is organized into the following sections:

\begin{enumerate}[label=\textbf{\Alph*},leftmargin=*]
	\item \hyperref[sec:app_stability_examples]{\textbf{Stability Examples}}: Illustrates the effect of $\alpha$ upon computing the \CST and \BCST on different 2-dimensional toy datasets.
	
	\item \hyperref[sec:app_applications]{\textbf{Applications}}: Explores further the applications mentioned in the main paper, namely trajectory inference of single-cell data and 3D plant skeletonization, and provides further examples.
	
	\item \hyperref[sec:app_CST_MCCNF]{\textbf{Reinterpreting CST as a Minimum Concave Cost Flow Problem}}: Presents how the \CST can be interpreted as an instance within the broader category of problems posed by the Minimum Concave Cost Network Flow (MCCNF) problems. In addition, it also discusses the relation to the Branched Optimal Transport problem
	
	\item \hyperref[sec:app_CST_MRCT_equivalence]{\textbf{Equivalence of the CST Problem with $\alpha=1$ and the Minimum Routing Cost Tree Problem}}: Demonstrates the equivalence between the Minimum Routing Cost Tree (MRCT) and the \CST when $\alpha=1$.

	\item \hyperref[sec:app_CSTlimits]{\textbf{Limit cases of the \CST/\BCST problems beyond the range $\alpha\in[0,1]$}}: Proves the statements regarding the limiting trees of both the \CST and \BCST as $\alpha$ approaches negative and positive infinity, and as the number of terminals, $N$, approaches infinity.
	
	\item \hyperref[sec:app_num_topos]{\textbf{Exploring the Number of Derivable Topologies from \CST and \BCST Topologies}}: Analytically illustrates the quantity of \BCST topologies that can be derived from a single \CST topology, and vice versa.

	\item \hyperref[sec:app_angles]{\textbf{Branching Angles at the Steiner Points in the \BCST problem}}: Analytically examines the angles formed by the edges meeting at Steiner points in the \BCST problem.
	
	\item \hyperref[sec:infeasibility-of-degree-4-steiner-point-for-alpha1]{\textbf{Infeasibility of Degree-4 Steiner Points in the Plane for $\alpha=1$}}: Proves the infeasibility of degree-4 Steiner points in the 2-dimensional Euclidean \BCST problem for $\alpha=1$.
	
	\item \hyperref[sec:app_IRLS]{\textbf{Iteratively Reweighted Least Square for the Geometric Optimization of the Steiner Points}}: Extends the iteratively reweighted least square approach to compute the optimal positions of the Steiner points presented in \citep{smith_how_1992,lippmann_theory_2022} to the \BCST case.
	
	\item \hyperref[sec:complexity-mstreg-heuristic]{\textbf{Complexity mSTreg heuristic}}: Analyzes the complexity of the mSTreg heuristic.
	
	\item \hyperref[sec:app_effect_freq_sampling]{\textbf{Effect of Additional Intermediate Points in the mSTreg heuristic}}: Examines the impact of introducing additional intermediate points, sampled along the edges of the tree, during the \mST step in the \mSTreg heuristic.
	
	\item \hyperref[sec:app_SP_removal_strategies]{\textbf{Strategies to Transform a Full Tree Topology into a \CST Topology}}: Analyzes different strategies to transform a \BCST topology into a \CST topology in order to optimize the \CST problem using the \mSTreg heuristic.
	
	\item \hyperref[sec:app_toydata_experiments]{\textbf{Further Details on the Brute Force Experiment}}: Provides additional details on the brute force experiment discussed in Section \ref{sec:benchmark} and examines the behavior of the mSTreg heuristic concerning $\alpha$.
	
	\item \hyperref[sec:selection-of-alpha]{\textbf{Selection of $\alpha$}}: Offers empirical insights to guide the selection of the optimal value for $\alpha$.
	
	\item \hyperref[sec:app_implementation_details]{\textbf{Implementation Details}}: Provides further details regarding the implementation.
\end{enumerate}

\section{Stability Examples}\label{sec:app_stability_examples}
\begin{figure}[h!]
    \centering
    \begin{subfigure}{0.5\linewidth}
        \centering
        \includegraphics[width=\linewidth]{Stabilty_figures_paper/uniform/overlayed_uniform_n=1000_CST_alpha=0.00.png}
         \caption{\mST (\CST, $\alpha=0.0$)}
		\label{sfig1:app_uniform_CST_noiserobustness_toydata}
	\end{subfigure}%
    \begin{subfigure}{0.5\linewidth}
    \centering
		\includegraphics[width=\linewidth]{Stabilty_figures_paper/uniform/overlayed_uniform_n=1000_BCST_alpha=0.00.png}
		\caption{Steiner tree (\BCST, $\alpha=0$))}
		\label{sfig2:app_uniform_CST_noiserobustness_toydata}
	\end{subfigure}
    \begin{subfigure}{0.5\linewidth}
        \centering
        \includegraphics[width=\linewidth]{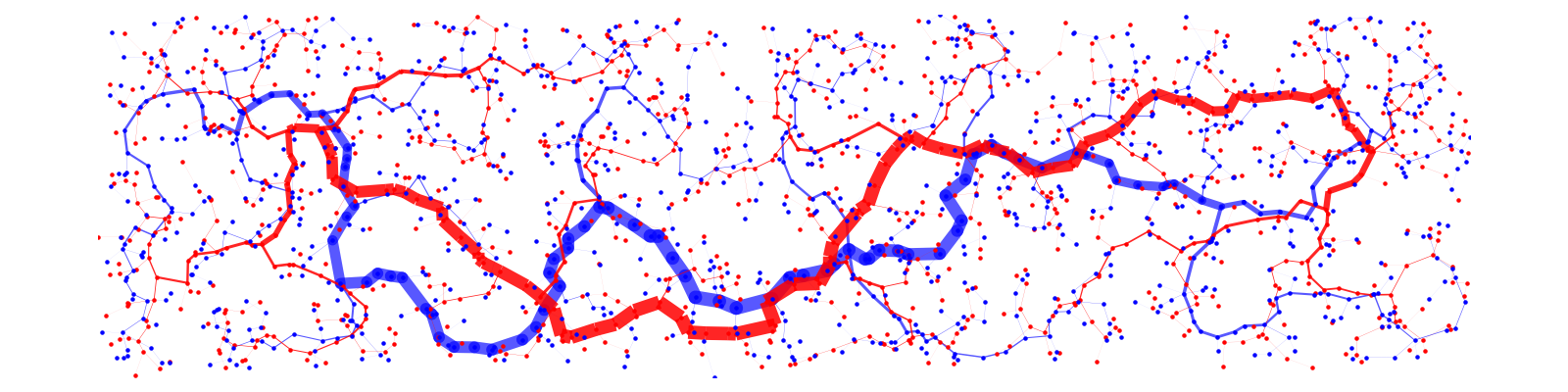}
         \caption{\CST, $\alpha=0.2$}
		\label{sfig3:app_uniform_CST_noiserobustness_toydata}
	\end{subfigure}%
    \begin{subfigure}{0.5\linewidth}
    \centering
		\includegraphics[width=\linewidth]{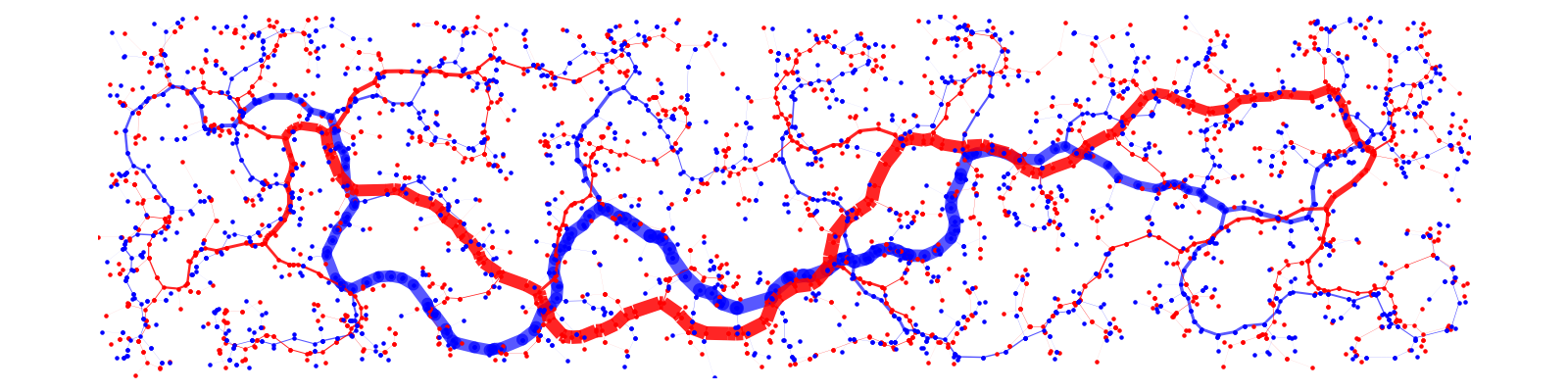}
		\caption{\BCST, $\alpha=0.2$}
		\label{sfig4:app_uniform_CST_noiserobustness_toydata}
	\end{subfigure}
    \begin{subfigure}{0.5\linewidth}
        \centering
        \includegraphics[width=\linewidth]{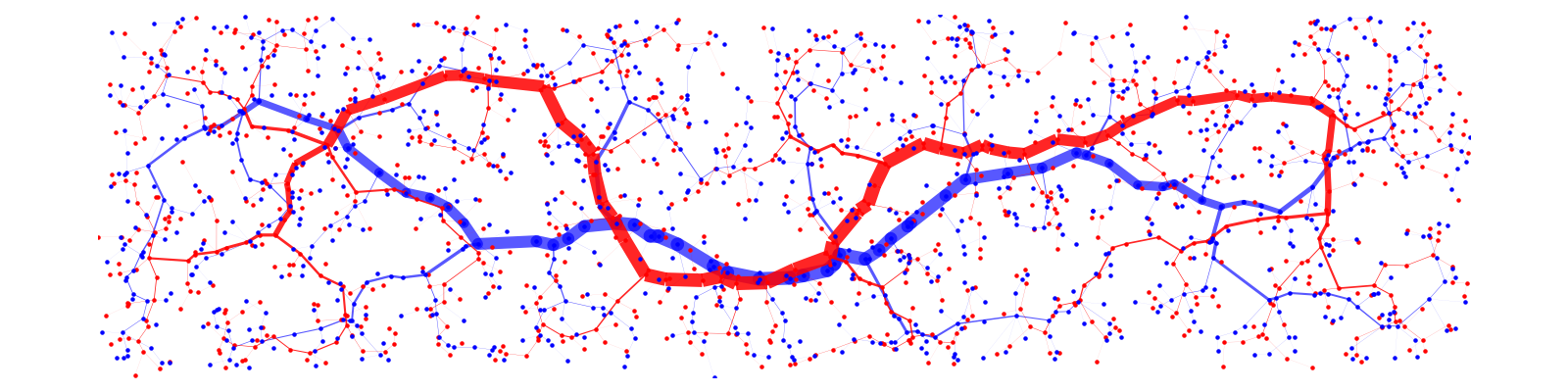}
         \caption{\CST, $\alpha=0.4$}
		\label{sfig3:app_uniform_CST_noiserobustness_toydata}
	\end{subfigure}%
    \begin{subfigure}{0.5\linewidth}
    \centering
		\includegraphics[width=\linewidth]{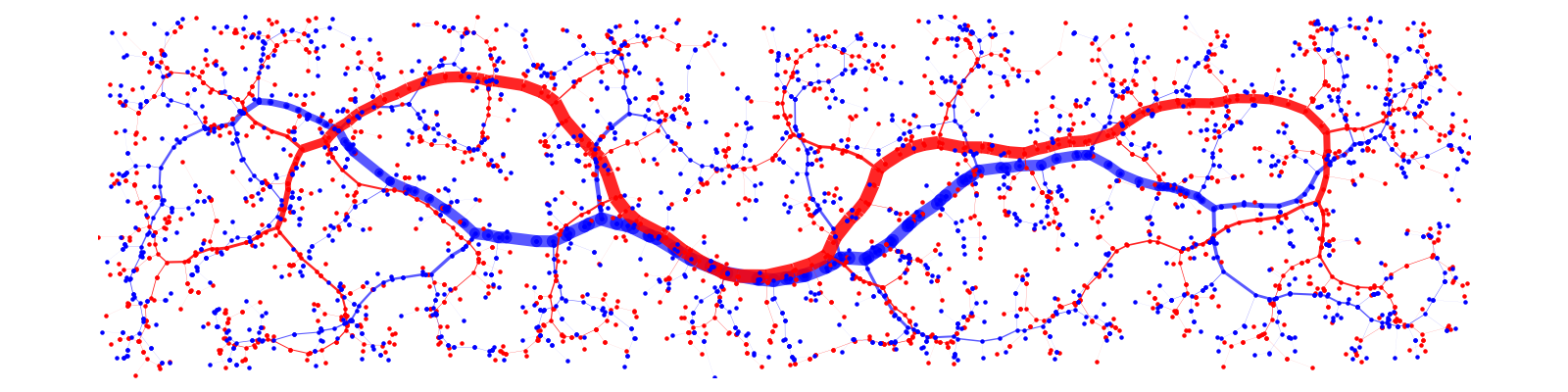}
		\caption{\BCST, $\alpha=0.4$}
		\label{sfig4:app_uniform_CST_noiserobustness_toydata}
	\end{subfigure}
    \begin{subfigure}{0.5\linewidth}
        \centering
        \includegraphics[width=\linewidth]{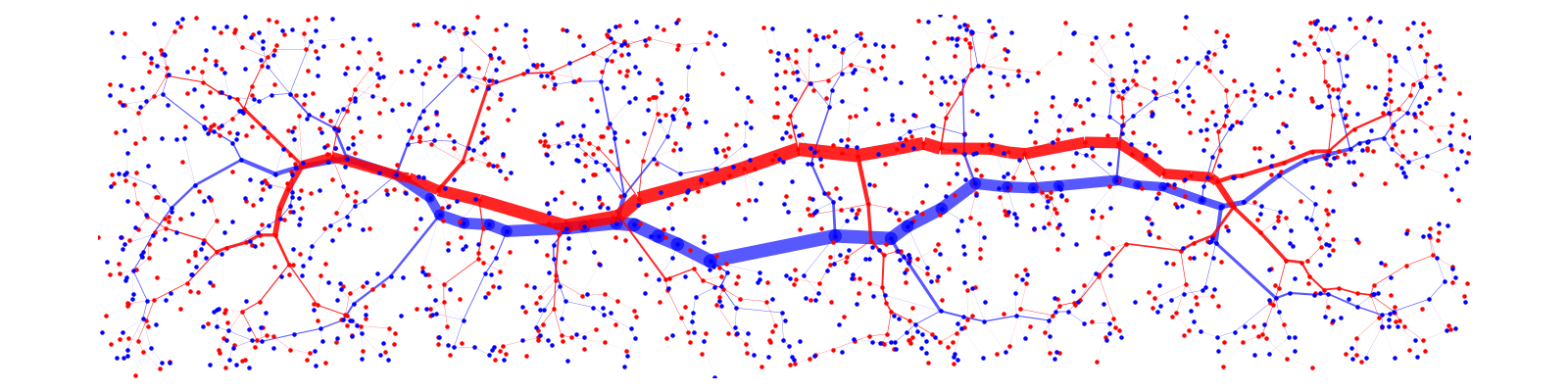}
         \caption{\CST, $\alpha=0.6$}
		\label{sfig5:app_uniform_CST_noiserobustness_toydata}
	\end{subfigure}%
    \begin{subfigure}{0.5\linewidth}
    \centering
		\includegraphics[width=\linewidth]{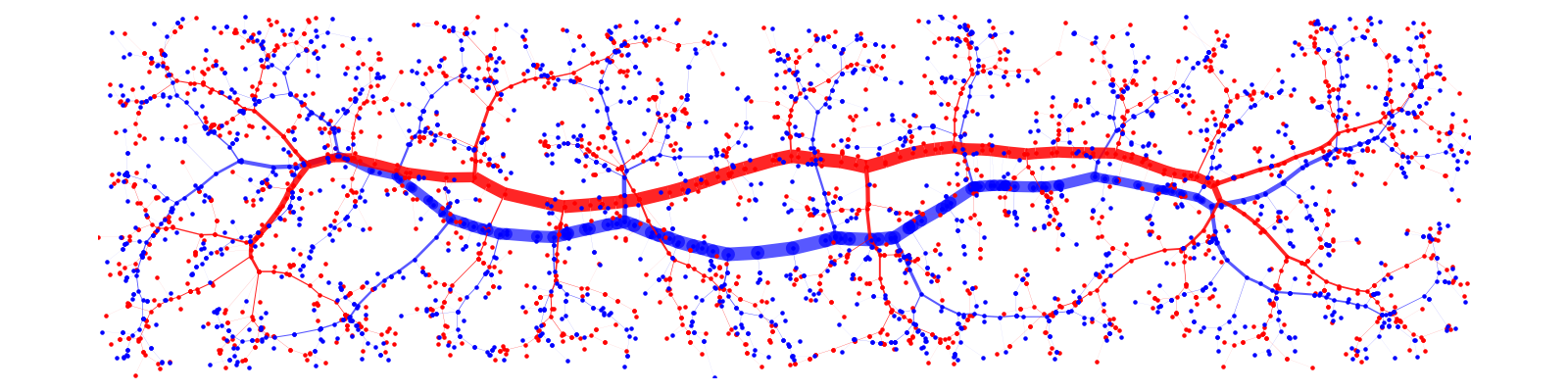}
		\caption{\BCST, $\alpha=0.6$}
		\label{sfig6:app_uniform_CST_noiserobustness_toydata}
	\end{subfigure}
    \begin{subfigure}{0.5\linewidth}
        \centering
        \includegraphics[width=\linewidth]{Stabilty_figures_paper/uniform/overlayed_uniform_n=1000_CST_alpha=0.80.png}
         \caption{\CST, $\alpha=0.8$}
		\label{sfig7:app_uniform_CST_noiserobustness_toydata}
	\end{subfigure}%
    \begin{subfigure}{0.5\linewidth}
    \centering
		\includegraphics[width=\linewidth]{Stabilty_figures_paper/uniform/overlayed_uniform_n=1000_BCST_alpha=0.80.png}
		\caption{\BCST, $\alpha=0.8$}
		\label{sfig8:app_uniform_CST_noiserobustness_toydata}
	\end{subfigure}
    \begin{subfigure}{0.5\linewidth}
        \centering
        \includegraphics[width=\linewidth]{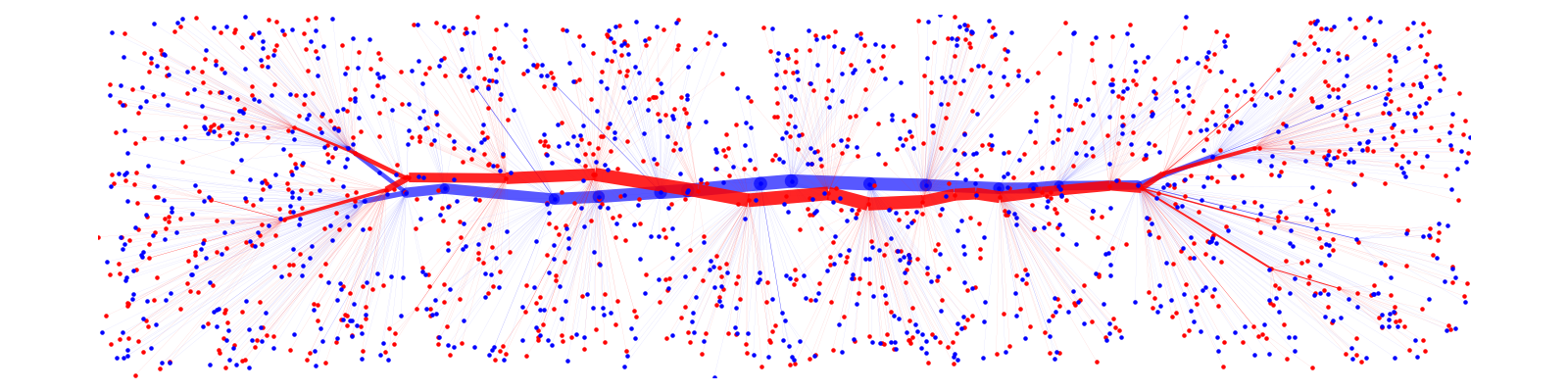}
         \caption{\CST, $\alpha=1.0$}
		\label{sfig9:app_uniform_CST_noiserobustness_toydata}
	\end{subfigure}%
    \begin{subfigure}{0.5\linewidth}
    \centering
		\includegraphics[width=\linewidth]{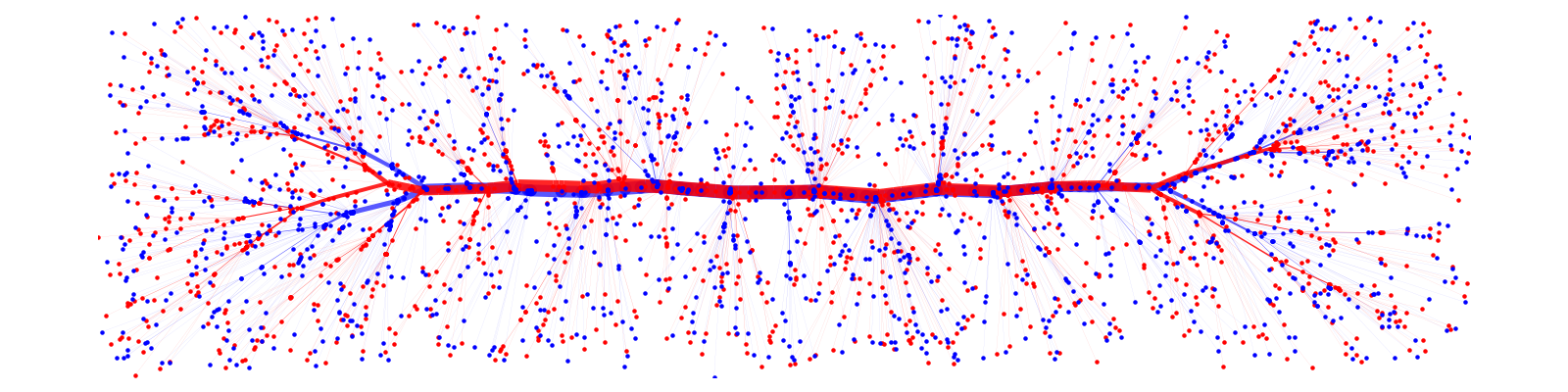}
		\caption{\BCST, $\alpha=1.0$}
		\label{sfig10:app_uniform_CST_noiserobustness_toydata}
	\end{subfigure}

  \caption[(B)CST Examples from Rectangle Uniform Distribution]{\textbf{(B)CST Examples from Rectangle Uniform Distribution}. \CST and \BCST are computed for two perturbed instances generated by adding zero-centered Gaussian noise to points derived from a common sample uniformly taken within a rectangle. (B)\CST for higher $\alpha$ values are more robust to noise and adhere to large scale structure in the data better.  The width of each edge is proportional to its centrality. All trees except  for the \mST were computed using the heuristic proposed in Section \ref{sec:heuristic}.}
	\label{fig:app_uniform_CST_noiserobustness_toydata}	
\end{figure}

\begin{figure}[h!]
    \centering
    \begin{subfigure}{0.25\linewidth}
        \centering
        \includegraphics[width=\linewidth]{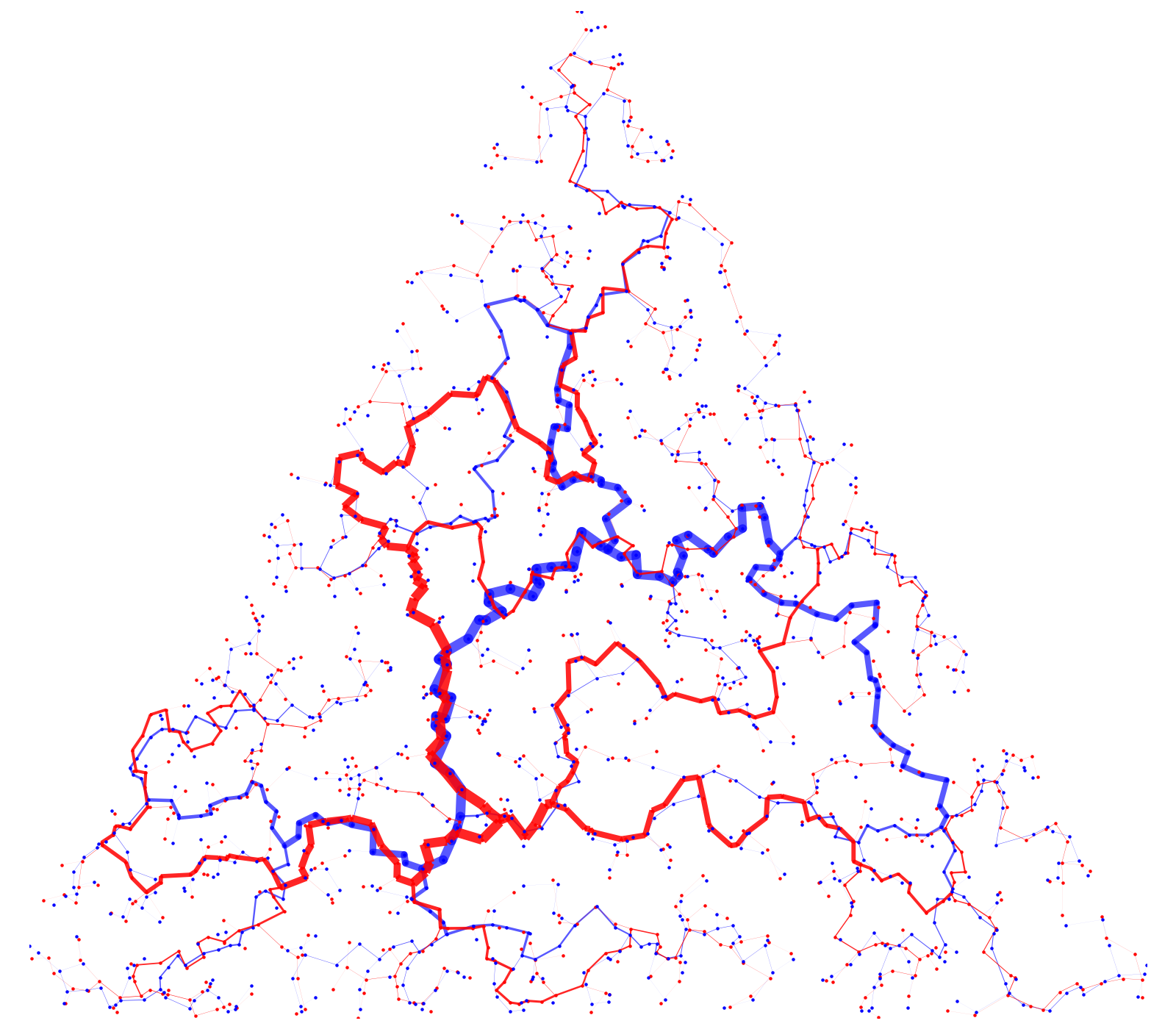}
         \caption{\mST \\(\CST, $\alpha=0.0$)}
		\label{sfig1:app_triangle_uniform_CST_noiserobustness_toydata}
	\end{subfigure}%
    \begin{subfigure}{0.25\linewidth}
    \centering
		\includegraphics[width=\linewidth]{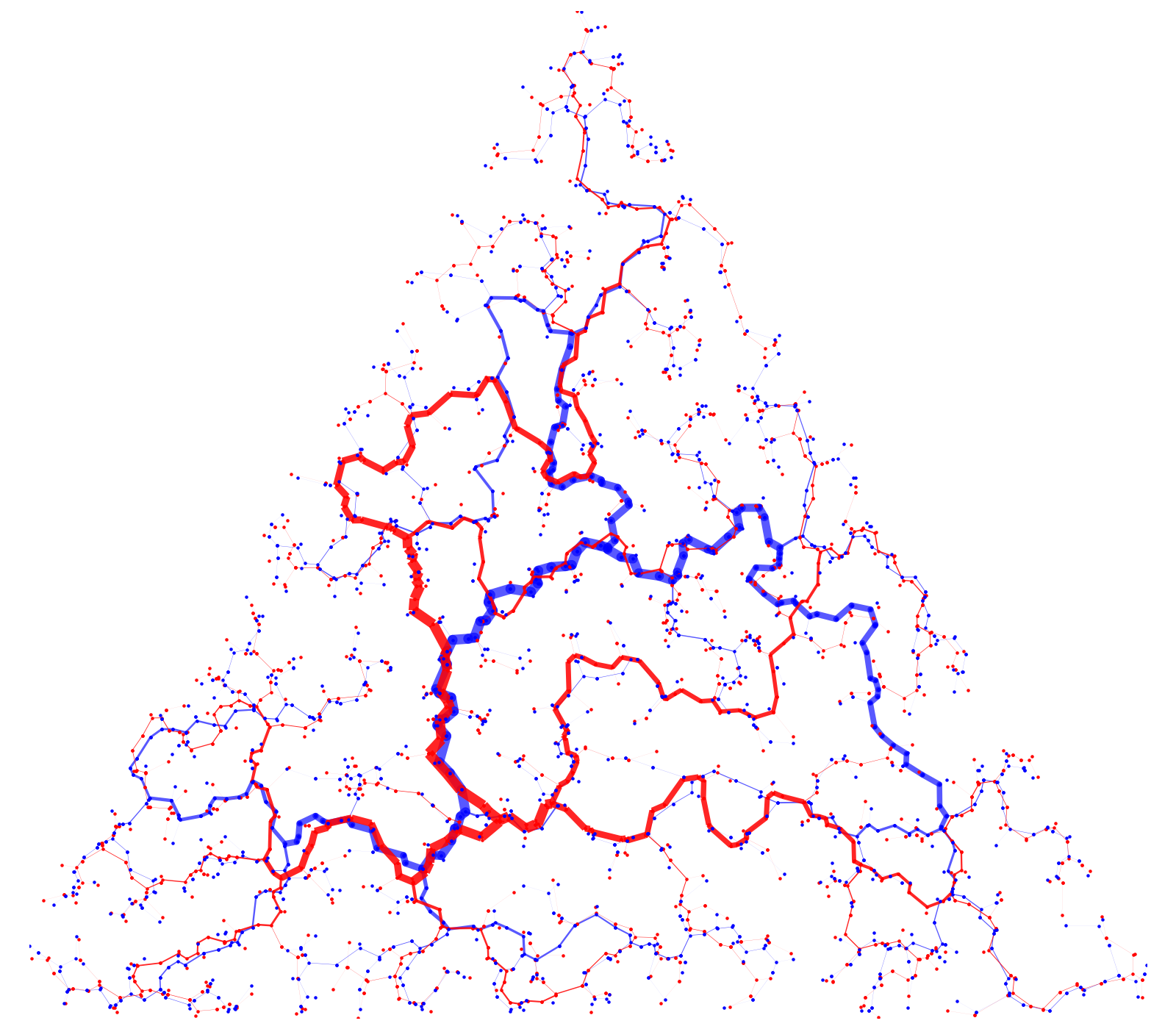}
        \captionsetup{justification=centering} 
		\caption{Steiner tree \newline (\BCST, $\alpha=0$)}
		\label{sfig2:app_triangle_uniform_CST_noiserobustness_toydata}
	\end{subfigure}%
    \begin{subfigure}{0.25\linewidth}
        \centering
        \includegraphics[width=\linewidth]{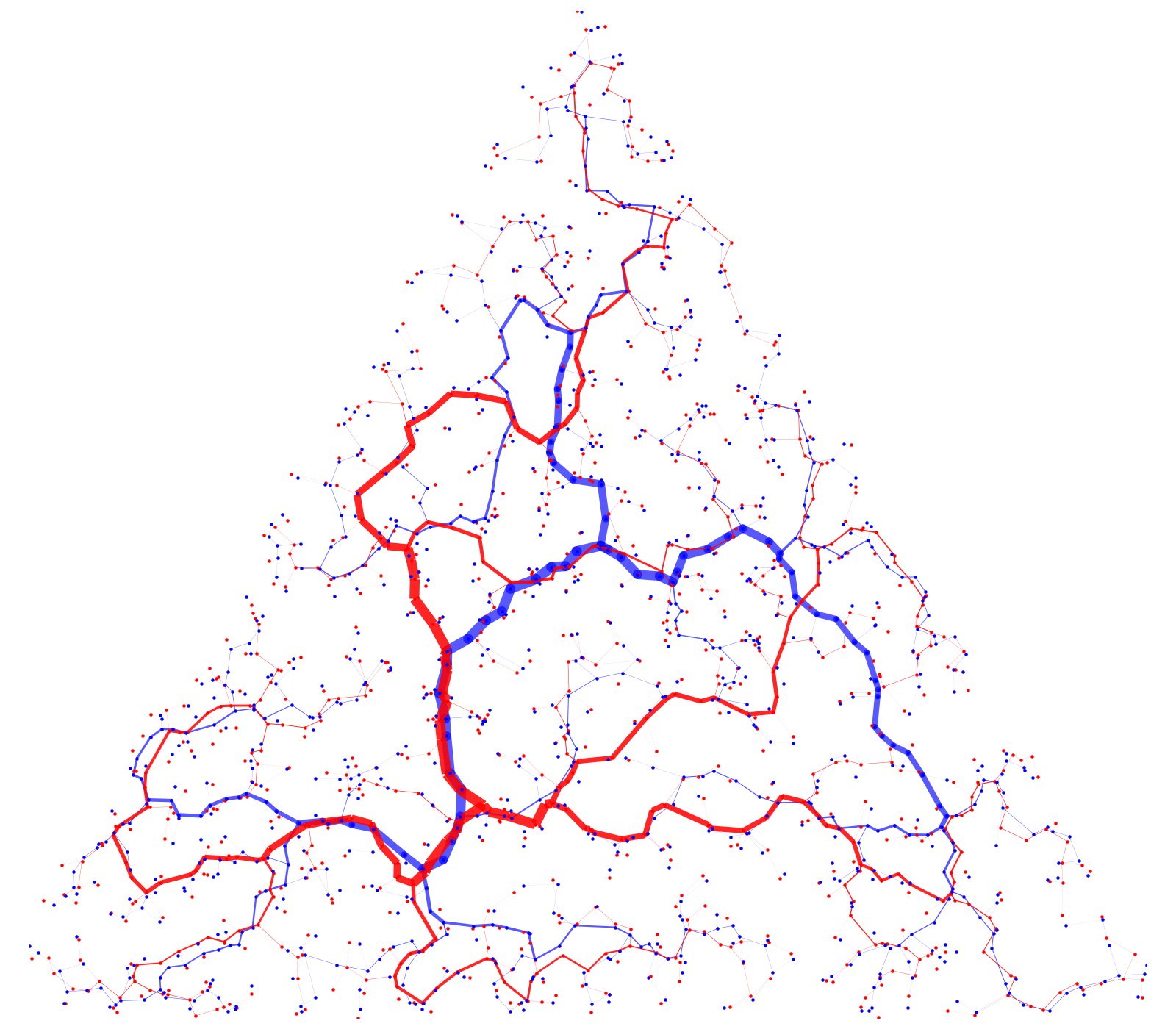}
         \caption{\CST, $\alpha=0.2$ \\\phantom{ $\alpha=0.2$}}
		\label{sfig3:app_triangle_uniform_CST_noiserobustness_toydata}
	\end{subfigure}%
    \begin{subfigure}{0.25\linewidth}
    \centering
		\includegraphics[width=\linewidth]{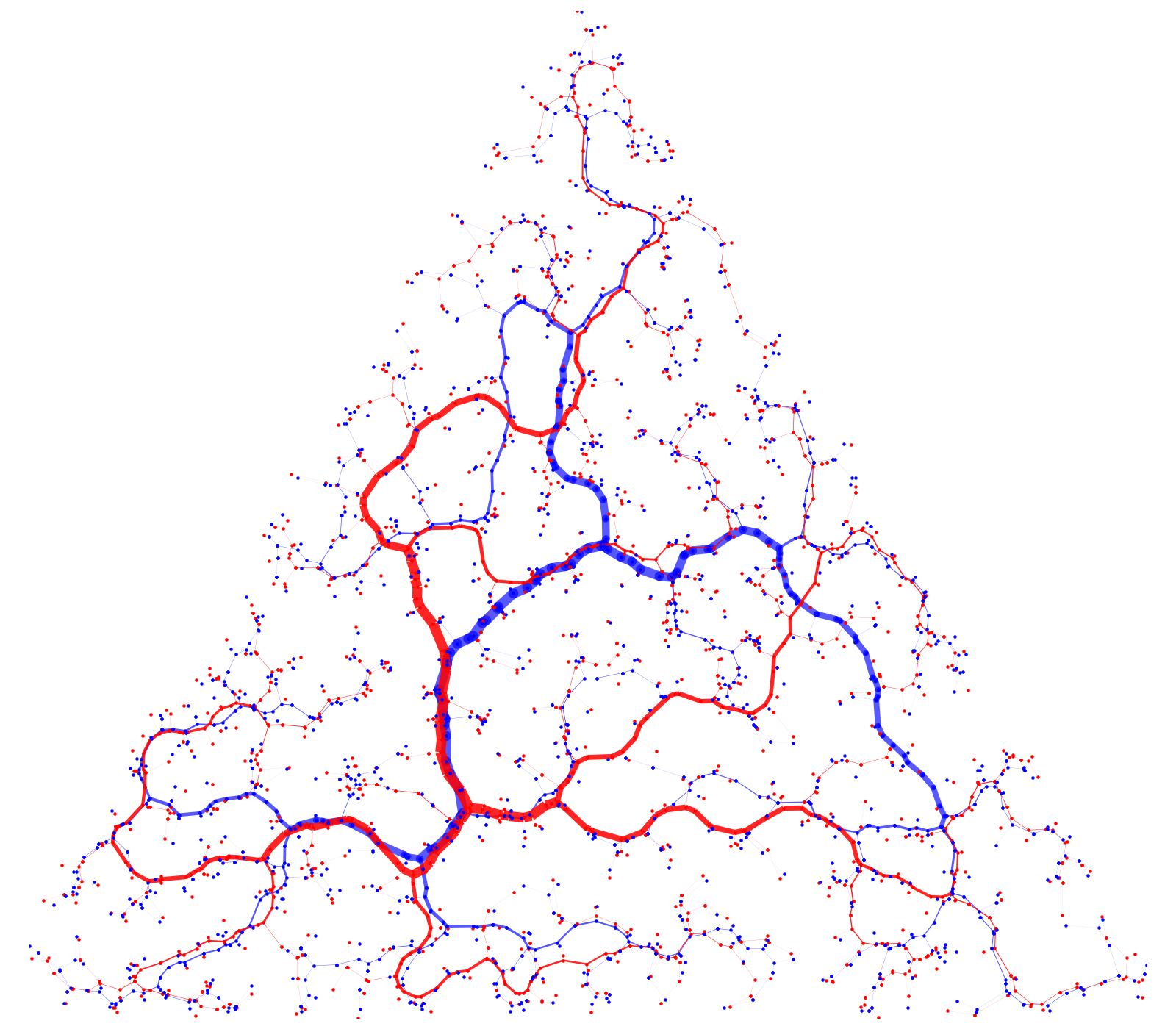}
		\caption{\BCST, $\alpha=0.2$\\\phantom{$\alpha=0.2$}}
		\label{sfig4:app_triangle_uniform_CST_noiserobustness_toydata}
	\end{subfigure}
    \begin{subfigure}{0.25\linewidth}
        \centering
        \includegraphics[width=\linewidth]{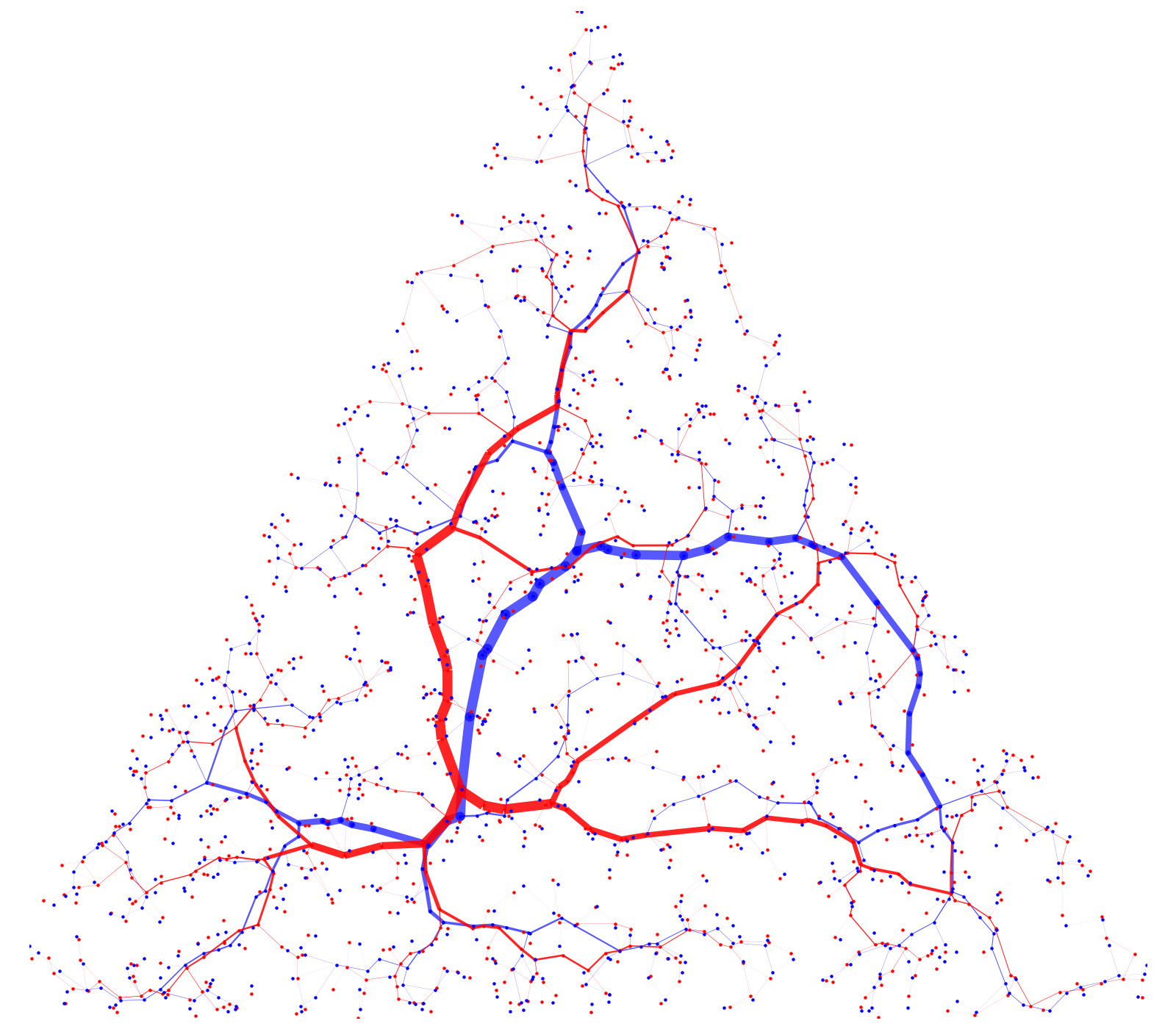}
         \caption{\CST, $\alpha=0.4$}
		\label{sfig5:app_triangle_uniform_CST_noiserobustness_toydata}
	\end{subfigure}%
    \begin{subfigure}{0.25\linewidth}
    \centering
		\includegraphics[width=\linewidth]{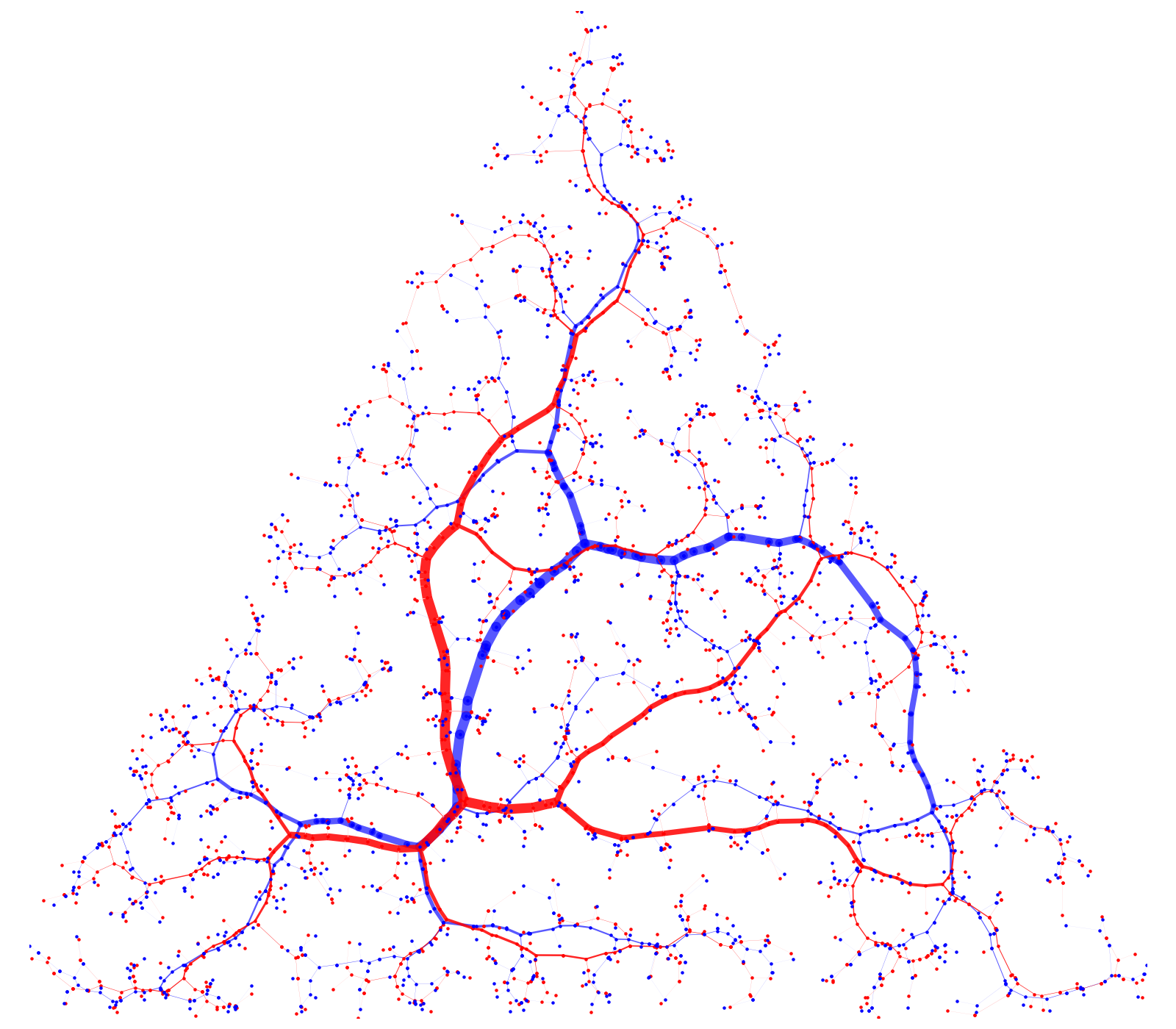}
		\caption{\BCST, $\alpha=0.4$}
		\label{sfig6:app_triangle_uniform_CST_noiserobustness_toydata}
	\end{subfigure}%
    \begin{subfigure}{0.25\linewidth}
        \centering
        \includegraphics[width=\linewidth]{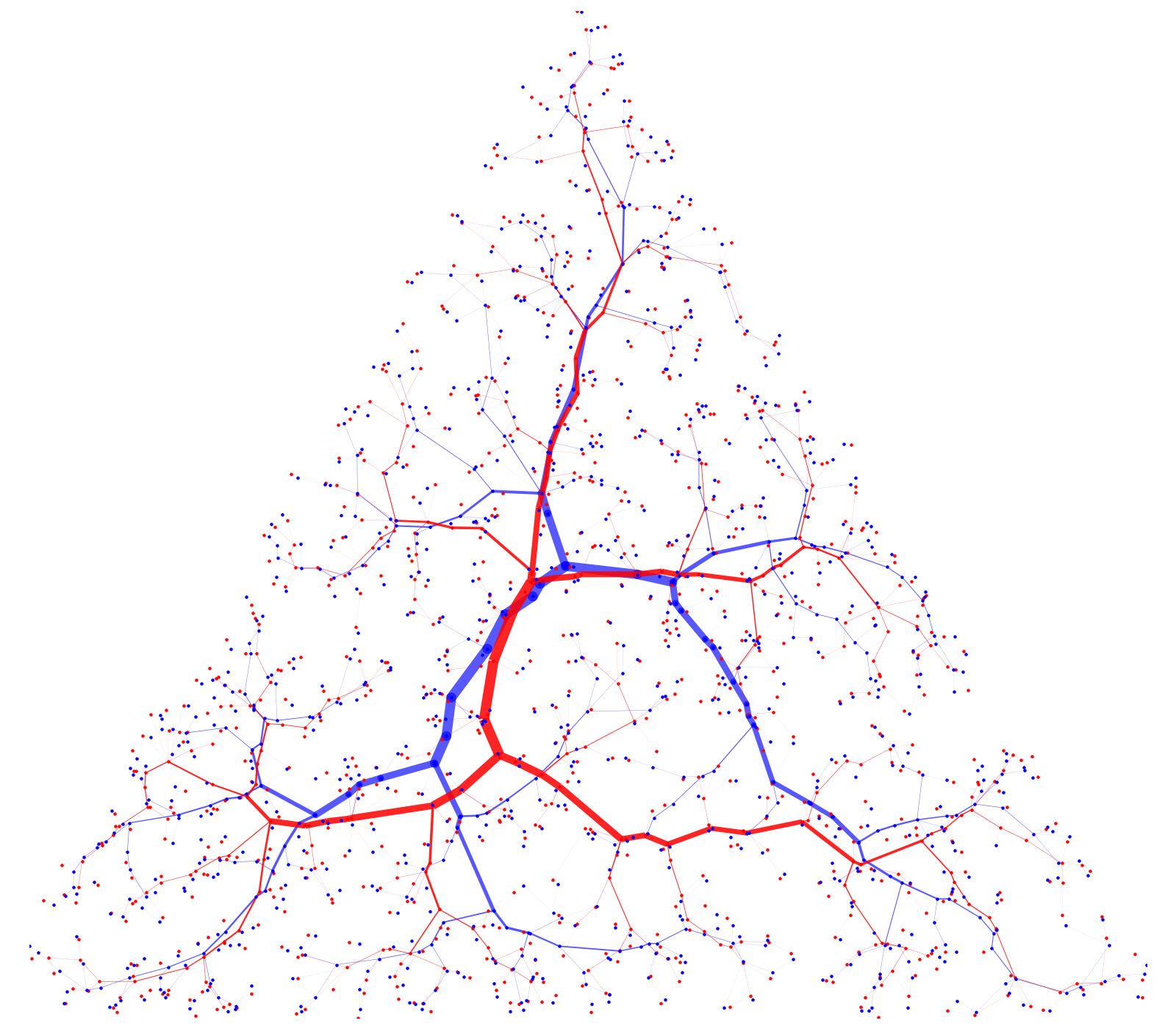}
         \caption{\CST, $\alpha=0.6$}
		\label{sfig7:app_triangle_uniform_CST_noiserobustness_toydata}
	\end{subfigure}%
    \begin{subfigure}{0.25\linewidth}
    \centering
		\includegraphics[width=\linewidth]{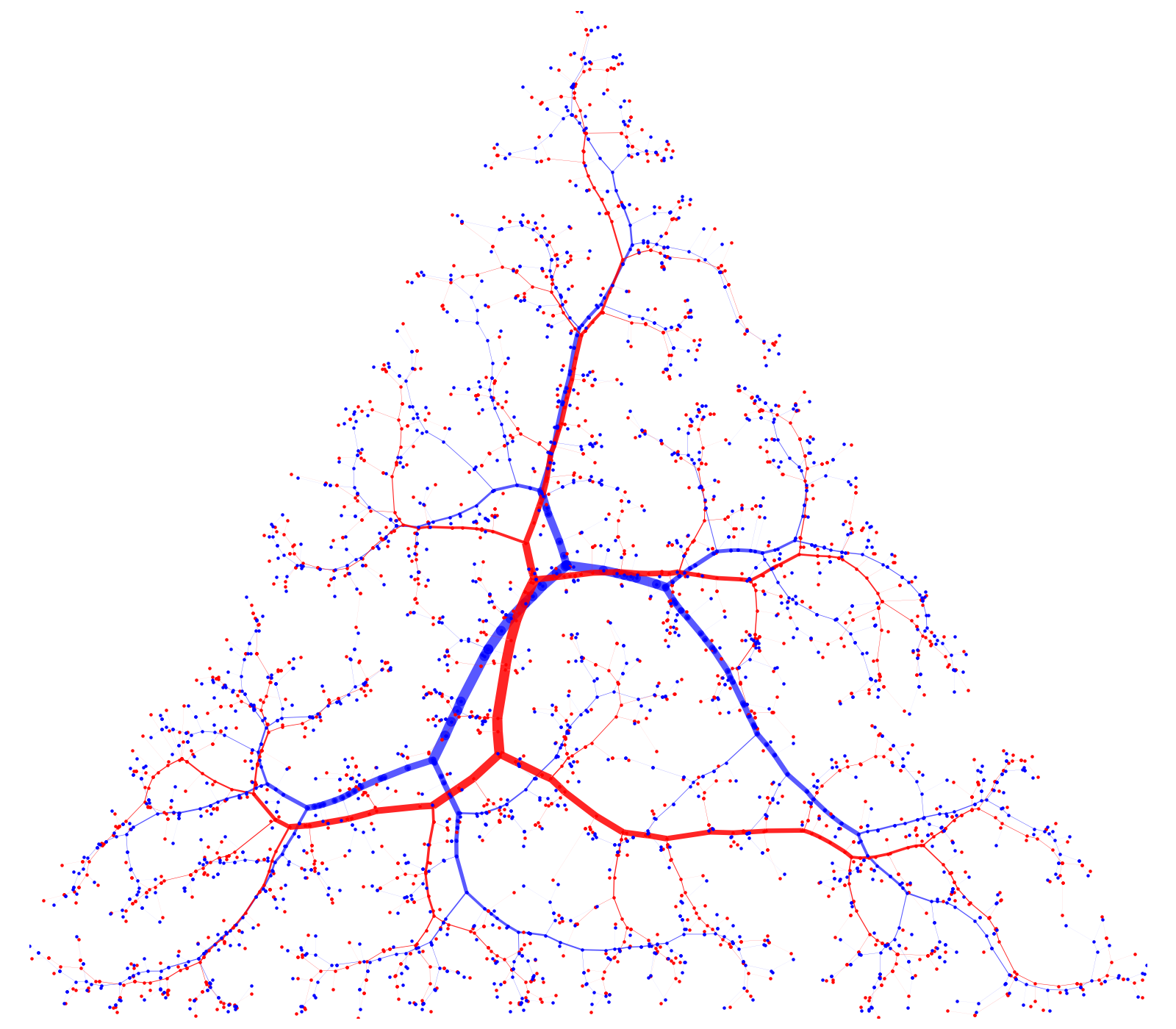}
		\caption{\BCST, $\alpha=0.6$}
		\label{sfig8:app_triangle_uniform_CST_noiserobustness_toydata}
	\end{subfigure}
    \begin{subfigure}{0.25\linewidth}
        \centering
        \includegraphics[width=\linewidth]{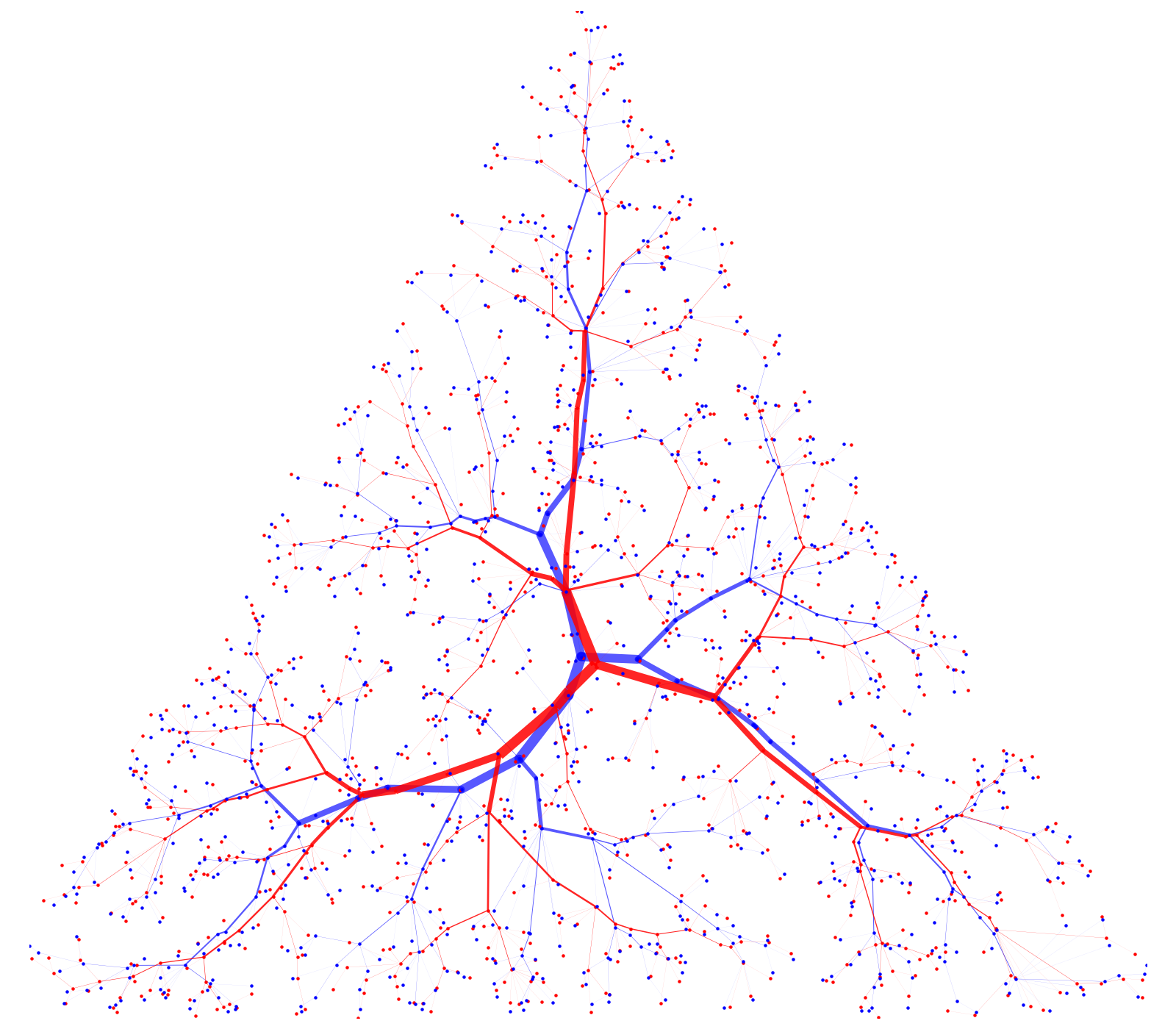}
         \caption{\CST, $\alpha=0.8$}
		\label{sfig9:app_triangle_uniform_CST_noiserobustness_toydata}
	\end{subfigure}%
    \begin{subfigure}{0.25\linewidth}
    \centering
		\includegraphics[width=\linewidth]{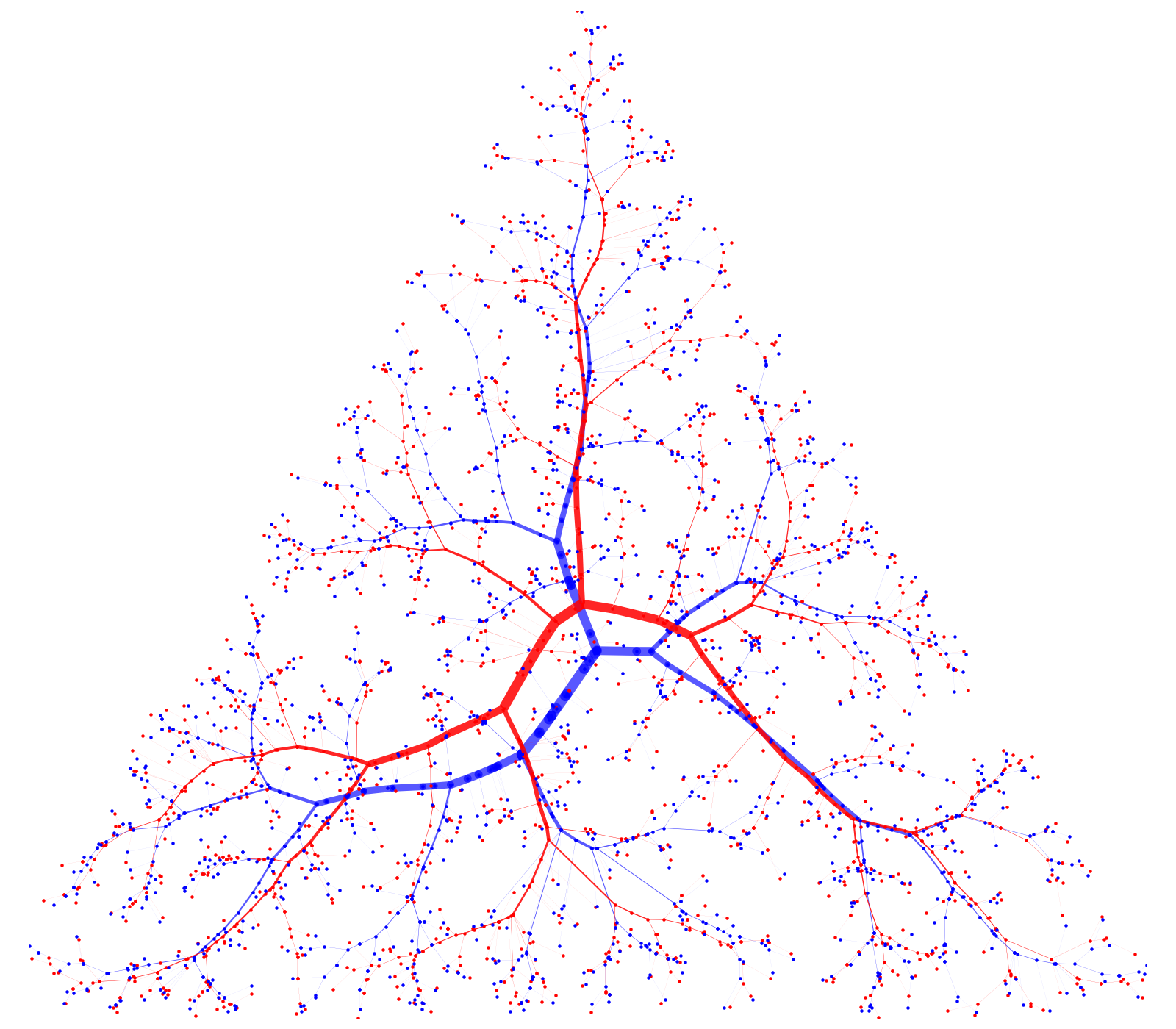}
		\caption{\BCST, $\alpha=0.8$}
		\label{sfig10:app_triangle_uniform_CST_noiserobustness_toydata}
	\end{subfigure}%
    \begin{subfigure}{0.25\linewidth}
        \centering
        \includegraphics[width=\linewidth]{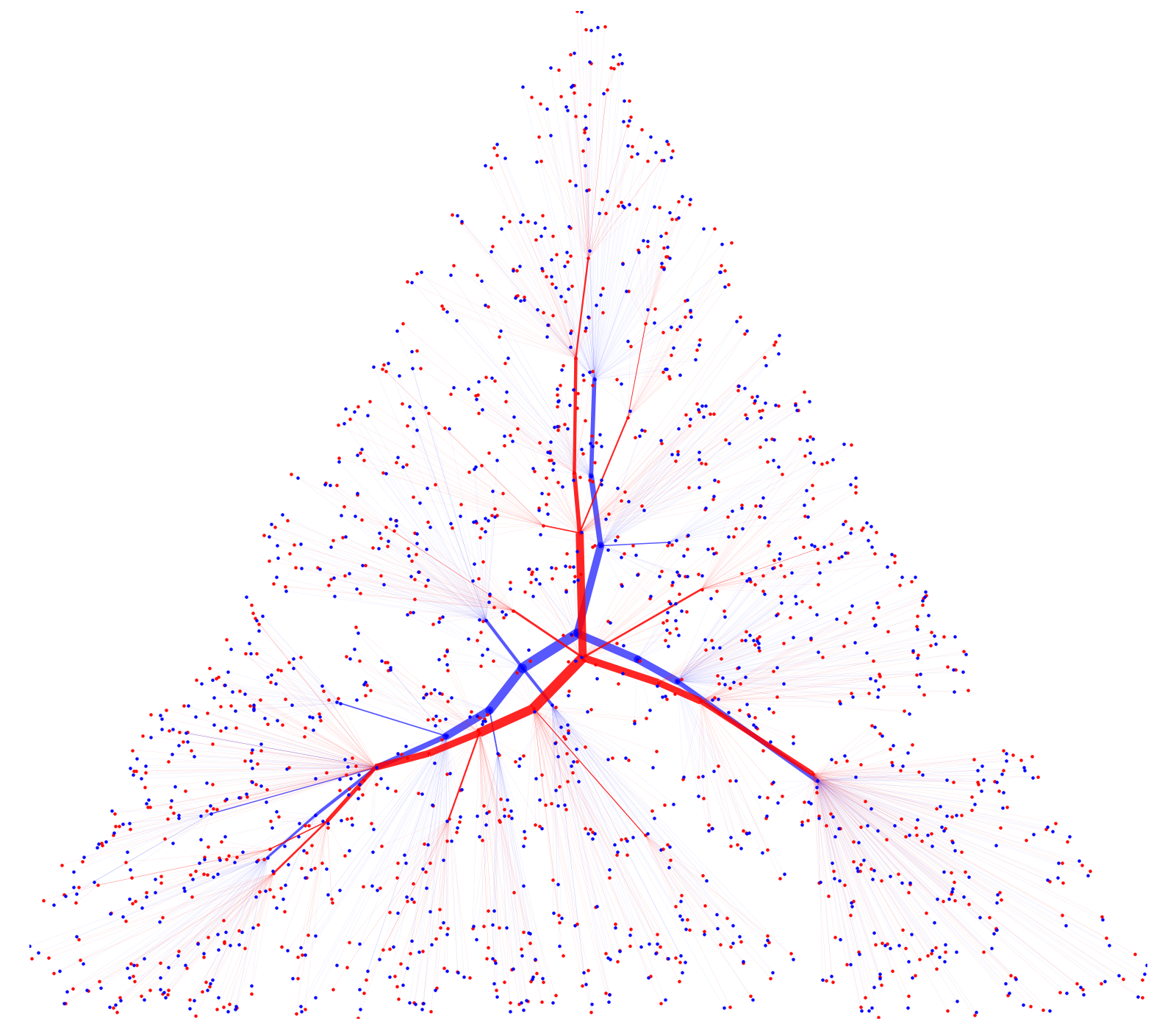}
         \caption{\CST, $\alpha=1.0$}
		\label{sfig11:app_triangle_uniform_CST_noiserobustness_toydata}
	\end{subfigure}%
    \begin{subfigure}{0.25\linewidth}
    \centering
		\includegraphics[width=\linewidth]{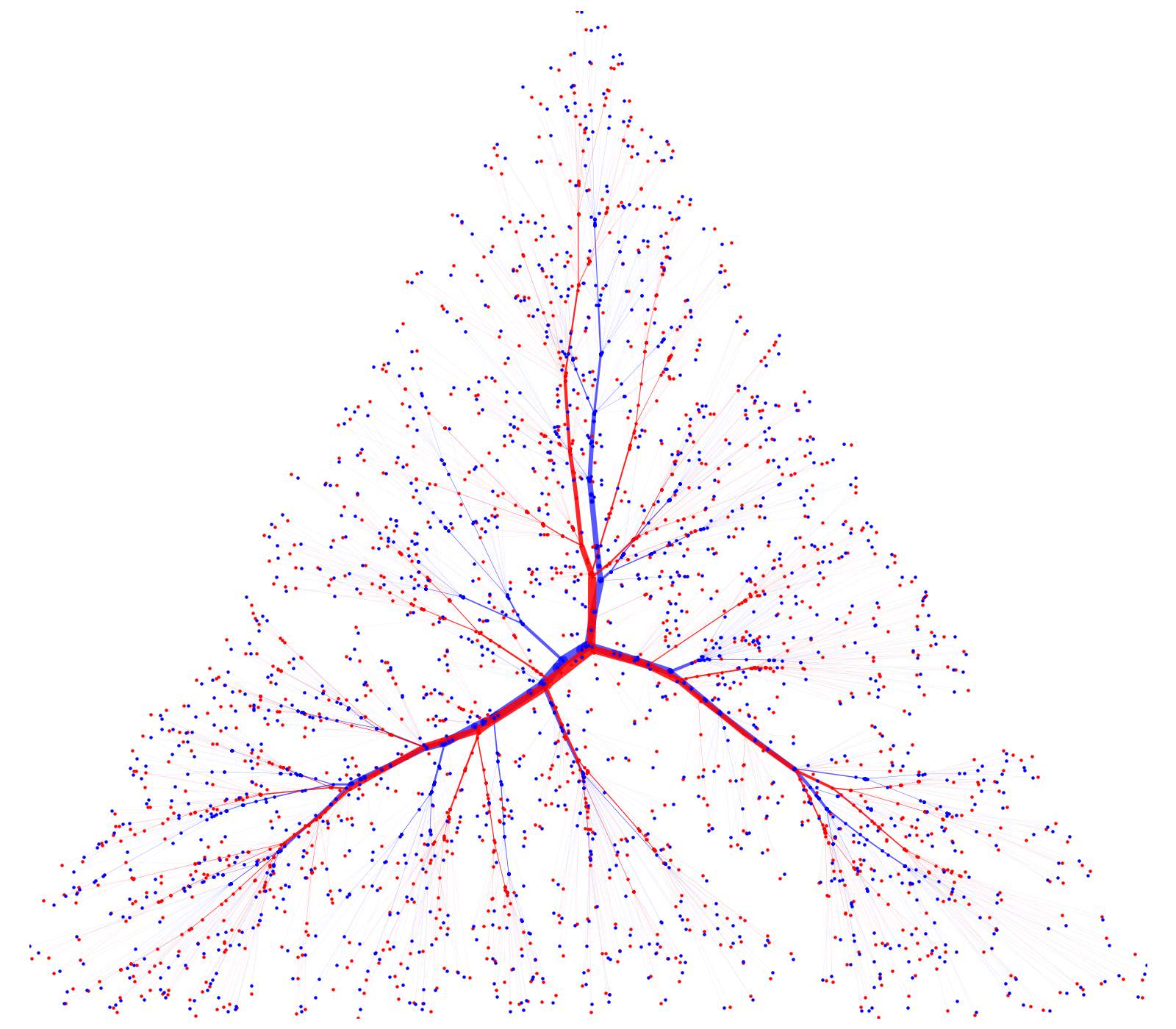}
		\caption{\BCST, $\alpha=1.0$}
		\label{sfig12:app_triangle_uniform_CST_noiserobustness_toydata}
	\end{subfigure}

  \caption[(B)CST Examples from Triangle Uniform Distribution]{\textbf{(B)CST Examples from Triangle Uniform Distribution}. \CST and \BCST are computed for two perturbed instances generated by adding zero-centered Gaussian noise to points derived from a common sample uniformly taken within a triangle. (B)\CST for higher $\alpha$ values are more robust to noise and adhere to large scale structure in the data better.  The width of each edge is proportional to its centrality. All trees except  for the \mST were computed using the heuristic proposed in Section \ref{sec:heuristic}.}
	\label{fig:app_triangle_uniform_CST_noiserobustness_toydata}	
\end{figure}

\begin{figure}[h!]
    \centering
    \begin{subfigure}{0.25\linewidth}
        \centering
        \includegraphics[width=\linewidth]{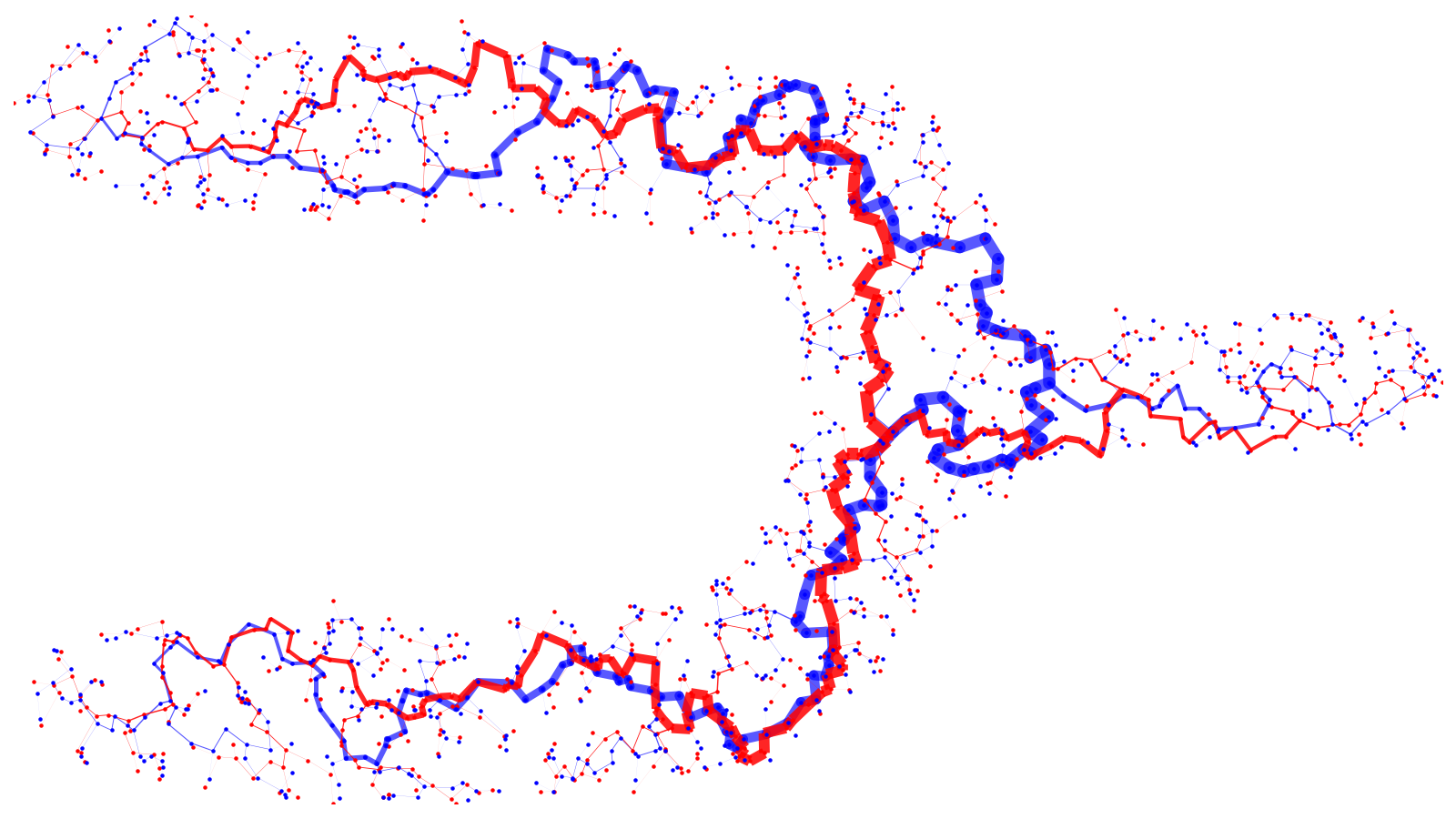}
         \caption{\mST\\(\CST, $\alpha=0.0$)}
		\label{sfig1:app_non_convex_CST_noiserobustness_toydata}
	\end{subfigure}%
    \begin{subfigure}{0.25\linewidth}
    \centering
		\includegraphics[width=\linewidth]{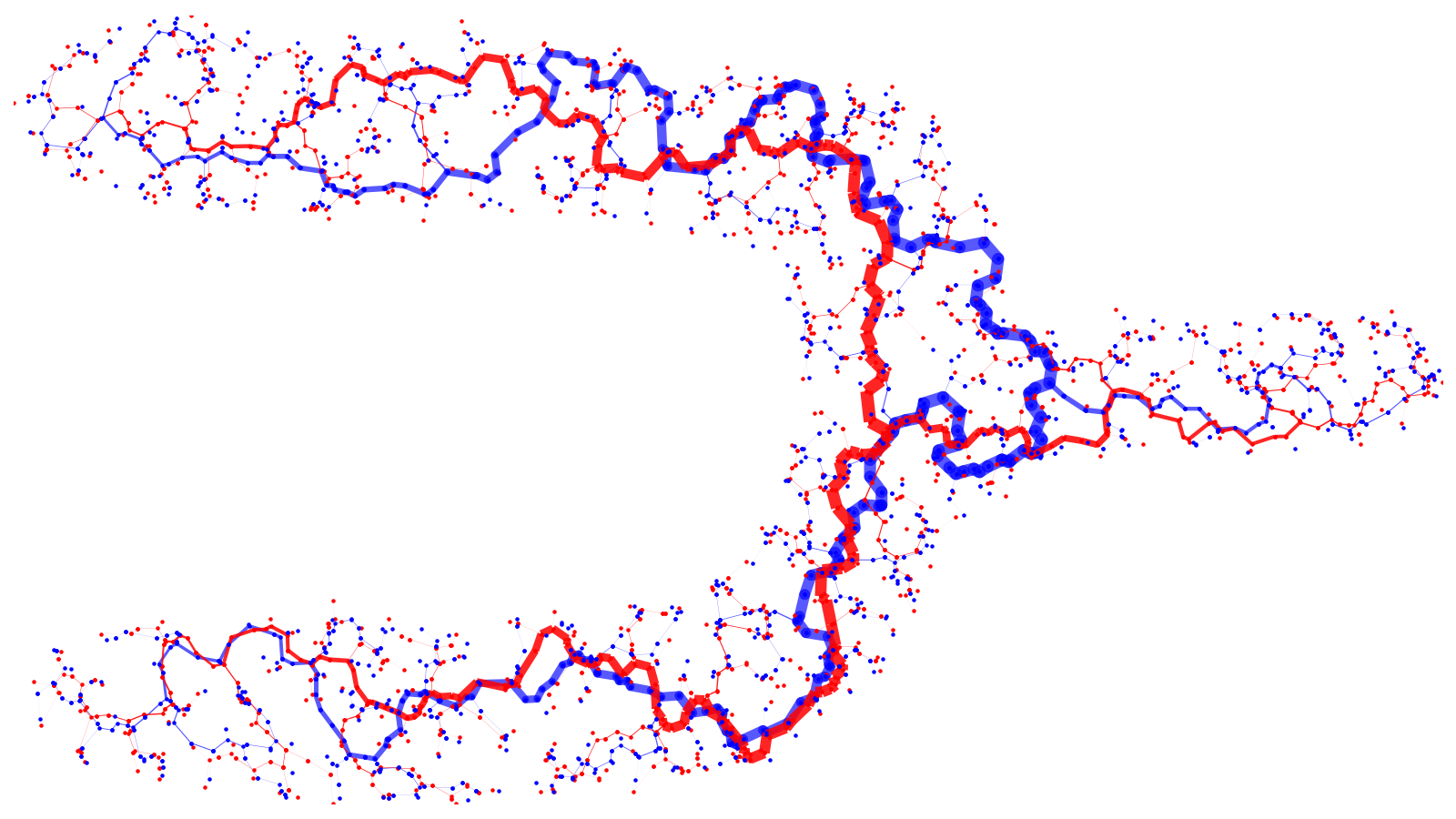}
        \captionsetup{justification=centering} 
		\caption{Steiner tree\\ (\BCST, $\alpha=0.0$)}
		\label{sfig2:app_non_convex_CST_noiserobustness_toydata}
	\end{subfigure}%
    \begin{subfigure}{0.25\linewidth}
        \centering
        \includegraphics[width=\linewidth]{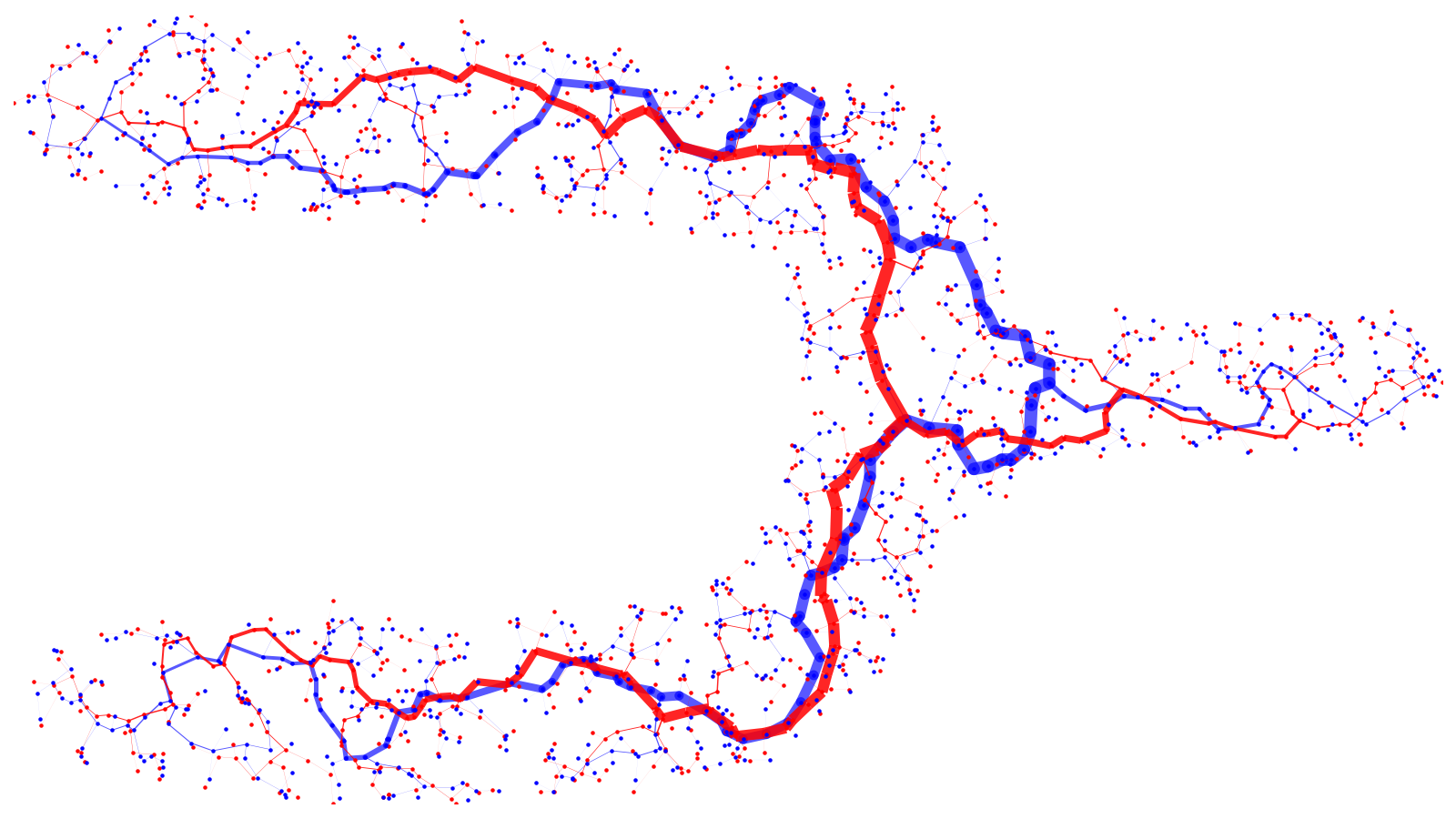}
         \caption{\CST, $\alpha=0.2$\\\phantom{$\alpha=0.2$}}
		\label{sfig3:app_non_convex_CST_noiserobustness_toydata}
	\end{subfigure}%
    \begin{subfigure}{0.25\linewidth}
    \centering
		\includegraphics[width=\linewidth]{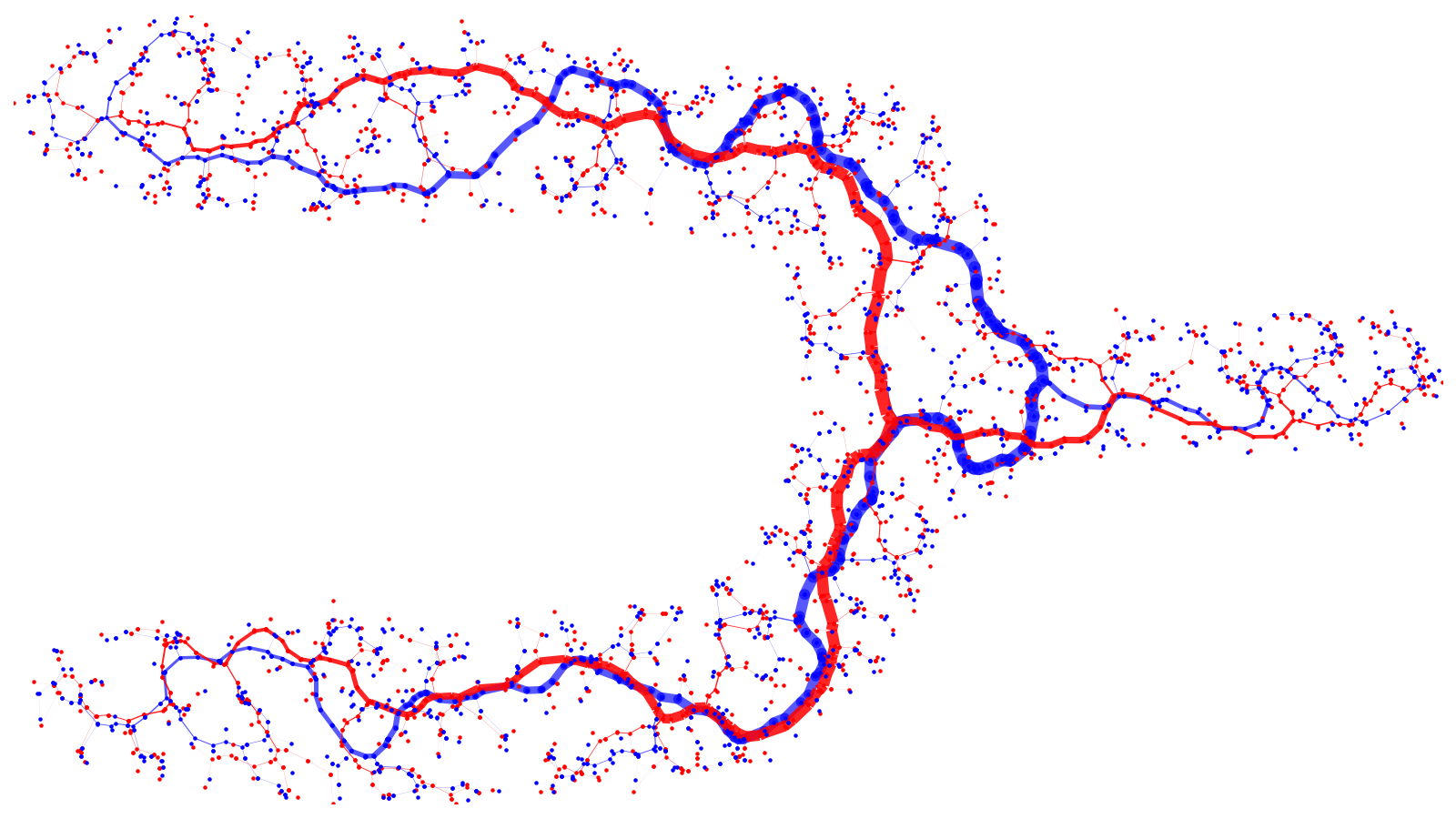}
		\caption{\BCST, $\alpha=0.2$\\\phantom{$\alpha=0.2$}}
		\label{sfig4:app_non_convex_CST_noiserobustness_toydata}
	\end{subfigure}
    \begin{subfigure}{0.25\linewidth}
        \centering
        \includegraphics[width=\linewidth]{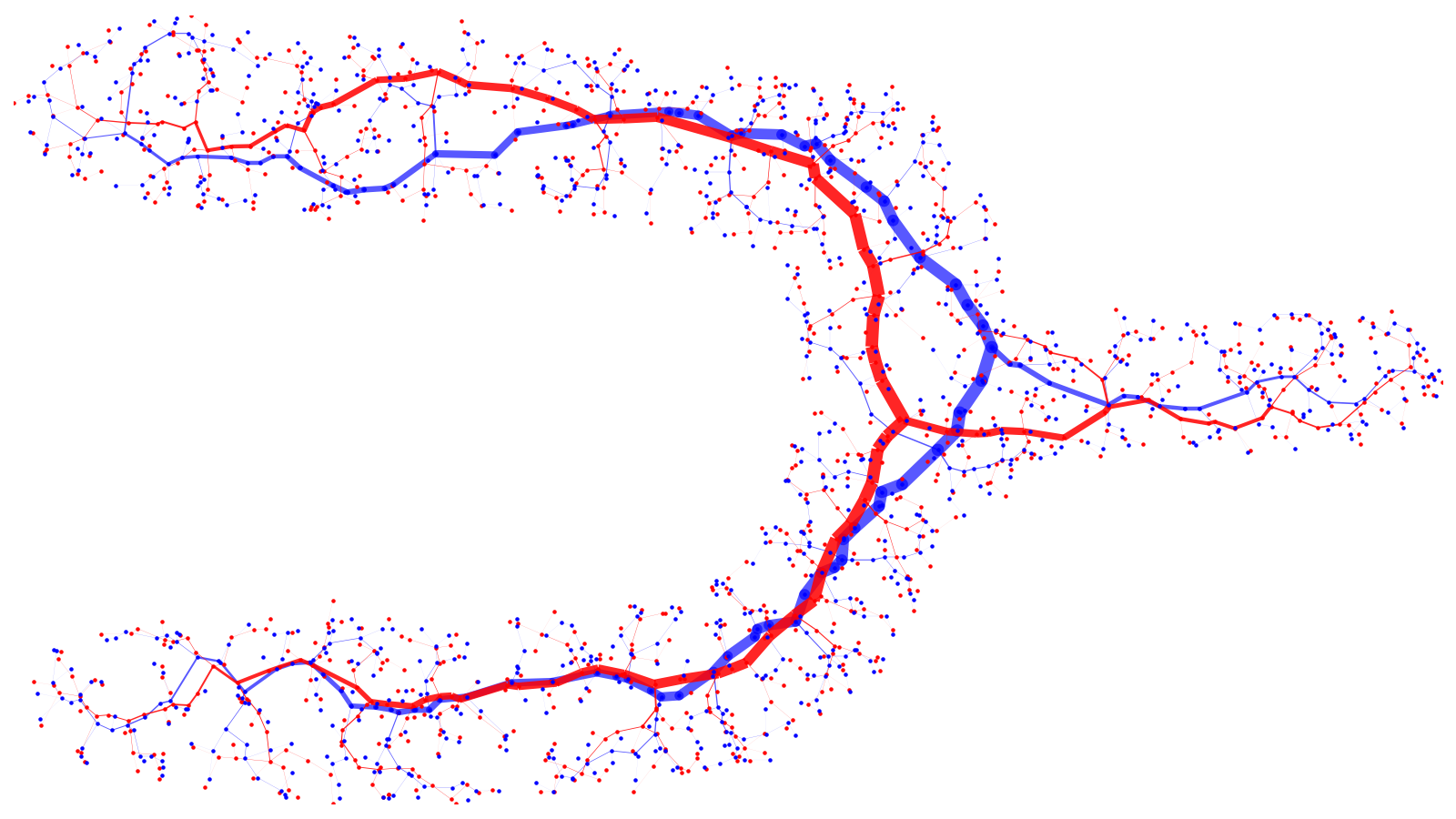}
         \caption{\CST, $\alpha=0.4$}
		\label{sfig5:app_non_convex_CST_noiserobustness_toydata}
	\end{subfigure}%
    \begin{subfigure}{0.25\linewidth}
    \centering
		\includegraphics[width=\linewidth]{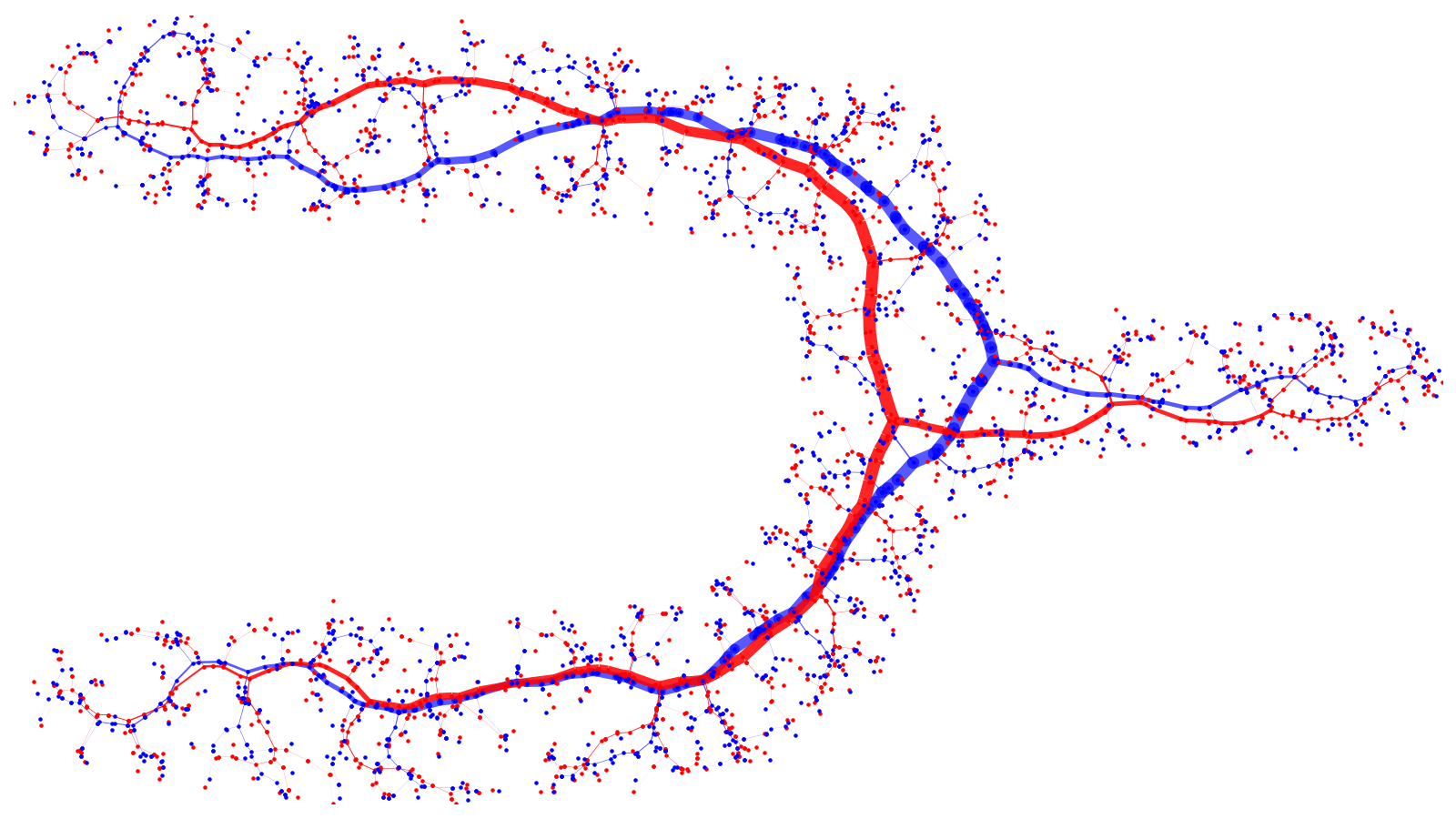}
		\caption{\BCST, $\alpha=0.4$}
		\label{sfig6:app_non_convex_CST_noiserobustness_toydata}
	\end{subfigure}%
    \begin{subfigure}{0.25\linewidth}
        \centering
        \includegraphics[width=\linewidth]{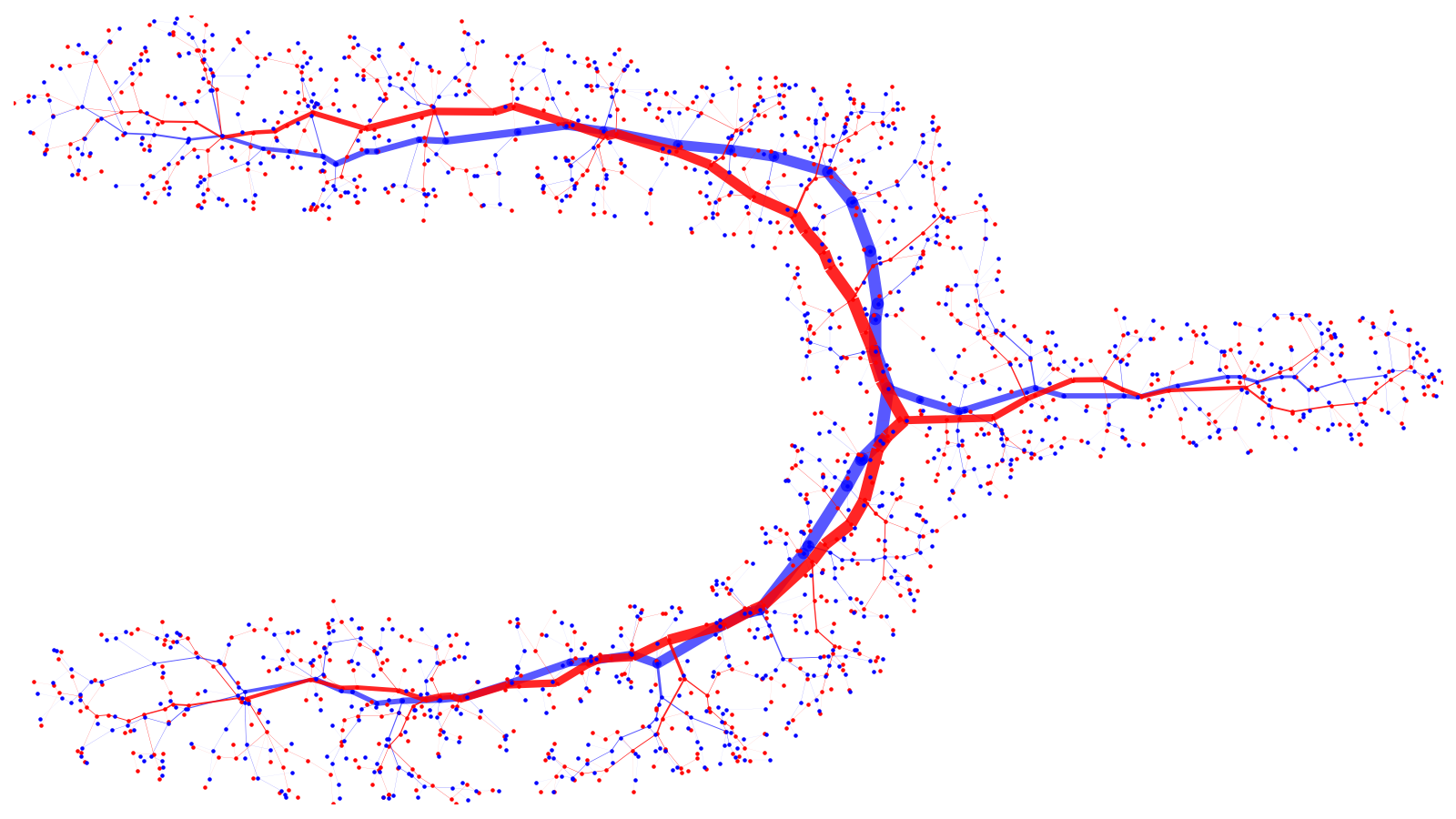}
         \caption{\CST, $\alpha=0.6$}
		\label{sfig7:app_non_convex_CST_noiserobustness_toydata}
	\end{subfigure}%
    \begin{subfigure}{0.25\linewidth}
    \centering
		\includegraphics[width=\linewidth]{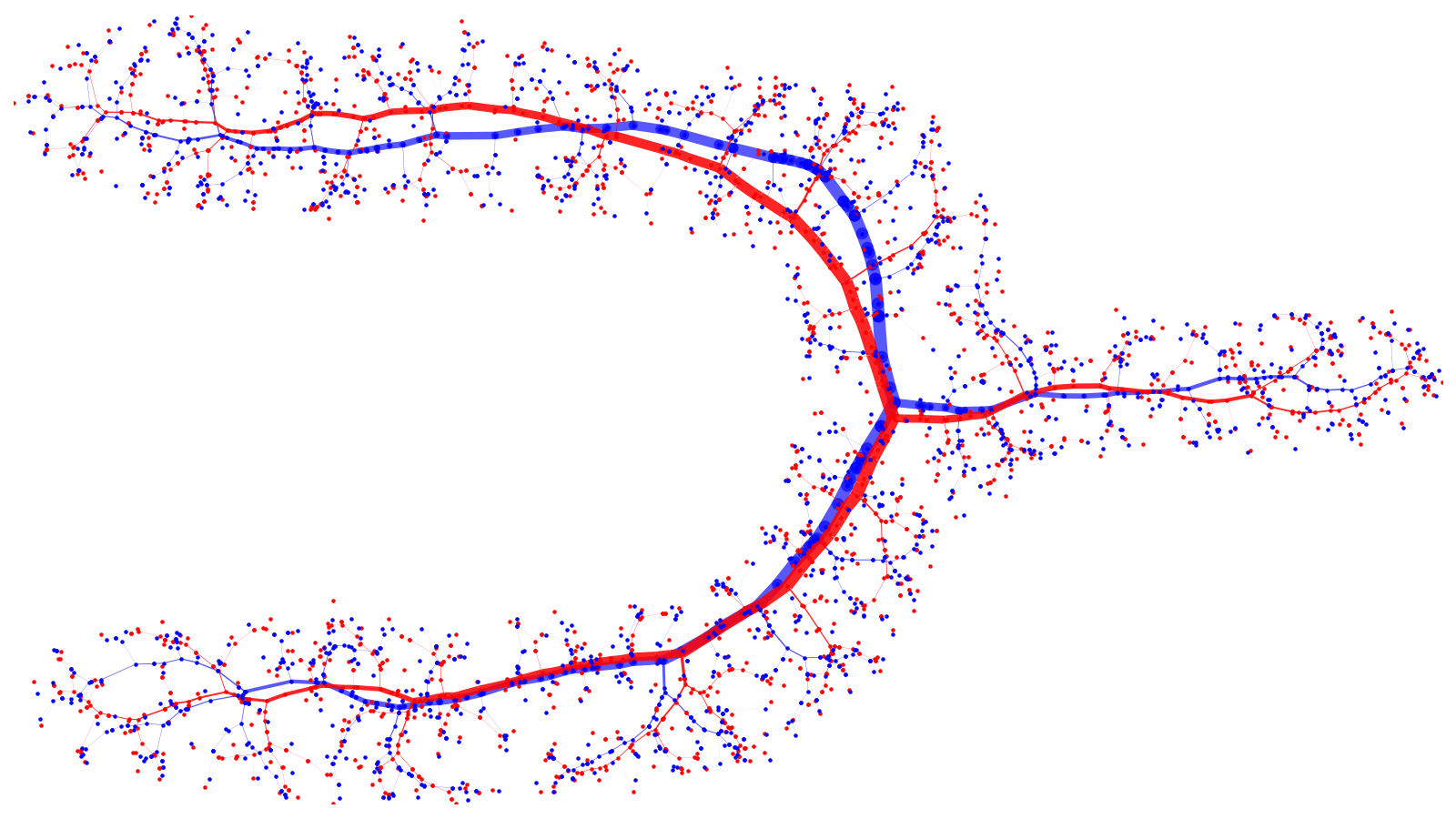}
		\caption{\BCST, $\alpha=0.6$}
		\label{sfig8:app_non_convex_CST_noiserobustness_toydata}
	\end{subfigure}
    \begin{subfigure}{0.25\linewidth}
        \centering
        \includegraphics[width=\linewidth]{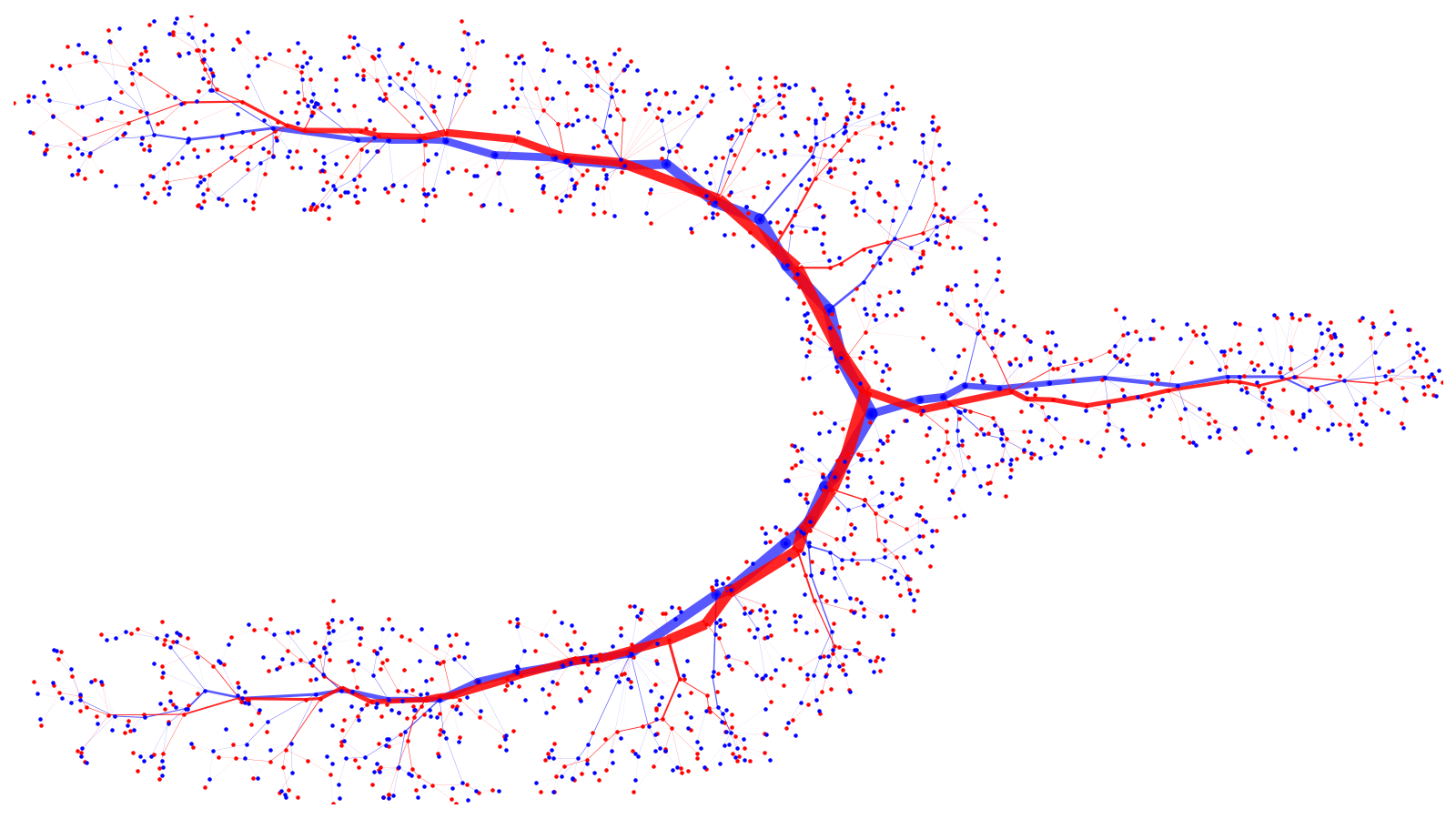}
         \caption{\CST, $\alpha=0.8$}
		\label{sfig9:app_non_convex_CST_noiserobustness_toydata}
	\end{subfigure}%
    \begin{subfigure}{0.25\linewidth}
    \centering
		\includegraphics[width=\linewidth]{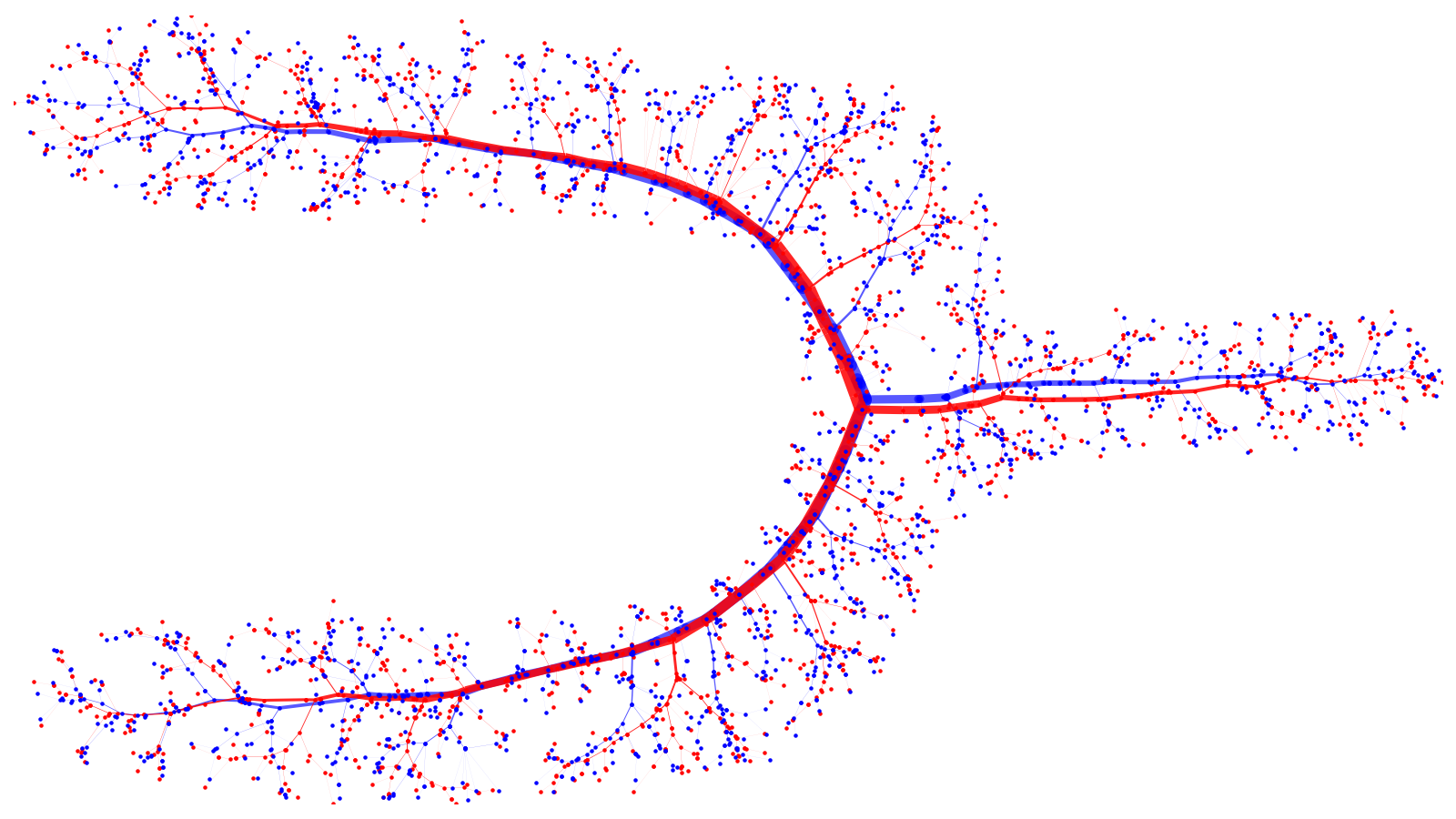}
		\caption{\BCST, $\alpha=0.8$}
		\label{sfig10:app_non_convex_CST_noiserobustness_toydata}
	\end{subfigure}%
    \begin{subfigure}{0.25\linewidth}
        \centering
        \includegraphics[width=\linewidth]{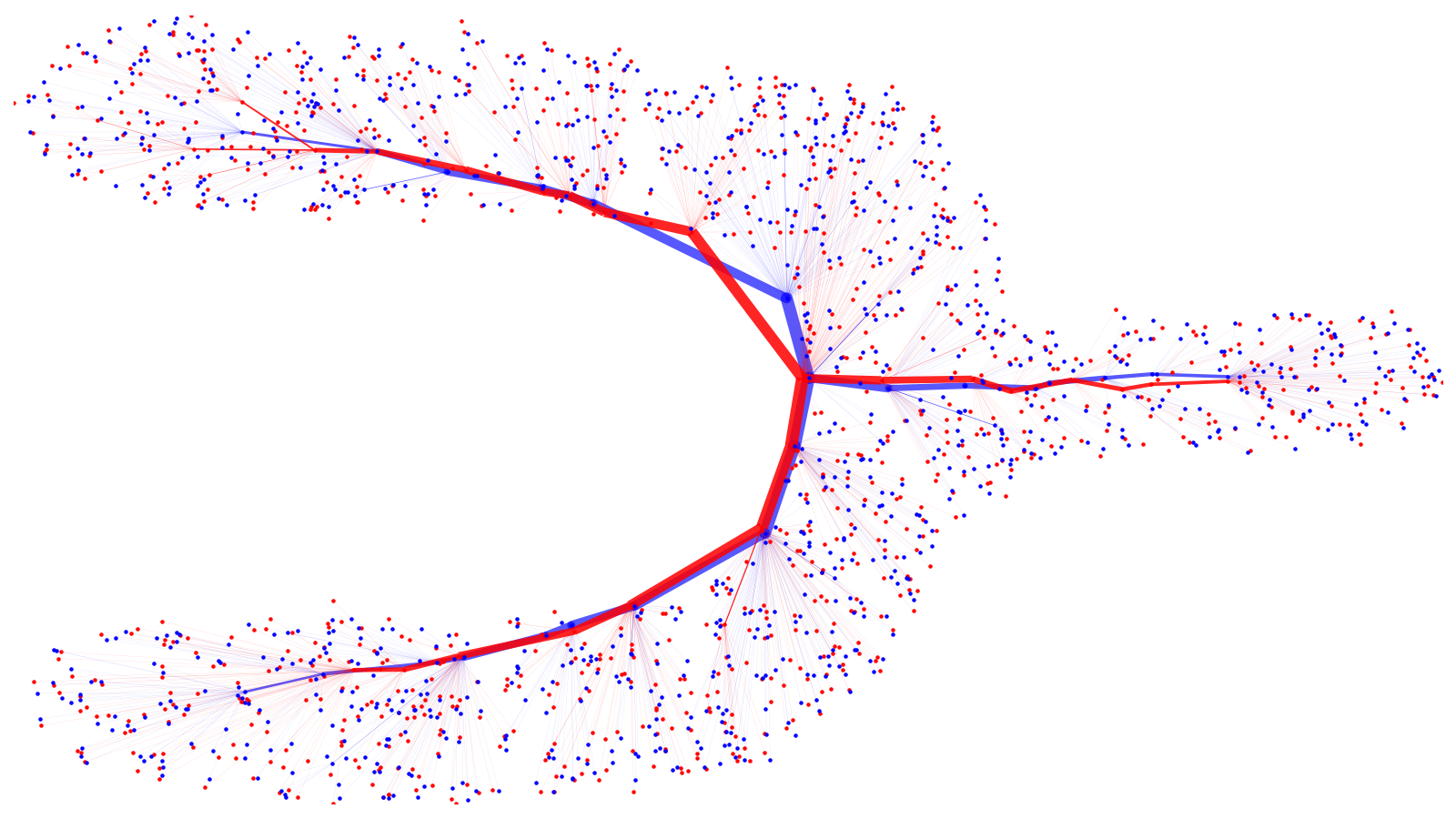}
         \caption{\CST, $\alpha=1.0$}
		\label{sfig11:app_non_convex_CST_noiserobustness_toydata}
	\end{subfigure}%
    \begin{subfigure}{0.25\linewidth}
    \centering
		\includegraphics[width=\linewidth]{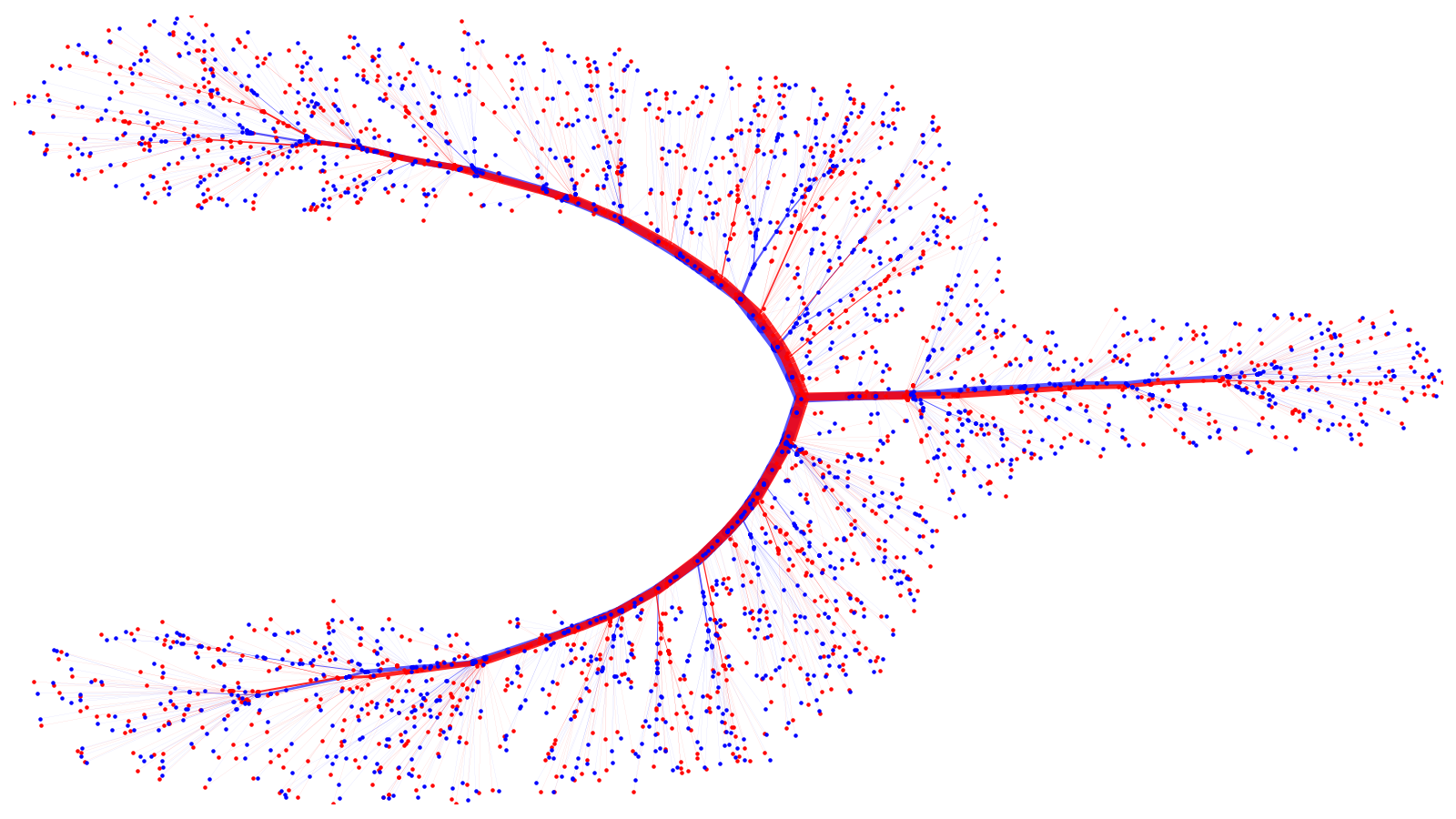}
		\caption{\BCST, $\alpha=1.0$}
		\label{sfig12:app_non_convex_CST_noiserobustness_toydata}
	\end{subfigure}

  \caption[(B)CST Examples from Non-Convex Uniform Distribution]{\textbf{(B)CST Examples from Non-Convex Uniform Distribution}.\CST and \BCST are computed for two perturbed instances generated by adding zero-centered Gaussian noise to points derived from a common sample uniformly taken within a non convex shape. (B)\CST for higher $\alpha$ values are more robust to noise and adhere to large scale structure in the data better.  The width of each edge is proportional to its centrality. All trees except  for the \mST were computed using the heuristic proposed in Section \ref{sec:heuristic}.}
	\label{fig:app_non_convex_CST_noiserobustness_toydata}	
\end{figure}

In this section, we show how stable the \BCST and \CST for different $\alpha$ values are. We sample 1000 points uniformly from uniform distributions over different supports and perturb them by adding zero centered Gaussian noise. We generate two perturbations and show how the tree structure evolves across different values. See \figurename{s} \ref{fig:app_uniform_CST_noiserobustness_toydata} 
 \ref{fig:app_triangle_uniform_CST_noiserobustness_toydata} and \ref{fig:app_non_convex_CST_noiserobustness_toydata}. As we increase the value of $\alpha$, the trees exhibit a more pronounced "star-shaped" pattern and enhanced stability. The parameter $\alpha$ provides a trade-off mechanism between preserving the structure of the data and ensuring the stability of the resulting tree.

\newpage
\newpage

\section{Applications}\label{sec:app_applications}
\subsection{Single cell transcriptomic data}
Single-cell transcriptomics analyzes the gene expression levels of individual cells in a particular population by counting the RNA transcripts of  genes at a given time. The high dimensional single cell RNA-sequencing data can be used to model the gene expression dynamics of a cell population as well as the cell differentiation process. The reconstruction of these trajectories can help discover which genes are critical to understand the underlying biological process. It is often assumed that these trajectories can be represented as trees \citep{saelens_comparison_2019,street_slingshot_2018}, and therefore the (B)\CST can be applied to model such trajectories.

In this section we will give additional details on the example shown in the main paper regarding the Paul dataset. In addition, we show another example of the performance of the (B)CST with a different dataset, namely the Setty dataset \citep{setty2019characterization}

\paragraph{Paul dataset} The Paul datset consists of gene expressions measurements of cells of mouse bone marrow \citet{paul_transcriptional_2015}. The original dataset is formed by 2730 cells each with 3451 gene measurements. The data is preprocessed using the recipe described in \citet{zheng_massively_2017}, which reduces the dimensionality to 1000 by selecting the most relevant genes. We further reduce the dimensionality of the data, by applying PCA with 50 principal components. Finally, we apply the corresponding spanning tree algorithm. For visualization purposes, we used the PAGA algorithm~\citep{wolf_paga_2019}, one of the best algorithms for single cell trajectory inference \citep{saelens_comparison_2019}. PAGA was designed to faithfully represent the trajectories. Thus, if a spanning tree aligns well with the embedding, this is an indication that the tree approximates the trajectory well. \figurename{} \ref{fig:stability_paul} in the main part shows the mST, Steiner tree ($\alpha=0$) and CST, BCST (at $\alpha=0.5$) and (B)\MRCT ($\alpha=1$) of the original sample and a perturbed sample with 50\% of the cells randomly sampled. The \mST and Steiner tree miss the highlighted bifurcation and they are more sensitive to the noise. The \CST and \BCST are robust to the perturbation and align well with the PAGA embedding, though the \CST may not reconstruct well the finer details. The addition of \BPs enables the \BCST to follow the trajectory more closely. The \MRCT results in a star tree, and its branched version nearly resembles a star tree as well. In high-dimensional data, such as the Paul dataset with 50 dimensions, the star-shaped tendency becomes more prominent. For $\alpha=1$, most of the intricate structure of the data is lost. Therefore, for higher dimensions lower $\alpha$ values become more relevant.

\paragraph{Setty dataset:} 
Extending our analysis, we applied the same experiment from Section \ref{sec:real_word_data} to a different real dataset: the human bone marrow dataset from \citet{setty2019characterization} referred here as the Setti dataset. After applying the same preprocessing as the one used for the Paul dataset, this dataset consists of 5780 samples within a 50-dimensional space, as opposed to the unprocessed data, which existed in a 14651-dimensional space. We computed \BCST, \CST, Steiner tree and \mST on both the original dataset and a downsampled version (25\% of the original), comparable in size to the Paul dataset's downsampled version. For visualization purposes we used the TSNE projection of the data provided by the scvelo library. \footnote{\url{https://scvelo.readthedocs.io/en/stable/scvelo.datasets.bonemarrow/}} \figurename{} \ref{fig:stability_setty} presents the
results, showing a consistent pattern akin to the Paul dataset. \BCST and \CST exhibited superior robustness compared to \mST and Steiner tree, with \BCST providing enhanced trajectory modeling.\footnote{In this particular case, we have excluded the $\alpha=1.00$ scenario. This is because, similar to the Paul dataset, both the \CST and \BCST algorithms yield star-shaped graphs under this condition.} This additional example reinforces the efficacy and robustness of our proposed methods.

\def \bottomvspace{-0.0}
\def \topvspace{-0.0}
\def \scalaabox{.75}
\begin{figure}[t!]
	\centering
	\scalebox{\scalaabox}{\begin{subfigure}{0.33\linewidth}
			\centering
			\vspace*{\topvspace cm}
			\includegraphics[width=1\linewidth]{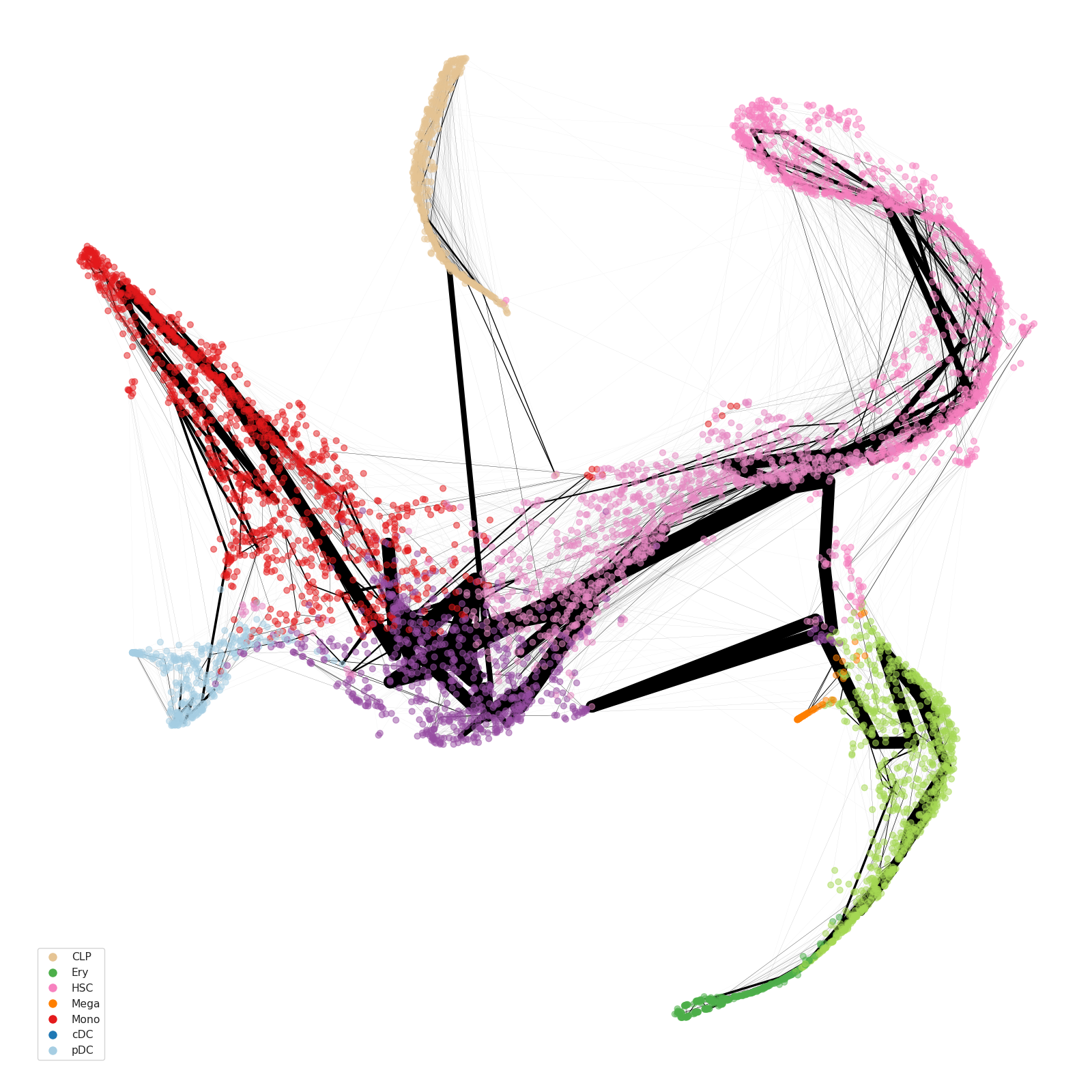}
					\vspace{\bottomvspace cm}
			\caption{Original; \mST}
			\label{sfig1:stability_setty}
		\end{subfigure}%
		\begin{subfigure}{0.33\linewidth}
			\centering
			\vspace*{\topvspace cm}
			\includegraphics[width=1\linewidth]{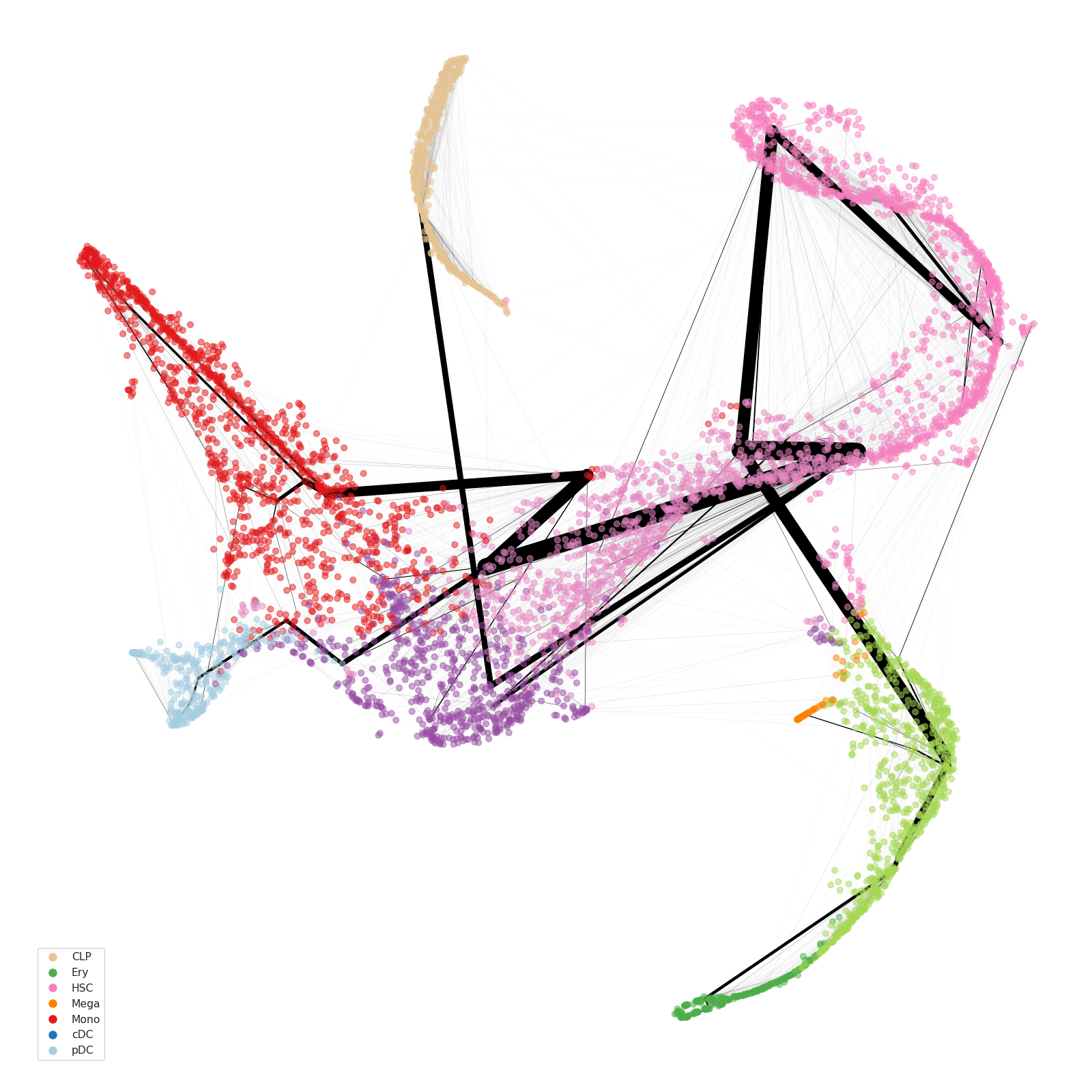}
			\vspace{\bottomvspace cm}
			\caption{Original; \CST $\alpha=0.40$}
			\label{sfig2:stability_setty}
		\end{subfigure}
		\begin{subfigure}{0.33\linewidth}
			\centering
			\vspace*{\topvspace cm}
			\includegraphics[width=1\linewidth]{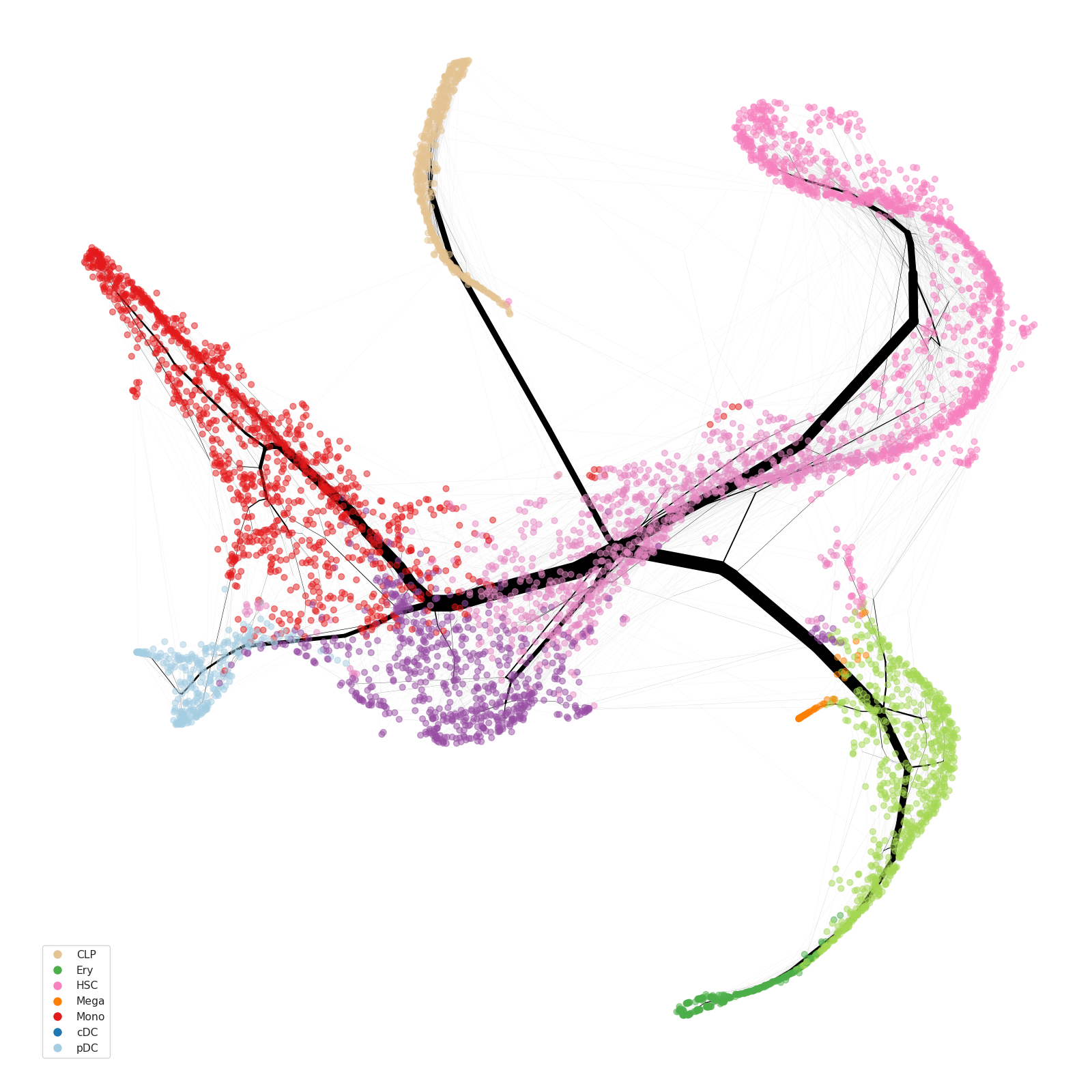}
			\vspace{\bottomvspace cm}
			\caption{Original; \BCST $\alpha=0.40$}
			\label{sfig3:stability_setty}
	\end{subfigure}}
	\scalebox{\scalaabox}{\begin{subfigure}{0.33\linewidth}
			\centering
			\vspace*{\topvspace cm}
			\includegraphics[width=1\linewidth]{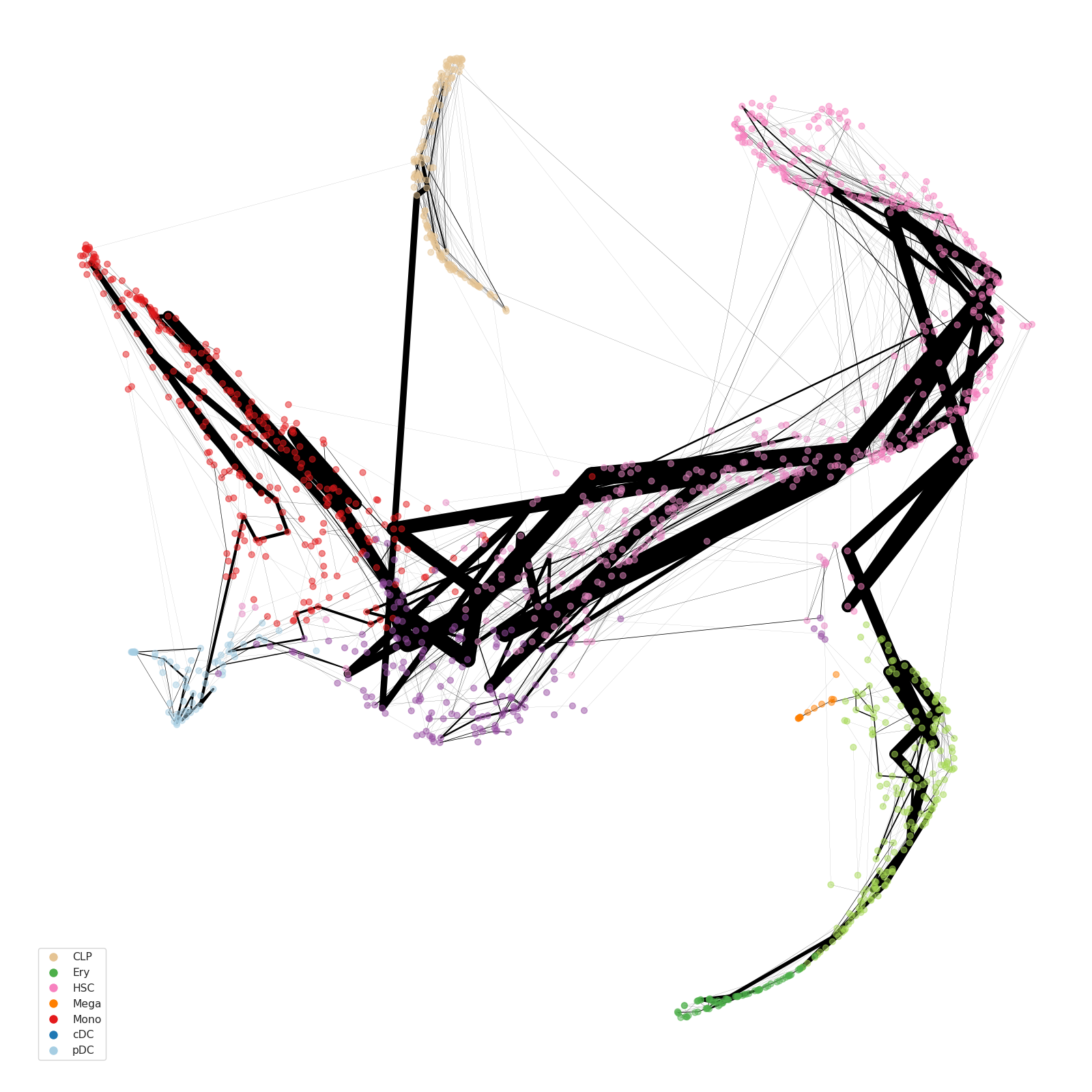}
			\vspace{\bottomvspace cm}
			\caption{Subsample; \mST}
			\label{sfig4:stability_setty}
		\end{subfigure}%
		\begin{subfigure}{0.33\linewidth}
			\centering
			\vspace*{\topvspace cm}
			\includegraphics[width=1\linewidth]{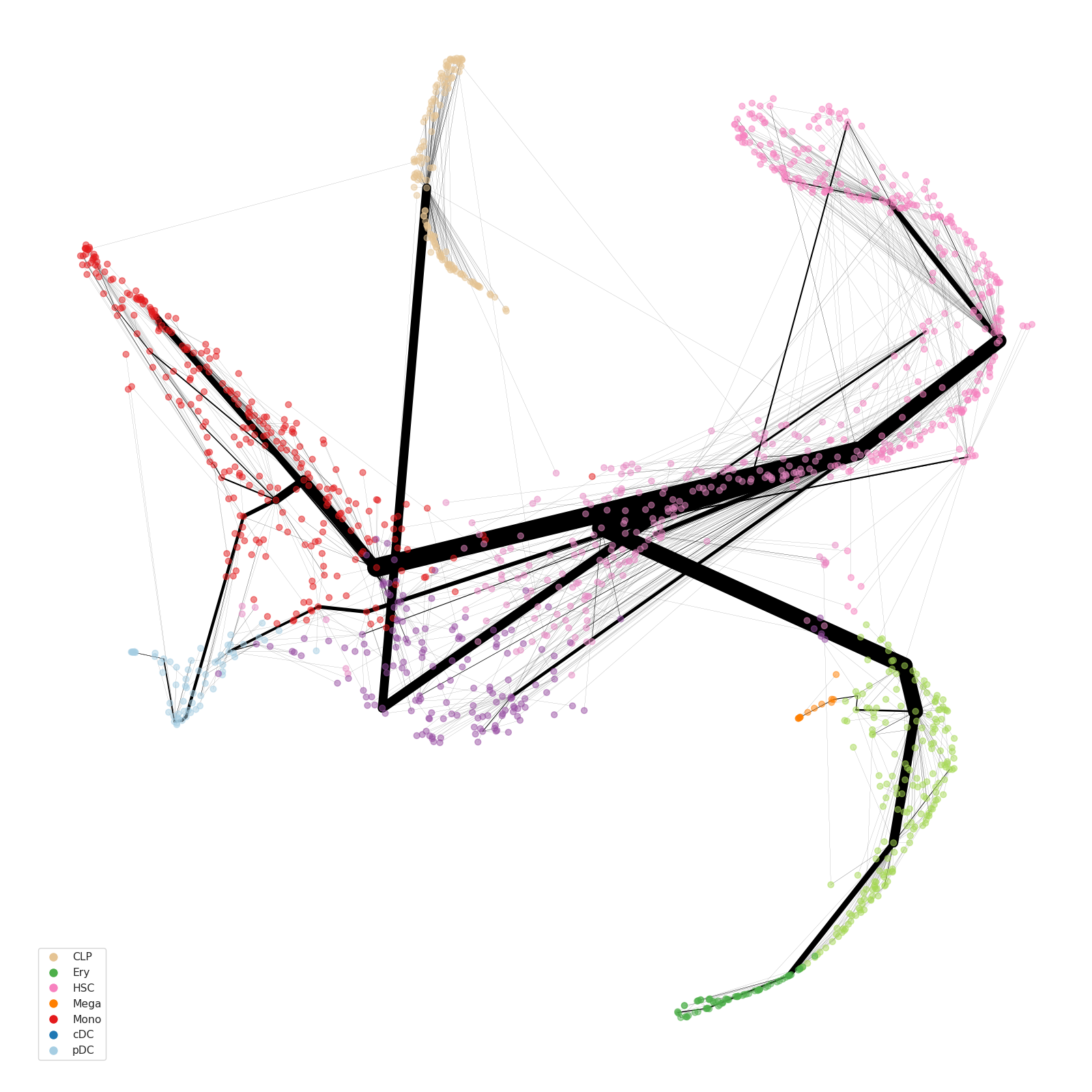}
			\vspace{\bottomvspace cm}
			\caption{Subsample; \CST $\alpha=0.40$}
			\label{sfig5:stability_setty}
		\end{subfigure}%
		\begin{subfigure}{0.33\linewidth}
			\centering
			\vspace*{\topvspace cm}
			\includegraphics[width=1\linewidth]{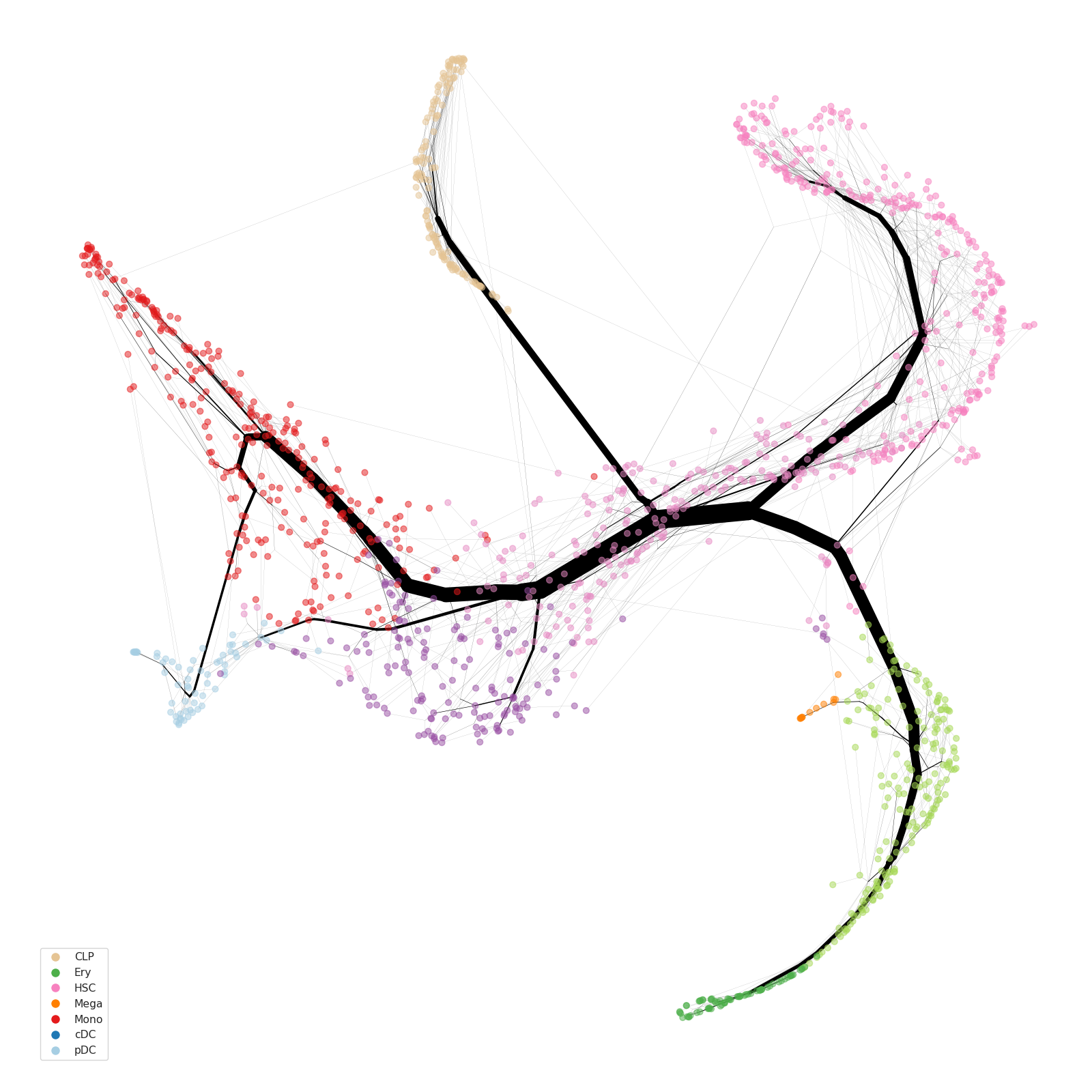}
			\vspace{\bottomvspace cm}
			\caption{Subsample; \BCST $\alpha=0.40$}
			\label{sfig6:stability_setty}
	\end{subfigure}}
	\caption[mST, CST and BCST of the the Setty dataset]{\textbf{mST, CST and BCST of the the Setty dataset \citep{setty2019characterization}}. We applied the algorithms on the original data (top) and perturbed version where 25\% of the points have been randomly removed (bottom). We use the TSNE algorithm to visualize the data in 2D, though the trees were computed on a 50 dimensional PCA projection of the preprocessed data.  Colors correspond to different ground truth cell populations. The width of the edges is proportional to their centralities. The \mST fails to accurately represent the trajectory and proves to be highly susceptible to noise. The \CST and \BCST are more faithful and robust, though the \BCST performance is superior.}
\label{fig:stability_setty}
\end{figure}

\subsection{3D Plant skeletonization}

Plant skeletonization is a foundational technique for elucidating growth patterns, branching hierarchies, and responses to environmental factors in plants. It simplifies complex plant structures into skeletal representations, often described by spanning trees. In this section, we demonstrate how the \BCST can model a plant's skeleton using a point cloud of its surface.\footnote{Note that we do not test the \CST in this context, as the ideal skeleton should closely follow the object's centerline. While the \BCST has Steiner points, which naturally align with the center of the surrounding points to minimize the distance, the \CST lacks this flexibility. Consequently, the backbone of the \CST can not align with the centerline, given that the terminals lie on the surface.}  We provide additional examples beyond those presented in the main paper.

For the skeletonization process, we utilized the 4D Plant Registration Dataset.\footnote{Data accessible at \url{https://www.ipb.uni-bonn.de/data/4d-plant-registration/}} It consists of 3D point cloud data that captures the surface of different plants at different growth stages. Specifically, our analysis focused on the point clouds of the tomato plant on days 5, 8, and 13. From each point cloud, we subsample uniformly at random 5000 points and then we compute the \BCST.

\begin{figure}[h!]
	\centering
	\caption*{Tomato plant 2, day 5.}
    \begin{subfigure}{0.25\linewidth}
        \centering
        \includegraphics[width=\linewidth]{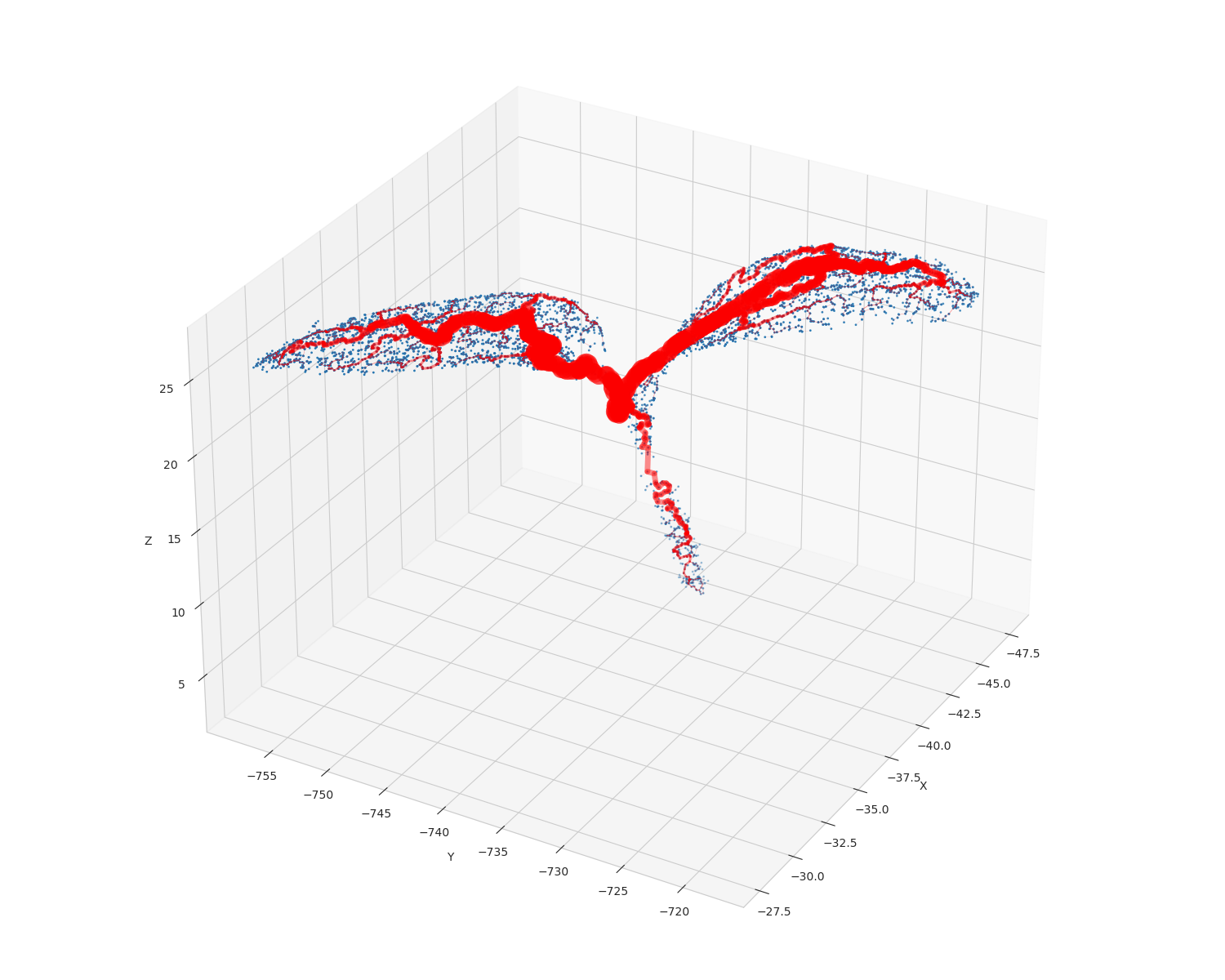}
        \caption{\BCST $\alpha=0.00$}
		\label{sfig1:tomato_plant2}
	\end{subfigure}%
    \begin{subfigure}{0.25\linewidth}
    \centering
		\includegraphics[width=\linewidth]{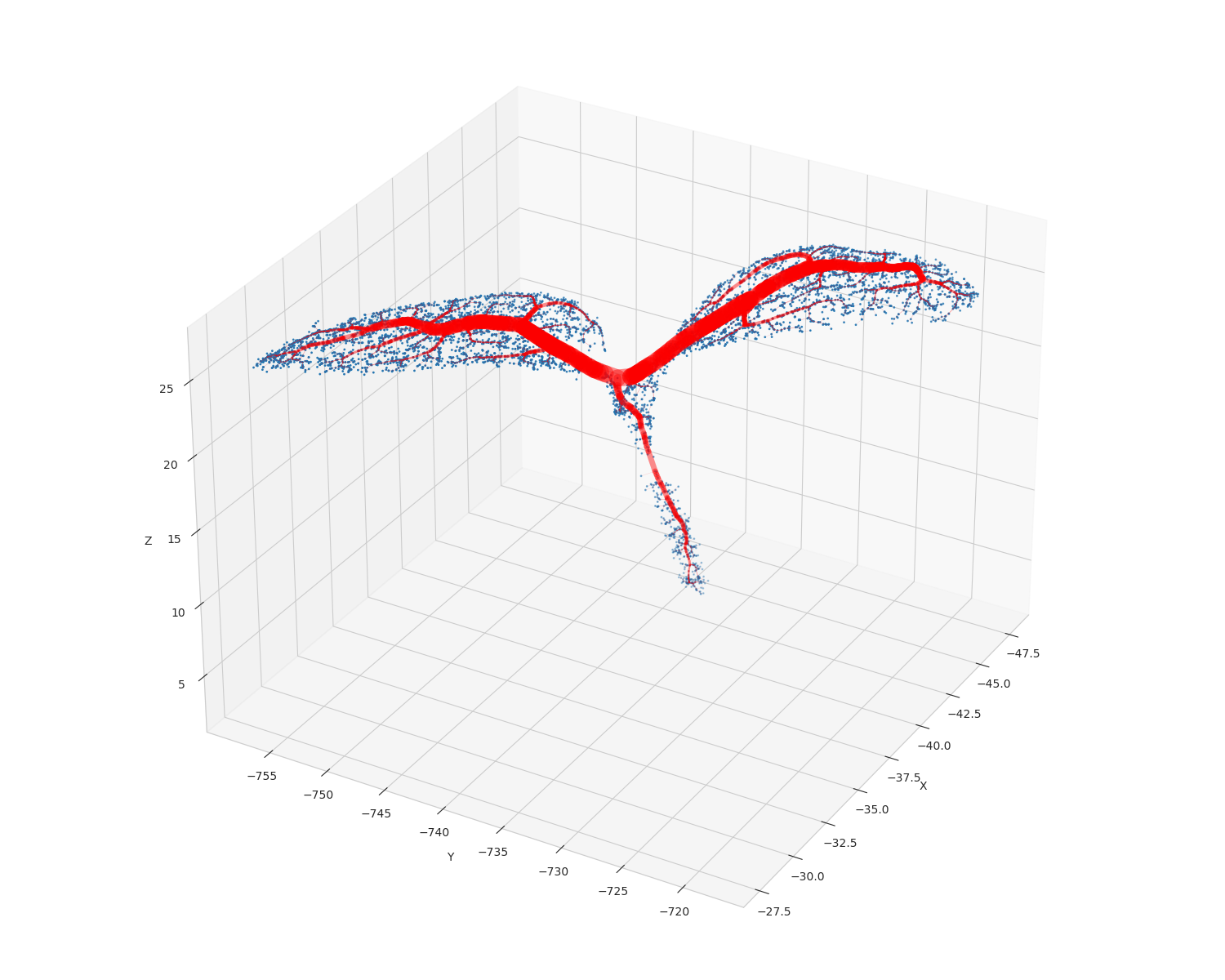}
		\caption{\BCST $\alpha=0.50$}
		\label{sfig2:tomato_plant2}
	\end{subfigure}%
    \begin{subfigure}{0.25\linewidth}
        \centering
        \includegraphics[width=\linewidth]{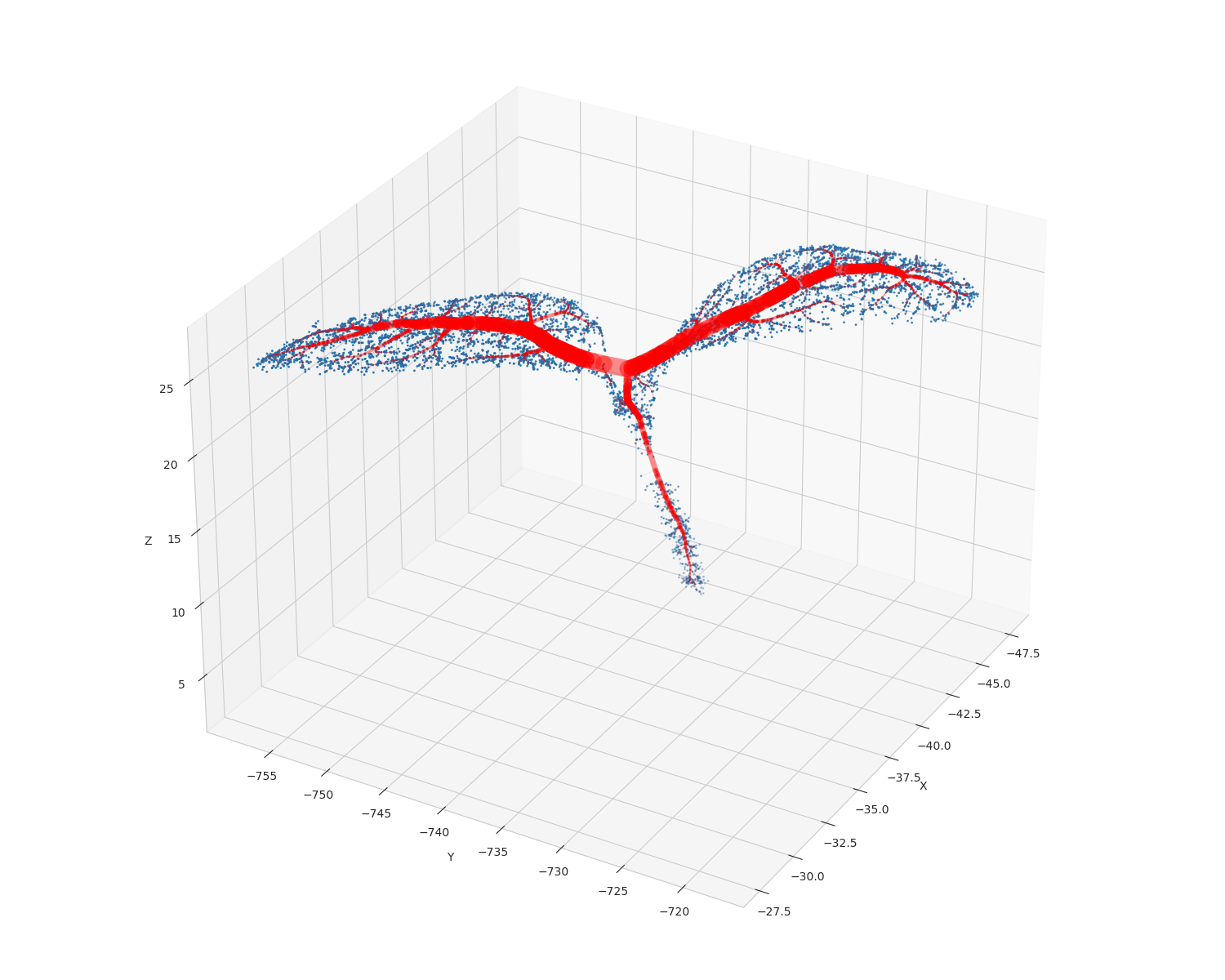}
        \caption{\BCST $\alpha=0.70$}
        \label{sfig3:tomato_plant2}
	\end{subfigure}%
    \begin{subfigure}{0.25\linewidth}
    	\centering
    	\includegraphics[width=\linewidth]{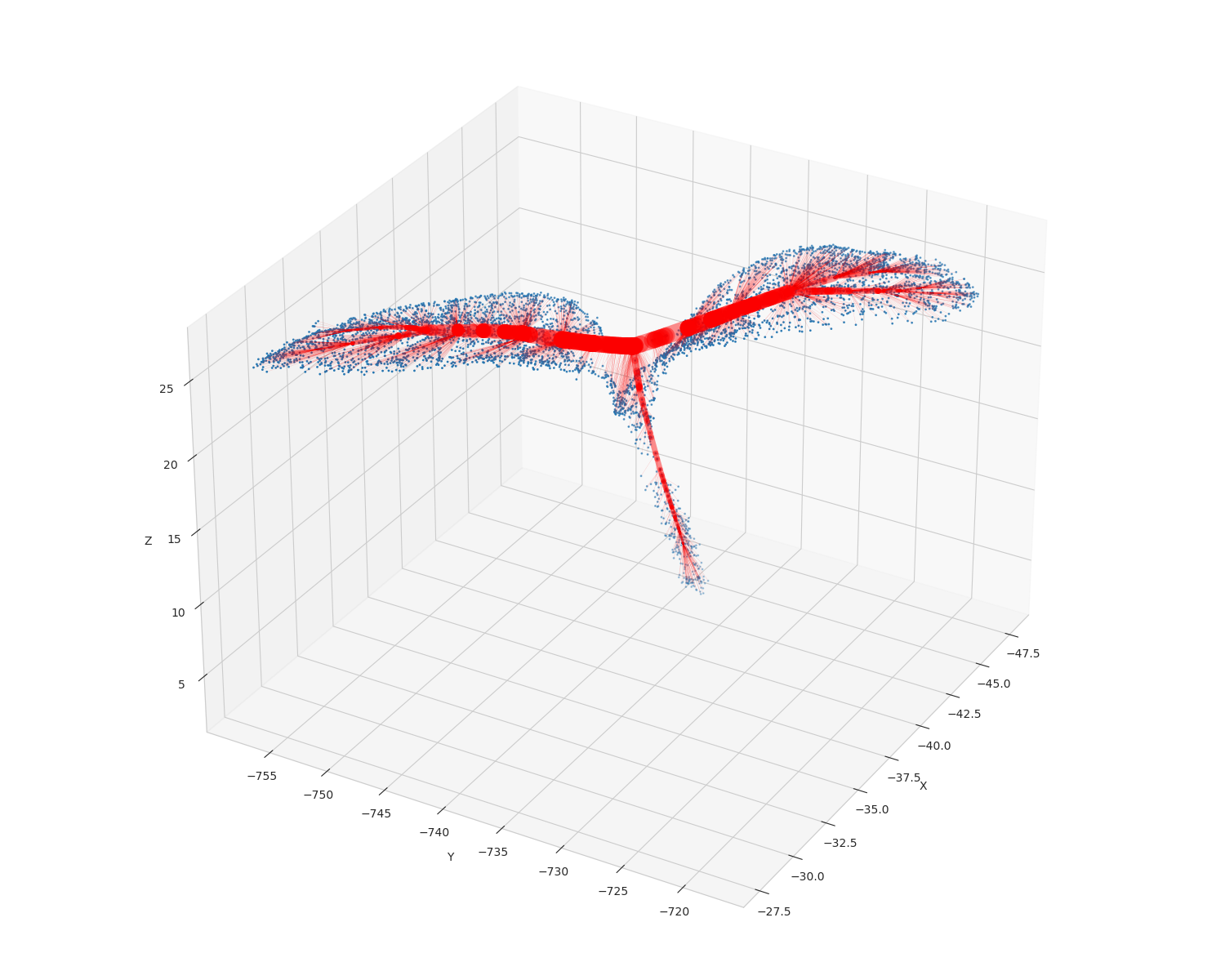}
    	\caption{\BCST $\alpha=1.00$}
    	\label{sfig4:tomato_plant2}
    \end{subfigure}%

	\centering
	\caption*{Tomato plant 2, day 8.}
	\begin{subfigure}{0.25\linewidth}
		\centering
		\includegraphics[width=\linewidth]{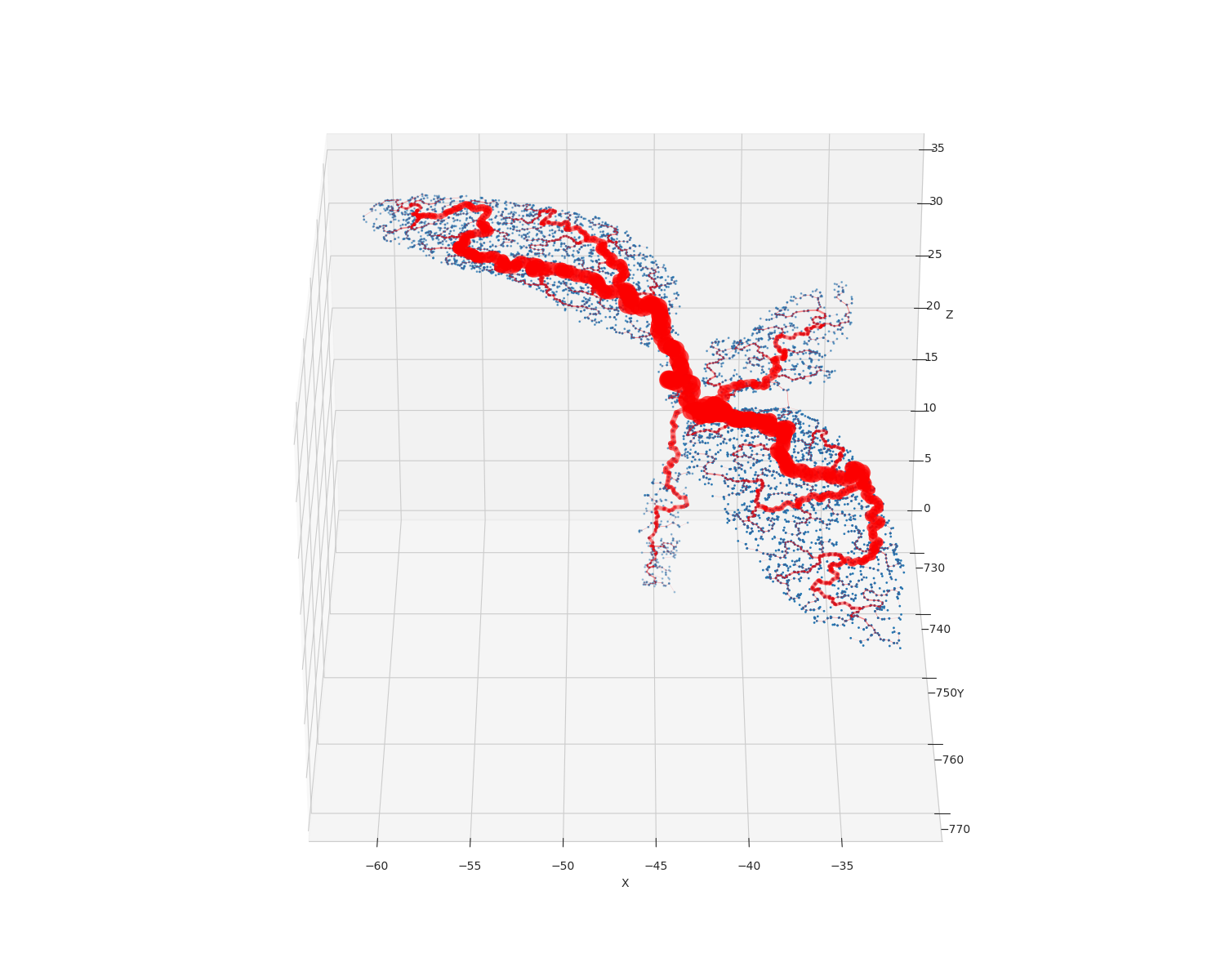}
		\caption{\BCST $\alpha=0.00$}
		\label{sfig5:tomato_plant2}
	\end{subfigure}%
	\begin{subfigure}{0.25\linewidth}
		\centering
		\includegraphics[width=\linewidth]{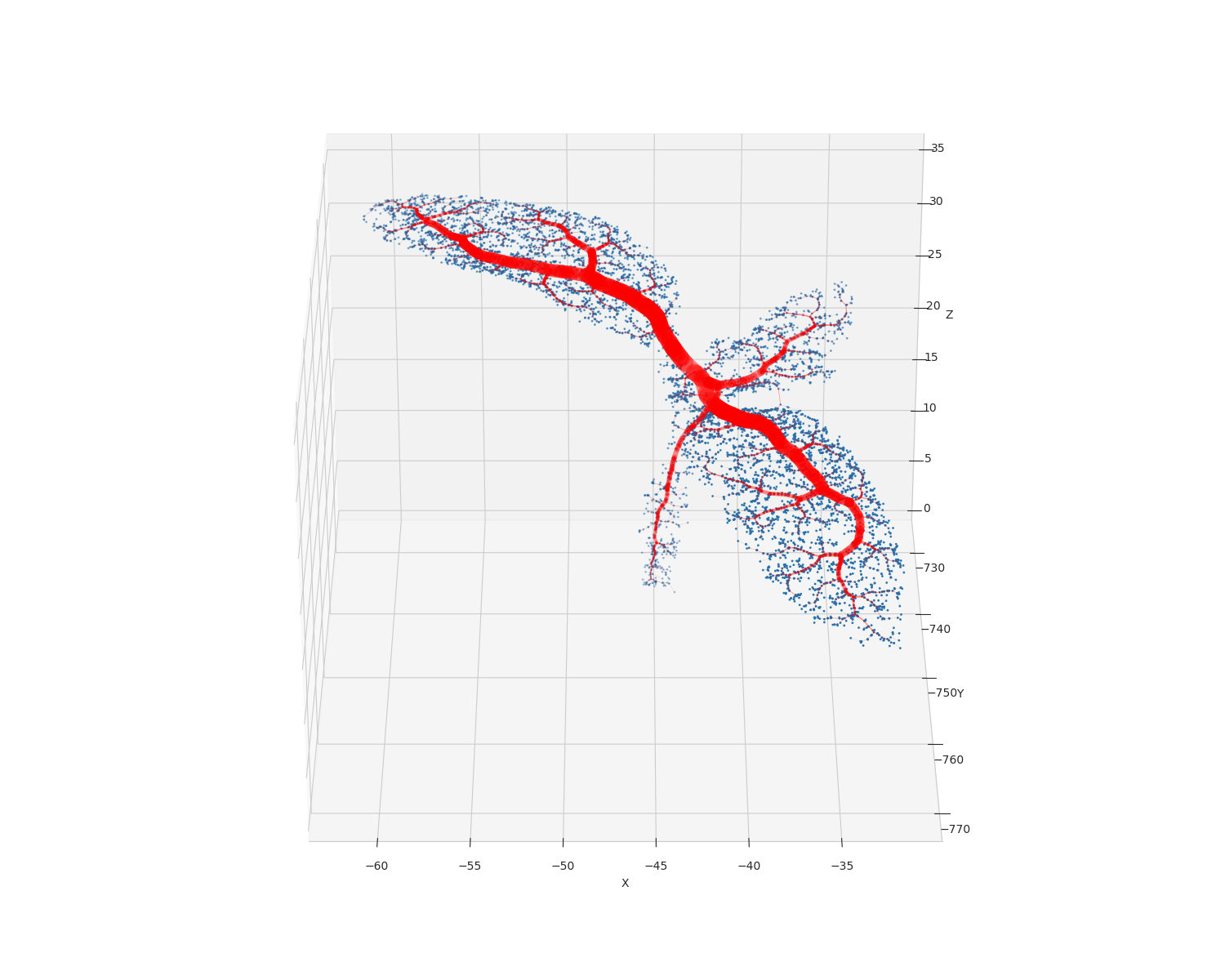}
		\caption{\BCST $\alpha=0.50$}
		\label{sfig6:tomato_plant2}
	\end{subfigure}%
	\begin{subfigure}{0.25\linewidth}
		\centering
		\includegraphics[width=\linewidth]{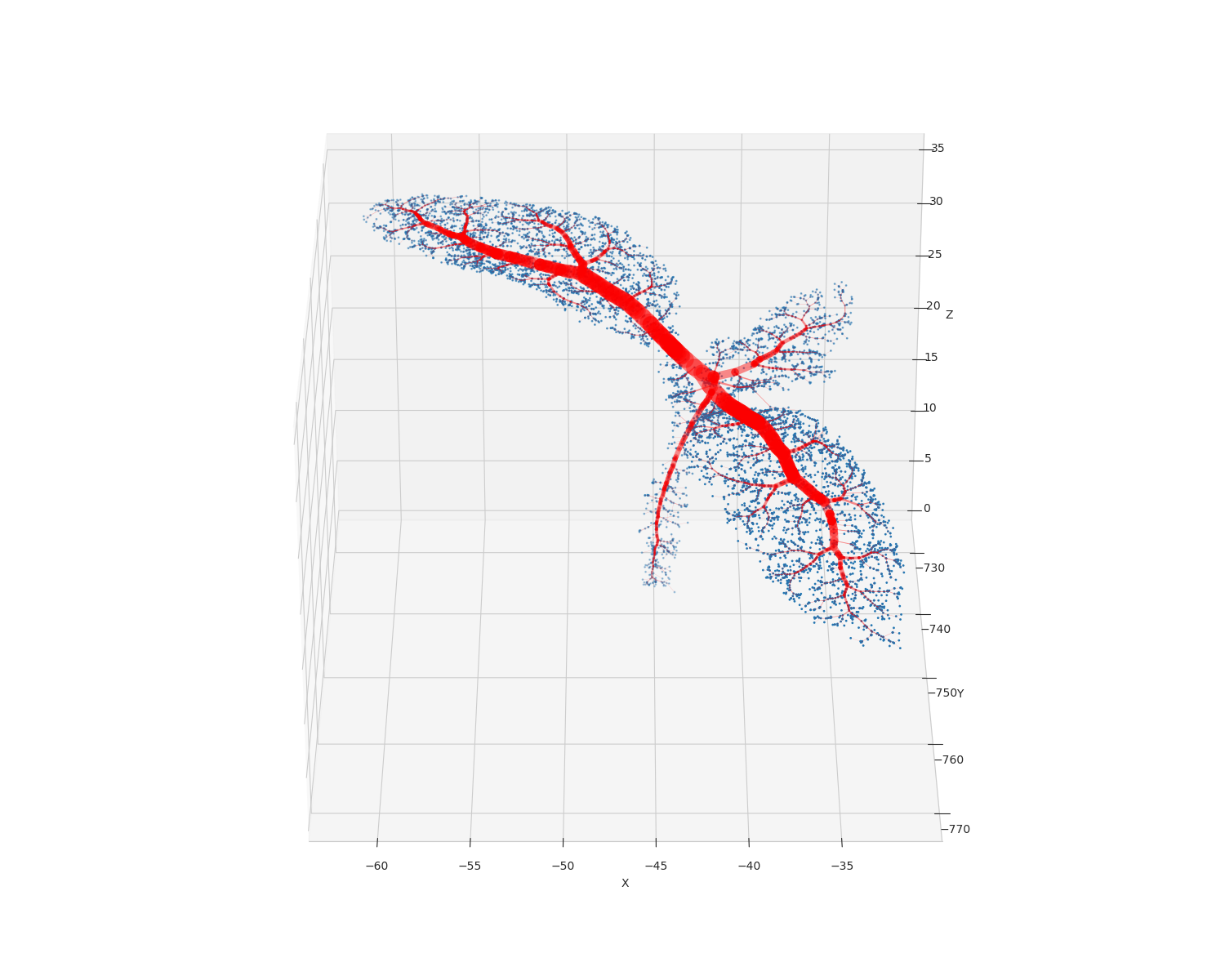}
		\caption{\BCST $\alpha=0.70$}
		\label{sfig7:tomato_plant2}
	\end{subfigure}%
	\begin{subfigure}{0.25\linewidth}
		\centering
		\includegraphics[width=\linewidth]{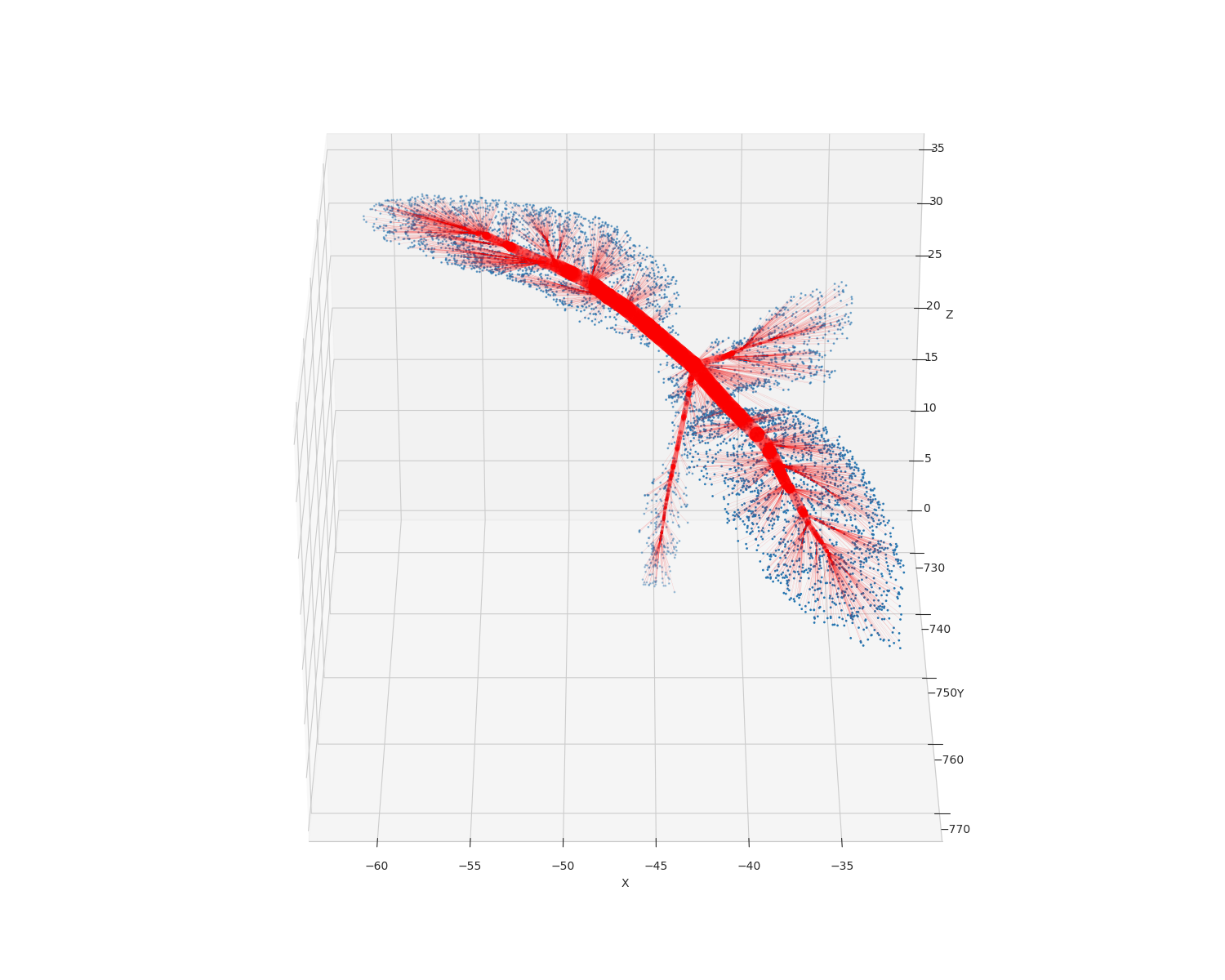}
		\caption{\BCST $\alpha=1.00$}
		\label{sfig8:tomato_plant2}
	\end{subfigure}%
	\centering
	\caption*{Tomato plant 2, day 13.}
	\begin{subfigure}{0.25\linewidth}
		\centering
		\includegraphics[width=\linewidth]{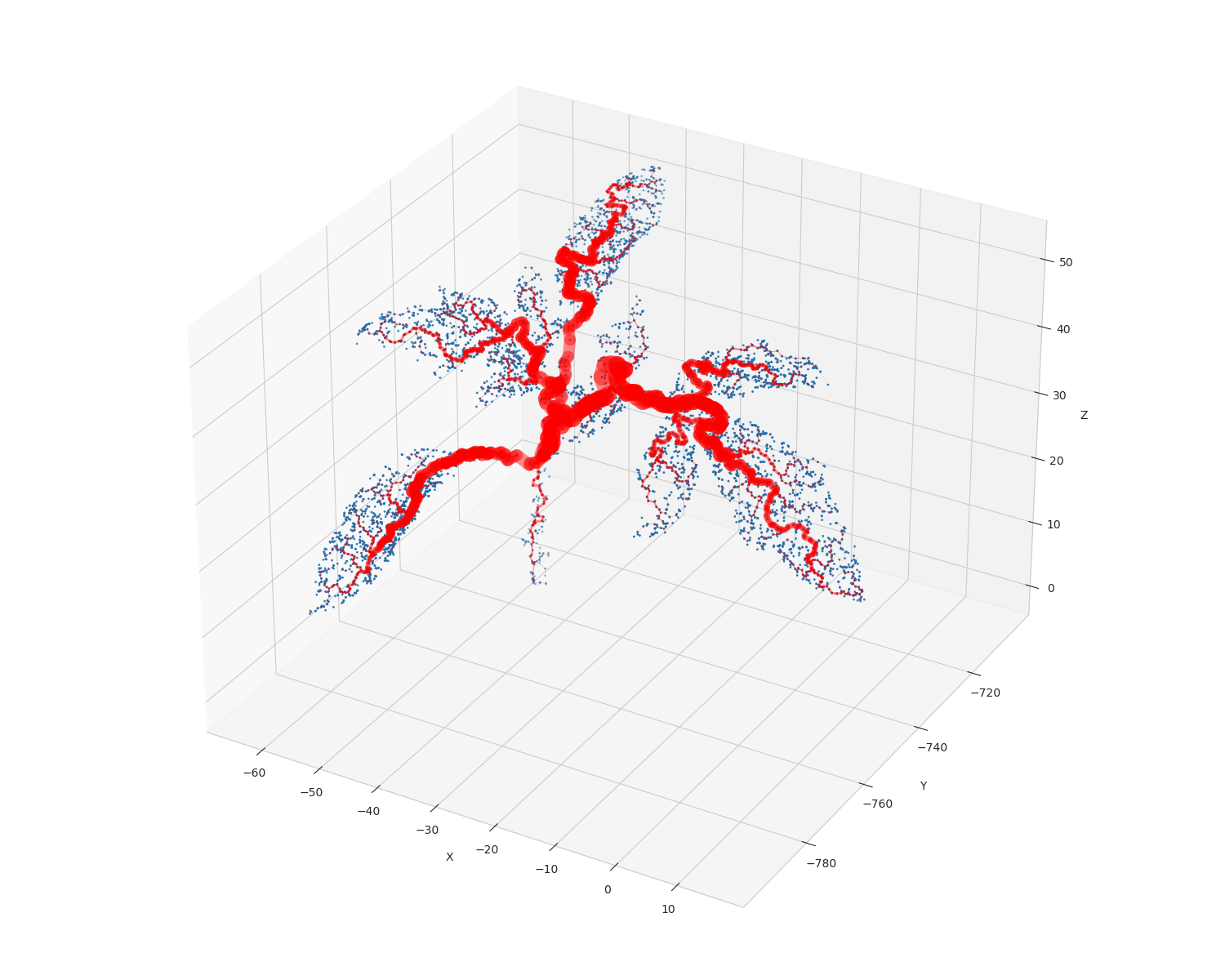}
		\caption{\BCST $\alpha=0.00$}
		\label{sfig9:tomato_plant2}
	\end{subfigure}%
	\begin{subfigure}{0.25\linewidth}
		\centering
		\includegraphics[width=\linewidth]{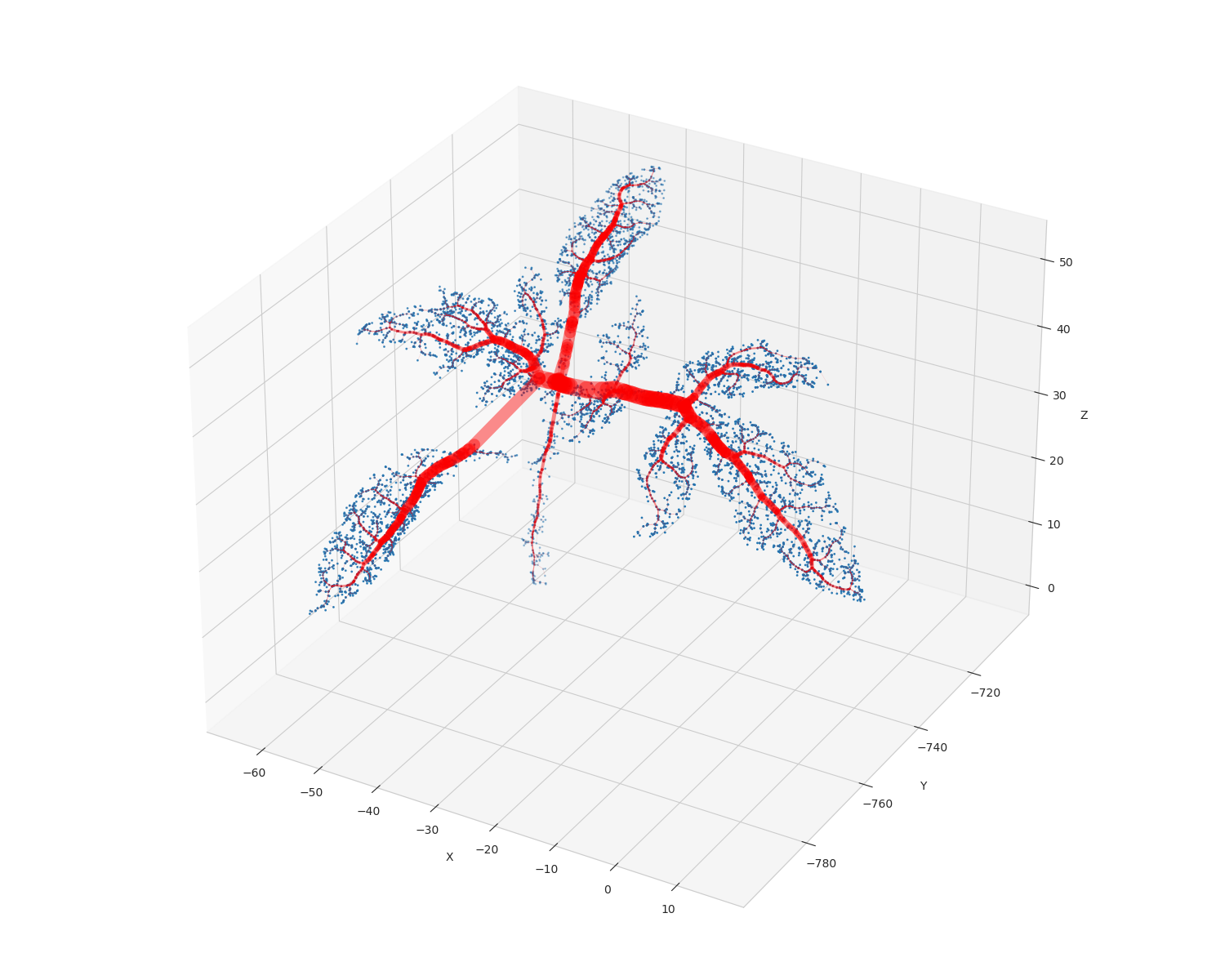}
		\caption{\BCST $\alpha=0.50$}
		\label{sfig10:tomato_plant2}
	\end{subfigure}%
	\begin{subfigure}{0.25\linewidth}
		\centering
		\includegraphics[width=\linewidth]{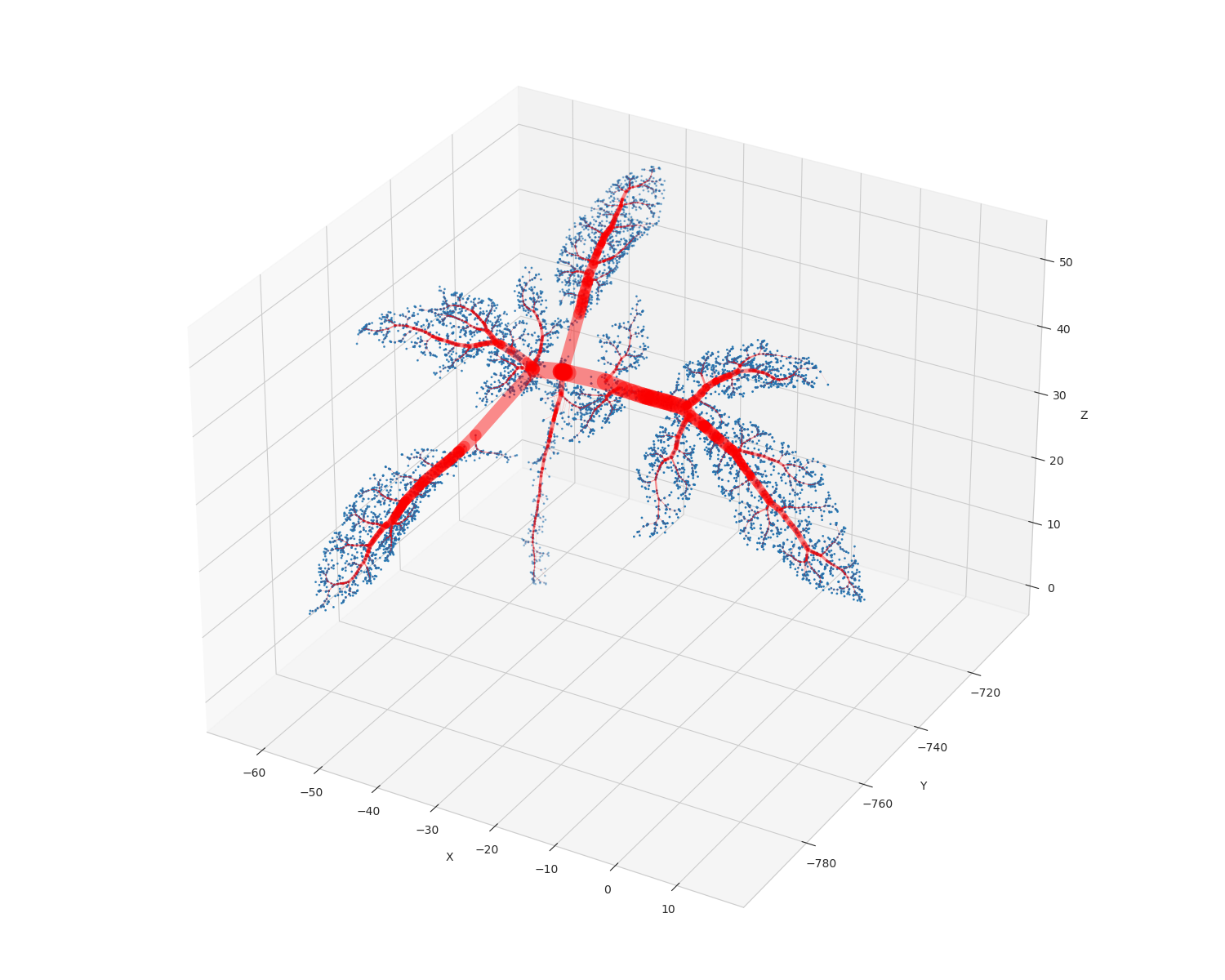}
		\caption{\BCST $\alpha=0.70$}
		\label{sfig11:tomato_plant2}
	\end{subfigure}%
	\begin{subfigure}{0.25\linewidth}
		\centering
		\includegraphics[width=\linewidth]{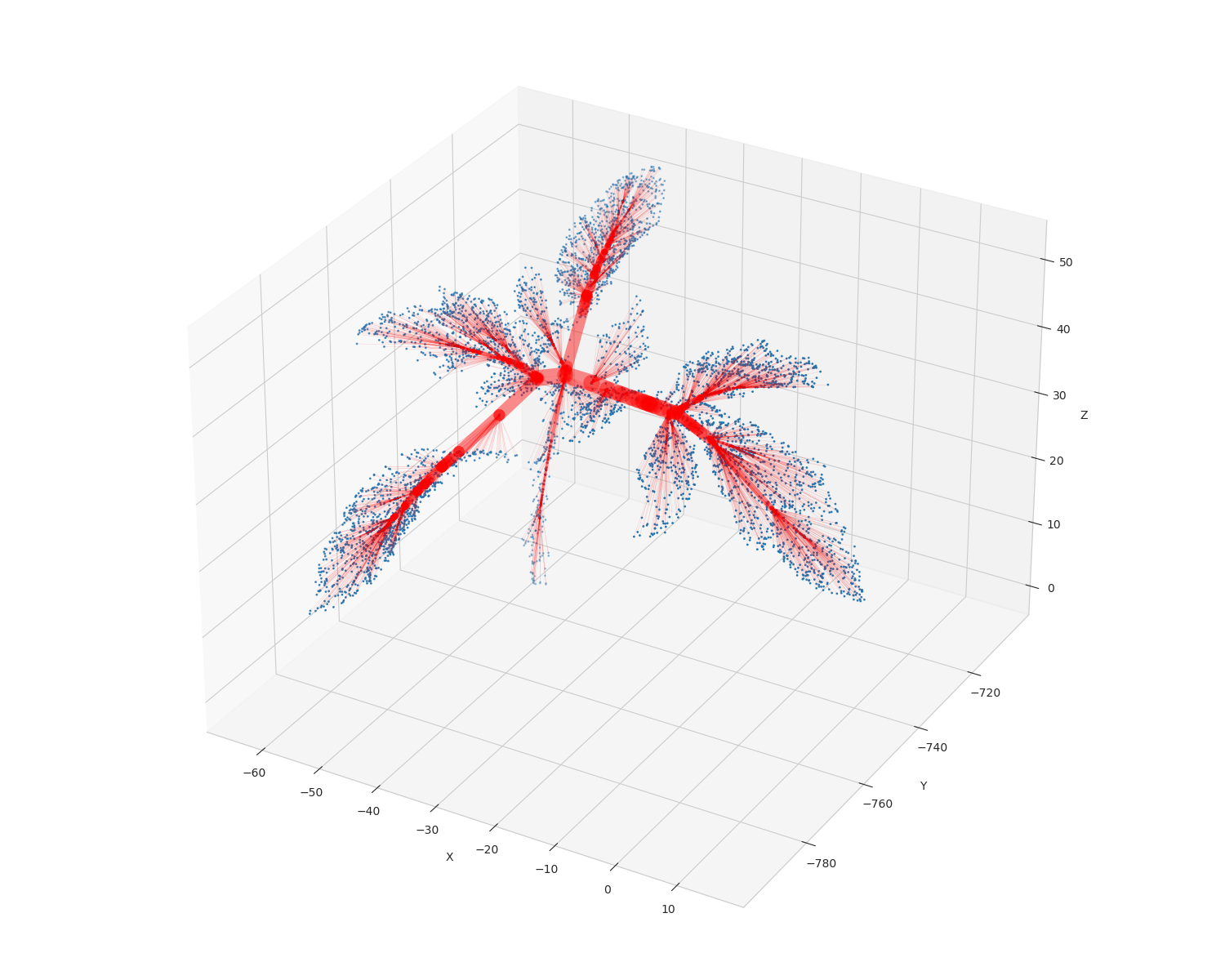}
		\caption{\BCST $\alpha=1.00$}
		\label{sfig12:tomato_plant2}
	\end{subfigure}%

	\caption[\BCST tomato plant skeletonization at different growth stages]{Skeletons at different $\alpha$ values of 3D point clouds of a tomato plant at different growth stages. The skeletons are modeled using the \BCST with varying $\alpha$ values. With $\alpha=0.00$, the tree branches exhibit greater irregularity, while at $\alpha=1.00$ the finer details are missed. Intermediate $\alpha$ values offer a more faithful representation of the plant's structure. Nonetheless these may present some slight deviation, where some modeled branches do not align through the center of the actual branches (see day 13). This effect is alleviated once prior information concerning the root's location is added (see \figurename{} \ref{fig:tomato_plant2_prior}).}
\label{fig:tomato_plant2}	
\end{figure}

\figurename{} \ref{fig:tomato_plant2} presents the \BCST results computed for different $\alpha$ values. When $\alpha=0.00$, the tree branches exhibit greater irregularity, while at $\alpha=1.00$, the finer details are obscured. Intermediate $\alpha$ values offer a more faithful representation of the plant's structure. However, some modeled branches deviate from the data, creating shortcut connections between points. This issue can be readily addressed by incorporating prior information about the root's position within the plant. By augmenting the point density near the root through virtual point replication (creating five times as many virtual points as there are original points at the root location), we encourage the branches to closely follow the natural branch density, resulting in a more faithful representation of the data. That is, if the original point cloud contains 5000 points, we introduce 25000 virtual points at the root location.

\begin{figure}[h!]
	\centering
	\caption*{Tomato plant 2, day 5.}
    \begin{subfigure}{0.25\linewidth}
        \centering
        \includegraphics[width=\linewidth]{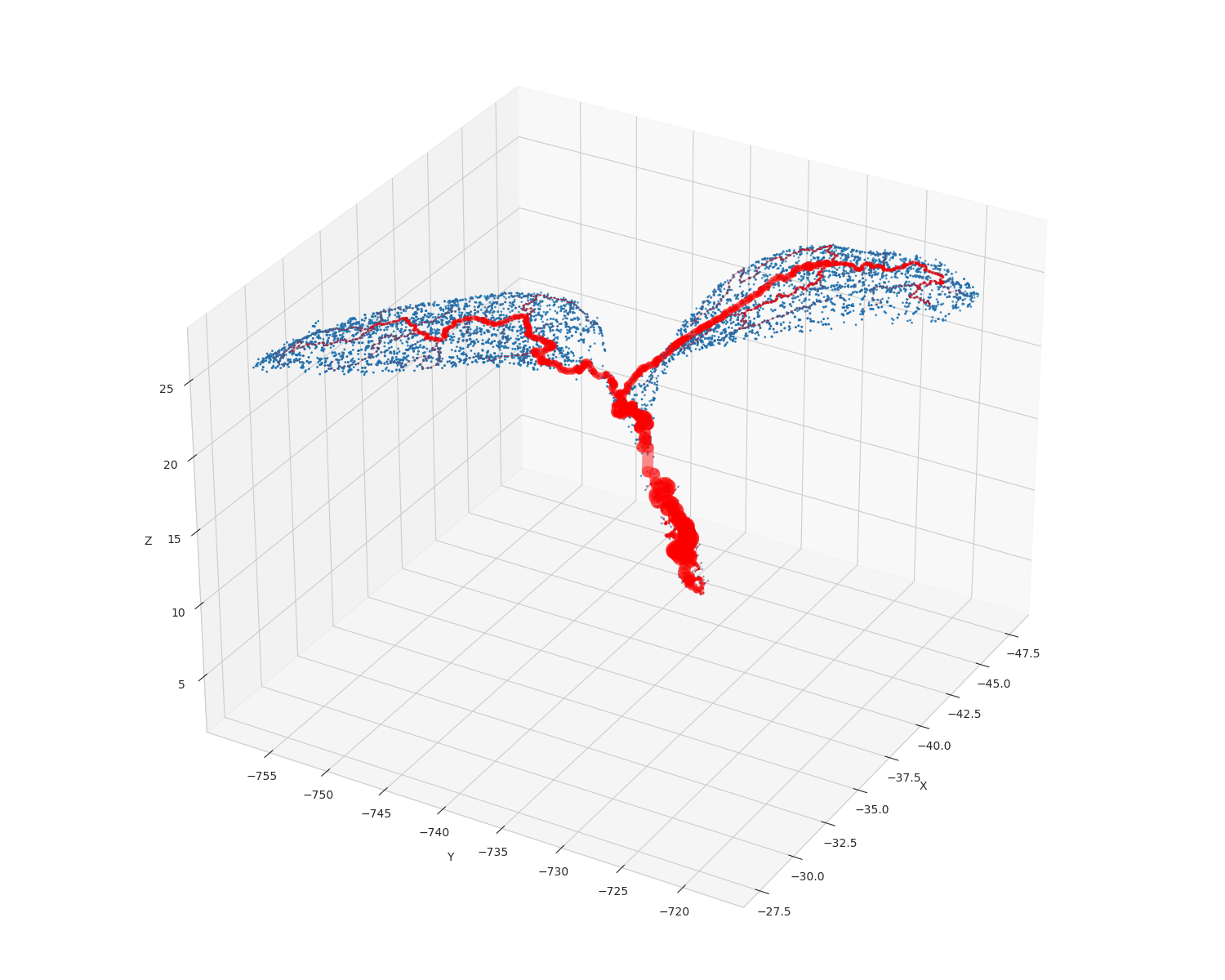}
        \caption{\BCST $\alpha=0.25$}
		\label{sfig1:tomato_plant2_prior}
	\end{subfigure}%
    \begin{subfigure}{0.25\linewidth}
    \centering
		\includegraphics[width=\linewidth]{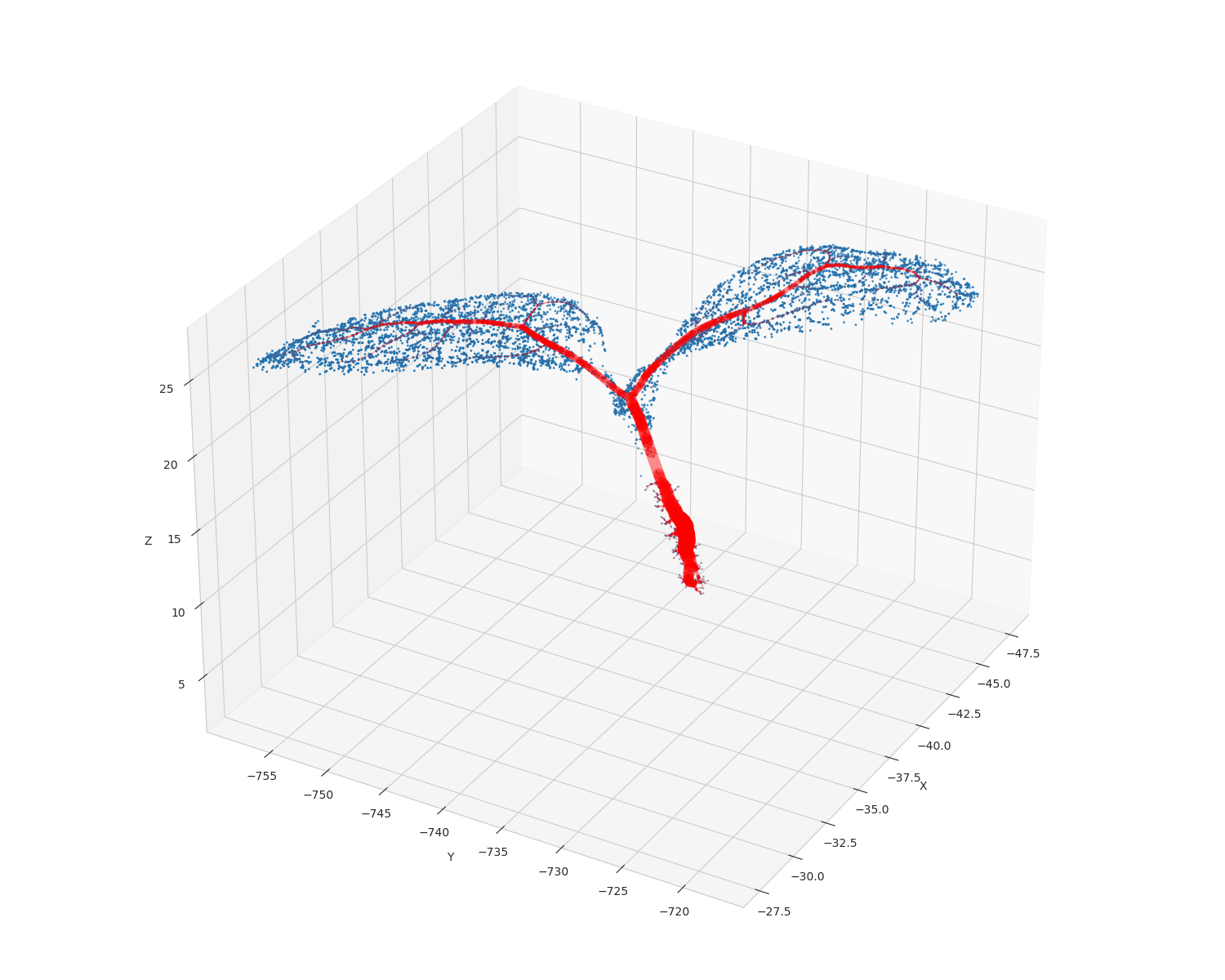}
		\caption{\BCST $\alpha=0.50$}
		\label{sfig2:tomato_plant2_prior}
	\end{subfigure}%
    \begin{subfigure}{0.25\linewidth}
        \centering
        \includegraphics[width=\linewidth]{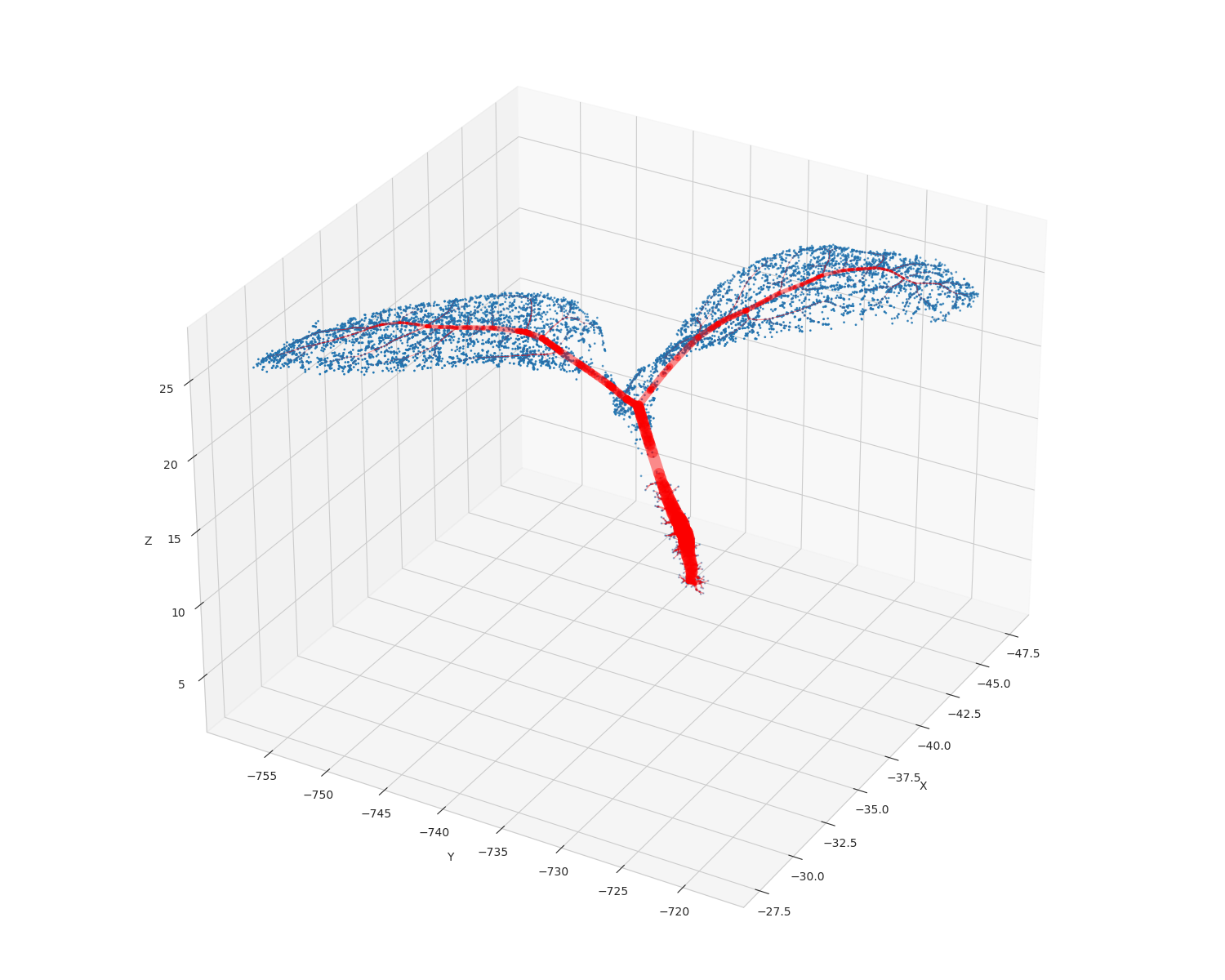}
        \caption{\BCST $\alpha=0.70$}
        \label{sfig3:tomato_plant2_prior}
	\end{subfigure}%
    \begin{subfigure}{0.25\linewidth}
    	\centering
    	\includegraphics[width=\linewidth]{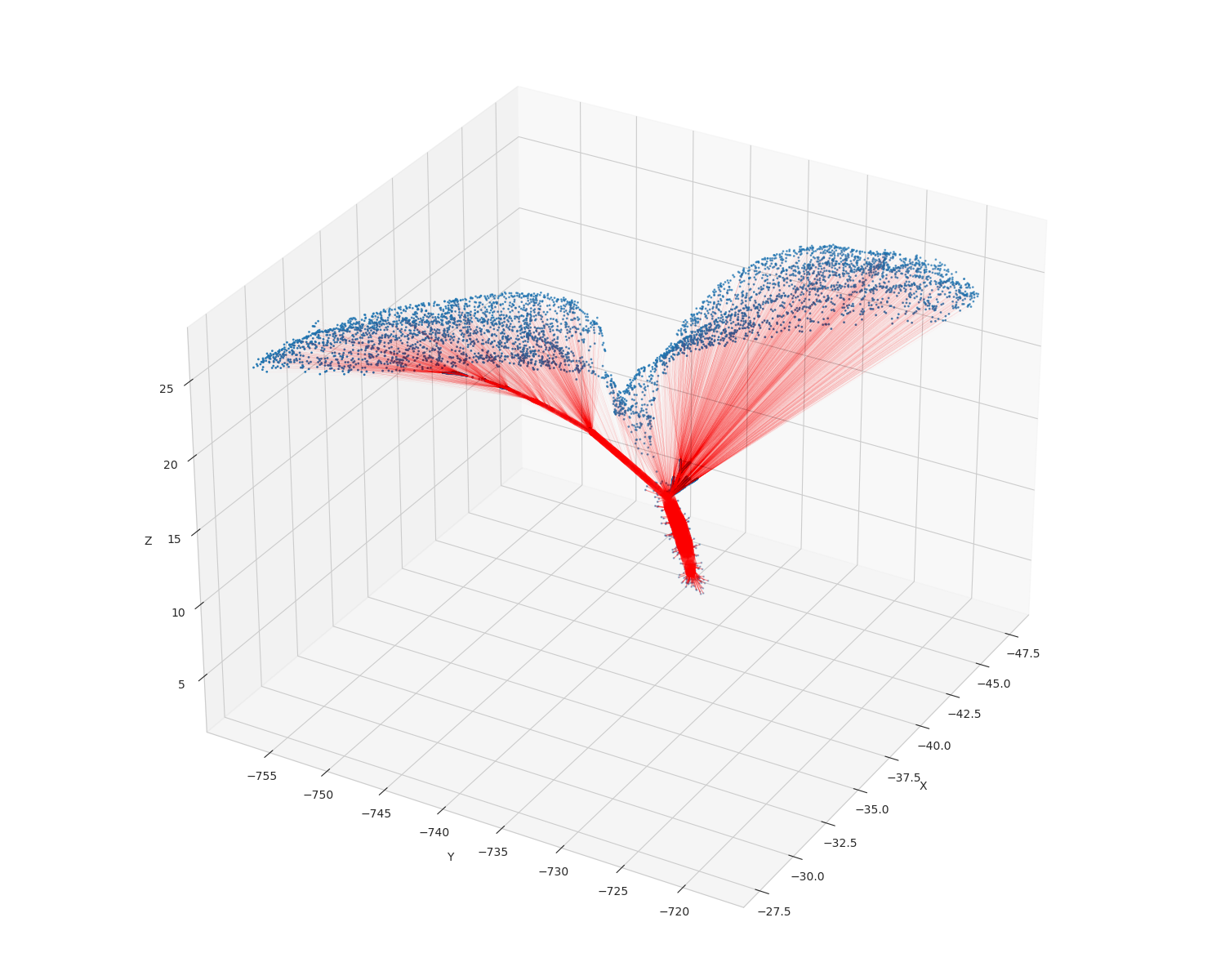}
    	\caption{\BCST $\alpha=1.00$}
    	\label{sfig4:tomato_plant2_prior}
    \end{subfigure}%

	\centering
	\caption*{Tomato plant 2, day 8.}
	\begin{subfigure}{0.25\linewidth}
		\centering
		\includegraphics[width=\linewidth]{Figures/plant_skeleton/tomato_plant2/n=5000/tomato_plant2_day8_n=5000_BCST_0.00_prior}
		\caption{\BCST $\alpha=0.00$}
		\label{sfig5:tomato_plant2_prior}
	\end{subfigure}%
	\begin{subfigure}{0.25\linewidth}
		\centering
		\includegraphics[width=\linewidth]{Figures/plant_skeleton/tomato_plant2/n=5000/tomato_plant2_day8_n=5000_BCST_0.50_prior}
		\caption{\BCST $\alpha=0.50$}
		\label{sfig6:tomato_plant2_prior}
	\end{subfigure}%
	\begin{subfigure}{0.25\linewidth}
		\centering
		\includegraphics[width=\linewidth]{Figures/plant_skeleton/tomato_plant2/n=5000/tomato_plant2_day8_n=5000_BCST_0.70_prior}
		\caption{\BCST $\alpha=0.70$}
		\label{sfig7:tomato_plant2_prior}
	\end{subfigure}%
	\begin{subfigure}{0.25\linewidth}
		\centering
		\includegraphics[width=\linewidth]{Figures/plant_skeleton/tomato_plant2/n=5000/tomato_plant2_day8_n=5000_BCST_1.00_prior}
		\caption{\BCST $\alpha=1.00$}
		\label{sfig8:tomato_plant2_prior}
	\end{subfigure}%
	\centering
	\caption*{Tomato plant 2, day 13.}
	\begin{subfigure}{0.25\linewidth}
		\centering
		\includegraphics[width=\linewidth]{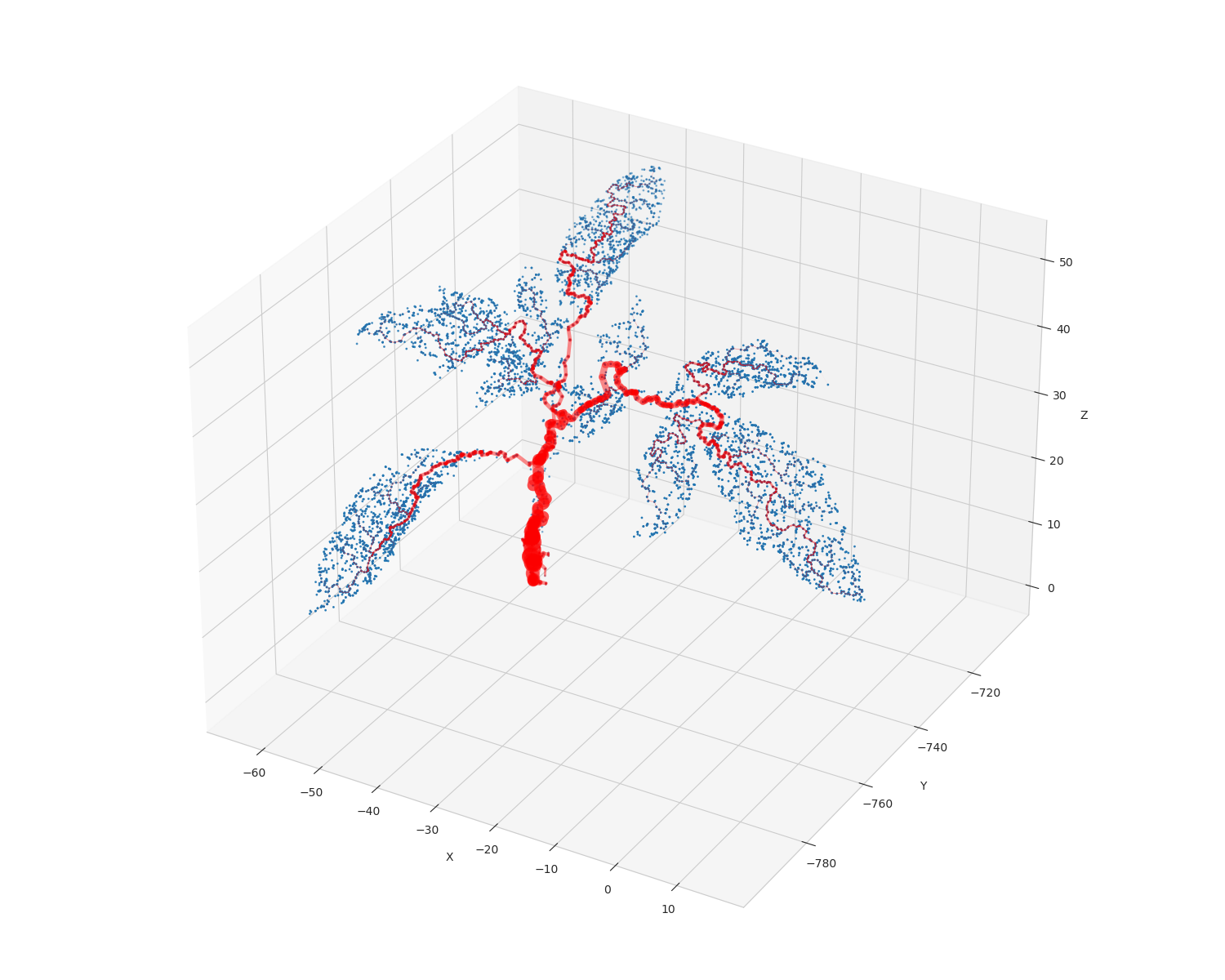}
		\caption{\BCST $\alpha=0.00$}
		\label{sfig9:tomato_plant2_prior}
	\end{subfigure}%
	\begin{subfigure}{0.25\linewidth}
		\centering
		\includegraphics[width=\linewidth]{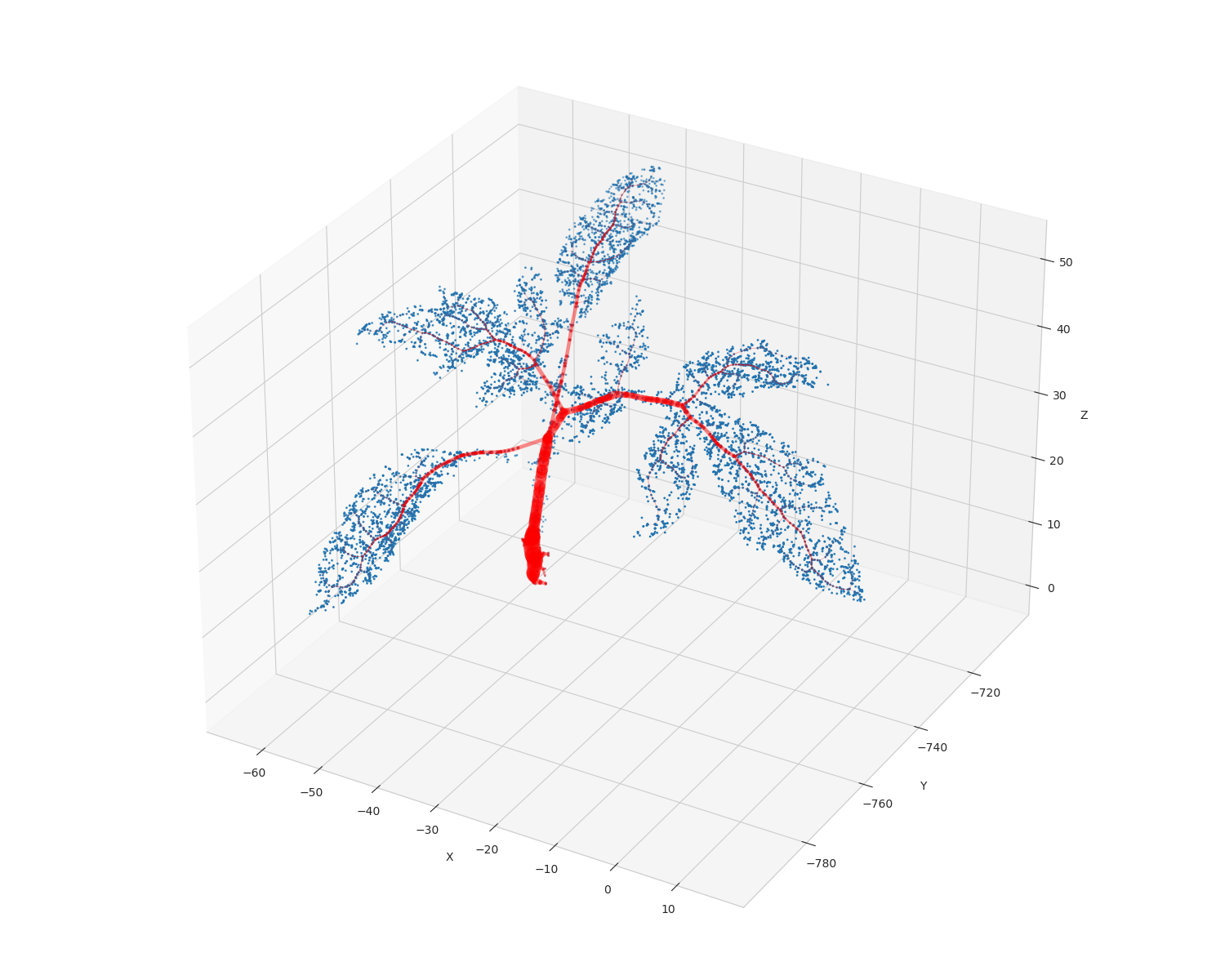}
		\caption{\BCST $\alpha=0.50$}
		\label{sfig10:tomato_plant2_prior}
	\end{subfigure}%
	\begin{subfigure}{0.25\linewidth}
		\centering
		\includegraphics[width=\linewidth]{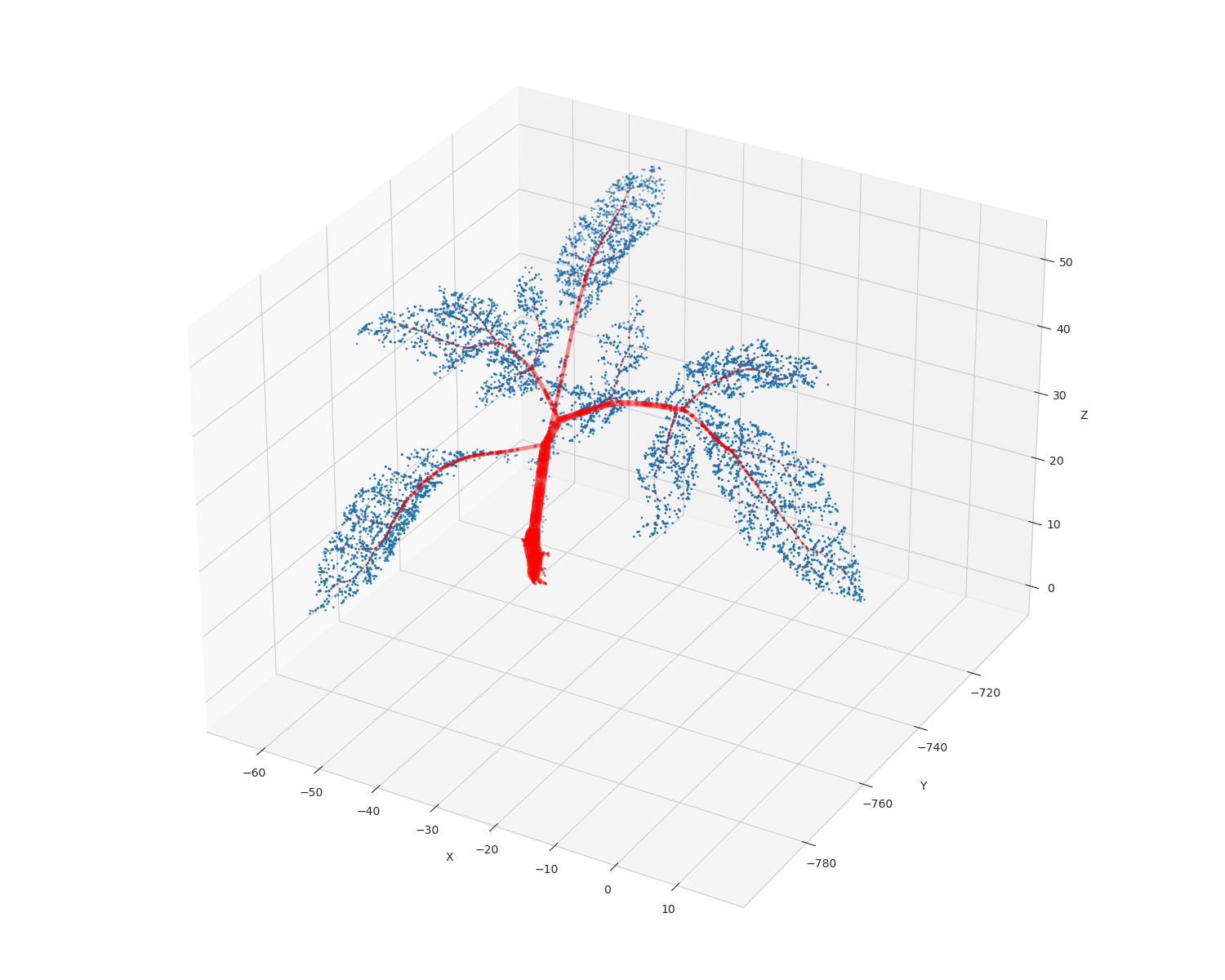}
		\caption{\BCST $\alpha=0.70$}
		\label{sfig11:tomato_plant2_prior}
	\end{subfigure}%
	\begin{subfigure}{0.25\linewidth}
		\centering
		\includegraphics[width=\linewidth]{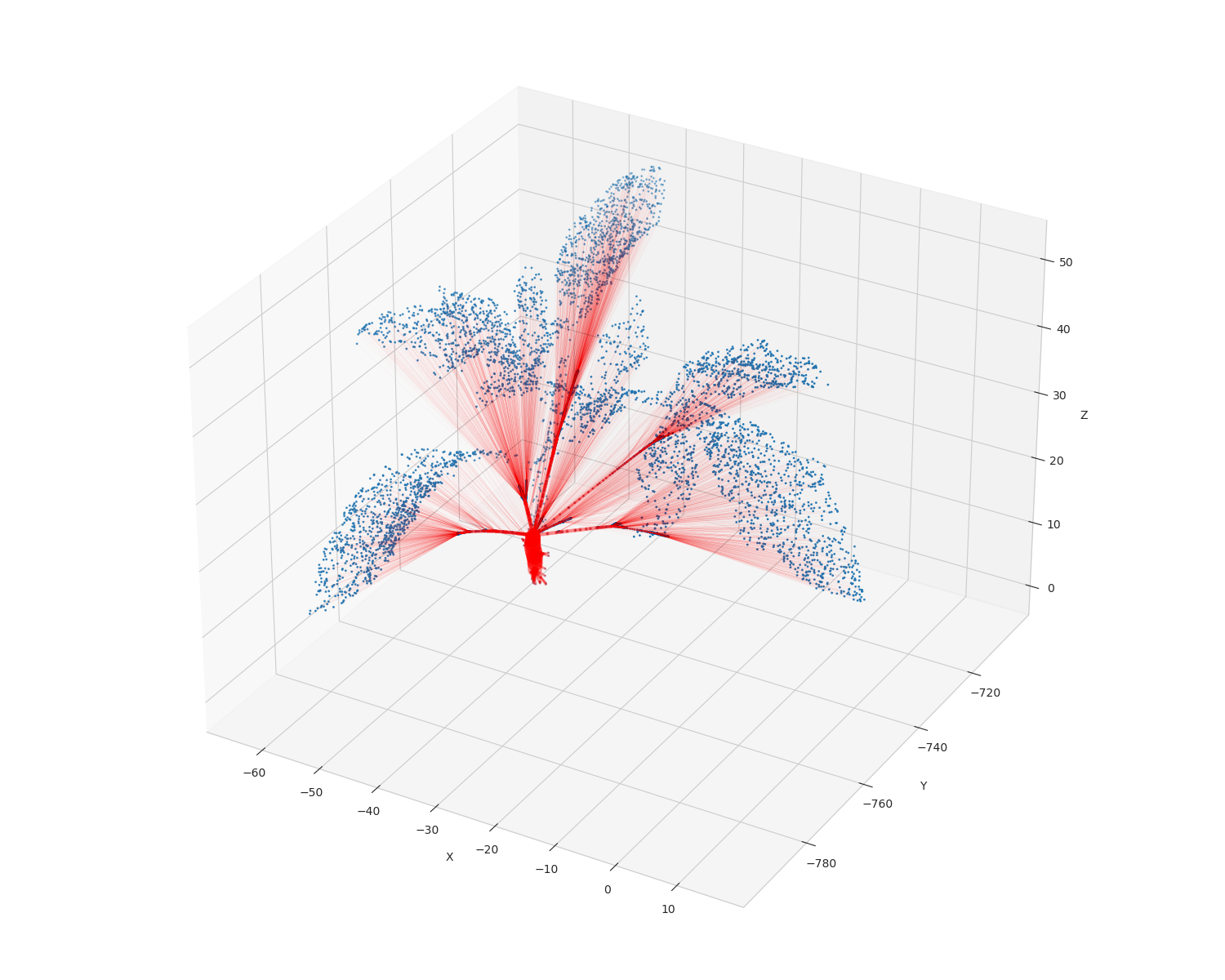}
		\caption{\BCST $\alpha=1.00$}
		\label{sfig12:tomato_plant2_prior}
	\end{subfigure}%

	\caption[\BCST tomato plant skeletonization using root location at different growth stages]{Skeletons at different $\alpha$ values of 3D point clouds of a tomato plant at different growth stages. The skeletons are modeled using the \BCST with varying $\alpha$ values, incorporating prior information about the root's location. In contrast to \figurename{} \ref{fig:tomato_plant2}, we have enhanced the density of points at the root location,  by virtually augmenting the number of points at the root's coordinates. This enhancement results in a more accurate and faithful representation of the plant's skeleton.}
\label{fig:tomato_plant2_prior}	
\end{figure}

\figurename{} \ref{fig:tomato_plant2_prior} displays the results after incorporating this prior information for the point cloud of the plant at different growth stages.\footnote{Day 8 of \figurename{} \ref{fig:tomato_plant2_prior} corresponds to Figure \ref{fig:plant_skeleton} in the main paper} Notably, the model exhibits greater fidelity for intermediate $\alpha$ values when the prior is applied.


\section{Reinterpreting CST as a Minimum Concave Cost Flow}\label{sec:app_CST_MCCNF}
In this section, we will pose the central spanning tree (\CST) problem as a minimum concave cost network flow (MCCNF) problem. The MCCNF problem minimizes the transportation cost of a commodity from sources to sinks. Here, the edge costs are modeled by concave functions that depend on the edge flow. Formally, given a demand vector $\mu\in\mathbb{R}^N$ with $\sum_{i=1}^N \mu_i=0$ and a network $G=(V,E)$ with $N$ nodes, we define the MCCNF problem as
\begin{equation}
	\label{eq:MCCNF}
	\begin{aligned}
		&\min_{f} \sum_{{ij}\in E} C_{ij}(f_{ij}),\  \text{ subject to } \\
		&\sum_{(i,j) \in E} f_{ij} -\sum_{(j,i) \in E} f_{ji} =\mu_i, \qquad \forall i\in V\\
		&f_{ij}\geq 0
	\end{aligned}.
\end{equation}
In the equation, $f_{ij}$ represents the flow associated with edge $(i,j)$ and $C_{ij}$ is a concave function dependent on $f_{ij}$ which determines the cost of the edge $(i,j)$. Note that the network defined by the flow, i.e. by the edges with $f_{ij}>0$, does not necessarily have to be a tree. We will refer to nodes with negative demands as sources and nodes with positive demands as sinks.

To be able to represent the \CST problem as an MCCNF, we need to identify the terms $m_e$ as flows. Since the function $\left(m_e(1-m_e)\right)^\alpha c_e$ is concave for $\alpha \in [0,1]$, it will follow that the \CST is an instance of the MCCNF problem.

Next, we will show how the $m_e$ can be interpreted as the flow along an edge of a particular single source flow problem. Consider a graph with $N$ nodes, where there is one source node $s$ with a mass $(N-1)/N$ that needs to be transported to the rest of the nodes. Each sink node has a demand of $1/N$ mass. Thus, \eqref{eq:MCCNF} becomes
\begin{equation}
	\label{eq:MCCNF_CST}
	\begin{aligned}
		&\min_{x} \sum_{{ij}\in E} c_{ij}\big(f_{ij}(1-f_{ij})\big)^{\alpha},\  \text{ subject to } \\
		&\sum_{(j,s) \in E} f_{js} - \sum_{(s,j) \in E} f_{sj} =\frac{N-1}{N}, \qquad\\
		&\sum_{(i,j) \in E} f_{ij} -\sum_{(j,i) \in E} f_{ji} =\frac{1}{N}, \qquad \forall i\neq s\\
		&f_{ij}\geq 0
	\end{aligned}
\end{equation}
For single source flow problems, it is well-known that the optimal solutions can only form trees \citep{zangwill_minimum_1968}. We will show that for this particular problem, $f_{ij}$ is equal to $m_{ij}$ for any tree. Recall that for a given tree $\tree$, the value $m_e$ associated with an edge $e$ was defined as the number of nodes that lie on one of the sides of $e$ divided by the total number of nodes. Although for our purposes the chosen side of the edge is arbitrary, since the objective function is symmetrized thanks to being multiplied by $(1-m_e)$, that is not the case when we want to interpret it as a flow. However, which side to choose will be canonically determined by the flow direction. 

Let $\tree$ be a feasible solution of \eqref{eq:MCCNF_CST}, i.e. a tree. The flow $f_{ij}$ at edge $(i,j)$ of $\tree$ indicates the outgoing mass that is transported through the edge. This mass is equal to the sum of the demands of the nodes that lie in the side of the edge $(i,j)$ indicated by the flow. Given that each node has a demand of $1/N$, then the flow $f_{ij}$ is equal to $1/N$ multiplied by the number of nodes in the side in question, that is $f_{ij}=m_{ij}$.

\subsection{Relation to the Branched Optimal Transport Problem}\label{subsec:app_relation_BOT_BCST}
The branched or irrigation optimal transport (BOT) problem is also a particular MCCNF instance. In the BOT problem, the nodes are embedded in a Euclidean space and also allows for the inclusion of additional Steiner points. It is an extension of the optimal transport problem, distinguished by its diminishing costs which lead to a branching effect by promoting the joint transportation of mass.

Formally, the BOT problem minimizes 

\begin{equation}
	\label{eq:BOT}
	\begin{aligned}
		&\underset{x_B,m_E}{\min} \
		\sum_{(i,j) \, \in \, E} m_{ij}^\alpha \left\|  x_i -  x_j \right\|_2 , \text{ subject to } \\
		&\sum_{(i,j) \in E} m_{ij} -\sum_{(j,i) \in E} m_{ji} =\mu_i, \qquad \forall i\in V\cup B\\
		&m_{ij}\geq 0
	\end{aligned},
\end{equation}
where $m_{ij}$ is the flow transported along an edge $(i,j)$ and $B$ and $x_B$ are the set of \BPs and their coordinates, respectively. As before, the vector $\mu$ represents the demands of the nodes. The demands of the \BPs are set to zero.

In the scenario where there is a single source and all nodes share the same demand, the \BCST and BOT problems differ only in the factors that multiply the distances. In the BOT problem, these factors correspond to $m_{ij}^\alpha$, representing the mass transported along an edge raised to the power of $\alpha$. In the \BCST problem, the factors are given by $\big(m_{ij}(1-m_{ij})\big)^\alpha$, representing the centralities of the edges raised to the power of $\alpha$. It is worth mentioning, that both problems converge to the Steiner tree problem when $\alpha = 0$. 

The primary distinction between the two problems lies in the selection of a source node. In the BOT problem, the selection of the source node determines the optimal topology of the network. However, that is not the case for the \BCST problem. Indeed, for the \BCST problem, the specific source node chosen is irrelevant due to the symmetrization effect caused by the term $m_e(1-m_e)$. In other words, the location of the source determines the edge orientation, which then defines the value of $m_e$. Nonetheless, this effect is nullified when multiplied by $(1-m_e)$. This independence of the source node choice makes the \BCST a more natural extension of the Steiner tree problem, since it does not require the choice of sources and sinks, unlike the BOT problem.

\section{Equivalence of the CST Problem with $\alpha=1$ and the Minimum Routing Cost Tree Problem}\label{sec:app_CST_MRCT_equivalence}
As mentioned in the main text, the term $m_{ij}(1-m_{ij})$ is proportional to the betweenness centrality of the edge $(i, j)$. This centrality quantifies the number of shortest paths that traverse the given edge. Thus, the multiplication of the length of each edge by its frequency in a shortest path is a rearrangement of the sum over all shortest path costs, i.e. the \MRCT cost. Formally,
\begin{equation}
	\begin{aligned}
		\sum_{i,j\in V\times V}d_{\tree}(i,j)&=\sum_{i,j\in V\times V}\sum_{(u,v)\in \Path_{ij}} ||x_u-x_v||\\
		&=\sum_{(u,v)\in \tree}\underbrace{\big|\{\Path_{ij} \ :\ (u,v)\in \Path_{ij}, \ i,j\in V\times V\}\big|}_{(u,v)\text{ betweeness centrality}} ||x_u-x_v||\\
		&\propto \sum_{(u,v)\in \tree}m_{uv}(1-m_{uv})||x_u-x_v||,
	\end{aligned}
\end{equation}
where $d_{\tree}(i,j)$ is the shortest path distance  in tree $\tree$ between $i$ and $j$ realized by the path $\Path_{ij}$.

\section{Limit Cases of the \CST/\BCST Problems Beyond the Range $\alpha\in[0,1]$}\label{sec:app_CSTlimits}
In this section, we investigate the topologies of the limit cases of the \CST as $\alpha$ approaches $\pm\infty$.  
We will use the following notation.
\begin{itemize}
	\item $N$ will represent the number of terminals.
	\item For a given tree $\tree$ containing edge $(x,y)$, $m^{\tree}_{xy}$ indicates the proportion of nodes (normalized by $N$) that are reachable from $x$, once edge $(x,y)$ is removed from $\tree$. That is, the normalized number of nodes that lie in the side of $x$.
	\item  For a given tree $\tree$, the term $\mathcal{N}_x$ denotes the set of neighbors of $x$ in $\tree$
\end{itemize}

\subsection{Proof \thref{lem:strong_triangle_inequality_(B)CST}}\label{sec:proof-threflemstrongtriangleinequalitybcst}  \thref{lem:strong_triangle_inequality_(B)CST} states that when a ``stronger" variant of the triangle inequality holds, then the optimum solution of the B\CST problem is a star tree. We divide the proof in two lemmas, \thref{lem:strong_triangle_inequality_CST} that proves it for the \CST problem; and \thref{lem:strong_triangle_inequality_BCST} for the \BCST case.
\begin{lemma}\thlabel{lem:strong_triangle_inequality_CST}
	Given a complete graph $G$ with $N$ nodes, let $c_{ij}$ be the edge-costs of any pair of nodes $i,j$ in the graph. If there exists 
	$$t\leq \min_{\ell\in[2,N/2]} \frac{\left(\frac{\ell(N-\ell)}{N-1}\right)^{\alpha}-1}{\ell-1}$$
	such that 
	\begin{equation}\label{eq:strong_triangle_inequality_CST_lemma}
	c_{kv}+tc_{uv}\geq c_{ku}
	\end{equation}
	for all triangles in the graph, then there exists an optimum $\CST$ evaluated at $\alpha$ which is a star tree.
\end{lemma}
\begin{proof}
	We will show that given a tree $\tree$, we can always increase the degree of a particular node without increasing the \CST-cost. Without loss of generality, we can assume $n\geq 4$, otherwise, any possible tree is a star-tree and the result holds trivially.
	
	Let us assume that $\tree$ is not a star-tree. Thus, there exist at least two nodes, $u$ and $v$, with degrees higher than 1, such that they are adjacent to each other. Moreover, we can assume that one of them, say $v$, is an extreme inner node, meaning that all its neighbors (except $u$) have degree $1$. Without loss of generality, we can assume that $\ell\coloneqq\left|\mathcal{N}_v\right|\leq N/2$, where $\mathcal{N}_v$ is the set of neighbors of $v$ in $\tree$. Otherwise, we could have chosen a different extreme inner node. Note that the centrality of the edge $(u,v)$ is $m_{uv}(1-m_{uv})=\frac{\ell(N-\ell)}{N^2}$.
	
	We will show that the topology $\tree'$ which connects all $k\in \mathcal{N}_v\backslash\{u\}$ to $u$ instead of $v$ has a lower \CST cost. The only edge centralities affected by this change are those associated with the edges $(u,v)$, $(u,k)$, and $(k,v)$ for all $k\in \mathcal{N}_v\backslash\{u\}$. To compare the costs of the topologies, it suffices to focus on these specific edges.
	
	First, let's determine the values of the centralities for the edges in both trees:
	\begin{itemize}
		\item  Normalized centrality of edge $(u,v)$ in $\tree$:
		$$m^{\tree}_{uv}(1-m^{\tree}_{uv})=\frac{\ell(N-\ell)}{N^2}$$
		\item  Normalized centrality of edge $(u,v)$ in $\tree'$: 
		$$m^{\tree'}_{uv}(1-m^{\tree'}_{uv})=\frac{N-1}{N^2}$$
		Note that $v$ has become a leaf of $\tree'$.
		\item Normalized centrality of edge $(k,v)$ in $\tree$ and centrality of edge $(k,u)$ in $\tree'$ for all $k~\in~\mathcal{N}_v\backslash\{u\}$: 
		$$m^{\tree}_{kv}(1-m^{\tree}_{kv})=\frac{N-1}{N^2}=m^{\tree'}_{ku}(1-m^{\tree'}_{ku})$$
		These nodes are leaves in both trees.
	\end{itemize}
	
	The difference between the costs of the topologies is 
	\footnotesize\begin{equation}
	\begin{aligned}\label{eq:tree_diff_CST}
	\operatorname{CST}(\tree)-\operatorname{CST}(\tree')=&c_{uv}\left(\frac{\ell(N-\ell)}{N^2}\right)^{\alpha}+\sum_{\substack{k\in\mathcal{N}_v \\k\neq u}}c_{kv}\left(\frac{N-1}{N^2}\right)^{\alpha}\\
	&-c_{uv}\left(\frac{N-1}{N^2}\right)^{\alpha}
	-\sum_{\substack{k\in\mathcal{N}_v \\k\neq u}}c_{ku}\left(\frac{N-1}{N^2}\right)^{\alpha}\\
	=&c_{uv}\left(\left(\frac{\ell(N-\ell)}{N^2}\right)^{\alpha}-\left(\frac{N-1}{N^2}\right)^{\alpha}\right)+\sum_{\substack{k\in\mathcal{N}_v \\k\neq u}}(c_{kv}-c_{ku})\left(\frac{N-1}{N^2}\right)^{\alpha}\\
	=&\sum_{\substack{k\in\mathcal{N}_v \\k\neq u}}\left[\frac{c_{uv}\left(\left(\frac{\ell(N-\ell)}{N^2}\right)^{\alpha}-\left(\frac{N-1}{N^2}\right)^{\alpha}\right)}{\ell-1}+(c_{kv}-c_{ku})\left(\frac{N-1}{N^2}\right)^{\alpha}\right]\\
	=&\left(\frac{N-1}{N^2}\right)^\alpha\sum_{\substack{k\in\mathcal{N}_v \\k\neq u}}\left[\frac{c_{uv}\left(\left(\frac{\ell(N-\ell)}{N-1}\right)^{\alpha}-1\right)}{\ell-1}+(c_{kv}-c_{ku})\right]
	\end{aligned}
	\end{equation}
	\normalsize
	Thus the decrease in  cost will positive if each term in the summand of the last equality of \eqref{eq:tree_diff_CST} is positive, namely
	\begin{equation}\label{eq:strong_triangle_inequality_CST}
		c_{kv}+\frac{\left(\left(\frac{\ell(N-\ell)}{N-1}\right)^{\alpha}-1\right)}{\ell-1}c_{uv}\geq c_{ku},
	\end{equation}
	which holds by assumption. Therefore, $\tree'$ will have a lower cost. Repeating this process, we can always decrease the cost till we form a star tree.
\end{proof}

The next result extends \thref{lem:strong_triangle_inequality_CST} to be applicable the \BCST problem. In contrast to the \CST case, the \BCST involves Steiner points, which must be treated differently. \thref{lem:strong_triangle_inequality_BCST} shows that if the ``strong" triangle inequality holds, we can collapse sequentially all Steiner points while decreasing the \BCST cost of a tree $\tree$.

\begin{lemma}\thlabel{lem:strong_triangle_inequality_BCST}
	Consider a solution $\tree$ of the \BCST problem with $N$ terminals. Let $c_{ij}$ be the edge-costs of any pair of nodes $i,j$ in the graph (Steiner or terminals). If there exists 
	$$t\leq \min_{\ell\in[2,N/2]} \frac{\left(\frac{(\ell)(N-\ell)}{N-1}\right)^{\alpha}-1}{\ell-1}$$
	such that 
	\begin{equation}\label{eq:strong_triangle_inequality_BCST_lemma}
	c_{kv}+tc_{uv}\geq c_{ku}
	\end{equation}
	for all triangles, then there exists an star tree with lower cost.
\end{lemma}
\begin{proof}
	Analogously to \thref{lem:strong_triangle_inequality_CST} we will show that given a tree $T$, we can always increase the degree of a particular node without increasing the \BCST-cost.

	Let us assume that $\tree$ is not a star-tree. Thus, there exist at least two nodes, $u$ and $v$, with degree higher than 1 which are adjacent to each other. Moreover, we can assume that one of them, say $v$, is an extreme inner node, meaning that all its neighbors (except $u$) have degree $1$. Without loss of generality, we can assume that $m_{uv}\coloneqq\frac{\ell}{N^2}\leq\frac{1}{2N}$. Otherwise, we could have chosen a different extreme inner node. 
	
	If $v$ is a terminal node, we can apply the same reasoning as in \thref{lem:strong_triangle_inequality_CST} to increase the degree of $u$. Let us assume then that $v$ is a Steiner point. In this case, we will construct a new  topology $\tree'$ by collapsing $v$ with $u$. This implies that the edge $(u,v)$ will disappear and that all $k\in \mathcal{N}_v\backslash\{u\}$ will be connected to $u$. In this case the normalized centralities of the edges are not changed.
	\begin{itemize}[leftmargin=*]
		\item  Normalized centrality of edge $(u,v)$ in $\tree$:
		$$m^{\tree}_{uv}(1-m^{\tree}_{uv})=\frac{\ell(N-\ell)}{N^2}$$
		\item Edge $(u,v)$ is not anymore in $\tree'$: 
		\item Normalized centrality of edge $(k,v)$ in $\tree$ and centrality of edge $(k,u)$ in $\tree'$ for all $k~\in~\mathcal{N}_v\backslash\{u\}$: 
		$$m^{\tree}_{kv}(1-m^{\tree}_{kv})=\frac{N-1}{N^2}=m^{\tree'}_{ku}(1-m^{\tree'}_{ku})$$
		These nodes are leaves in both trees.
	\end{itemize}
	
	The difference between the costs of the topologies is 
	\footnotesize\begin{equation}
		\begin{aligned}\label{eq:tree_diff_BCST}
			\operatorname{CST}(\tree)-\operatorname{CST}(\tree')=&c_{uv}\left(\frac{\ell(N-\ell)}{N^2}\right)^{\alpha}+\sum_{\substack{k\in\mathcal{N}_v \\k\neq u}}c_{kv}\left(\frac{N-1}{N^2}\right)^{\alpha}-\sum_{\substack{k\in\mathcal{N}_v \\k\neq u}}c_{ku}\left(\frac{N-1}{N^2}\right)^{\alpha}\\
			=&c_{uv}\left(\frac{\ell(N-\ell)}{N^2}\right)^{\alpha}+\sum_{\substack{k\in\mathcal{N}_v \\k\neq u}}(c_{kv}-c_{ku})\left(\frac{N-1}{N^2}\right)^{\alpha}\\
			=&\sum_{\substack{k\in\mathcal{N}_v \\k\neq u}}\left[\frac{c_{uv}(\left(\frac{\ell(N-\ell)}{N^2}\right)^{\alpha}}{\ell-1}+(c_{kv}-c_{ku})\left(\frac{N-1}{N^2}\right)^{\alpha}\right]\\
			=&\left(\frac{N-1}{N^2}\right)^\alpha\sum_{\substack{k\in\mathcal{N}_v \\k\neq u}}\left[\frac{c_{uv}\left(\left(\frac{\ell(N-\ell)}{N-1}\right)^{\alpha}\right)}{\ell-1}+(c_{kv}-c_{ku})\right]
		\end{aligned}
	\end{equation}
	\normalsize
	Thus the decrease in  cost will positive if each term in the summand of the last equality of \eqref{eq:tree_diff_BCST} is positive, namely
	\begin{equation}\label{eq:strong_triangle_inequality_BCST}
		c_{kv}+\frac{\left(\frac{\ell(N-\ell)}{N-1}\right)^{\alpha}}{\ell-1}c_{uv}\geq c_{ku},
	\end{equation}
	which holds by assumption, since
	
	$$\frac{\left(\frac{\ell(N-\ell)}{N-1}\right)^{\alpha}}{\ell-1}\geq \min_{\ell\in[2,N/2]}\frac{\left(\frac{\ell(N-\ell)}{N-1}\right)^{\alpha}}{\ell-1}>\min_{\ell\in[2,N/2]} \frac{\left(\frac{\ell(N-\ell)}{N-1}\right)^{\alpha}-1}{\ell-1}.$$
	
	Therefore, $\tree'$ will have a lower cost. Repeating this process, we can always decrease the cost till we form a star tree.
\end{proof}

\begin{remark}
	Note that \thref{lem:strong_triangle_inequality_CST} and \thref{lem:strong_triangle_inequality_BCST} state only a sufficient condition, which means that the optimum can be a star tree even if the strong triangle inequality does not hold. Additionally, it is worth to highlight that \thref{lem:strong_triangle_inequality_CST} also holds true for the \CST problem even when the nodes lack embedding in any specific space, allowing for edge costs with arbitrary values. 
\end{remark}

\subsection{Proof $h_1(\ell,N,\alpha)>1$ as $N$ Approaches Infinity, for $\alpha>1$}\label{sec:proof-h1ellnalpha1-as-n-approaches-infinity-and-alpha1}

Recall that $h_1$ is defined as 
$$h_1(\ell,N,\alpha)\coloneqq\frac{\left(\left(\frac{\ell(N-\ell)}{N-1}\right)^{\alpha}-1\right)}{\ell-1}=\frac{\left(1+\frac{(\ell-1)(N-\ell-1)}{N-1}\right)^{\alpha}-1}{\ell-1}.$$
We will show that for high enough $N$, $\min_{\ell\in[2,N/2]}h_1(\ell,N,\alpha)>1$. By leveraging the Mathematica software \citep{Mathematica}, we can establish that the function $h_1(\ell,N,\alpha)$ exhibits concavity concerning $\ell$ within the interval $[2,N/2]$ under the conditions $\alpha>1$ and $N>3$. Consequently, for fixed values of $\alpha>1$ and $N$, the minimum of $h$ is achieved either at $\ell=2$ or at $\ell=N/2$. Next we show that as $N$ tends to infinity, both evaluations tend towards a value greater than 1.
\begin{itemize}[leftmargin=*]
	\item When $\ell=2$ we have
	\begin{equation}\label{eq:xtrem0}
	h_1(2,N,\alpha)=\left(1+\frac{N-3}{N-1}\right)^\alpha-1\xrightarrow{N\to\infty}2^\alpha -1>1
	\end{equation}
	\item When $\ell=N/2$ we have
	\begin{equation}\label{eq:xtrem1}
	h_1(N/2,N,\alpha)=\frac{\left(1+\frac{\left(N/2-1\right)^2}{N-1}\right)^\alpha-1}{N/2-1}\xrightarrow{N\to\infty}\infty
	\end{equation}
\end{itemize}
Therefore, $h_1(\ell,N,\alpha)>1$ as $N$ approaches infinity and $\alpha>1$.

Combining this inequality with \thref{lem:strong_triangle_inequality_(B)CST}, we conclude that the optimum (B)\CST will be a tree as $N$ approaches infinity and $\alpha>1$.

\subsection{Computation $\alpha^{\ast}(N)$}\label{sec:computation-alphaastn}

Given $N$, recall that $\alpha^{\ast}(N)$ is the minimum $\alpha$ at which $h_1(\ell,N,\alpha)>1$ for all $\ell\in[2,N/2]$. Since $h_1$ is concave with respect to $\ell$ when $\alpha>1$, our focus narrows down to investigating the cases $\ell=2$ and $\ell=N/2$ --the values where the minima can be attained.
\begin{itemize}[leftmargin=*]
	\item When $\ell=2$ we have
	\begin{equation}\label{eq:min_xtrem}
	\small \begin{aligned}
	h_1(2,N,\alpha)=\left(1+\frac{N-3}{N-1}\right)^\alpha-1>1&\iff \left(1+\frac{N-3}{N-1}\right)^\alpha>2\\
	&\iff \alpha \log\left(1+\frac{N-3}{N-1}\right)>\log(2)\\
	&\iff\alpha >\frac{\log(2)}{\log\left(1+\frac{N-3}{N-1}\right)}
	\end{aligned}
	\end{equation}
	\item When $\ell=N/2$ we have
	\begin{equation}\label{eq:min_xtrem}
	\begin{aligned}
	h_1(N/2,N,\alpha)=\frac{\left(1+\frac{\left(N/2-1\right)^2}{N-1}\right)^\alpha-1}{N/2-1}>1&\iff \left(1+\frac{\left(N/2-1\right)^2}{N-1}\right)^\alpha>N/2\\
	&\iff \alpha \log\left(1+\frac{\left(N/2-1\right)^2}{N-1}\right)>\log(N/2)\\
	&\iff\alpha >\frac{\log(N/2)}{\log\left(1+\frac{\left(N/2-1\right)^2}{N-1}\right)}
	\end{aligned}
	\end{equation}
\end{itemize}
Therefore,

\begin{equation}
\alpha^{\ast}(N)\coloneqq \max\left(\frac{\log(2)}{\log\left(1+\frac{N-3}{N-1}\right)},\frac{\log(N/2)}{\log\left(1+\frac{\left(N/2-1\right)^2}{N-1}\right)}\right)
\end{equation}

\subsection{Proof \thref{th:tree_opt_alpha_-inf_reformulation}}\label{sec:proof-threfthtreeoptalpha-infreformulation}
In this section we prove \thref{th:tree_opt_alpha_-inf_reformulation} which states that if a variant of the triangle inequality holds, then the optimum (B)\CST tree will be a path as $\alpha$ approaches infinity. First, \thref{lem:tree_opt_alpha_-inf} will show that if the triangle inequality holds strictly, then for $\alpha$ negative enough the optimum $\CST$ will be a path. \thref{cor:tree_opt_alpha_-inf} demonstrates that when the nodes are embedded in a geodesic space, the triangle inequality does not need to hold strictly for \thref{lem:tree_opt_alpha_-inf} to be true. \thref{th:tree_opt_alpha_-inf_reformulation} will also be derived as corollary (\thref{cor:tree_opt_alpha_-inf_reformulation}). 

\begin{lemma}\thlabel{lem:tree_opt_alpha_-inf}
	Let $G$ be a complete graph with edge-costs satisfying the condition of the strict triangle inequality for every triplet of nodes $(u, v, k)$, defined as
	$$c_{uv}+c_{kv}<c_{ku}$$
	As the parameter $\alpha$ approaches negative infinity ($\alpha \to -\infty$), there exists a Hamiltonian path $\tree_{\star}$ in $G$ with a lower \CST cost than any other tree $\tree$ that is not a path.
\end{lemma}
\begin{proof}
	We will show that for any node $v$ with degree higher than $2$, we can always decrease its degree such that the \CST cost of $\tree$ decreases. By iteratively applying this process, we ensure that the degrees of all nodes will eventually be reduced to at most 2, culminating in the formation of a path.
	
	Let $v$ be a node with degree higher than $3$. Let $(k,v)$ and $(u,v)$ be the two edges adjacent to $v$ and assume w.l.o.g that $m^{\tree}_{uv}=\min_{i\in\mathcal{N}_v}m_{iv}$. Since $m^{\tree}_{kv}\geq m^{\tree}_{uv}$, we have
	$$\left(m^{\tree}_{kv}\big(1-m^{\tree}_{kv}\big)\right)\geq\left(m^{\tree}_{uv}\big(1-m^{\tree}_{uv}\big)\right).$$
	We will now demonstrate that the modified topology $\tree'$, where node $k$ is connected to node $u$ instead of node $v$, results in a lower \CST cost. The only edge centralities affected by this change are those associated with edges $(u,v)$, $(u,k)$, and $(k,v)$. To compare the costs of the topologies, it suffices to focus on these specific edges.
	
	First, we determine the values of the centralities for these edges in both trees.
	\begin{itemize}
		\item   Normalized centrality of edge $(u,v)$ in tree $\tree$: 
		$$m^{\tree}_{uv}(1-m^{\tree}_{uv})$$
		\item   Normalized centrality of edge $(u,v)$ in tree $\tree'$: 
		$$m^{\tree'}_{uv}(1-m^{\tree'}_{uv})=\left(m^{\tree}_{uv}+m^{\tree}_{kv}\right)\left(1-m^{\tree}_{uv}-m^{\tree}_{kv}\right)$$
		The equality is due to the fact that once $k$ is a neighbor of $u$, all nodes that were in the same side as $k$ will be now in the same side as $u$.
		\item Normalized centrality edge $k,v$ in $\tree$ and normalized centrality edge $k,u$ in $\tree'$: 
		$$m^{\tree}_{kv}(1-m^{\tree}_{kv})=m^{\tree'}_{ku}(1-m^{\tree'}_{ku})$$
		Both $u$ and $v$ lie in the same side of the edges, hence the equality of their normalized centralities.
	\end{itemize}
	Note that $m^{\tree'}_{uv}(1-m^{\tree'}_{uv})>m^{\tree}_{uv}(1-m^{\tree}_{uv})$. Otherwise, it would imply that
	\begin{equation*}
	\begin{aligned}
	m^{\tree'}_{uv}(1-m^{\tree'}_{uv})&<&m^{\tree}_{uv}(1-m^{\tree}_{uv})\iff\\
	\left(m^{\tree}_{uv}+m^{\tree}_{kv}\right)\left(1-m^{\tree}_{uv}-m^{\tree}_{kv}\right)&<&m^{\tree}_{uv}(1-m^{\tree}_{uv})\iff\\
	\min\left(m^{\tree}_{uv}+m^{\tree}_{kv},1-m^{\tree}_{uv}-m^{\tree}_{kv}\right)&<&\min\left(m^{\tree}_{uv},1-m^{\tree}_{uv}\right)=m^{\tree}_{uv}
	\end{aligned}
	\end{equation*}
	Trivially, $m^{\tree}_{uv}+m^{\tree}_{kv}<m^{\tree}_{uv}$ leads to a contradiction since $m^{\tree}_{kv}>0$. Thus, the only possibility is
	\begin{eqnarray*}
		1-m^{\tree}_{uv}-m^{\tree}_{kv}<m_{uv}^{\tree}&\iff& 1-m^{\tree}_{kv}<2m^{\tree}_{uv}\\&\iff& 1-m^{\tree}_{kv}=m^{\tree}_{uv}+\sum_{\substack{i\in\mathcal{N}_v\\i\neq k, i\neq u}}m_{iv}^{\tree}<2m^{\tree}_{uv}\\&\iff&\sum_{\substack{i\in\mathcal{N}_v\\i\neq k, i\neq u}}m_{iv}^{\tree}<m^{\tree}_{uv}
	\end{eqnarray*}
	which is also a contradition since by assumption $m^{\tree}_{uv}=\min_{i\in\mathcal{N}_v} m_{iv}^{\tree}$. 
	Now we are able to show that the cost of $\tree'$ is lower than the one of $\tree$
	\scriptsize
	\begin{equation*}
	\begin{aligned}\label{eq:tree_opt_alpha_-inf}
	&\operatorname{CST}(\tree)-\operatorname{CST}(\tree')=c_{uv}\left(m^{\tree}_{uv}\big(1-m^{\tree}_{uv}\big)\right)^{\alpha}+c_{kv}\left(m^{\tree}_{kv}\big(1-m^{\tree}_{kv}\big)\right)^{\alpha}\\
	&-c_{uv}\left(\left(m^{\tree}_{uv}+m^{\tree}_{kv}\right)\left(1-m^{\tree}_{uv}
	-m^{\tree}_{kv}\right)\right)^{\alpha}
	-c_{ku}\left(m^{\tree}_{kv}\big(1-m^{\tree}_{kv}\big)\right)^{\alpha}\\
	=&c_{uv}\Bigg(\left(m^{\tree}_{uv}\big(1-m^{\tree}_{uv}\big)\right)^{\alpha}-\left(\left(m^{\tree}_{uv}+m^{\tree}_{kv}\right)\left(1-m^{\tree}_{uv}-m^{\tree}_{kv}\right)\right)^{\alpha}\Bigg)+(c_{kv}-c_{ku})\left(m^{\tree}_{kv}\big(1-m^{\tree}_{kv}\big)\right)^{\alpha}\\
	=&\left(m^{\tree}_{uv}\big(1-m^{\tree}_{uv}\big)\right)^{\alpha}\left(c_{uv}-c_{uv}\left(\underbrace{\frac{\left(m^{\tree}_{uv}+m^{\tree}_{kv}\right)\left(1-m^{\tree}_{uv}-m^{\tree}_{kv}\right)}{\left(m^{\tree}_{uv}\big(1-m^{\tree}_{uv}\big)\right)}}_{>1}\right)^{\alpha}
	+(c_{kv}-c_{ku})\left(\underbrace{\frac{m^{\tree}_{kv}\big(1-m^{\tree}_{kv}\big)}{m^{\tree}_{uv}\big(1-m^{\tree}_{uv}\big)}}_{\geq1}\right)^{\alpha}\right).\\
	\end{aligned}
	\end{equation*}
	\normalsize
	We can differentiate two cases. 
	If $m^{\tree}_{kv}\big(1-m^{\tree}_{kv}\big)= m^{\tree}_{uv}\big(1-m^{\tree}_{uv}\big)$ then 
	\begin{equation}
	\label{eq2:tree_opt_alpha_-inf}
	\frac{\operatorname{CST}(\tree)-\operatorname{CST}(\tree')}{\left(m^{\tree}_{uv}\big(1-m^{\tree}_{uv}\big)\right)^{\alpha}}\xrightarrow{\alpha\to-\infty}(c_{uv}+c_{kv}-c_{ku})>0,
	\end{equation}
	where we have used the strict triangle inequality. Otherwise, the limit tends to
	$$        \frac{\operatorname{CST}(\tree)-\operatorname{CST}(\tree')}{\left(m^{\tree}_{uv}\big(1-m^{\tree}_{uv}\big)\right)^{\alpha}}\xrightarrow{\alpha\to-\infty}c_{uv}>0.$$

	Hence, for sufficiently negative values of $\alpha$, the difference $\CST(\tree)-\CST(\tree')$ will be positive. By repeating this process, we can continue reducing the degree of nodes with degree higher than $2$ until all nodes have degree at most $2$. This process will eventually lead to the formation of a Hamiltonian path with a lower cost than the original tree $\tree$.
\end{proof}

\begin{remark}
	Due to  equation \eqref{eq2:tree_opt_alpha_-inf}, \thref{lem:tree_opt_alpha_-inf} required the triangle inequality to hold strictly. However, the strict triangle inequality is not an indispensable for the validity of  \thref{lem:tree_opt_alpha_-inf}. \thref{cor:tree_opt_alpha_-inf} demonstrates that, when the nodes of the graph are embedded in a geodesic metric space (e.g. Euclidean space), the strict triangle inequality becomes unnecessary. This result extends also to the \BCST problem. 
	
	Nevertheless, the scenario portrayed in \figurename{} \ref{fig:counterexample_hamilitonian} serves as an example where the strict triangle inequality is not satisfied, leading to a non-optimal Hamiltonian path. This illustrates that while the strict triangle inequality may be dispensable under certain conditions, there are instances of arbitrary graphs, as demonstrated in the figure, where it cannot be abandoned. 
\end{remark}

\tikzset{
	dot/.style 2 args={fill, circle, inner sep=0pt, label={#1:\scriptsize #2}},	
	fulldot/.style 2 args={circle,draw,minimum size=0.3cm,inner sep=0pt, label={#1:\scriptsize #2}},
	main node/.style={circle,draw,minimum size=0.4cm,inner sep=0pt]},
	mini node/.style={circle,draw,minimum size=0.3cm,inner sep=0pt]}
}
\def \scale{2}

\begin{figure*}[h!]
	\centering
	
	\begin{subfigure}{0.33\textwidth}
		\centering
		\begin{tikzpicture}[]
				\node[main node,opacity=.5,black,fill=cyan,text opacity=1] (1) at (\scale*0,\scale*0) {\small 1};
				\node[main node,opacity=.5,black,fill=cyan,text opacity=1] (2) at (\scale *1,\scale*0) {\small $2$};
				\node[main node,opacity=.5,black,fill=cyan,text opacity=1] (3) at (\scale* -0.5,\scale*-0.866) {\small $3$};
				\node[main node,opacity=.5,black,fill=cyan,text opacity=1] (4) at (\scale*-0.5,\scale*0.866) {\small $4$};

				\path[-,draw,line width=1pt]		
				
				(1) edge node[above] {1} (2)
				
				(1) edge node[right] {1} (3)
				
				(1) edge node[left] {1} (4)
    
                (2) edge node[sloped,below] {2} (3)
				
				(2) edge node[sloped, above] {2} (4)
				
				(3) edge node[left] {2} (4);
				

		\end{tikzpicture}
		\caption{Graph with edge costs}
		\label{sfig1:counterexample_hamilitonian}
	\end{subfigure}%
	\begin{subfigure}{.33\textwidth}
		\centering
		\begin{tikzpicture}[]
				\node[main node,opacity=.5,black,fill=cyan,text opacity=1] (1) at (\scale*0,\scale*0) {\small 1};
				\node[main node,opacity=.5,black,fill=cyan,text opacity=1] (2) at (\scale *1,\scale*0) {\small $2$};
				\node[main node,opacity=.5,black,fill=cyan,text opacity=1] (3) at (\scale* -0.5,\scale*-0.866) {\small $3$};
				\node[main node,opacity=.5,black,fill=cyan,text opacity=1] (4) at (\scale*-0.5,\scale*0.866) {\small $4$};

				\path[-,draw,line width=1pt]		
				
				(1) edge node[above] {1} (2)
				
				(1) edge node[right] {1} (3)
				
				(3) edge node[left] {2} (4);
				

		\end{tikzpicture}
		\caption{Hamiltonian path}
		\label{sfig2:counterexample_hamilitonian}
	\end{subfigure}%
	\begin{subfigure}{.33\textwidth}
		\centering
        \begin{tikzpicture}[]
				\node[main node,opacity=.5,black,fill=cyan,text opacity=1] (1) at (\scale*0,\scale*0) {\small 1};
				\node[main node,opacity=.5,black,fill=cyan,text opacity=1] (2) at (\scale *1,\scale*0) {\small $2$};
				\node[main node,opacity=.5,black,fill=cyan,text opacity=1] (3) at (\scale* -0.5,\scale*-0.866) {\small $3$};
				\node[main node,opacity=.5,black,fill=cyan,text opacity=1] (4) at (\scale*-0.5,\scale*0.866) {\small $4$};

				\path[-,draw,line width=1pt]		
				
				(1) edge node[above] {1} (2)
				
				(1) edge node[right] {1} (3)
				
				(1) edge node[left] {1} (4);
				

		\end{tikzpicture}
		
		\caption{Optimal \CST solution}
		\label{sfig3:counterexample_hamilitonian}
	\end{subfigure}%
	\caption[Necessity of the Triangle Inequality for Path Graph to be (B)\CST Optimum as $\alpha\to-\infty$]{\textbf{Necessity of the Triangle Inequality for Path Graph to be (B)\CST Optimum as $\alpha\to-\infty$}. \ref{sfig1:counterexample_hamilitonian}) Graph with edge costs depicted, where the triangle inequality is not strictly satisfied. \ref{sfig2:counterexample_hamilitonian}) Optimal hamiltonian path with \CST cost  equal to $\frac{3^\alpha \cdot3+4^\alpha}{16^{\alpha}}$. \ref{sfig3:counterexample_hamilitonian}) Optimal \CST with cost equal to $\frac{3\cdot 3^\alpha}{16^{\alpha}}$. Thus, if the triangle inequality does not strictily hold, the Hamiltonian path will not necessarily be optimal even for sufficiently negative $\alpha$ values}
	\label{fig:counterexample_hamilitonian}	
\end{figure*}%

\begin{corollary}\thlabel{cor:tree_opt_alpha_-inf}
	Consider the \BCST and \CST problem where the nodes are embedded in a geodesic metric space. As $\alpha$ tends to negative infinity ($\alpha\to-\infty$) there exists a Hamiltonian path $\tree_{\star}$ with a lower \CST/\BCST cost than any other tree $\tree$ that is not a path.
\end{corollary}
\begin{proof}
	The reasoning aligns with the exposition in \thref{lem:tree_opt_alpha_-inf}, proving that for any node $v$ with degree exceeding $2$, we can always decrease its degree such that the \CST/\BCST cost of $\tree$ decreases. Through the iterative application of this process, the degrees of all nodes are systematically decreased, ultimately converging to a state where each node has at most a degree of 2, thereby resulting in the formation of a path.
	
	Consider a node $v$ with a degree higher than 2. To apply the logic presented in Theorem \ref{lem:tree_opt_alpha_-inf}, it is essential to ensure the ability to select two neighbors such that the triangle inequality holds strictly. Let $u$, $k$, and $\ell$ represent three distinct neighbors of $v$, and assume that the triangle inequality is an equality between each pair of neighbors and the node $v$. In other words, we have
	\begin{eqnarray}
	&c_{uv}+c_{vk}=c_{uk},\label{eq:triangle_inequality1}\\
	&c_{uv}+c_{v\ell}=c_{u\ell},\label{eq:triangle_inequality2}\\
	&c_{\ell v}+c_{kv}=c_{\ell k}.\label{eq:triangle_inequality3}
	\end{eqnarray}
	Starting with \eqref{eq:triangle_inequality1}, we conclude that $v$ lies on the geodesic path between $u$ and $k$. Similarly, from \eqref{eq:triangle_inequality2} and \eqref{eq:triangle_inequality3}, we deduce that $v$ is positioned between $u$ and $\ell$ and also between $\ell$ and $k$. Consequently, $\ell$ is established to be between $u$ and $k$, as it is situated between $u$ and $v$ and also between $k$ and $v$.
	
	As a result, we can infer that $c_{\ell k} + c_{\ell u} = c_{uk}$. Summing \eqref{eq:triangle_inequality1} and \eqref{eq:triangle_inequality2} and utilizing the derived equality $c_{\ell k} + c_{\ell u} = c_{uk}$ (owing to the position of $\ell$ between $u$ and $k$), we obtain:
	\[c_{uv}+c_{vk}+c_{uv}+c_{v\ell}=c_{u\ell}+c_{uk}\rightarrow 2c_{uv}+c_{\ell k}=c_{\ell k}\rightarrow c_{uv}=0\]
	
	The deduction that $c_{uv} = 0$ implies that nodes $u$ and $v$ occupy the same position and can effectively be considered as a single node. Consequently, we can systematically remove neighbors of $v$ by repeating this process until $v$ attains a degree of 2.
	
	Alternatively, if the triangle inequality must hold strictly for a pair of nodes $u, k$, and node $v$, we can, w.l.o.g., assume that node $u$ is chosen such that $m_{uv} = \min_{i\in\mathcal{N}_v}m_{iv}$. In this scenario, applying the same reasoning as the one presented in \thref{lem:tree_opt_alpha_-inf}, we demonstrate that the modified topology $\tree'$ --where node $k$ is connected to node $u$ instead of node $v$-- results in a lower \BCST/\CST cost.
	
\end{proof}

\thref{lem:tree_opt_alpha_-inf} resembles \thref{lem:strong_triangle_inequality_CST} in the sense that both lemmas require the satisfaction of a weighted triangle inequality. The following corollary reformulates \thref{lem:tree_opt_alpha_-inf}, mirroring the structure found in \thref{lem:strong_triangle_inequality_CST}, and accentuates the correlation between the weighted triangle inequality and the number of terminals denoted by $N$.

\begin{corollary}[\thref{th:tree_opt_alpha_-inf_reformulation}]\thlabel{cor:tree_opt_alpha_-inf_reformulation}
	Given a complete graph $G$ with $N$ nodes, let $c_{ij}$ be the edge-costs of any pair of nodes $(i,j)$. If there exists 
	$$t\leq \min_{\substack{1 \leq s\leq N-3\\1\leq \ell \leq \min(s,(N-s)/2-1)}} \frac{\left(\ell(N-\ell)\right)^{\alpha}-\left((\ell+s)(N-\ell-s)\right)^\alpha}{\left(s(N-s)\right)^\alpha}$$
	such that 
	\[c_{kv}+tc_{uv}\geq c_{ku}\]
	for all triangles in the graph, then there exists an optimum $\CST$ evaluated at $\alpha$ which is a Hamiltonian path.
\end{corollary}
\begin{proof}
	\thref{lem:tree_opt_alpha_-inf} holds if
	\begin{eqnarray}\label{eq1:cor:tree_opt_alpha_-inf_reformulation}
	\nonumber &\left(c_{uv}-c_{uv}\left(\frac{\left(m^{\tree}_{uv}+m^{\tree}_{kv}\right)\left(1-m^{\tree}_{uv}-m^{\tree}_{kv}\right)}{\left(m^{\tree}_{uv}\big(1-m^{\tree}_{uv}\big)\right)}\right)^{\alpha}
	+(c_{kv}-c_{ku})\left(\frac{m^{\tree}_{kv}\big(1-m^{\tree}_{kv}\big)}{m^{\tree}_{uv}\big(1-m^{\tree}_{uv}\big)}\right)^{\alpha}\right)>0\\ &\iff c_{uv}\left(\frac{\left(m^{\tree}_{uv}\big(1-m^{\tree}_{uv}\big)\right)^{\alpha}-\left(\left(m^{\tree}_{uv}+m^{\tree}_{kv}\right)\left(1-m^{\tree}_{uv}-m^{\tree}_{kv}\right)\right)^{\alpha}}{\left(m^{\tree}_{kv}\left(1-m^{\tree}_{kv}\right)\right)^{\alpha}}\right)+c_{kv}>c_{ku}
	\end{eqnarray}
	where it is assumed that $m_{uv}\min_{i\in\mathcal{N}_v} m_{iv}$. Thus, let $m_{uv}=\frac{\ell}{N}$ and $m_{kv}=\frac{s}{N}$. Substituting these values into \eqref{eq1:cor:tree_opt_alpha_-inf_reformulation}, we derive the following inequality
	\[c_{kv}+\frac{\left(\ell(N-\ell)\right)^{\alpha}-\left((\ell+s)(N-\ell-s)\right)^\alpha}{\left(s(N-s)\right)^\alpha}c_{uv}\geq c_{ku}.\]
	
	Note that $\ell\leq \min(s,(N-s)/2-1)$ since $m_{uv}\min_{i\in\mathcal{N}_v} m_{iv}$. Thus, $m_{uv}=\frac{\ell}{N}\leq m_{kv}=\frac{s}{N}$. On the other hand, since  in the proof of \thref{lem:tree_opt_alpha_-inf}, $v$ has at least three neighbors ($u$, $k$ and say $p$), the $N-s$ nodes not lying in the side corresponding to $k$ of  edge $(k,v)$ must be distributed in the side of $v$. At exception of node $v$, one part of the remaining $N-s-1$ nodes will be in the side corresponding to $u$ of edge $(u,v)$ and the other part in the side corresponding to $p$ of edge $(p,v)$. Since $m_{uv}$ must be minimum, the side corresponding to $u$ of edge $(u,v)$ can at most be equal to $\frac{N-s-1}{2N}$.
	
	Notice also that $1\leq s\leq N-3$, since $v$ has degree at least three and therefore there are at least three nodes ($v$, $u$, $p$) lying in the same side of $v$ with respect to edge $(k,v)$. Thus, $m^{\tree}_{vk}=(1-m^{\tree}_{kv})=1-\frac{s}{N}\geq \frac{3}{N}$, or equivalently $s\leq N-3$.
\end{proof}

\section{Exploring the Number of Derivable Topologies from \CST and \BCST Topologies}\label{sec:app_num_topos}
\subsection{Number of BCST Topologies Derivable from a CST Topology}
In this section, we explicitly determine the number of topologies of the \BCST problem that can be derived from a single \CST topology. 

To derive a full tree topology $\tree_{\BCST}$ from a \CST topology $\tree_{\CST}$ with $N$ terminals, we need to add $N-2$ \BP. In particular, for each terminal node, $v$, with degree $d_v\geq 2$, we need to spawn $d_v-1$ \BP. Since for $k$ terminal nodes, there exist a total $(2k-5)!!$ of full tree topologies \citep{schroder_vier_1870}, there are $(2(d_v+1)-5)!!=(2d_v-3)!!$ ways to connect the added \BP to the neighbors of $v$ and $v$ itself. Thus the total number of full tree topologies is equal to the number of possible combinations of subtopologies engendered per terminal neighborhood for terminals with degree higher than 2. Formally, this number is equal
\begin{equation}
    \prod_{v\ :\ d_v\geq 2} (2d_v-3)!!.
\end{equation}
Note that, on the one hand, if all nodes have degree lower or equal than $2$, i.e. the tree is a path, then a single full tree topology can be derived. On the other hand, if the original $\tree_{\CST}$ is a star graph, then there is a single graph with degree higher than $2$, which is equal to $N-1$. Thus, the total number of topologies derived from it is equal to $(2N-5)!!$. This is the total number of possible full tree topologies, hence a star graph can generate any full tree topology. In general, the higher the degree of the nodes in $\tree_{\CST}$, the higher the number of derivable full tree topologies.

\subsection{Number of CST Topologies Derivable from a BCST Topology}\label{subsec:app_numtopos_CST}
In this case, we need to collapse each \BP to a terminal. The collapse process can be carried out sequentially, where each \BP is collapsed to one of its neighboring nodes, until no \BP remain. Naively, we might assume that there are $3^{N-2}$ possible topologies, given that each \BP has 3 neighbors available for collapse and there are $N-2$ \BPs. However, this is not the case because some combinations may result in non-valid topologies. For instance, if all \BPs collapse with neighbors that are also \BP, none of the \BP will be collapsed with a terminal node.


Before providing the formula for the number of $\CST$ topologies that can be derived from a full tree topology, let's introduce the All Minors Matrix Tree Theorem \citep{chaiken1982combinatorial}, which is necessary to derive the formula.  The Matrix Tree Theorem \citep{Tutte1984, kirchhoff1847ueber} states that the number of spanning trees of a given graph $G=(V,E)$ can be calculated from the determinant of a submatrix of its Laplacian. Recall that the Laplacian matrix of a graph $L$ is given by 
$$L=D-A, \ \text{ where } \  A_{ij}=\begin{cases}
    1 & \text{if } (i,j)\in E\\
    0 & \text{otherwise }\\
\end{cases} \qquad \& \qquad D_{ij}=\begin{cases}
    \sum_{k}A_{ik} & \text{if } i=j\\
    0 & \text{otherwise }\end{cases}.$$
In other words, $A$ is the adjacency matrix of the graph, and $D$ is the diagonal degree matrix.  We state a simplified version of the theorem without providing a proof.
\begin{theorem}\thlabel{thm:allminorsMTT}
    Given a graph $G=(V,E)$ and a subset $U$ of nodes in $G$, let $W=V\backslash U$. We define $L_{W,W}$ as the submatrix of the Laplacian matrix of $G$, which includes the rows and columns indexed by the nodes in $W$. In this context, the determinant of $L_{W,W}$, denoted as $\det (L_{W,W})$, provides a count of the number of spanning forests of $G$ that consist of $|U|$ disjoint trees, with the nodes in $U$ being disconnected across these trees.
\end{theorem}
\begin{proof}
See \citet{chaiken1982combinatorial}.
\end{proof}

Now we are ready to present the main result of this subsection. Consider a full tree topology $\tree_{\BCST}$. The number of topologies for the \CST problem that can be derived from $\tree_{\BCST}$ is given by
\begin{equation}
	\label{eq:number_topos_derived_from_fulltopo}
	\det (L_{\BPs,\BPs}),
\end{equation}
where $L_{\BPs,\BPs}$ represents the submatrix of the Laplacian matrix $L$ of $\tree_{\BCST}$. This submatrix is formed by selecting the rows and columns associated with the \BPs. By virtue of \thref{thm:allminorsMTT}, equation \eqref{eq:number_topos_derived_from_fulltopo} counts the number of spanning forests of the $\tree_{\BCST}$ which disconnect the terminal nodes. To demonstrate that this count of forests coincides with the number of topologies that can be derived from the full tree topology $\tree_{\BCST}$, we will establish a bijection.

Indeed, if we have a forest that disconnects all the terminals, each \BP within the forest must belong to a component with a single terminal. In this scenario, we can unambiguously collapse each \BP to its corresponding terminal. Once we have collapsed the \BP, we still need to reconnect the terminals between them to form a valid \CST topology. Now, notice that in the original full  tree  topology $\tree_{\BCST}$, each terminal is uniquely adjacent to a \BP. We can connect the terminals between them based on the collapse process of the \BP. Specifically, a terminal $v_t$ is connected to another terminal $u_t$ if the neighboring \BP of $v_t$ in the original $\tree_{\BCST}$ has been collapsed to $u_t$. Similarly, we can reverse these steps to map a \CST topology back to a unique forest that disconnects the terminals. \figurename{} \ref{fig:equivalence_forest_CSTtopo} illustrates the individual steps of this bijection using two examples. We have proven the following theorem

\begin{theorem}\thlabel{th:num_derivable_topos_from_BCST}
	Let $\tree_{\BCST}$ be a full tree topology with $N$ terminals. Consider the Laplacian matrix $L$ of $\tree_{\BCST}$. The number of $\CST$ topologies that can be derived from $\tree_{\BCST}$ is equal to the determinant of $L_{\BPs,\BPs}$, which is the submatrix of the Laplacian obtained by selecting the rows and columns indexed by the \BPs. Hence, the number of $\CST$ topologies derived from $\tree_{\BCST}$ can be calculated as $\det L_{\BPs,\BPs}$.
\end{theorem}

\tikzset{
	dot/.style 2 args={fill, circle, inner sep=0pt, label={#1:\scriptsize #2}},	
	fulldot/.style 2 args={circle,draw,minimum size=0.3cm,inner sep=0pt, label={#1:\scriptsize #2}},
	main node/.style={circle,draw,minimum size=0.4cm,inner sep=0pt]},
	mini node/.style={circle,draw,minimum size=0.3cm,inner sep=0pt]},
}
\def \scale{1 }
\def \scalemini{0.4*\scale}
\def \dispx{2.75}
\def \dispy{2}
\def \startarrow{0.58}
\def \endarrow{0.7}
\def \escalabox{1}
\begin{figure*}[h!]
	\centering
 \begin{subfigure}{1\textwidth}
		\centering
		\scalebox{\escalabox}{
            \begin{tikzpicture}[]
				\def \scalegraph{2}

				\node[main node,opacity=.5,black,fill=cyan,text opacity=1] (1) at (\scalegraph*-1.5,\scalegraph*0.75) {1};
				\node[main node,opacity=.5,black,fill=cyan,text opacity=1] (2) at (\scalegraph*-1.5,\scalegraph*-0.75) {\small $2$};
				\node[main node,opacity=.5,black,fill=cyan,text opacity=1] (3) at (\scalegraph*1.5,\scalegraph*0.75) {\small $3$};
				\node[main node,opacity=.5,black,fill=cyan,text opacity=1] (4) at (\scalegraph*1.5,-\scalegraph*0.75) {\small $4$};

                \node[main node,opacity=.5,black,fill=red,text opacity=1] (5) at (\scalegraph*-0.5,\scalegraph*0) {\small $5$};
    
				\node[main node,opacity=.5,black,fill=red,text opacity=1] (6) at (\scalegraph*0.5,\scalegraph*0) {\small $6$};	
				
				\path[-,draw,line width=1pt]		
				
				(1) edge node[below] {} (5)		
				
				(2) edge node[below] {} (5)		
				
				(3) edge node[below] {} (6)		
				
				(4) edge node[below] {} (6)
				
				(5) edge node[below] {} (6);
			
		\end{tikzpicture}}
		\caption{Original full tree topology $\tree_{\BCST}$}
		\label{sfig1:equivalence_forest_CSTtopo}
	\end{subfigure}
	\begin{subfigure}{1\textwidth}
		\centering
		\scalebox{\escalabox}{
            \begin{tikzpicture}[]
				\def \scalegraph{1}
				\def \shiftxstepA{5}
				\def \shiftxstepB{2*\shiftxstepA}
				\def \shiftystepA{0}
				\def \shiftystepB{2*\shiftystepA}
                \def \sepxgraphs{\scalegraph*2}

				\node[main node,opacity=.5,black,fill=cyan,text opacity=1] (1) at (\scalegraph*-1.5,\scalegraph*0.75) {1};
				\node[main node,opacity=.5,black,fill=cyan,text opacity=1] (2) at (\scalegraph*-1.5,\scalegraph*-0.75) {\small $2$};
				\node[main node,opacity=.5,black,fill=cyan,text opacity=1] (3) at (\scalegraph*1.5,\scalegraph*0.75) {\small $3$};
				\node[main node,opacity=.5,black,fill=cyan,text opacity=1] (4) at (\scalegraph*1.5,-\scalegraph*0.75) {\small $4$};

                \node[main node,opacity=.5,black,fill=red,text opacity=1] (5) at (\scalegraph*-0.5,\scalegraph*0) {\small $5$};
    
				\node[main node,opacity=.5,black,fill=red,text opacity=1] (6) at (\scalegraph*0.5,\scalegraph*0) {\small $6$};	

                \node[opacity=.0,text opacity=1] (inv1) at (\scalegraph*0,-\scalegraph*1.5) {\scriptsize  Spanning Forest Separating the Terminals};	
				
				\path[-,draw,line width=1pt]		
				
				(1) edge node[below] {} (5)
				
				(5) edge node[below] {} (6);


				\draw [latex-latex,line width=1pt](\scalegraph*1.5 +0.3*\sepxgraphs,\scalegraph*0) -- (\scalegraph*1.5 +0.7*\sepxgraphs,\scalegraph*0);
				
				\node[main node,opacity=.5,black,fill=cyan,text opacity=1] (1) at (\scalegraph*-1.5+\scalegraph*\shiftxstepA,\scalegraph*0.75+\scalegraph*\shiftystepA) {\tiny 1$|$5$|$6};
				\node[main node,opacity=.5,black,fill=cyan,text opacity=1] (2) at (\scalegraph*-1.5+\scalegraph*\shiftxstepA,\scalegraph*-0.75+\scalegraph*\shiftystepA) {\small $2$};
				\node[main node,opacity=.5,black,fill=cyan,text opacity=1] (3) at (\scalegraph*1.5+\scalegraph*\shiftxstepA,\scalegraph*0.75+\scalegraph*\shiftystepA) {\small $3$};
				\node[main node,opacity=.5,black,fill=cyan,text opacity=1] (4) at (\scalegraph*1.5+\scalegraph*\shiftxstepA,-\scalegraph*0.75+\scalegraph*\shiftystepA) {\small $4$};	

                \node[opacity=.0,text opacity=1] (inv2) at (\scalegraph*0+\scalegraph*\shiftxstepA,-\scalegraph*1.5) {\scriptsize \BP Collapsed with Terminals};


				\draw [latex-latex,line width=1pt](\scalegraph*1.5 +0.3*\sepxgraphs+\scalegraph*\shiftxstepA,\scalegraph*0) -- (\scalegraph*1.5 +0.7*\sepxgraphs+\scalegraph*\shiftxstepA,\scalegraph*0);
				
                \node[main node,opacity=.5,black,fill=cyan,text opacity=1] (1) at (\scalegraph*-1.5+\scalegraph*\shiftxstepB,\scalegraph*0.75+\scalegraph*\shiftystepB) {\tiny 1$|$5$|$6};
				\node[main node,opacity=.5,black,fill=cyan,text opacity=1] (2) at (\scalegraph*-1.5+\scalegraph*\shiftxstepB,\scalegraph*-0.75+\scalegraph*\shiftystepB) {\small $2$};
				\node[main node,opacity=.5,black,fill=cyan,text opacity=1] (3) at (\scalegraph*1.5+\scalegraph*\shiftxstepB,\scalegraph*0.75+\scalegraph*\shiftystepB) {\small $3$};
				\node[main node,opacity=.5,black,fill=cyan,text opacity=1] (4) at (\scalegraph*1.5+\scalegraph*\shiftxstepB,-\scalegraph*0.75+\scalegraph*\shiftystepB) {\small $4$};

                \node[opacity=.0,text opacity=1,align=center] (inv3) at (\scalegraph*0+\scalegraph*\shiftxstepB,-\scalegraph*1.5) {\scriptsize Interconnected Terminals Based\\\scriptsize on Neighboring \BP};	
				
				\path[-,draw,line width=1pt]		
				
				(1) edge node[below] {} (2)
				
				(1) edge node[below] {} (3)
				
				(1) edge node[below] {} (4);

		\end{tikzpicture}}
		\caption{Example 1, Bijection Steps}
		\label{sfig2:equivalence_forest_CSTtopo}
	\end{subfigure}
    \begin{subfigure}{1\textwidth}
		\centering
		\scalebox{\escalabox}{
            \centering
            \begin{tikzpicture}[]
				\def \scalegraph{1}
				\def \shiftxstepA{5}
				\def \shiftxstepB{2*\shiftxstepA}
				\def \shiftystepA{0}
				\def \shiftystepB{2*\shiftystepA}
                \def \sepxgraphs{\scalegraph*2}

				\node[main node,opacity=.5,black,fill=cyan,text opacity=1] (1) at (\scalegraph*-1.5,\scalegraph*0.75) {1};
				\node[main node,opacity=.5,black,fill=cyan,text opacity=1] (2) at (\scalegraph*-1.5,\scalegraph*-0.75) {\small $2$};
				\node[main node,opacity=.5,black,fill=cyan,text opacity=1] (3) at (\scalegraph*1.5,\scalegraph*0.75) {\small $3$};
				\node[main node,opacity=.5,black,fill=cyan,text opacity=1] (4) at (\scalegraph*1.5,-\scalegraph*0.75) {\small $4$};

                \node[main node,opacity=.5,black,fill=red,text opacity=1] (5) at (\scalegraph*-0.5,\scalegraph*0) {\small $5$};
    
				\node[main node,opacity=.5,black,fill=red,text opacity=1] (6) at (\scalegraph*0.5,\scalegraph*0) {\small $6$};	

                \node[opacity=.0,text opacity=1] (inv1) at (\scalegraph*0,-\scalegraph*1.5) {\scriptsize Spanning Forest Separating the Terminals};	
				
				\path[-,draw,line width=1pt]		
				
				(1) edge node[below] {} (5)
				
				(4) edge node[below] {} (6);


				\draw [latex-latex,line width=1pt](\scalegraph*1.5 +0.3*\sepxgraphs,\scalegraph*0) -- (\scalegraph*1.5 +0.7*\sepxgraphs,\scalegraph*0);
				
				\node[main node,opacity=.5,black,fill=cyan,text opacity=1] (1) at (\scalegraph*-1.5+\scalegraph*\shiftxstepA,\scalegraph*0.75+\scalegraph*\shiftystepA) {\tiny 1$|$5};
				\node[main node,opacity=.5,black,fill=cyan,text opacity=1] (2) at (\scalegraph*-1.5+\scalegraph*\shiftxstepA,\scalegraph*-0.75+\scalegraph*\shiftystepA) {\small $2$};
				\node[main node,opacity=.5,black,fill=cyan,text opacity=1] (3) at (\scalegraph*1.5+\scalegraph*\shiftxstepA,\scalegraph*0.75+\scalegraph*\shiftystepA) {\small $3$};
				\node[main node,opacity=.5,black,fill=cyan,text opacity=1] (4) at (\scalegraph*1.5+\scalegraph*\shiftxstepA,-\scalegraph*0.75+\scalegraph*\shiftystepA) {\tiny 4$|$6};	

                \node[opacity=.0,text opacity=1] (inv2) at (\scalegraph*0+\scalegraph*\shiftxstepA,-\scalegraph*1.5) {\scriptsize \BP Collapsed with Terminals};


				\draw [latex-latex,line width=1pt](\scalegraph*1.5 +0.3*\sepxgraphs+\scalegraph*\shiftxstepA,\scalegraph*0) -- (\scalegraph*1.5 +0.7*\sepxgraphs+\scalegraph*\shiftxstepA,\scalegraph*0);
				
                \node[main node,opacity=.5,black,fill=cyan,text opacity=1] (1) at (\scalegraph*-1.5+\scalegraph*\shiftxstepB,\scalegraph*0.75+\scalegraph*\shiftystepB) {\tiny 1$|$5};
				\node[main node,opacity=.5,black,fill=cyan,text opacity=1] (2) at (\scalegraph*-1.5+\scalegraph*\shiftxstepB,\scalegraph*-0.75+\scalegraph*\shiftystepB) {\small $2$};
				\node[main node,opacity=.5,black,fill=cyan,text opacity=1] (3) at (\scalegraph*1.5+\scalegraph*\shiftxstepB,\scalegraph*0.75+\scalegraph*\shiftystepB) {\small $3$};
				\node[main node,opacity=.5,black,fill=cyan,text opacity=1] (4) at (\scalegraph*1.5+\scalegraph*\shiftxstepB,-\scalegraph*0.75+\scalegraph*\shiftystepB) {\tiny 4$|$6};

                \node[opacity=.0,text opacity=1,align=center] (inv3) at (\scalegraph*0+\scalegraph*\shiftxstepB,-\scalegraph*1.5) {\scriptsize Interconnected Terminals Based\\\scriptsize on Neighboring \BP};	
				
				\path[-,draw,line width=1pt]		
				
				(1) edge node[below] {} (2)
				
				(1) edge node[below] {} (4)
				
				(3) edge node[below] {} (4);

		\end{tikzpicture}}
		\caption{Example 2, Bijection Steps}
		\label{sfig3:equivalence_forest_CSTtopo}
	\end{subfigure}%
    \caption[Bijection Between Derivable \CST Topologies from a Full Tree Topology and Their Terminal-Separating Forests]{\textbf{Bijection Between Derivable \CST Topologies from a Full Tree Topology and Their Terminal-Separating Forests}. Figures \ref{sfig2:equivalence_forest_CSTtopo}) and \ref{sfig3:equivalence_forest_CSTtopo}) illustrate two examples of the relationship between a spanning forest and a derived \CST topology of a full tree topology depicted in Figure \ref{sfig1:equivalence_forest_CSTtopo}.}
    \label{fig:equivalence_forest_CSTtopo}
 \end{figure*}

\section{Branching Angles at the Steiner Points in the \BCST Problem}\label{sec:app_angles}

In this section, we formulate the branching angles in terms of the centralities of the edges for a given topology of the \BCST problem. As stated in Section \ref{sec:branching-angles-at-the-steiner-points}, it is sufficient to study the geometric optimization of 3 nodes connected by a single \BP, with minimization objective given by
\begin{equation}\label{eq:objective_3terminals}
    C(b)=\centflow_0 ||b-a_0||+\centflow_1 ||b-a_1||+\centflow_2 ||b-a_2||.
\end{equation}
Recall that node $b$ represents the Steiner point whose coordinates need to be optimized, nodes $\{a_i\}$ are the terminals with fixed positions and $\centflow_i\coloneqq m_{ba_i}(1-m_{ba_i})$ are the centralities of the edges $(b,a_i)$ (see \figurename{} \ref{fig:angles_SP}).

\tikzset{
	dot/.style 2 args={fill, circle, inner sep=0pt, label={#1:\scriptsize #2}},	
	fulldot/.style 2 args={circle,draw,minimum size=0.3cm,inner sep=0pt, label={#1:\scriptsize #2}},
	main node/.style={circle,draw,minimum size=0.4cm,inner sep=0pt]},
	mini node/.style={circle,draw,minimum size=0.3cm,inner sep=0pt]},
    invisible/.style={circle,minimum size=0.001cm,inner sep=0pt]},
}
\def \scale{2}

\begin{figure*}[h!]
	\centering

    \begin{tikzpicture}[]

        \node[invisible,text opacity=0] (i) at (\scale*-0.,\scale*1) {\small $1$};

        \node[main node,opacity=.5,black,fill=cyan,text opacity=1] (0) at (\scale*-
                      0.,\scale*-1) {\small $a_0$};
        \node[main node,opacity=.5,black,fill=cyan,text opacity=1] (1) at (\scale* -0.8,\scale*0.866) {\small $a_1$};
        \node[main node,opacity=.5,black,fill=cyan,text opacity=1] (2) at (\scale *1,\scale*0.75) {\small $a_2$};

        \node[main node,opacity=.5,black,fill=red,text opacity=1] (b) at (\scale*0,\scale*0) {\small b};
        
        \path[-,draw,line width=1pt]		
        
            (b) edge node[right] {$\centflow_0$} (0)
            
            (b) edge node[left] {$\centflow_1$} (1)
            
            (b) edge node[right] {$\centflow_2$} (2);

        \path[dashed,draw,line width=1pt]		
        
            (b) edge node[above] {} (i);

        \draw  pic["$\theta_1$", draw, -, angle eccentricity=1.2, angle radius=1.2cm]
        {angle=i--b--1};
        \draw  pic["$\theta_2$", draw, -, angle eccentricity=1.2, angle radius=1cm]
        {angle=2--b--i};


    \end{tikzpicture}
    \caption[Branching Angles at Steiner Point]{\textbf{Branching Angles at Steiner Point}. The symbols $\centflow_i$ represent the normalized centralities of the edges, that is $\centflow_i\coloneqq m_{ba_i}(1-m_{ba_i})$.}
    \label{fig:angles_SP}
\end{figure*}

We will reproduce the arguments exposed for the BOT problem in \citet{bernot_optimal_2008} and \citet{lippmann_theory_2022} to determine the angles $\theta_1$ and $\theta_2$. 
We will differentiate two cases: when the \BP does not coincide with any other terminal node; and when $b$ collapses with one of the terminals.

\subsection{Steiner Point $b$ Does Not Collapse with a Terminal}
In this case, the function is differentiable with respect to $b$, and therefore we just need to see where the gradient of equation \eqref{eq:objective_3terminals} is equal to zero. The formula for the gradient is as follows
\begin{equation}\label{eq:grad_objective_3terminals}
    \nabla_b C(b)= \centflow_0^\alpha n_0+\centflow_1^\alpha n_1+\centflow_2^\alpha n_2,
\end{equation}
where $n_i=\frac{b-a_i}{||b-a_i||}$. By applying the dot product to $\nabla_b C(b)$ with each $n_i$ and setting it equal to zero, we derive the following equalities:%
\begin{eqnarray*}
    \langle\nabla_b C(b),n_0\rangle=0 &\rightarrow& \centflow_0^\alpha +\centflow_1^\alpha \underbrace{\langle n_1,n_0\rangle}_{-\cos(\theta_1)}+\centflow_2^\alpha \underbrace{\langle n_2,n_0\rangle}_{-\cos(\theta_2)}=0\\
    \langle\nabla_b C(b),n_1\rangle=0 &\rightarrow& \centflow_0^\alpha\underbrace{\langle n_0,n_1\rangle}_{-\cos(\theta_1)} +\centflow_1^\alpha+\centflow_2^\alpha\langle n_2,n_1\rangle=0\\
    \langle\nabla_b C(b),n_2\rangle=0 &\rightarrow& \centflow_0^\alpha \underbrace{\langle n_0,n_2\rangle}_{-\cos(\theta_2)}+\centflow_1^\alpha \langle n_1,n_2\rangle+\centflow_2^\alpha=0
\end{eqnarray*}%
Solving the linear system we obtain that the angles satisfy
\begin{equation}\label{eq:optimal_angles}
    \begin{aligned}
        \cos(\theta_1)&=&\frac{\centflow_0^{2\alpha}+\centflow_1^{2\alpha}-\centflow_2^{2\alpha}}{2\centflow_0^{\alpha}\cdot\centflow_1^{\alpha}},\\
        \cos(\theta_2)&=&\frac{\centflow_0^{2\alpha}+\centflow_2^{2\alpha}-\centflow_1^{2\alpha}}{2\centflow_0^{\alpha}\cdot\centflow_2^{\alpha}},\\
        \cos(\theta_1+\theta_2)&=&\frac{\centflow_0^{2\alpha}-\centflow_1^{2\alpha}-\centflow_2^{2\alpha}}{2\centflow_1^{\alpha}\cdot\centflow_2^{\alpha}}.
    \end{aligned}
\end{equation}

\subsection{Steiner Point $b$ Collapses with a Terminal} \label{subsec:Vbranching}
In this case, in order to determine the optimality angles, we will use the subdifferential argument applied in \citet{lippmann_theory_2022}. W.l.o.g. we will assume that $b$ collapses with terminal $a_0$. 

The subdifferential of a convex function $h:\mathbb{R}^n\to \mathbb{R}$ at $x$ is defined as the following set of vectors 
\[\partial g(x)\coloneqq\{v \ : \ h(z)\geq h(x) +\langle v,z-x\rangle, \ \forall z\in\mathbb{R}^n\}.\]
In other words, $\partial g(x)$ comprises all vectors $v$ such that the line passing through $h(x)$ in the direction of $v$ lies below the function $h$ at all points. Each of these vectors is called a subgradient of $h$ at $x$. When a function is differentiable at $x$, the subdifferential only contains the gradient of the function at $x$.

Fermat rule states that a convex function attains its minimum at $x$ if and only if $0\in \partial g(x)$. Furthermore, the subdifferential of two convex functions is equal to the union of the pairwise sums of their subgradients. In other words, for $g(x)=g_1(x)+g_2(x)$ then 
\begin{equation}\label{eq:sum_subdifferential}
    \partial g(x)=\{v_1 + v_2 \ : \ v_1\in \partial g_1(x), \ v_2\in \partial g_2(x)\}.  
\end{equation}

We can apply Fermat's rule to determine when the minimum is attained at $b=a_0$. For the function $g(x)=w\cdot ||x-a||$, the subdifferential is given by 
\[\partial g(x)=\begin{cases}
    \{v \ : \ ||v||\leq w\}, \quad &\text{ if } x=a\\
    \left\{w\frac{x-a}{||x-a||}\right\}, &\text{ otherwise } 
\end{cases}.\]
Thus, applying equation \eqref{eq:sum_subdifferential}, the subdifferential of $C(b)$ at $b=a_0$ is given by
\[\partial C(a_0)=\left\{v+\centflow_1^\alpha\frac{b-a_1}{||b-a_1||}+\centflow_2^\alpha\frac{b-a_2}{||b-a_2||} \ : \ ||v||\leq \centflow_0^{\alpha} \right\}.\]
In order for $b$ to be optimal at $a_0$, zero has to belong to $\partial C(a_0)$, which is true if and only if 
\begin{equation}
    \begin{aligned}
    &\left|\left|\centflow_1^\alpha\frac{b-a_1}{||b-a_1||}+\centflow_2^\alpha\frac{b-a_2}{||b-a_2||}\right|\right|\leq \centflow_0^\alpha\\
    \iff& \left|\left|\frac{b-a_1}{||b-a_1||}+\frac{b-a_2}{||b-a_2||}\right|\right|^2=\centflow_1^{2\alpha}+\centflow_2^{2\alpha}+2\centflow_1^{\alpha}\centflow_2^{\alpha}\cos(\gamma)\leq \centflow_0^{2\alpha}
    \end{aligned}
\end{equation}
where $\gamma$ is the angle of the terminal triangle at $a_0$, that is $\gamma\coloneqq\angle a_1a_0a_2$. Isolating $\gamma$, we get
\begin{equation}
    \label{eq:branching angle}
    \gamma\geq \arccos\left(\frac{\centflow_0^{2\alpha}-\centflow_1^{2\alpha}-\centflow_2^{2\alpha}}{2\centflow_1^{\alpha}\cdot\centflow_2^{\alpha}}\right)=\theta_1+\theta_2.
\end{equation}
Thus $b$ will collapse to $a_0$ if the angle $\angle a_1a_0a_2$ is greater than the optimal angle given by \eqref{eq:optimal_angles}. In such cases, the resulting branching is referred to as a $V$-branching.
\begin{remark}\thlabel{rem:sol_depend_on_flows}
It is worth noting that the reasoning presented in this section remains independent of the weighting factors, which, in our case, were set equal to the normalized edge centralities powered to $\alpha$. As a result, this finding holds true for any weights and can be used to determine an arbitrary weighted geometric median of three points.  Furthermore, we emphasize that the position of the \BP, $b$, depends exclusively on the angles and weighting factors and not on the distances between the terminal nodes.
\end{remark}

\section{Infeasibility of Degree-4 Steiner Points in the Plane for $\alpha=1$}\label{sec:infeasibility-of-degree-4-steiner-point-for-alpha1}
\tikzset{
	dot/.style 2 args={fill, circle, inner sep=0pt, label={#1:\scriptsize #2}},	
	fulldot/.style 2 args={circle,draw,minimum size=0.3cm,inner sep=0pt, label={#1:\scriptsize #2}},
	main node/.style={circle,draw,minimum size=0.4cm,inner sep=0pt]},
	mini node/.style={circle,draw,minimum size=0.3cm,inner sep=0pt]},
    invisible/.style={circle,minimum size=0.001cm,inner sep=0pt]},
}
\def \scale{1.8}

\begin{figure*}[h!]
	\centering
    \begin{subfigure}{0.5\textwidth}
    	\centering

        \begin{tikzpicture}[]
            \node[main node,opacity=.5,black,fill=cyan,text opacity=1] (0) at (\scale*-
                          1.2,\scale*-0.8) {\small $a_3$};
            \node[main node,opacity=.5,black,fill=cyan,text opacity=1] (1) at (\scale* -1.2,\scale*0.8) {\small $a_1$};
            \node[main node,opacity=.5,black,fill=cyan,text opacity=1] (2) at (\scale *1.2,\scale*0.8) {\small $a_2$};
    
            \node[main node,opacity=.5,black,fill=cyan,text opacity=1] (3) at (\scale *1.2,-\scale*0.8) {\small $a_4$};
    
            \node[main node,opacity=.5,black,fill=red,text opacity=1] (b) at (\scale*0,\scale*0) {\small $b$};

            \node[invisible] (i1) at (\scale*-1.2,\scale*0) {};
            \node[invisible] (i2) at (\scale*1.2,\scale*0) {};
            
            \path[-,draw,line width=1pt]		
            
                (b) edge node[right] {} (0)
                
                (b) edge node[left] {} (1)
                
                (b) edge node[right] {} (2)
                
                (b) edge node[right] {} (3);

            \path[dashed,draw,line width=1pt]		
                (b) edge node[right] {} (i1)
                
                (b) edge node[right] {} (i2);
            \draw  pic["$\gamma$", draw, -, angle eccentricity=1.2, angle radius=1cm]
            {angle=2--b--1};
            \draw  pic["$\theta_1$", draw, -, angle eccentricity=1, angle radius=1.2cm,left]
            {angle=1--b--i1};
            \draw  pic["$\theta_2$", draw, -, angle eccentricity=1, angle radius=1.2cm,right,]
            {angle=i2--b--2};
        \end{tikzpicture}
        \caption{}
        \label{sfig1:degree4_alph1_approach}
    \end{subfigure}\hfill
    \begin{subfigure}{0.5\textwidth}
        \centering
        \begin{tikzpicture}[]

            \node[main node,opacity=.5,black,fill=cyan,text opacity=1] (0) at (\scale*-
                          1.2,\scale*-0.8) {\small $a_3$};
            \node[main node,opacity=.5,black,fill=cyan,text opacity=1] (1) at (\scale* -1.2,\scale*0.8) {\small $a_1$};
            \node[main node,opacity=.5,black,fill=cyan,text opacity=1] (2) at (\scale *1.2,\scale*0.8) {\small $a_2$};
    
            \node[main node,opacity=.5,black,fill=cyan,text opacity=1] (3) at (\scale *1.2,-\scale*0.8) {\small $a_4$};
    
            \node[main node,opacity=.5,black,fill=red,text opacity=1] (b1) at (\scale*-0.4,\scale*0) {\small $b_1$};

            \node[main node,opacity=.5,black,fill=red,text opacity=1] (b2) at (\scale*0.4,\scale*0) {\small $b_2$};

            \node[invisible] (i1) at (\scale*-1.2,\scale*0) {};
            \node[invisible] (i2) at (\scale*1.2,\scale*0) {};

            \node[invisible] (ia0) at (\scale*-1.2+-\scale*0.3,\scale*-0.8) {};
            \node[invisible] (ia1) at (\scale*-1.2+-\scale*0.3,\scale*0.8) {};
            \node[invisible] (ia2) at (\scale*1.2+\scale*0.3,\scale*0.8) {};
            \node[invisible] (ia3) at (\scale*1.2+\scale*0.3,-\scale*0.8) {};
            
            \path[-,draw,line width=1pt]		
            
                (b1) edge node[right] {} (0)
                
                (b1) edge node[left] {} (1)
                
                (b2) edge node[right] {} (2)
                
                (b2) edge node[right] {} (3)
                
                (b2) edge node[above ] {$\leftarrow \cdot \rightarrow$} (b1);
                
            \path[dashed,draw,line width=1pt]		
                (b1) edge node[right] {} (i1)
                
                (b2) edge node[right] {} (i2);

            \path[->,draw,line width=0.5pt]		
                (0) edge node[right] {} (ia0)
                
                (1) edge node[right] {} (ia1)
                
                (2) edge node[right] {} (ia2)
                
                (3) edge node[right] {} (ia3);

            \draw  pic["$\theta_1$", draw, -, angle eccentricity=1, angle radius=0.8cm,left]
            {angle=1--b1--i1};
            \draw  pic["$\theta_2$", draw, -, angle eccentricity=1, angle radius=0.8cm,right,]
            {angle=i2--b2--2};

        \end{tikzpicture}
        
        \caption{}
        \label{sfig2:degree4_alph1_approach}
    \end{subfigure}
    \caption[Splitting Collapsed \BP While Preserving Optimal Angles]{\textbf{Splitting Collapsed \BP While Preserving Optimal Angles}. \ref{sfig1:degree4_alph1_approach}) illustrates the collapsed solution of a 4-terminal configuration. \ref{sfig2:degree4_alph1_approach}) demonstrates that it is possible to move jointly the terminal points $\{a_1,a_3\}$ in a specific but opposite direction to the one of the terminals $\{a_2,a_4\}$, resulting in the splitting of the collapsed \BP $b$ into two distinct \BPs, $b_1$ and $b_2$. Remarkably, this split can be executed while preserving the angles $\theta_1$ and $\theta_2$. Importantly, these angles must correspond to the optimal angles given by \eqref{eq:optimal_angles}.}
    \label{fig:degree4_alph1_approach}
\end{figure*}

As stated in Section \ref{sec:infeasibility-of-degree-4-steiner-points-in-the-plane}, the optimal position of the \BPs is continuously dependent on the terminal positions and solely relies on the branching angles, as shown in Section \ref{sec:branching-angles-at-the-steiner-points}. Consequently, assuming that there exists a configuration such that the \BPs collapse, it is possible to find terminal positions that lead to an unstable collapse of the \BPs. Here, instability refers to a configuration where an infinitesimal translation of the terminals results in the splitting of the \BPs. This scenario is depicted in \figurename{} \ref{fig:degree4_alph1_approach}. In such cases, the angles realized by the terminals and the \BPs will reach the upper bounds specified by \eqref{eq:branching angle}. Therefore, the angles depicted in \figurename{} \ref{sfig1:degree4_alph1_approach} fulfill the condition 
\begin{equation}\label{eq:equality_angles_approach_alpha=1}
    \gamma=\pi -\theta_1-\theta_2,
\end{equation}
where the angles satisfy
\begin{eqnarray}
    \cos(\gamma)&=&\frac{F\left(m_{a_1b}+m_{a_2b}\right)^{2\alpha}-F\left(m_{a_1b}\right)^{2\alpha}-F\left(m_{a_2b}\right)^{2\alpha}}{2F\left(m_{a_1b}\right)^{\alpha}F\left(m_{a_2b}\right)^{\alpha}}\\
    \cos(\theta_1)&=&\frac{F\left(m_{a_3b}+m_{a_1b}\right)^{2\alpha}+F\left(m_{a_1b}\right)^{2\alpha}-F\left(m_{a_3b}\right)^{2\alpha}}{2F\left(m_{a_3b}+m_{a_1b}\right)^{\alpha}F\left(m_{a_1b}\right)^{\alpha}}\\
    \cos(\theta_2)&=&\frac{F\left(m_{a_2b}+m_{a_4b}\right)^{2\alpha}+F\left(m_{a_2b}\right)^{2\alpha}-F\left(m_{a_4b}\right)^{2\alpha}}{2F\left(m_{a_2b}+m_{a_4b}\right)^{\alpha}F\left(m_{a_2b}\right)^{\alpha}}
\end{eqnarray}%
We can manipulate \eqref{eq:equality_angles_approach_alpha=1} in the following way
\begin{eqnarray}\label{eq:aux_equation2solve}
        \nonumber\gamma=\pi -\theta_1-\theta_2\iff& \cos(\gamma-\pi)=\cos(-\theta_1-\theta_2)\\
        \iff& -\cos(\gamma)=\cos(\theta_1+\theta_2)\\
        \nonumber\underbrace{\iff}_{*}&-\cos(\gamma)=\cos(\theta_1)\cos(\theta_2)-\sqrt{\left(1-\cos(\theta_1)^2\right)\left(1-\cos(\theta_2)^2\right)}
\end{eqnarray}
where in (*) we have used the fact that $$\cos(x+y)=\cos(x)\cos(y)-\sin(x)\sin(y)=\cos(x)\cos(y)-\sqrt{(1-\cos(x)^2)(1-\cos(y)^2)}.$$
If we square both sides of \ref{eq:aux_equation2solve} and equate to 0 we obtain.
\begin{equation}\label{eq:equation2solve_alpha1}
    \left(\cos(\gamma)+\cos(\theta_1)\cos(\theta_2)\right)^2-\left(1-\cos(\theta_1)^2\right)\left(1-\cos(\theta_2)^2\right)=0.
\end{equation}
Equation \eqref{eq:equation2solve_alpha1} depends on the variables $m_{a_1b}$, $m_{a_2b}$, $m_{a_3b}$, and $\alpha$\footnote{Since $\sum_{i=1}^4m_{a_i b}=1$, $m_{a_4b}$ can be expressed as $1-m_{a_1b}-m_{a_3b}-m_{a_4b}$.}. Equation \eqref{eq:equation2solve_alpha1} is generally too complex to be solved analytically. However, with the help of the Mathematica software \citep{Mathematica}, we have determined that for $\alpha=1$, the equality does not hold within the constraints of the problem, namely
$$\sum_{i=1}^4m_{a_i,b}=1 \qquad \text{ and } \qquad 0<m_{a_i,b}<1, \qquad \forall i.$$

To simplify the notation, let's denote $m_{a_ib}$ as $m_i$. For $\alpha=1$, when we expand equation \eqref{eq:equation2solve_alpha1}, we find that the numerator of the formula becomes a fourth-degree polynomial with respect to $m_1$. The four roots, $\{m_1^{(j)}\}$ of the polynomial are
\scriptsize
\begin{equation}\label{eq:m1_roots}
    \begin{aligned}
        m^{(1)}_{1}&=\frac{1}{2}\left(1-m_{2}-m_{3} \colorbox{yellow!50}{$-$}\sqrt{(-1+m_{2}+m_{3})^2-\frac{4}{3}\left(-2m_{2}-2m_{3}+3m_{2} m_{3}\colorbox{blue!10}{$-$}\sqrt{m_{2}^2-m_{2} m_{3} +m_{3}^2}\right)}\right)\\
        m^{(2)}_{1}&=\frac{1}{2}\left(1-m_{2}-m_{3}\colorbox{yellow!50}{$+$}\sqrt{(-1+m_{2}+m_{3})^2-\frac{4}{3}\left(-2m_{2}-2m_{3}+3m_{2} m_{3}\colorbox{blue!10}{$-$}2\sqrt{m_{2}^2-m_{2} m_{3} +m_{3}^2}\right)}\right)\\
        m^{(3)}_{1}&=\frac{1}{2}\left(1-m_{2}-m_{3}\colorbox{yellow!50}{$-$}\sqrt{(-1+m_{2}+m_{3})^2-\frac{4}{3}\left(-2m_{2}-2m_{3}+3m_{2} m_{3}\colorbox{blue!10}{$+$}2\sqrt{m_{2}^2-m_{2} m_{3} +m_{3}^2}\right)}\right)\\
        m^{(4)}_{1}&=\frac{1}{2}\left(1-m_{2}-m_{3}\colorbox{yellow!50}{$+$}\sqrt{(-1+m_{2}+m_{3})^2-\frac{4}{3}\left(-2m_{2}-2m_{3}+3m_{2} m_{3}\colorbox{blue!10}{$+$}2\sqrt{m_{2}^2-m_{2} m_{3} +m_{3}^2}\right)}\right)
    \end{aligned},
\end{equation}
\normalsize
where we have highlighted the difference between the roots. We will show that 
$$1<m_1^{(4)}+m_2+m_3<m_1^{(2)}+m_2+m_3 \qquad \text{ and } \qquad m_1^{(1)}<m_1^{(3)}<0,$$ 
which implies that the problem constraints are not satisfied, and therefore \BPs of degree 4 are not possible.

\paragraph{Claim 1:  $1<m_1^{(4)}+m_2+m_3\leq m_1^{(2)}+m_2+m_3$.} From \eqref{eq:m1_roots} it is clear that $m_1^{(4)}\leq m_1^{(2)}$. Thus, it is enough to prove the inequality for $m_1^{(4)}$:
\tiny
\begin{equation*}
    \begin{aligned}
        m_1^{(4)}+m_2+m_3=&\frac{1}{2}(1+m_{2}+m_{3}) +\frac{1}{2}\sqrt{(-1+m_{2}+m_{3})^2-\frac{4}{3}\left(-2m_{2}-2m_{3}+3m_{2} m_{3}-\sqrt{m_{2}^2-m_{2} m_{3} +m_{3}^2}\right)}\\ 
        =&\frac{1}{2}(1+m_{2}+m_{3}) +\frac{1}{2}\sqrt{1+\frac{2}{3}(m_2+m_3)+(m_2-m_3)^2-\frac{8}{3}\sqrt{\underbrace{m_2^2-m_2m_3+m_3^2}_{< (m_2+m_3)^2}}} \\
        >&\frac{1}{2}(1+m_{2}+m_{3}) +\frac{1}{2}\sqrt{1+\frac{2}{3}(m_2+m_3)+(m_2-m_3)^2-\frac{8}{3}(m_2+m_3)}\\
        =&\frac{1}{2}(1+m_{2}+m_{3}) +\frac{1}{2}\sqrt{1+(m_2-m_3)^2-\frac{1}{2}(m_2+m_3)}\\
        >&\frac{1}{2}\left(\underbrace{1+m_{2}+m_{3} +\sqrt{1-\frac{1}{2}(m_2+m_3)}}_{>2}\right)>1
    \end{aligned}
\end{equation*}
\normalsize
For the last inequality, we have used the fact $0<m2+m3<1$, that the function $g(x)=1+x+\sqrt{1-x/2}$ is increasing in $[0,1]$ and that $g(0)=2$.

\paragraph{Claim 2:  $m_1^{(1)}\leq m_1^{(3)}<0$.} From \eqref{eq:m1_roots} it is clear that $m_1^{(1)}\leq m_1^{(3)}$. Thus, it is enough to prove the inequality for $m_1^{(3)}$.
\scriptsize
\begin{align}
        \nonumber&m_1^{(3)}<0\\&\iff
        \nonumber\frac{1}{2}\left(1-m_{2}-m_{3}-\sqrt{(-1+m_{2}+m_{3})^2-\frac{4}{3}\left(-2m_{2}-2m_{3}+3m_{2} m_{3}+2\sqrt{m_{2}^2-m_{2} m_{3} +m_{3}^2}\right)}\right)<0\\
        &\label{eq:root_3_simplication}\iff \left(-2m_{2}-2m_{3}+3m_{2} m_{3}+2\sqrt{m_{2}^2-m_{2} m_{3} +m_{3}^2}\right)<0
\end{align}
\normalsize
Thus, we need to focus on inequality \ref{eq:root_3_simplication}. We will differentiate various cases:
\begin{itemize}[leftmargin=*]
    \item\textbf{ If $2/3> m_3\geq m_2$}:
    \begin{equation*}
    \begin{aligned}
        &\left(-2m_{2}-2m_{3}+3m_{2} m_{3}+2\sqrt{m_{2}^2-m_{2} m_{3} +m_{3}^2}\right) \\    &=m_2\underbrace{(2-3m_3)}_{>0\iff2/3> m3}+2\left(\underbrace{m_3-\sqrt{m_2^2-m2m3+m_3^2}}_{ \geq 0 \iff m_3\geq m_2}\right)>0
    \end{aligned}
\end{equation*}
For the second term, we have used the fact that 
\begin{equation*}
    \begin{aligned}
        m_3\geq\sqrt{m_2^2-m2m3+m_3^2}&\iff m_3^2\geq m_2^2-m_2m_3+m_3^2\iff m_2m_3\geq m_2^2\\
        &\iff m_3\geq m_2
    \end{aligned}
\end{equation*}
    \item \textbf{If $2/3\geq m_3\geq m_2$: }Anologous to previous case due to the symmetry of $m_2$ and $m_3$ in \eqref{eq:root_3_simplication}

    \item \textbf{If $\max(m_2,m_3)\geq 2/3$: } W.l.o.g~ we can assume $m_2\geq2/3$ and $m_3<1/3$ due to the symmetry between $m_2$ and $m_3$. For this case, we will find the roots with respect to $m_2$ of inequality \eqref{eq:root_3_simplication} and see that the constraints on $m_2$ do not hold. Indeed,
    \begin{eqnarray*}
        &-2m_{2}-2m_{3}+3m_{2} m_{3}+2\sqrt{m_{2}^2-m_{2} m_{3} +m_{3}^2}=0\Rightarrow\\
        &\left(-2m_{2}-2m_{3}+3m_{2} m_{3}\right)^2=4(m_{2}^2-m_{2} m_{3} +m_{3}^2)
    \end{eqnarray*}
    The roots $\{m_2^{(j)}\}$ of the polynomial are
    
    \begin{equation}\label{eq:m2_roots}
        \begin{aligned}
            m^{(1)}_{2}&=\frac{2m_3}{-5+6m_3\colorbox{red!30}{$+$}\sqrt{3}\sqrt{7-12m_3+6m^2_3}} \qquad  \& \qquad m_3\neq\frac{2-\sqrt{2}}{3},\\
            m^{(2)}_{2}&=\frac{2m_3}{-5+6m_3\colorbox{red!30}{$-$}\sqrt{3}\sqrt{7-12m_3+6m^2_3}} \qquad  \& \qquad m_3\neq\frac{2-\sqrt{2}}{3}\\
            m^{(3)}_{2}&=\frac{2(-3+2\sqrt{2})}{3(-2+3\sqrt{2})} \qquad \& \qquad m_3=\frac{2-\sqrt{2}}{3}.
        \end{aligned},
    \end{equation}
    The denominator of the root $m^{(1)}_{2}$ has negative sign for $0<m_3<1$, which leads to $m^{(2)}_{1}<0$, contradicting the initial constraints. Trivially, $m^{(3)}_{2}$ is also negative.

    The denominator of the root $m^{(2)}_{2}$ is negative for $0<m_3<\frac{2-\sqrt{2}}{3}$, resulting in $m^{(2)}_{2}<0$. When $\frac{2-\sqrt{2}}{3}<m_3<1/3$, the denominator becomes positive but remains lower than $2m_3$, thus $m^{(2)}_{2}>1$, which is also a contradiction.
\end{itemize}

We have ruled out all possible cases, thus we have proven the next theorem.

\begin{theorem}
Let $\alpha=1$. Given a set of terminals which lie in the plane, then the \BPs of the optimal solution of the \BCST problem will not contain \BPs of degree $4$ unless these collapse with a terminal.
\end{theorem}

\section{Iteratively Reweighted Least Square for the Geometric Optimization of the Steiner Points}\label{sec:app_IRLS}
In this section we review briefly the iteratively reweighted least square (IRLS) algorithm proposed in \cite{smith_how_1992}. This algorithm was initially developed for the geometric optimization of Steiner points (\BPs) and later adapted in \cite{lippmann_theory_2022} for the branched optimal transport (BOT) problem. We will show that the same algorithm can be adapted for the \BCST problem, since the algorithm is agnostic to the weighting factors multiplying the distances involved in the BOT and \BCST objectives, as defined in equations \eqref{eq:BOT} and \eqref{eq:BCST}, respectively. 

Consider the following minimization problem for a fixed tree topology
\begin{equation}\label{eq:objective_agnostic_weights}
    \min_{X_B}C(X)=\min_{X_B}\sum_{(i,j) \, \in \, E} w_{ij} \left\|  x_i -  x_j \right\|
\end{equation}
where $w_{ij}$ are arbitrary weights, $E$ is the set of edges of the tree, $X_B=\{x_{N+1},\dots,x_{2N-2}\}$ are the coordinates of the \BPs, which need to be optimized, and $X=\{x_{1},\dots,x_{2N-2}\}$ is the set of all coordinates (terminals and \BPs). Starting from arbitrary \BPs coordinates, denoted as $X^{(0)}$, the algorithm iteratively solves the following linear system of equations.
\begin{equation}\label{eq:iteration_IRLS}
    x_i^{(k+1)}=\frac{\displaystyle\sum_{j:(i,j)\in E}w_{ij}\frac{x_j^{(k+1)}}{||x_i^{(k)}-x_j^{(k)}||}}{\displaystyle\sum_{j:(i,j)\in E}\frac{w_{ij}}{||x_i^{(k)}-x_j^{(k)}||}}, \qquad \forall N+1\leq i\leq 2N-2.
\end{equation}
Note that only the coordinates corresponding to the \BPs are updated. The coordinates of the terminals are kept fixed and set equal to their original coordinates.

We will show that in each iteration the cost of the objective function decreases, i.e. $C(X^{(k+1)})<C(X^{(k)})$. As shown in \cite{smith_how_1992}, this implies that the $\lim_{k\to\infty}X^{(k)}=\arg\min C(X)$. 

The algorithm can be considered an IRLS approach because it reinterprets the cost function as a quadratic function. Indeed, $C(X)$ can be rewritten as 
\[C(X)=\sum_{(i,j) \, \in \, E} w_{ij} \left\|  x_i -  x_j \right\|=\sum_{(i,j) \, \in \, E} \underbrace{\frac{w_{ij}}{\left\|  x_i -  x_j \right\|}}_{W_{ij}(X)}\left\|  x_i -  x_j \right\|^2=\sum_{(i,j) \, \in \, E} W_{ij}(X) \left\|  x_i -  x_j \right\|^2\]

In concrete, the solution of the linear system \eqref{eq:iteration_IRLS} minimizes the following quadratic function
\[Q^{(k)}(X)=\sum_{(i,j) \, \in \, E} W_{ij}(X^{(k)}) \left\|  x_i -  x_j \right\|^2.\]
That is $Q^{(k)}(X)\geq Q^{(k)}(X^{(k+1)})$ $\forall X$. Moreover, note that $C(X^{(k)})=Q^{(k)}(X^{(k)})$. Now we can show that the cost $C$ decreases at each iteration:
\begin{equation}
    \begin{aligned}
        C(X^{(k)})=&Q^{(k)}(X^{(k)})\geq Q^{(k)}(X^{(k+1)})\\
        =&\sum_{(i,j) \, \in \, E}w_{ij}\frac{\left(\left|x_i^{(k)}-x_j^{(k)}\right|+\left|x_i^{(k+1)}-x_j^{(k+1)}\right|-\left|x_i^{(k)}-x_j^{(k)}\right|\right)^2}{\left|x_i^{(k)}-x_j^{(k)}\right|}\\
        =&C(X^{(k)})+2\left(C(X^{(k+1})-C(X^{(k)})\right)\\
        &+\sum_{(i,j) \, \in \, E}w_{ij}\frac{\left(\left|x_i^{(k+1)}-x_j^{(k+1)}\right|-\left|x_i^{(k)}-x_j^{(k)}\right|\right)^2}{\left|x_i^{(k)}-x_j^{(k)}\right|}\\&\iff\\
         C(X^{(k+1})\leq& C(X^{(k)})-\underbrace{\sum_{(i,j) \, \in \, E}\frac{w_{ij}}{2}\frac{\left(\left|x_i^{(k+1)}-x_j^{(k+1)}\right|-\left|x_i^{(k)}-x_j^{(k)}\right|\right)^2}{\left|x_i^{(k)}-x_j^{(k)}\right|}}_{\geq 0}\\
        \leq&C(X^{(k)})
    \end{aligned}    
\end{equation}

\section{Complexity mSTreg Heuristic}\label{sec:complexity-mstreg-heuristic}
\begin{figure}
	\centering
	\includegraphics[width=1\linewidth]{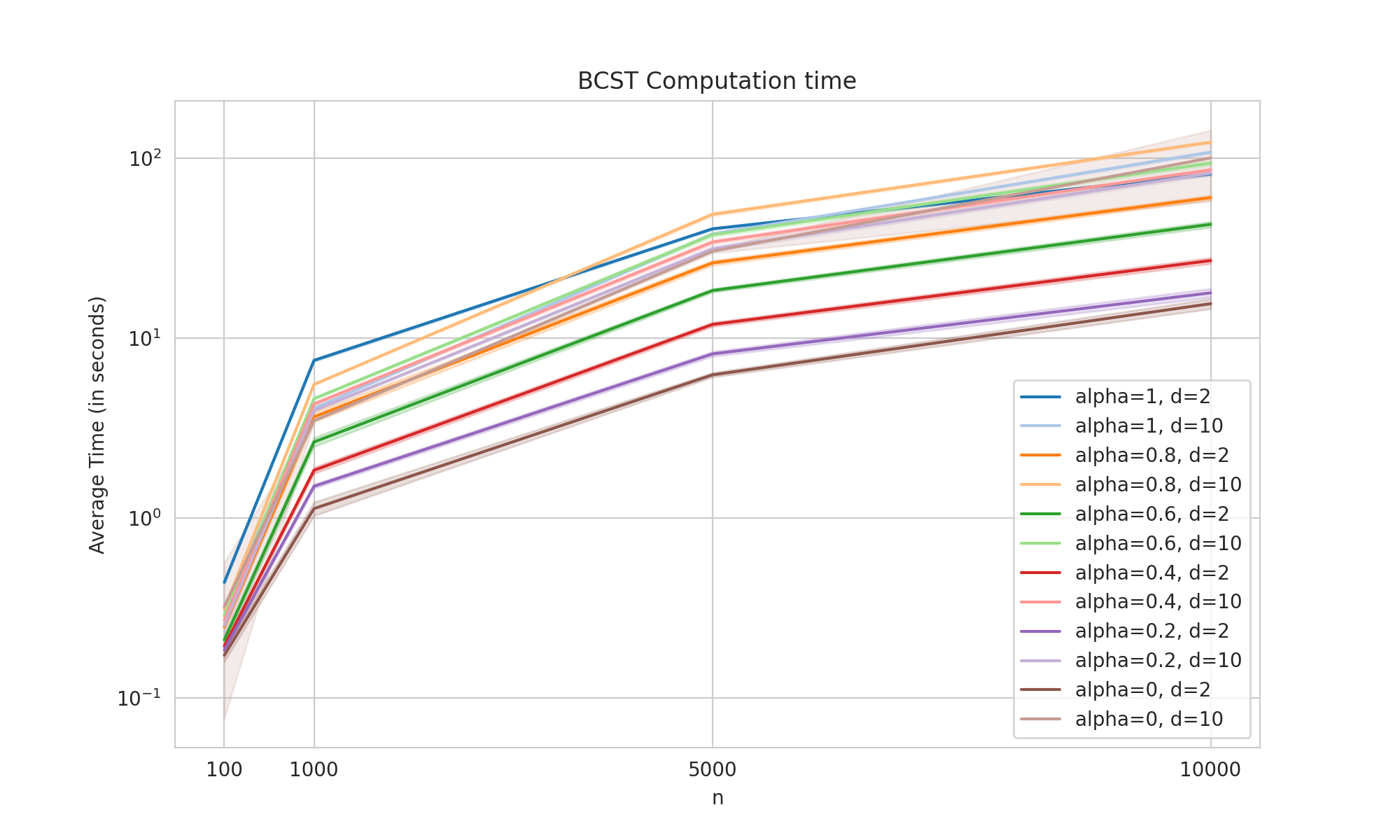}
	\caption[\BCST Time Complexity]{\textbf{\BCST time complexity}. Computation time of 20 iterations of the mSTreg heuristic averaged over 5 distinct instances with different numbers of terminals $n$, data dimensionality $d$, and $\alpha$ values.}
	\label{fig:time_vs_n}
\end{figure}

We will delve into the complexity of the two steps of our heuristic.
\begin{itemize}[leftmargin=*]
	\item \textbf{Complexity geometric optimization}: We approximate the optimal \BPs coordinates using the IRLS approach presented in Section \ref{sec:app_IRLS}. Each iteration of the IRLS requires $\mathcal{O}(nd)$ operations, where $n$ is the number of terminals and $d$ is the data dimensionality. Within each iteration, $d$ linear systems are solved. These can be solved in linear time and in parallel. The number of iterations needed for the IRLS to converge is not known a priori, however, \citet{lippmann_theory_2022} suggest that this number could scale on average like $\mathcal{O}(\log(n))$. Consequently, each geometric optimization step takes $\mathcal{O}(\log(n)nd)$.
	\item\textbf{Topology optimization step}: In the topology optimization step, we compute the minimum spanning tree (mST)	over the terminals and \BPs. Given a graph $G=(V,E)$, Kruskal's algorithm takes $\mathcal{O}(|E|\log|V|)$ operations to compute the mST. In a complete graph, this becomes $\mathcal{O}(n^2\log(n))$. However, in some situations, we may expedite the mST computation by computing the mST over a $k$-nearest neighbor (kNN) graph. Approximating a kNN graph with k-d trees can have a complexity of $\mathcal{O}(dn\log(n)^2)$. In this case, the number of edges in the graph would be $|E|\approx kn$. Hence, the overall mST complexity would be $\mathcal{O}(dn\log(n)^2+kn \log(n)) \approx \mathcal{O}(dn\log(n)^2)$.
\end{itemize}

Therefore, the heuristic's per-iteration complexity is approximately $\mathcal{O}(dn\log(n) + n^2\log(n))$ or $\mathcal{O}(dn\log(n)^2)$ if the mST is computed over a kNN graph. Throughout our experiments, a limit of 20 iterations was set, though practical convergence often demands fewer. In addition, we gauged the computational time of the heuristic by averaging its performance over 20 iterations across 5 distinct instances, varying $n$, $d$, and $\alpha$. Data was generated by sampling $n$ points from a $d$-dimensional unit cube. The performance times are presented in \figurename{} \ref{fig:time_vs_n}. The heuristic was executed on an Intel Xeon Gold 6254 CPU @ 3.10GHz.

\section{Effect of Additional Intermediate Points in the mSTreg Heuristic}\label{sec:app_effect_freq_sampling}

\begin{figure}
	\centering
    \includegraphics[width=1\linewidth]{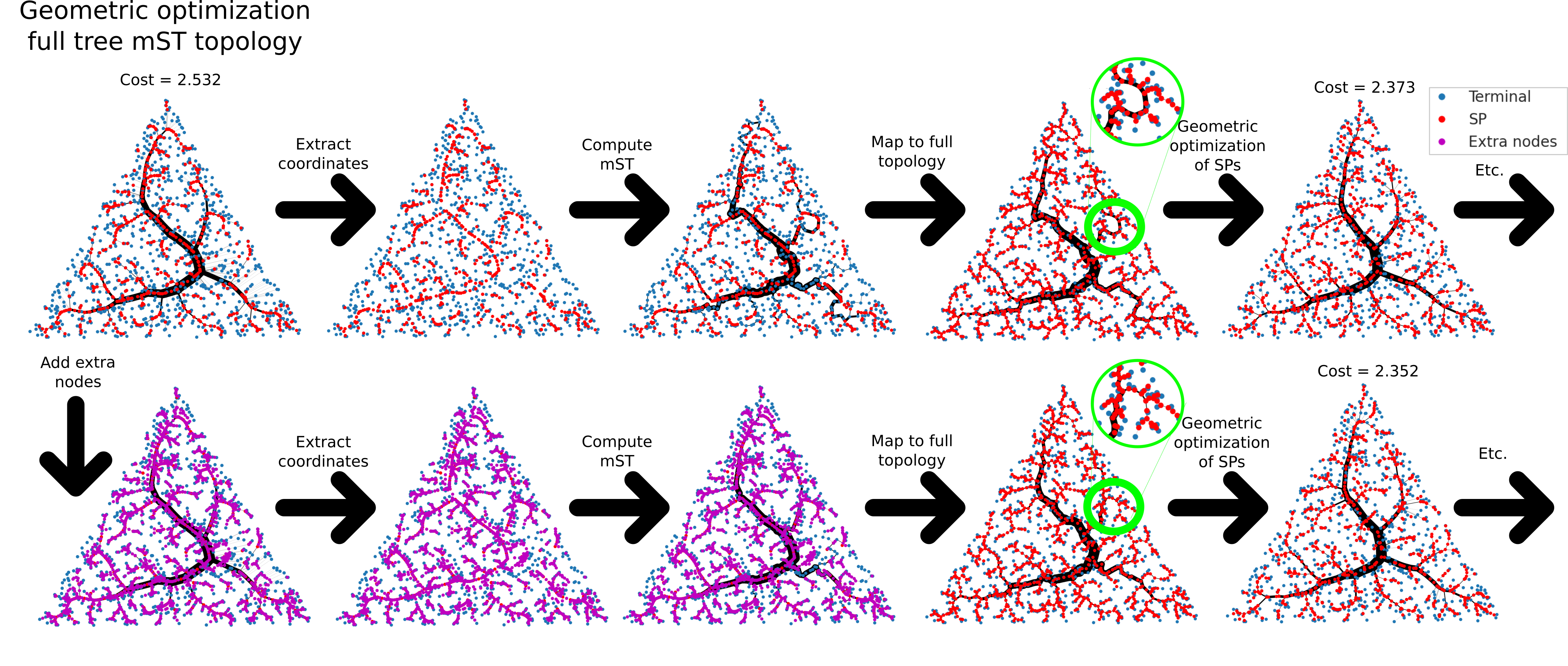}
	\caption[\mSTreg Heuristic with Additional Sampled Points]{\textbf{\mSTreg Heuristic with Additional Sampled Points}. Effect of adding extra points per edge (visualized in violet) in the \mST computation step of the \mSTreg heuristic. Top left: \BCST solution obtained once the \mST has been mapped to a full tree topology and its Steiner point coordinates have been optimized. Top row: Next steps of the \mSTreg heuristic without adding any extra point. Bottom row: Next steps of the \mSTreg heuristic once an extra point has been added at the middle of each edge (shown in violet). The addition of extra points may allow the \mST to more reliably follow the edges of the geometry-optimized tree from previous step. We zoom in to highlight an improvement in the topology resulting from the addition of these extra points. In this particular case, the cost obtained with the inclusion of the extra nodes is lower than the cost without them.}
	\label{fig:mSTreg_freqsampling}
\end{figure}

In Section \ref{sec:heuristic} we have described the \mSTreg heuristic as a solution approach for the \BCST problem. The algorithm can be summarized in two steps: 1) Optimization of the \BPs coordinates given a fixed topology; 2) topology update by computing the \mST over the terminals and \BPs. The motivation for the topology update is that the optimal positions of the \BPs may suggest a more desirable topology, since they may be biased to move closer to other nodes than the ones to which they are connected. Thus, we hope that the new topology, defined by the \mST over the \BPs and the terminals, interconnects such nodes.

However, there are instances where the \BPs may not be sufficiently close to each other, causing the \mST to fail in recovering the desired connections. the addition of intermediate nodes along the edges may address this problem, allowing the \mST to more reliably follow the edges of the geometry-optimized tree from the previous step. An illustrative example highlighting the benefits of this approach can be seen in \figurename{} \ref{fig:mSTreg_freqsampling}. In general, we have seen that adding between 1 and 3 nodes per edge can often yield improvements. However, the impact on the main backbone is minimal. In Algorithm \ref{alg:CST_mSTreg}, the number of intermediate points that are added along an edge is regulated by the \verb|sampling\_frequency| variable.

\section{Strategies to Transform a Full Tree Topology into a \CST Topology}\label{sec:app_SP_removal_strategies}
When using the \mSTreg heuristic described in Section \ref{sec:heuristic} for solving the \CST problem without branching points, we need to map from a full tree topology to a \CST topology. As shown in Section \ref{subsec:app_numtopos_CST}, this process is ambiguous and there may be an exponential number of derivable topologies with respect to the number of terminals. Hence to brute force the one which minimizes the \CST cost is out of reach. 

In this section, we describe some heuristic rules to transform a full tree topology into a \CST topology. In order to transform a full tree topology into a \CST topology, we collapse iteratively one \BP at a time with one of its neighbors until there are no more \BPs to collapse. The first factor to take into account is in which order the \BPs are collapsed. We consider two strategies: 1) collapse the \BP that is closest to a terminal (``Ordclosestterminal'') or 2) collapse the \BP with the closest neighbor, i.e. the one that minimizes the distance to one of each neighbors independently of if it is a terminal or a \BP (``Ordclosest''). In practice we did not see any big difference, though ``Ordclosest'' tends to be slightly better.

The second factor to take into account is to which neighbor should an \BP collapse. We again compare two different heuristics: 1) collapsing the \BP to the neighbor that minimally increases the \CST cost (``greedy''); 2) collapsing the \BP to the closest neighbor in terms of distance. We found empirically that the greedy approach yields significantly superior results.

Lastly, we conducted tests on updating the position of the collapsed \BP. When a \BP, denoted as $b_1$, is collapsed with a neighbor $b_2$, then the other neighbors of $b_1$ become neighbors of $b_2$. We observed that updating the position of $b_2$ to the weighted geometric median of its neighbors (including those inherited from $b_1$) yielded improved results compared to not updating the coordinates of $b_2$. The coordinates of $b_2$ were only updated when $b_2$ was an \BP. If $b_2$ happened to be a terminal, its position was kept fixed. To denote if a strategy updated the position or not we will use the expression ``update'' and ``no\_update'' respectively.

To evaluate the effectiveness of the strategies, we conducted a series of experiments by sampling 200 problem instances for each $N$ in the set $\{5,6,7,8,9\}$, where $N$ represents the number of terminals. For each instance, we applied the \mSTreg heuristic with different $\alpha$ values and utilized the aforementioned strategies to transform a full tree topology into a CST topology. \figurename{} \ref{fig:bruteforce_CST_comparison} shows the mean ranking positions obtained by the different strategies, once all feasible solutions have been sorted. The results confirm the observations that we already pointed out. For all of our experiments we used the combination that produced the best results i.e. ``update''+``greedy''+``Ordclosest''.

\def \bottomvspace{-.5}
\def \topvspace{0.1}

\begin{figure}
	\centering         
    \begin{subfigure}{0.33\linewidth}
    		\centering
    		\vspace*{\topvspace cm}
    		\includegraphics[width=1\linewidth]{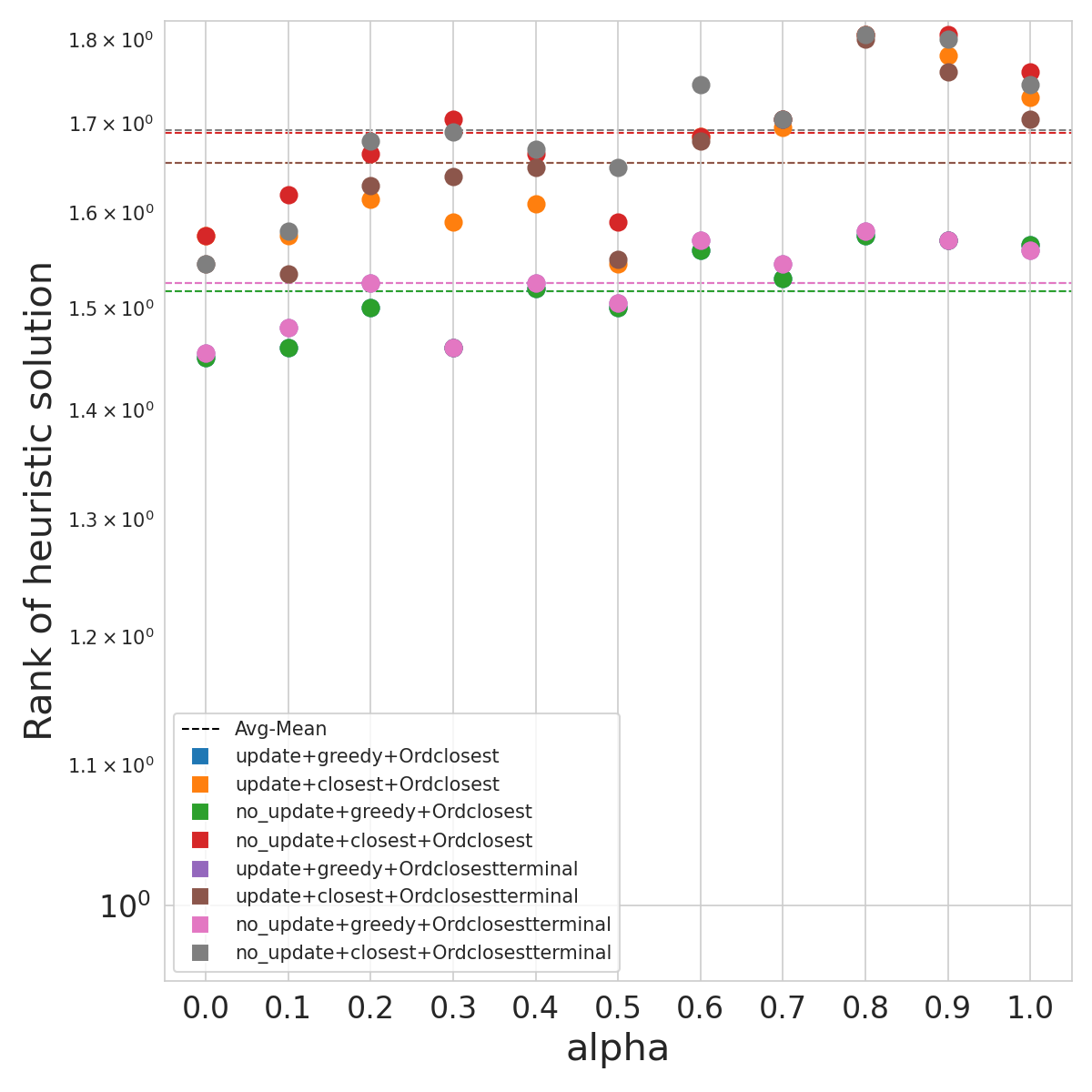}
    		\vspace{\bottomvspace cm}
    		\caption{N=5}
    		\label{sfig1:bruteforce_CST_comparison}
    	\end{subfigure}%
    	\begin{subfigure}{0.33\linewidth}
    		\centering
    		\vspace*{\topvspace cm}
    		\includegraphics[width=1\linewidth]{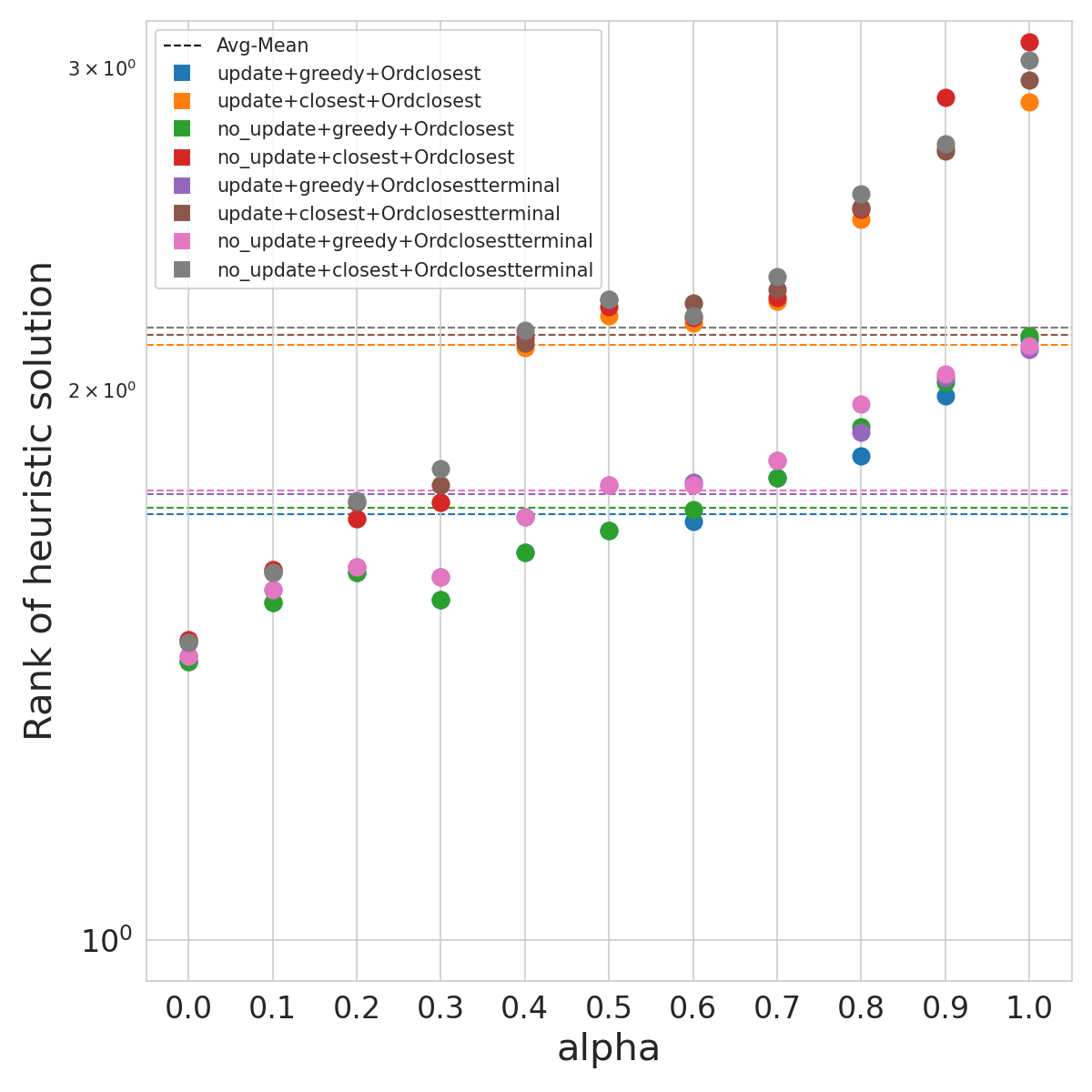}
    		\vspace{\bottomvspace cm}
    		\caption{N=6}
    		\label{sfig2:bruteforce_CST_comparison}
    	\end{subfigure}%
    	\begin{subfigure}{0.33\linewidth}
    		\centering
    		\vspace*{\topvspace cm}
    		\includegraphics[width=1\linewidth]{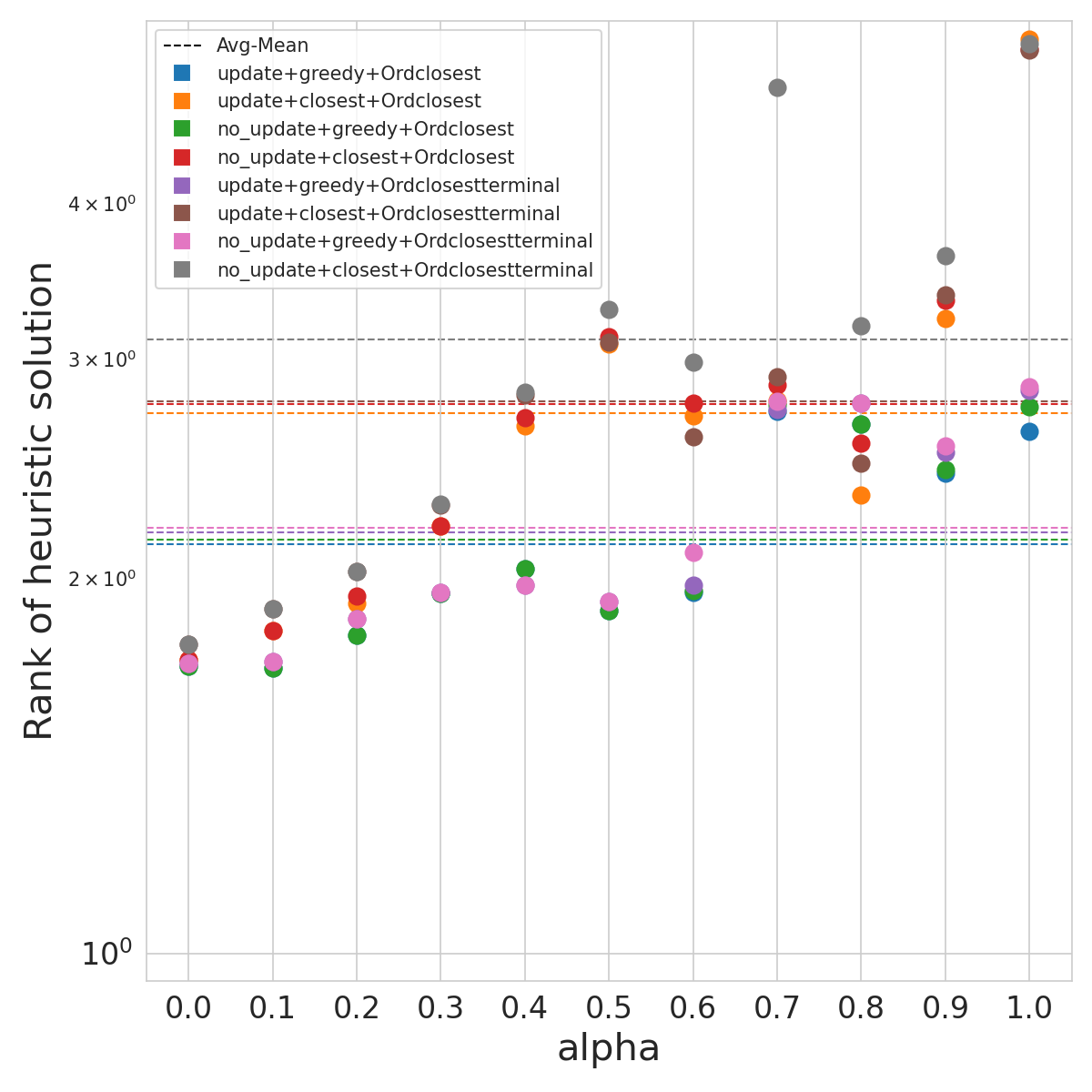}
    		\vspace{\bottomvspace cm}
    		\caption{N=7}
    		\label{sfig3:bruteforce_CST_comparison}
    	\end{subfigure}
        \begin{subfigure}{0.33\linewidth}
    		\centering
    		\vspace*{\topvspace cm}
    		\includegraphics[width=1\linewidth]{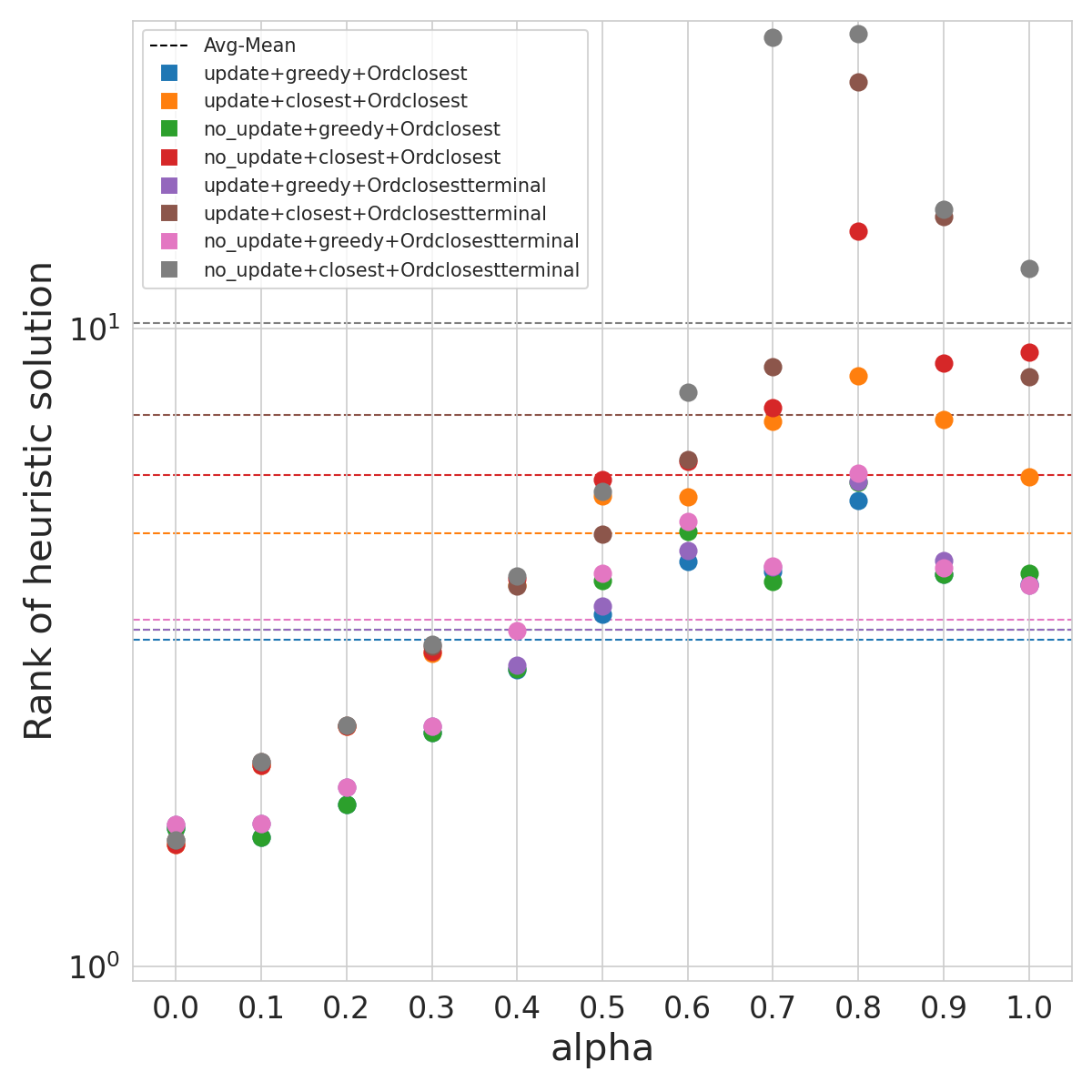}
    		\vspace{\bottomvspace cm}
    		\caption{N=8}
    		\label{sfig4:bruteforce_CST_comparison}
    	\end{subfigure}%
        \begin{subfigure}{0.33\linewidth}
    		\centering
    		\vspace*{\topvspace cm}
    		\includegraphics[width=1\linewidth]{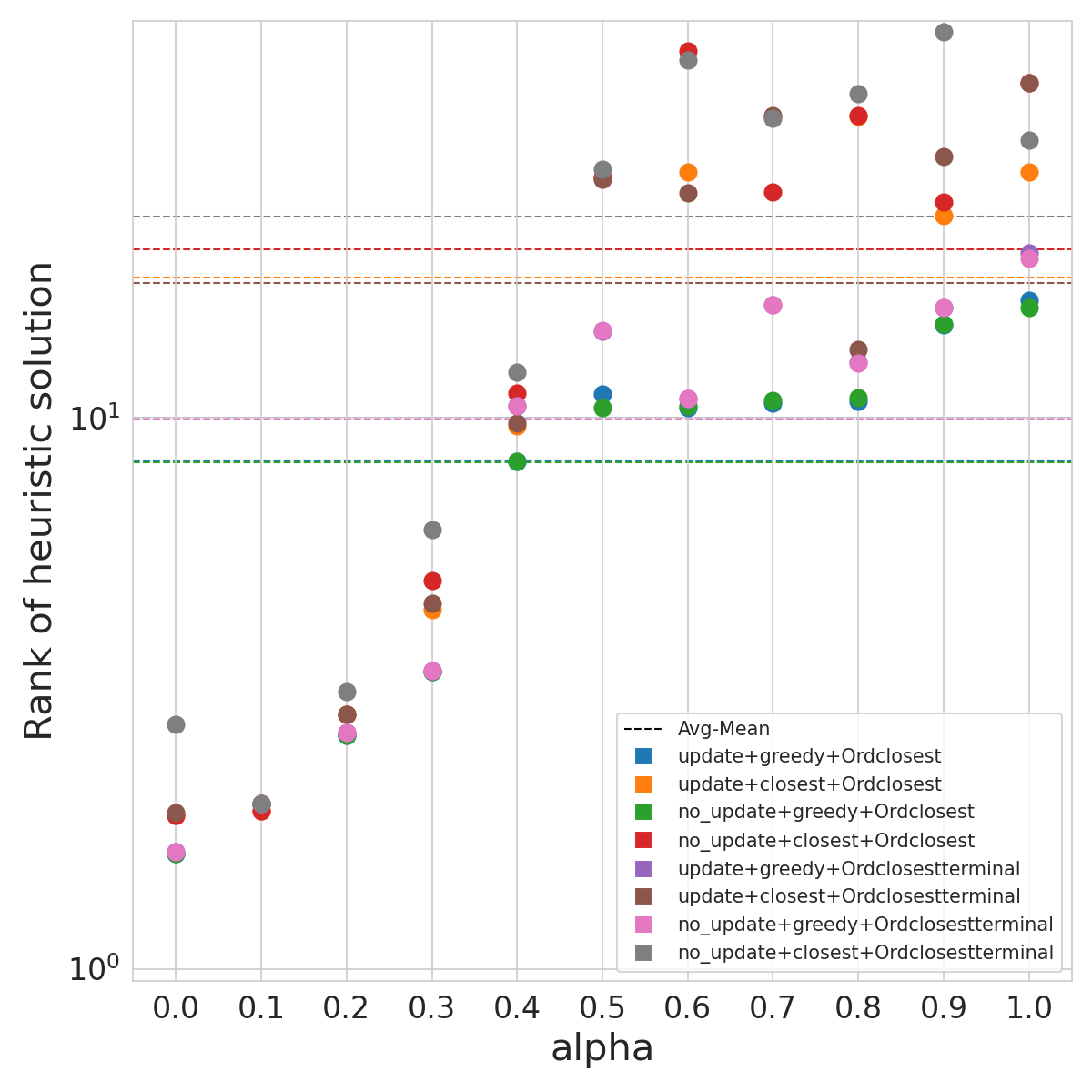}
    		\vspace{\bottomvspace cm}
    		\caption{N=9}
    		\label{sfig5:bruteforce_CST_comparison}
    	\end{subfigure}%
 
	 \caption[Performance \BPs Collapse Strategies]{\textbf{Performance \BPs Collapse Strategies}. Comparison of different collapse strategies to transform a full tree topology into a \CST topology. See Section \ref{sec:app_SP_removal_strategies} for a short description of the strategies. We plot the average sorted position of the heuristic for different number of terminals, $N$ and for different $\alpha$ values. We observe that the strategy combinations including the ``greedy'' collapse approach have significantly better results. The combinations  which update the position of the collapsed \BP (``update'') perform slightly better than the ones that do not (``no\_update''). Analogously, the strategies that order the \BPs based on the closeness to the neighbors (``Ordclosest'') is slightly better than ``Ordclosestterminal''.}
	\label{fig:bruteforce_CST_comparison}
\end{figure}

\section{Further Details on the Brute Force Experiment}\label{sec:app_toydata_experiments}
In this section, we analyze the behavior of the \mSTreg heuristic with respect to $\alpha$. To investigate this, we utilize the experiment described in Section \ref{sec:benchmark}, which compared the cost of the output tree generated by our heuristic with the optimal solution for different numbers of terminal nodes, denoted as N, while specifically examining the influence of $\alpha$. 

For each $N\in \{5,6,7,8,9\}$, we sample 200 problem instances. We computed the optimal \CST and \BCST topologies of all problems via brute-force and with the \mSTreg heuristic\footnote{As described in \ref{sec:app_SP_removal_strategies}, we can use different strategies to transform a full tree topology into a \CST topology, when solving the \CST problem with the \mSTreg heuristic. We used the one that updates the position of \BPs and collapses to the closest neighbor.} for all $\alpha \in \{0,0.1,\dots,0.9,1\}$. \figurename{s} \ref{fig:bruteforce_BCST_alpha} and \ref{fig:bruteforce_CST_alpha} show the relative error and how the heuristic solution ranks, when the costs of all topologies are sorted. When solving the \BCST, though there is not a clear trend, we can observe that for higher $N$ the heuristic tends to perform worse for higher $\alpha$ values, since on average the heuristic's solution ranking is higher. When solving the \CST problem this pattern can be more clearly seen.

\def \bottomvspace{-.5}
\def \topvspace{0.1}
\begin{figure}
	\centering
     
    \begin{minipage}[t]{1\textwidth}   
        \begin{subfigure}{0.2\linewidth}
            \centering
            \vspace*{\topvspace cm}
            \includegraphics[width=1\linewidth]{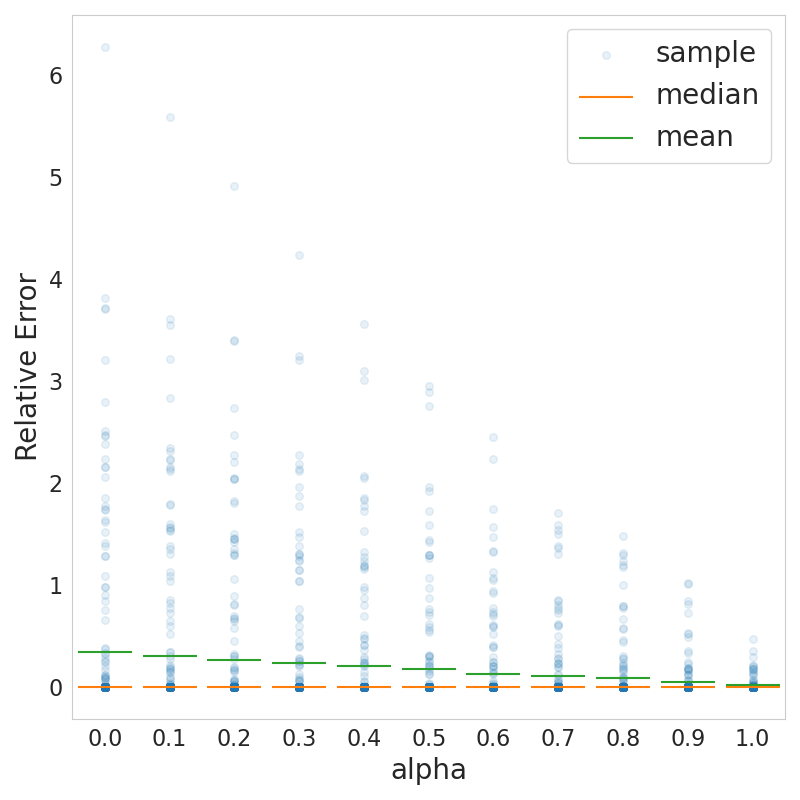}
            \vspace{\bottomvspace cm}
            \subcaption{N=5}
            \label{sfig1:bruteforce_BCST_alpha}
        \end{subfigure}%
        \begin{subfigure}{0.2\linewidth}
            \centering
            \vspace*{\topvspace cm}
            \includegraphics[width=1\linewidth]{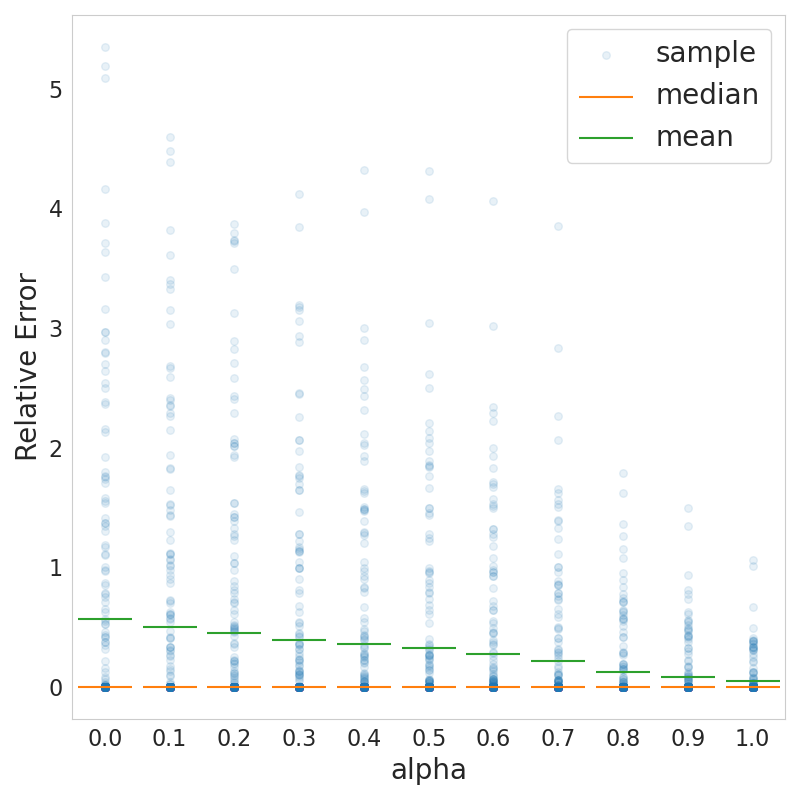}
            \vspace{\bottomvspace cm}
            \subcaption{N=6}
            \label{sfig2:bruteforce_BCST_alpha}
        \end{subfigure}%
        \begin{subfigure}{0.2\linewidth}
            \centering
            \vspace*{\topvspace cm}
            \includegraphics[width=1\linewidth]{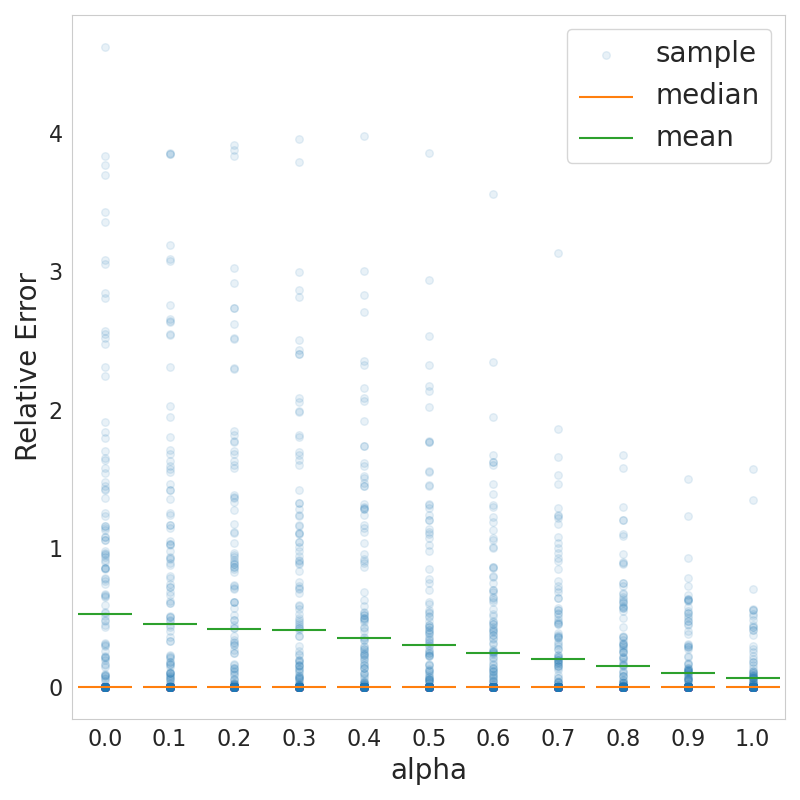}
            \vspace{\bottomvspace cm}
            \subcaption{N=7}
            \label{sfig3:bruteforce_BCST_alpha}
        \end{subfigure}%
        \begin{subfigure}{0.2\linewidth}
            \centering
            \vspace*{\topvspace cm}
            \includegraphics[width=1\linewidth]{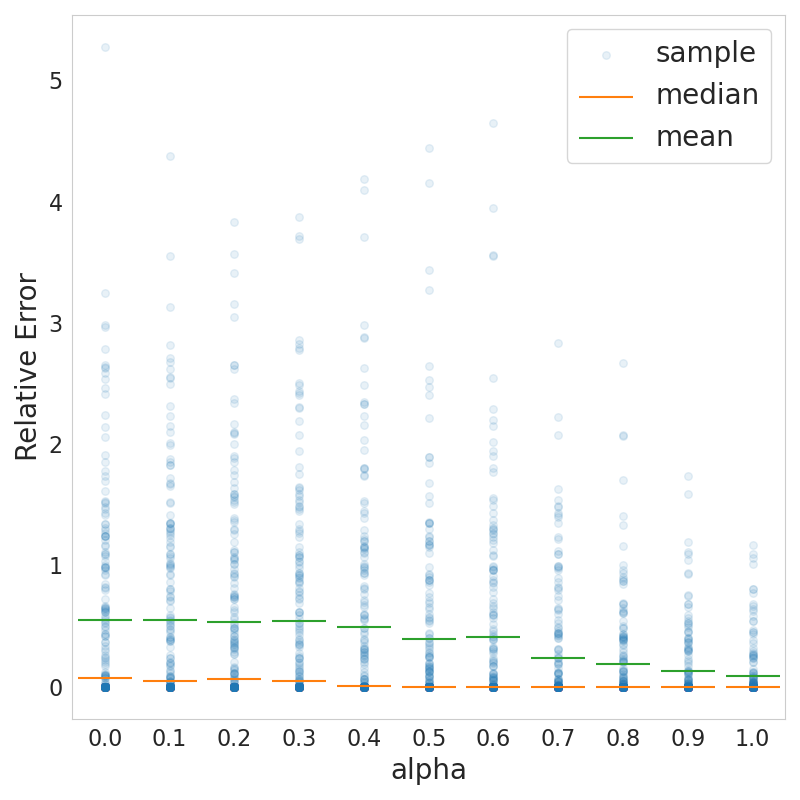}
            \vspace{\bottomvspace cm}
            \subcaption{N=8}
            \label{sfig4:bruteforce_BCST_alpha}
        \end{subfigure}%
        \begin{subfigure}{0.2\linewidth}
            \centering
            \vspace*{\topvspace cm}
            \includegraphics[width=1\linewidth]{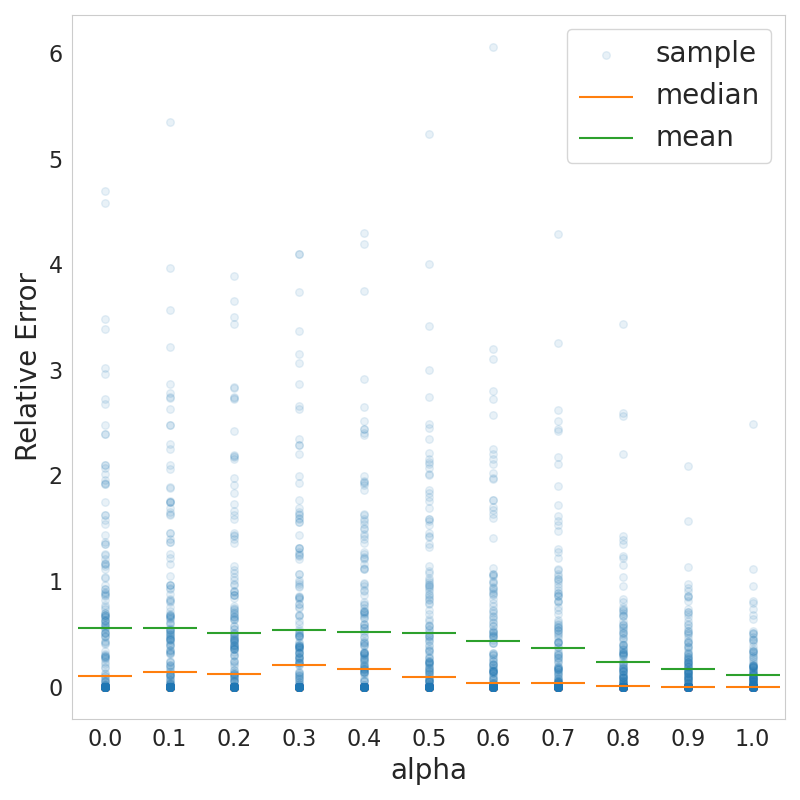}
            \vspace{\bottomvspace cm}
            \subcaption{N=9}
            \label{sfig5:bruteforce_BCST_alpha}
        \end{subfigure}%
	    \centering
        \caption*{\BCST relative error for different number of terminals N}
    \end{minipage}

    \begin{minipage}[t]{1\textwidth}   
        \begin{subfigure}{0.2\linewidth}
    		\centering
    		\vspace*{\topvspace cm}
    		\includegraphics[width=1\linewidth]{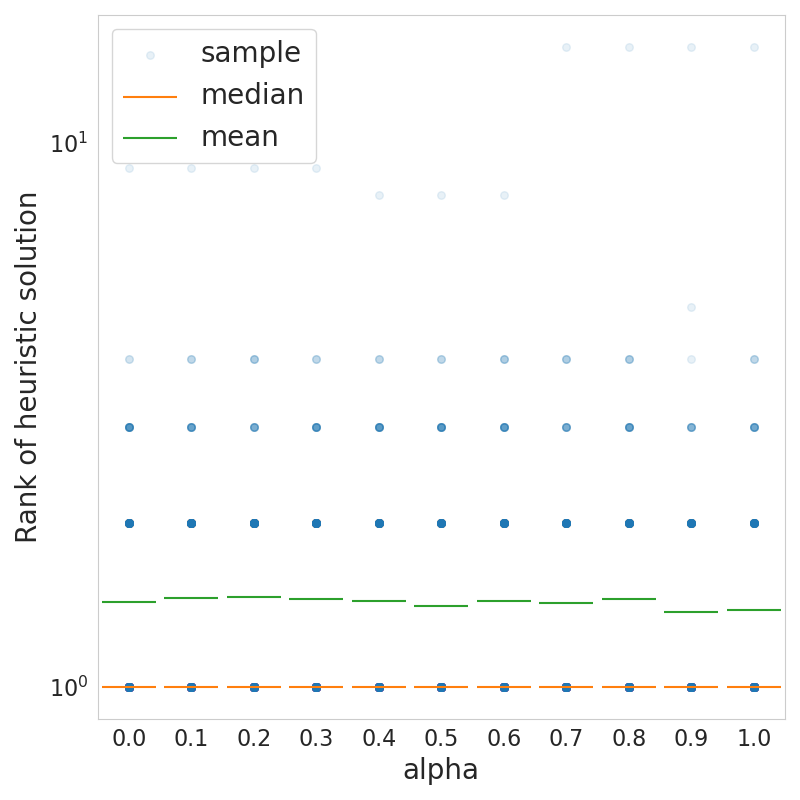}
    		\vspace{\bottomvspace cm}
    		\caption{N=5}
    		\label{sfig6:bruteforce_BCST_alpha}
    	\end{subfigure}%
    	\begin{subfigure}{0.2\linewidth}
    		\centering
    		\vspace*{\topvspace cm}
    		\includegraphics[width=1\linewidth]{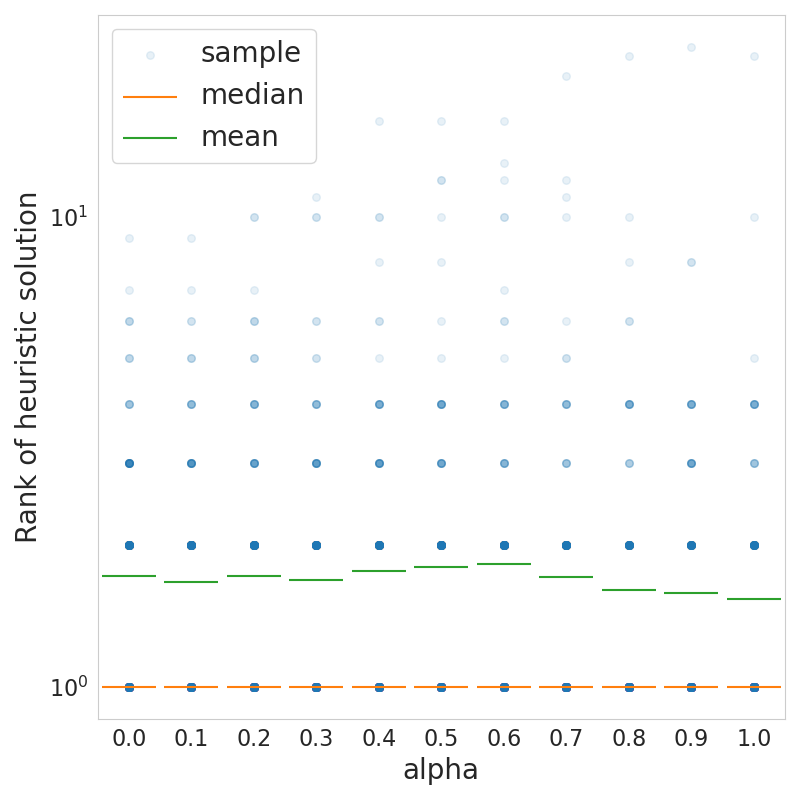}
    		\vspace{\bottomvspace cm}
    		\caption{N=6}
    		\label{sfig7:bruteforce_BCST_alpha}
    	\end{subfigure}%
    	\begin{subfigure}{0.2\linewidth}
    		\centering
    		\vspace*{\topvspace cm}
    		\includegraphics[width=1\linewidth]{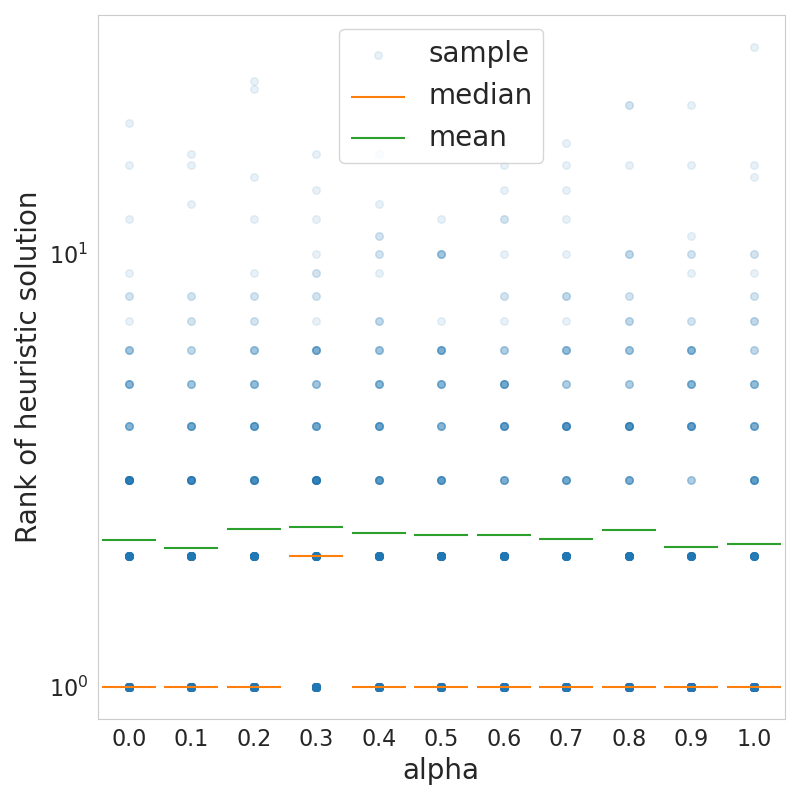}
    		\vspace{\bottomvspace cm}
    		\caption{N=7}
    		\label{sfig8:bruteforce_BCST_alpha}
    	\end{subfigure}%
        \begin{subfigure}{0.2\linewidth}
    		\centering
    		\vspace*{\topvspace cm}
    		\includegraphics[width=1\linewidth]{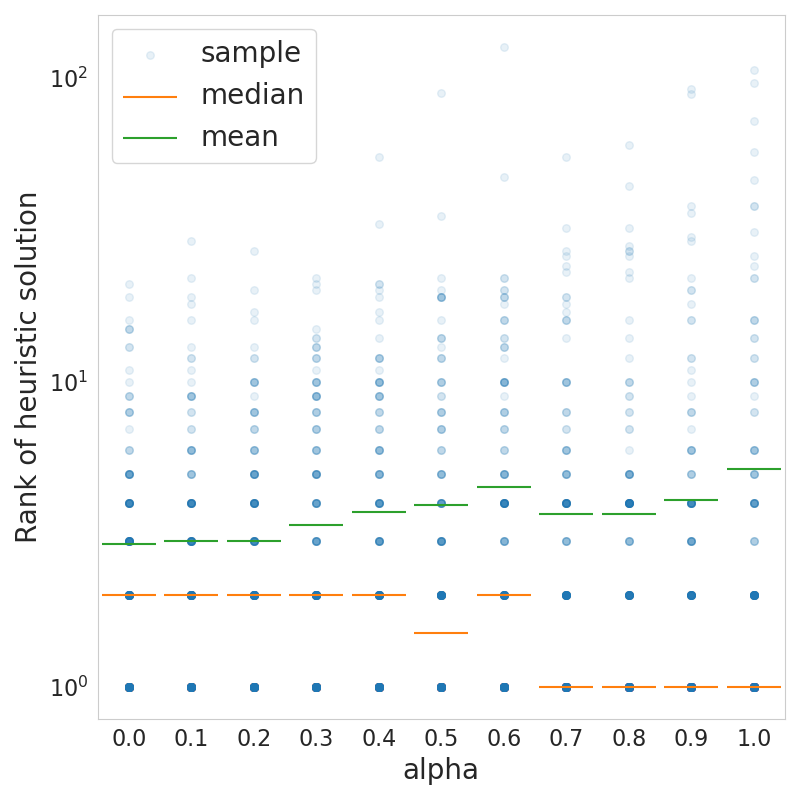}
    		\vspace{\bottomvspace cm}
    		\caption{N=8}
    		\label{sfig9:bruteforce_BCST_alpha}
    	\end{subfigure}%
        \begin{subfigure}{0.2\linewidth}
    		\centering
    		\vspace*{\topvspace cm}
    		\includegraphics[width=1\linewidth]{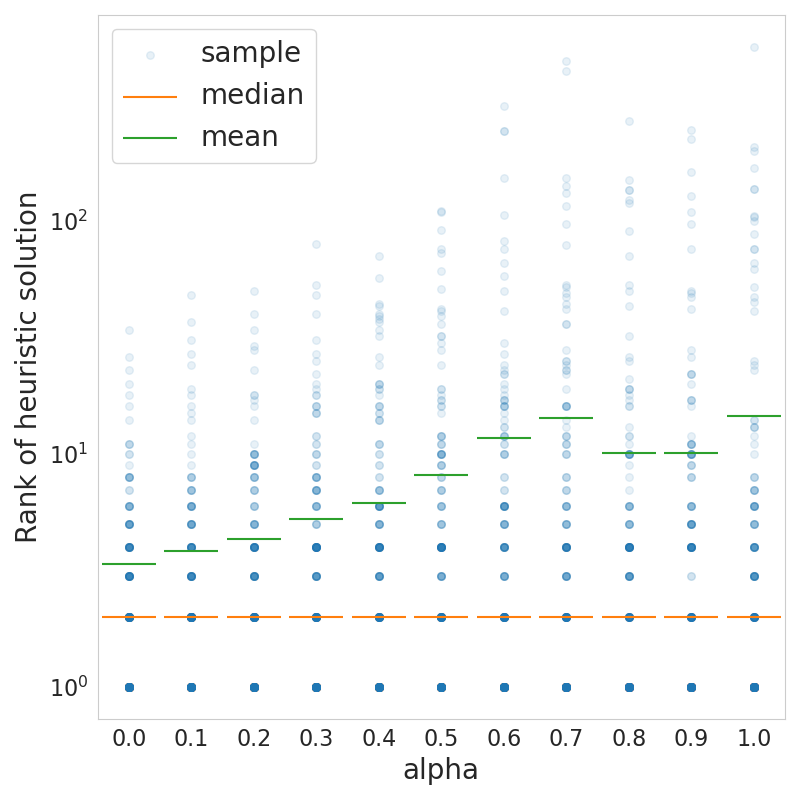}
    		\vspace{\bottomvspace cm}
    		\caption{N=9}
    		\label{sfig10:bruteforce_BCST_alpha}
    	\end{subfigure}%
    	\centering
        \caption*{\BCST rank of heuristic for different number of terminals N}
    \end{minipage}
 
	 \caption[\BCST Bruteforce Benchmark with Respect to $\alpha$]{\textbf{\BCST Bruteforce Benchmark with Respect to $\alpha$}. Relative cost errors between the \mSTreg heuristic and \BCST optimal solutions; and sorted position of the heuristic tree for different number of terminals, $N$. For each $N$ we uniformly sampled 200 different terminal configurations and we solved them for all $\alpha\in\{0.0,0.1,\dots,1.0\}$. Most runs ended up close to the global optimum. There is no clear pattern with respect to the performance of the heuristic with respect to the value of $\alpha$, though for higher number of terminals, it seems that the rank of our solution gets to be worse on average.}
	\label{fig:bruteforce_BCST_alpha}
\end{figure}

\def \bottomvspace{-.5}
\def \topvspace{0.1}
\begin{figure}
	\centering
     
    \begin{minipage}[t]{1\textwidth}   
        \begin{subfigure}{0.2\linewidth}
            \centering
            \vspace*{\topvspace cm}
            \includegraphics[width=1\linewidth]{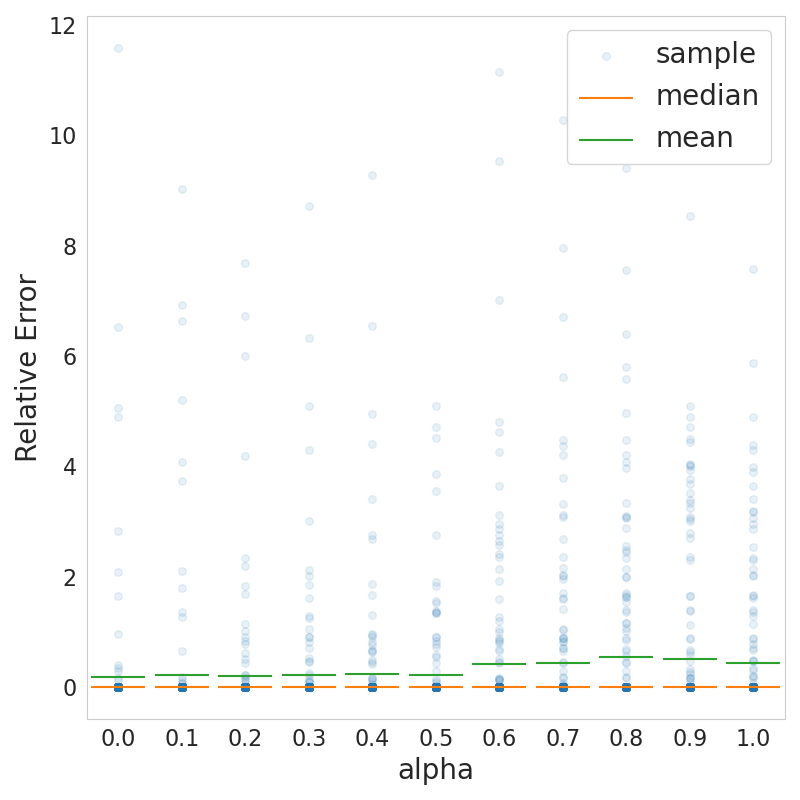}
            \vspace{\bottomvspace cm}
            \subcaption{N=5}
            \label{sfig1:bruteforce_CST_alpha}
        \end{subfigure}%
        \begin{subfigure}{0.2\linewidth}
            \centering
            \vspace*{\topvspace cm}
            \includegraphics[width=1\linewidth]{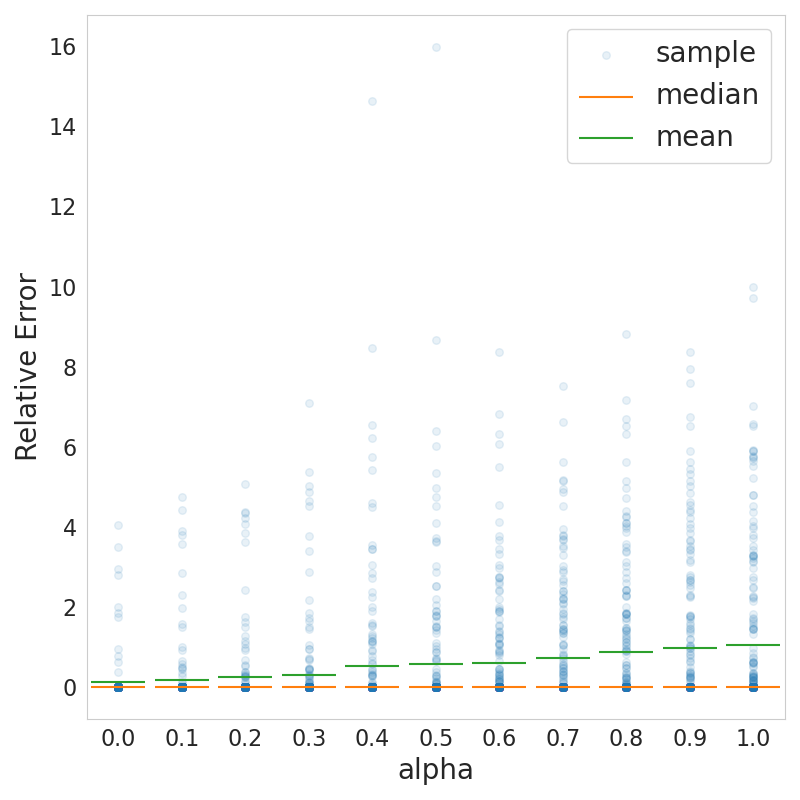}
            \vspace{\bottomvspace cm}
            \subcaption{N=6}
            \label{sfig2:bruteforce_CST_alpha}
        \end{subfigure}%
        \begin{subfigure}{0.2\linewidth}
            \centering
            \vspace*{\topvspace cm}
            \includegraphics[width=1\linewidth]{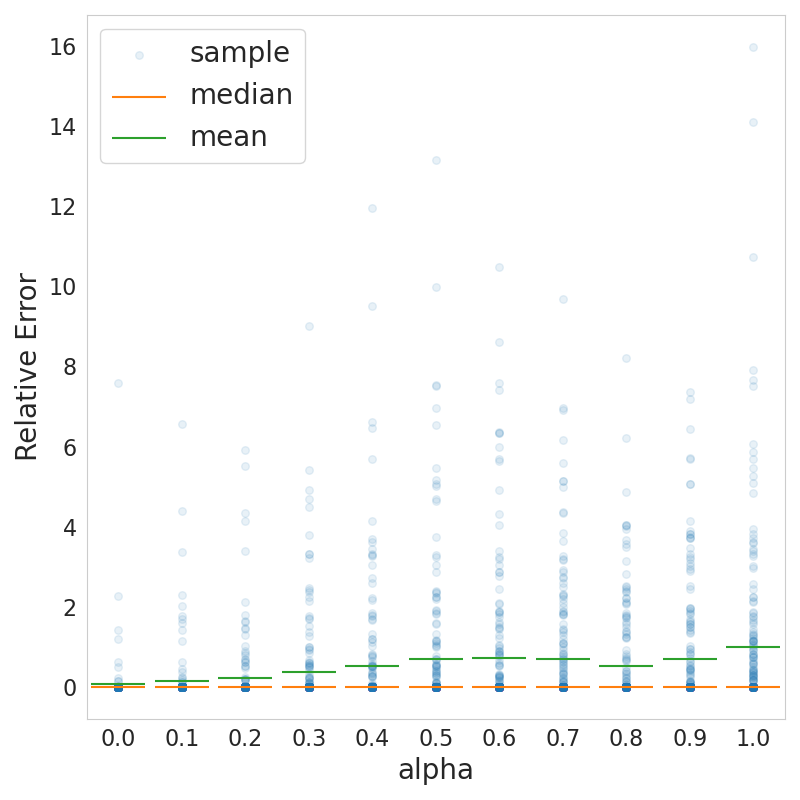}
            \vspace{\bottomvspace cm}
            \subcaption{N=7}
            \label{sfig3:bruteforce_CST_alpha}
        \end{subfigure}%
        \begin{subfigure}{0.2\linewidth}
            \centering
            \vspace*{\topvspace cm}
            \includegraphics[width=1\linewidth]{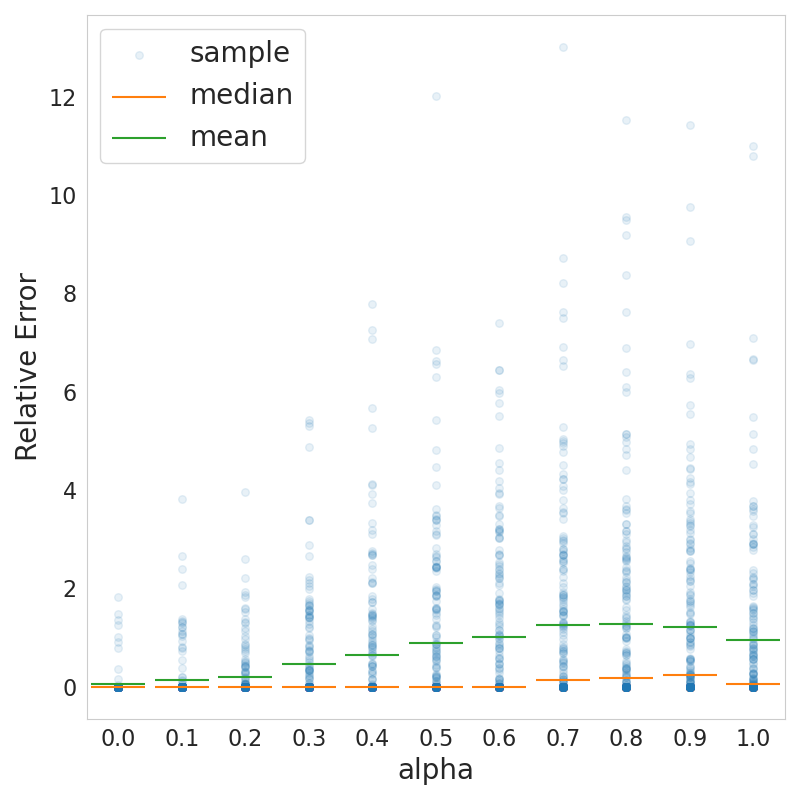}
            \vspace{\bottomvspace cm}
            \subcaption{N=8}
            \label{sfig4:bruteforce_CST_alpha}
        \end{subfigure}%
        \begin{subfigure}{0.2\linewidth}
            \centering
            \vspace*{\topvspace cm}
            \includegraphics[width=1\linewidth]{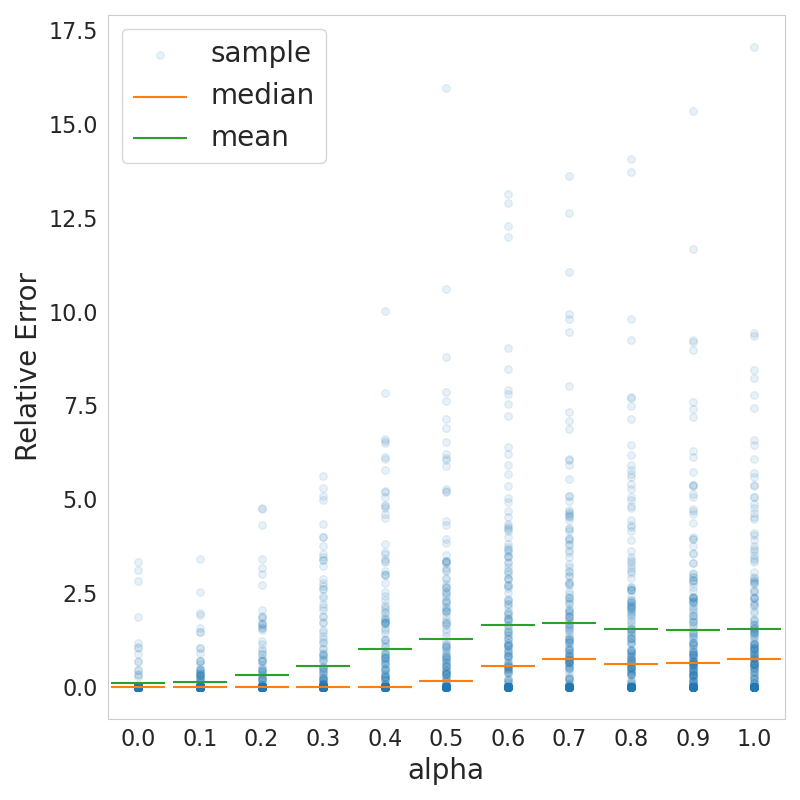}
            \vspace{\bottomvspace cm}
            \subcaption{N=9}
            \label{sfig5:bruteforce_CST_alpha}
        \end{subfigure}%
        \caption*{\CST relative error for different number of terminals N}
    \end{minipage}

    \begin{minipage}[t]{1\textwidth}   
        \begin{subfigure}{0.2\linewidth}
    		\centering
    		\vspace*{\topvspace cm}
    		\includegraphics[width=1\linewidth]{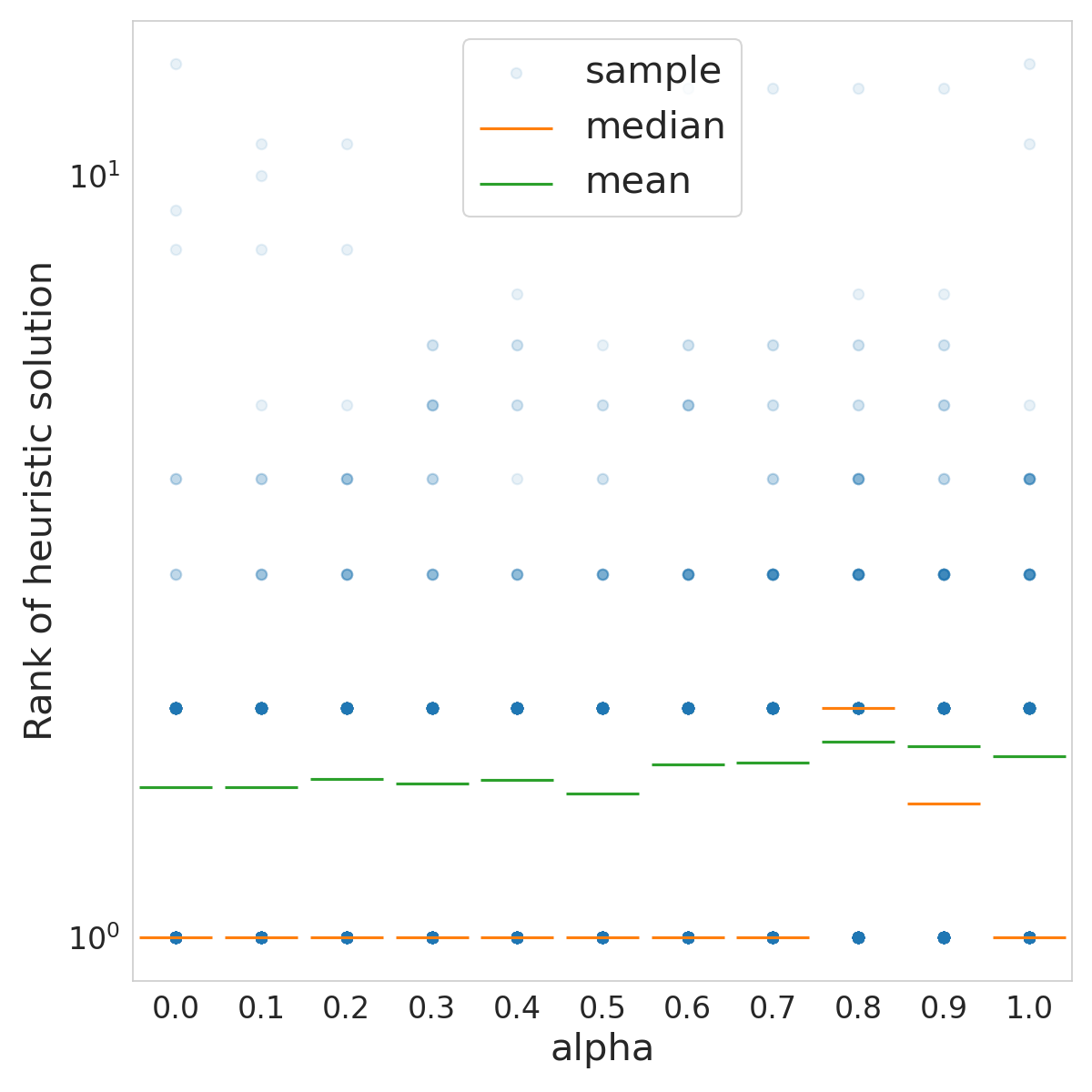}
    		\vspace{\bottomvspace cm}
    		\caption{N=5}
    		\label{sfig6:bruteforce_CST_alpha}
    	\end{subfigure}%
    	\begin{subfigure}{0.2\linewidth}
    		\centering
    		\vspace*{\topvspace cm}
    		\includegraphics[width=1\linewidth]{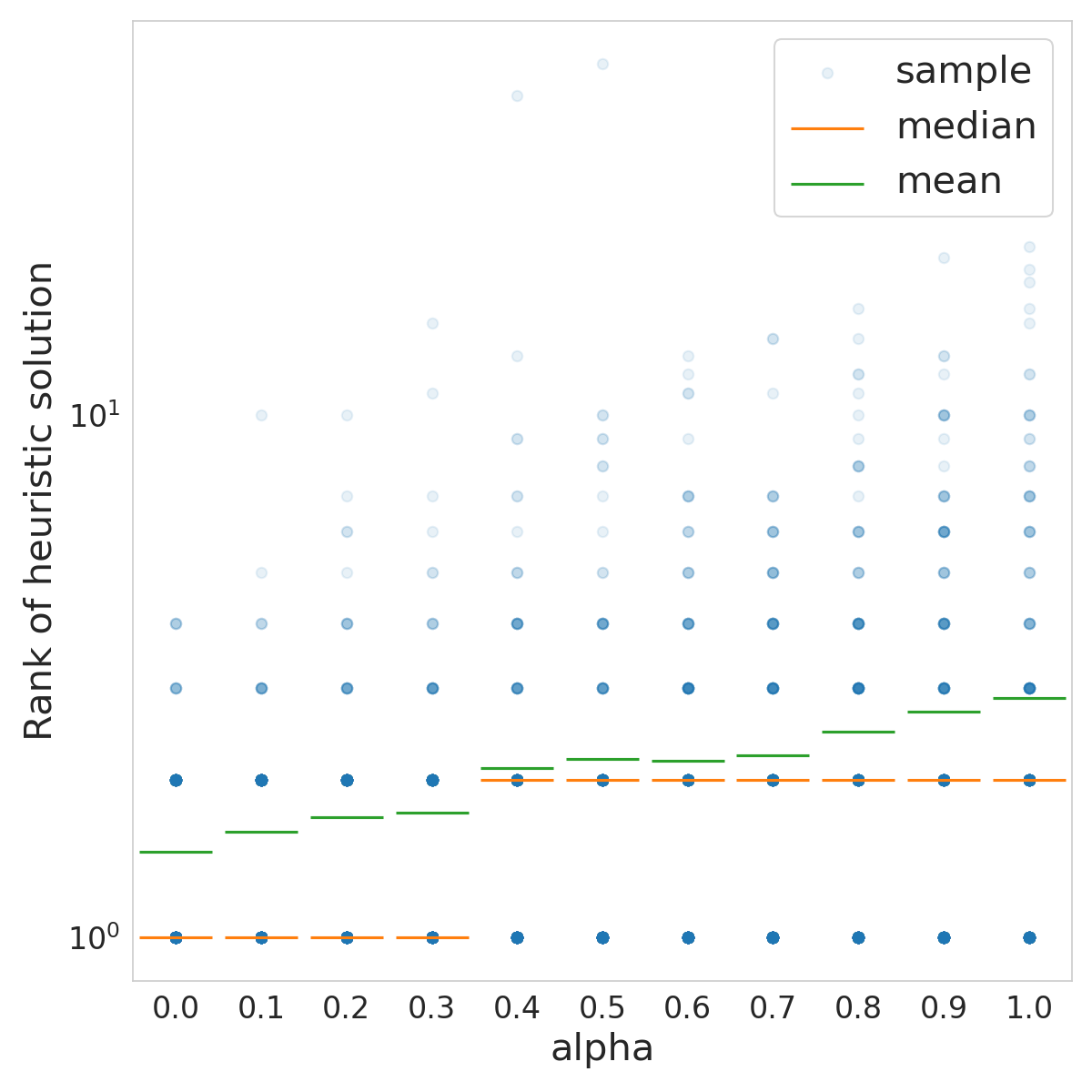}
    		\vspace{\bottomvspace cm}
    		\caption{N=6}
    		\label{sfig7:bruteforce_CST_alpha}
    	\end{subfigure}%
    	\begin{subfigure}{0.2\linewidth}
    		\centering
    		\vspace*{\topvspace cm}
    		\includegraphics[width=1\linewidth]{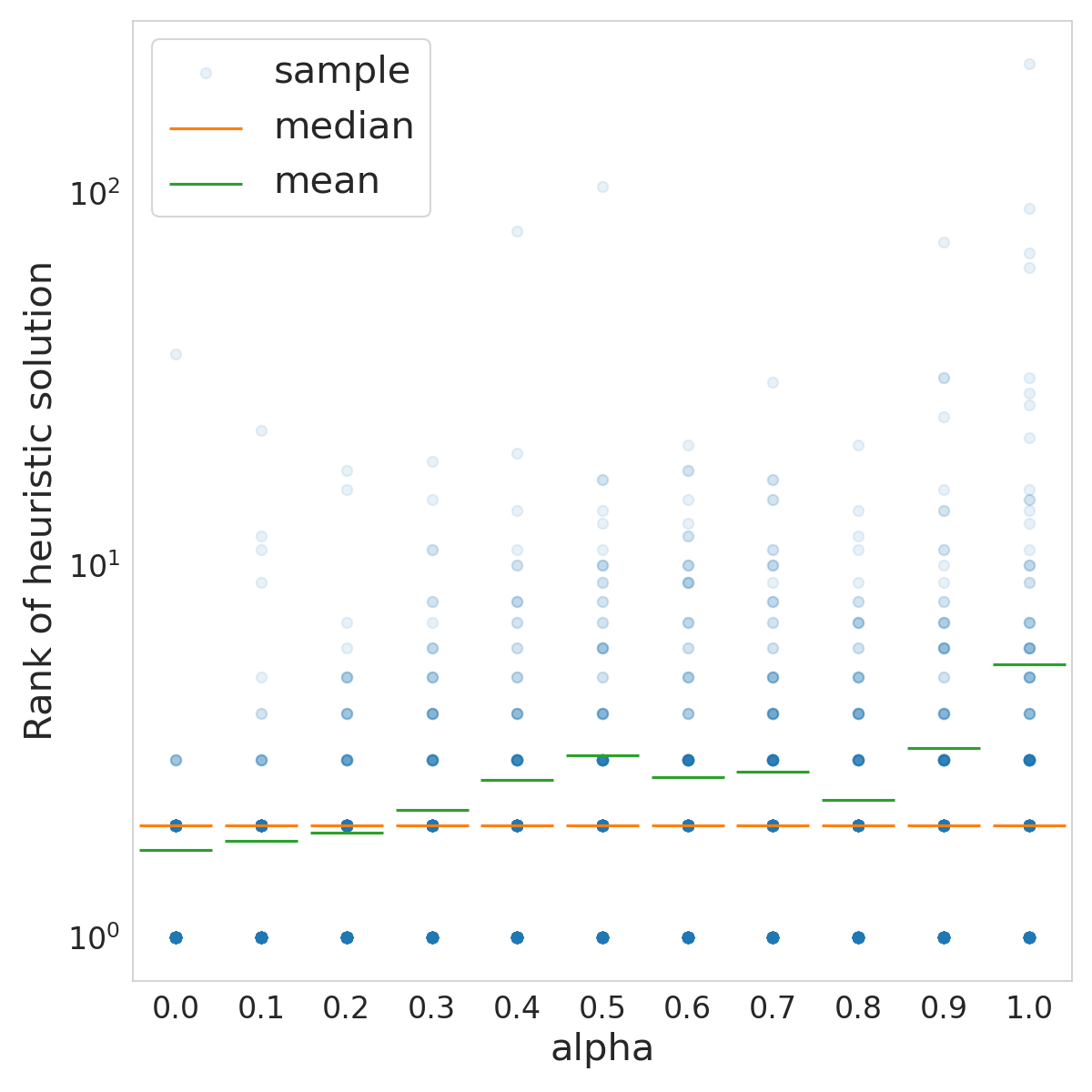}
    		\vspace{\bottomvspace cm}
    		\caption{N=7}
    		\label{sfig8:bruteforce_CST_alpha}
    	\end{subfigure}%
        \begin{subfigure}{0.2\linewidth}
    		\centering
    		\vspace*{\topvspace cm}
    		\includegraphics[width=1\linewidth]{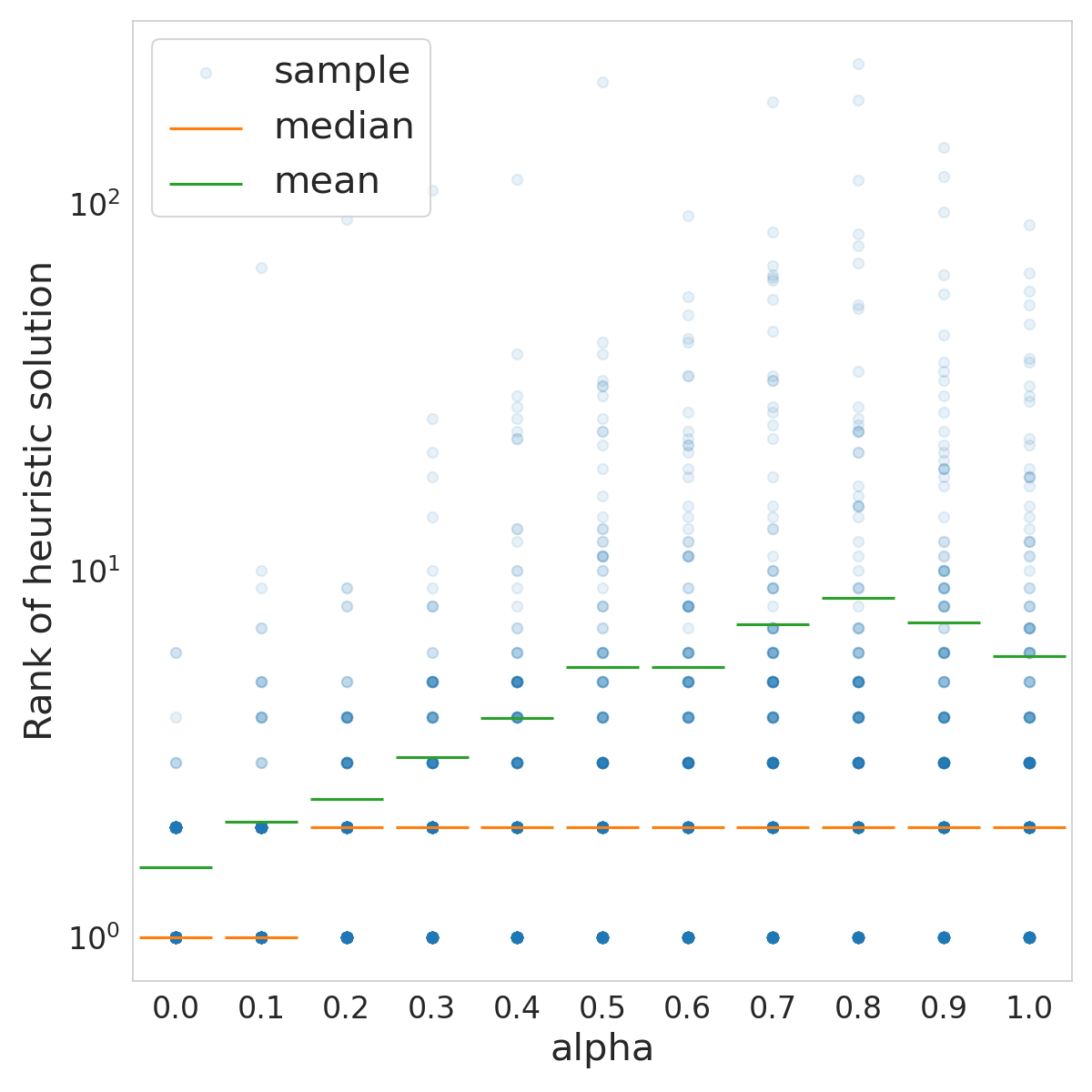}
    		\vspace{\bottomvspace cm}
    		\caption{N=8}
    		\label{sfig9:bruteforce_CST_alpha}
    	\end{subfigure}%
        \begin{subfigure}{0.2\linewidth}
    		\centering
    		\vspace*{\topvspace cm}
    		\includegraphics[width=1\linewidth]{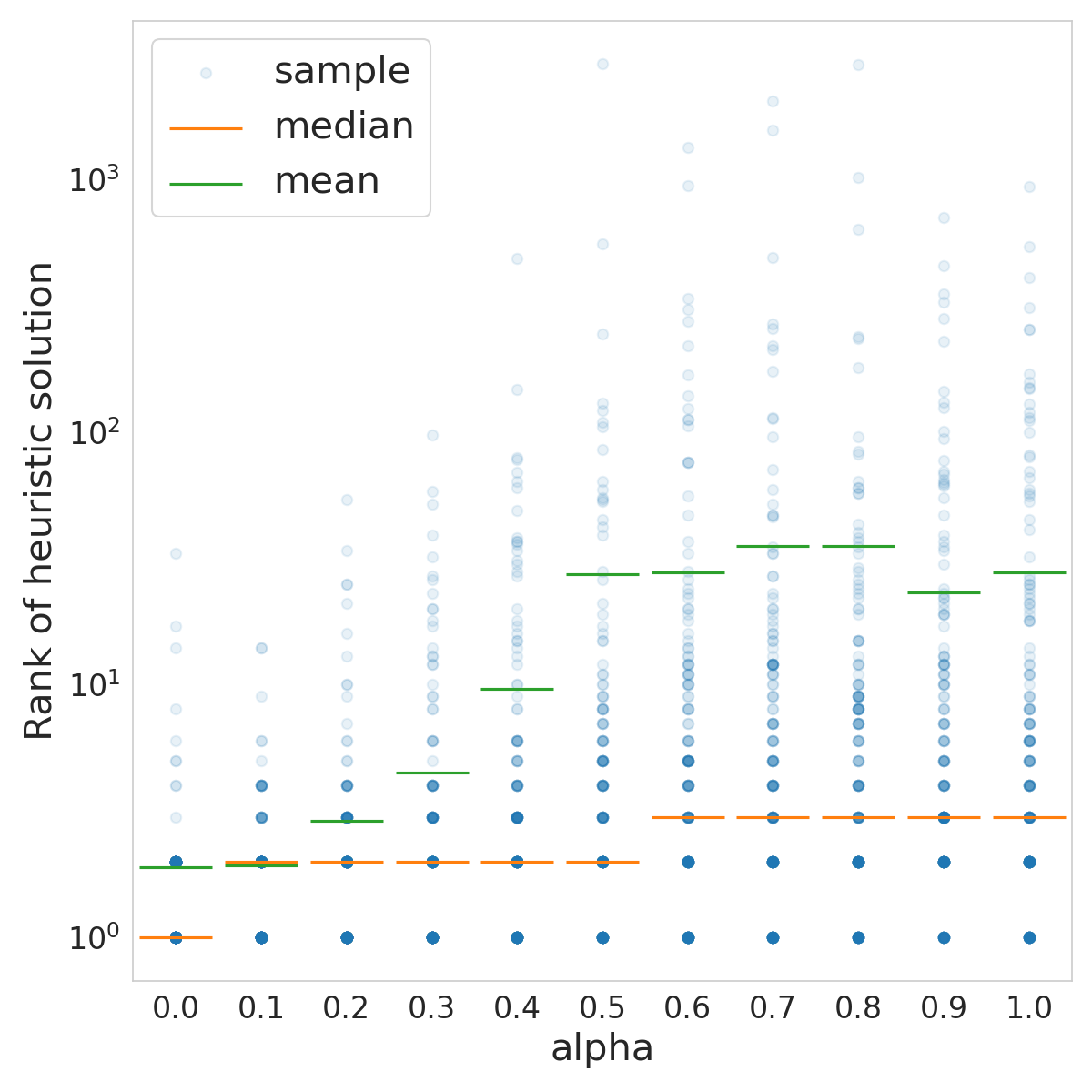}
    		\vspace{\bottomvspace cm}
    		\caption{N=9}
    		\label{sfig10:bruteforce_CST_alpha}
    	\end{subfigure}%
        \caption*{\CST rank of heuristic for different number of terminals N}
    \end{minipage}
 
	 \caption[\CST Bruteforce Benchmark with Respect to $\alpha$]{\textbf{\CST Bruteforce Benchmark with Respect to $\alpha$}.Relative cost errors between the \mSTreg heuristic and optimal \CST solutions; and sorted position of the heuristic tree for different number of terminals, $N$. For each $N$ we uniformly sampled 200 different terminal configurations and we solved them for all $\alpha\in\{0.0,0.1,\dots,1.0\}$. Most runs ended up close to the global optimum. There is no clear pattern with respect to the performance of the heuristic with respect to the value of $\alpha$, though for higher number of terminals, it seems that the rank of our solution gets to be worse on average.}
	\label{fig:bruteforce_CST_alpha}
\end{figure}

\section{Selection of $\alpha$}\label{sec:selection-of-alpha}
In this section, we present our practical insights into determining the optimal value for $\alpha$. It is crucial to emphasize that the choice of $\alpha$ is task-dependent and influenced by the desired level of structure preservation. Nevertheless, we share the observations derived from our empirical experiences.

For simpler examples, as the ones illustrated in Section \ref{sec:app_stability_examples}, we have consistently found that $\alpha$ values within the range of $[0.7, 1]$ yield high stability while preserving the primary data structure. In general, an increase in $\alpha$ correlates with a heightened inclination toward a star-shaped tree, in line with the limit case discussed in Appendix \ref{sec:limitCST_alpha>1_n_infty}. As pointed in \thref{rem:more_star_in_high_dim}, this pattern intensifies in higher dimensions, being noticeable with even modest $\alpha$ values ($\alpha \lesssim 1$), which results in a nearly star-shaped tree that compromises the preservation of data structure.

We refrained from further exploring the case of $\alpha>1$ due to both practical and theoretical observations pointing towards an excessively star-shaped tree. Indeed, as demonstrated in \ref{sec:limitCST_alpha>1_n_infty}, when the number of terminals, $N$, is sufficiently large and $\alpha>1$, the optimal solution results in a star-graph. The transition to a star-tree occurs quite early in this scenario. For instance, with moderate values of $N$ and $\alpha\gtrsim 1$ (e.g., $N=1000$, $\alpha=1.13$), an optimal star-tree emerges. Refer to \figurename{} \ref{fig:threshold_alpha} to observe how the value of $\alpha$ at which the optimal solution becomes a star-tree approaches $1$ as $N$ increases.

In summary, our empirical experience suggests that intermediate $\alpha$ values (around 0.5) effectively preserve the data structure while maintaining relative stability. This choice holds true for the applications highlighted in the thesis. We hope that by sharing our experiences, practitioners can better select an appropriate $\alpha$ for their respective applications.

\section{Implementation Details}\label{sec:app_implementation_details}
In this section, we explain some implementation details of the \mSTreg heuristic and also the parameters used for the different experiments.

In each iteration of the \mSTreg algorithm, it is necessary to compute the \mST. Since we are working with a complete graph, the computational complexity of the \mST computation is $O(N^2)$. To reduce this cost, we compute the \mST over a k-nearest neighbor (kNN) graph, where we set the value of $k$ to $\log(N)$. While the resulting \mST over the kNN graph may not always match the optimal \mST, in practice, they often yield similar results. It is worth noting that the introduction of additional nodes, as described in Section \ref{sec:app_effect_freq_sampling}, may provide more significant benefits when using the \mST computed over a kNN graph.

In Section \ref{sec:app_SP_removal_strategies} we have described different approaches to transform a full tree topology into a \CST tree topology. The strategy used to collapse the \BP nodes upon transforming a full tree topology into a \CST tree was the one that updates the collapsed \BP to the weighted geometric median, collapses greedily the \BPs and determines the \BP to be collapsed as the one with minimum distance to one of its neighbors (``update+greedy+Ordclosest'').

In all experiments, we set the \verb|sampling_frequency| variable of Algorithm \ref{alg:CST_mSTreg} equal to $3$, 
and we set the maximum number of iterations of the \mSTreg heuristic equal to 20.

\end{document}